%% file: PhDImp.tex
\documentclass[a4paper,10pt]{book}
\usepackage[latin1]{inputenc}
\usepackage[T1]{fontenc}

\usepackage[dvips]{graphicx,color}
\usepackage{colordvi}
\usepackage{amsmath}
\usepackage{palatino}
\usepackage{bbm}

\newcommand{\jbox}[1]{
\noindent\fbox{
\begin{minipage}{.96\textwidth}
\medskip

#1
\medskip

\end{minipage}
}
\bigskip
}

\bf 
\usepackage{fancyheadings}
\newcommand{\nb}{\Gray{\rule{1.6mm}{1.6mm}}}

\newcommand{\probreak}{\medskip}
\newcommand{\proof}{\noindent \begin{description}\item {\bf Proof:  } }
\newcommand{\proofend}{\hfill\fbox\\\bigskip\medskip \end{description}   }

\newcommand{\chapterbreak}{\bigskip\bigskip\bigskip}

\newcommand{\newsl}{\begin{center}
--------------------------------------
\end{center}}

\newcommand{\newfigure}[3]{
\bigskip\medskip\medskip

\centerline{
      \includegraphics[width=#1]{#2}
}
\vspace*{0.2cm}
\noindent {\small #3}

\smallskip
\hrule

\bigskip\medskip\medskip
}

\makeindex

\begin{document}


\pagestyle{empty}

~\newpage~\newpage

\vspace*{.5cm}

\hspace*{-6cm}\includegraphics[width=26cm]{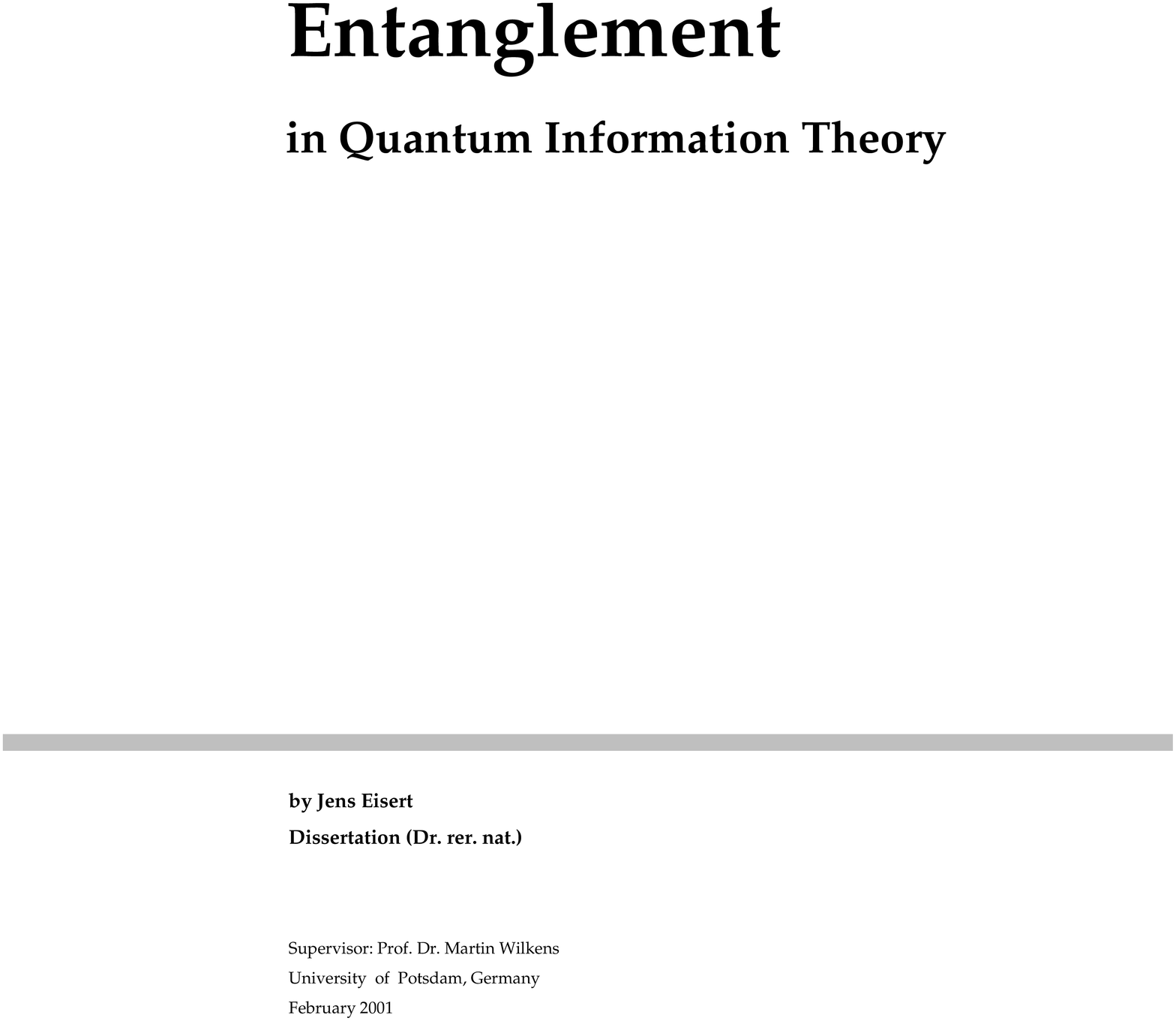}



~\newpage

\vspace*{19cm}
\begin{center}
\copyright \, February 2001, Jens  Eisert (covering
 April 1998 -- December 2000)\\
\medskip

MSc\ (Physics), University of Connecticut, Storrs, USA

Diploma (Physics), Albert Ludwig University, Freiburg, Germany
\end{center}

~\newpage
~\newpage


\pagestyle{fancyplain}
\renewcommand{\chaptermark}[1]{\markboth{#1}{}}
\renewcommand{\sectionmark}[1]{\markright{\thesection.\, #1}}
\lhead[\fancyplain{}{\bfseries\thepage}]%
   {\fancyplain{}{\bfseries\rightmark}}
	
\rhead[\fancyplain{}{\bfseries \,\, \leftmark}]%
   {\fancyplain{}{\bfseries\thepage}}%
\cfoot{}

\pagenumbering{roman}


\newpage

\chapter*{List of Publications}
\addcontentsline{toc}{chapter}{\protect\numberline{}List of Publications}

\medskip\medskip

\noindent {\bf  Journal Publications}

\bigskip

\begin{itemize}

\item[{[E1]}]
 J.\ Eisert and M.\ Wilkens,
``Catalysis of Entanglement Manipulation for Mixed States'',
{\it Physical Review Letters}\/ {\bf 85}, 437   (2000). 

(Lanl e-print quant-ph/9912080)

See Chapter 3.
 
\bigskip

\item[{[E2]}]
 J.\ Eisert, T.\ Felbinger, P.\ Papadopolous, M.B.\ Plenio, and
M.\ Wilkens,
``Classical Information and Distillable Entanglement'',
{\it Physical Review Letters}\/ {\bf 84}, 1611   (2000). 

(Lanl e-print quant-ph/9907021)

See Chapter 5.

\bigskip

\item[{[E3]}] J.\ Eisert and H.-J.\ Briegel,
``The Schmidt Measure as a Tool for Quantifying Multi-Particle Entanglement'',
{\it Physical Review A}\/ {\bf 64}, 022306 (2000). 

(Lanl e-print quant-ph/0007081)

See Section 2.3.




\bigskip

\item[{[E4]}] 

 J.\ Eisert, K.\ Jacobs, P.\ Papadopoulos, and M.B.\ Plenio,
``Optimal Local Implementation of Non-Local Quantum Gates'',
{\it Physical Review A}\/  {\bf 62}, 052317  (2000). 

(Lanl e-print quant-ph/0005101)

See Chapter 4.

\bigskip

\item[{[E5]}] J.\ Eisert and M.\ Wilkens,
``Quantum Games'',
{\it Journal of Modern Optics}\/ 
{\bf 47}, 2543  (2000).  

(Lanl e-print quant-ph/0004076)

See Chapter 6.

\bigskip

\item[{[E6]}]
 J.\ Eisert, M.\ Wilkens, and M.\ Lewenstein,
``Quantum Games and Quantum Strategies'',
{\it Physical Review Letters}\/ {\bf 83}, 3077   (1999). 

(Lanl e-print quant-ph/9806088)

See Chapter 6.

\bigskip

\item[{[E7]}] J.\ Eisert and M.B.\ Plenio,
``A Comparison of Entanglement Measures'',
{\it Journal of Modern Optics}\/ {\bf 46}, 145   (1999). 

(Lanl e-print quant-ph/9807034)

\bigskip

\item[{[E8]}] R.A.\ Gonzales, J.\ Eisert, I.\ Koltracht, M.\ Neumann, and 
G. Rawitscher,
``Integral Equation Method for the Continuous Spectrum Radial 
Schr{\"o}\-din\-ger Equation'',
{\it Journal of Computational Physics}\/ {\bf 134}, 134   (1997).

(Compare also lanl e-print nucl-th/9802022)

\bigskip

\item[{[E9]}] 
K.\ Audenaert, J.\ Eisert, E.\ Jane, M.B.\ Plenio, and
S. Virmani, 
``The Asymptotic Relative Entropy of Entanglement'',
published after submission as
{\it Physical Review Letters} {\bf 87},  217902 (2001). 

See Subsection 2.2.5.

\bigskip

\item[{[E10]}] 
J.\ Eisert, K.\ Audenaert,  and M.B.\ Plenio, 
``Entanglement Measures and Non-Local State Distinguishability'',
published after submission as
{\it Journal of Physics A} {\bf   36}, 5605 
(2003).

See Subsection 2.2.5 and Appendix C.


\end{itemize}

\bigskip\bigskip\bigskip

\noindent 
\hspace*{-.3cm}{\bf Miscellaneous}

\bigskip

\begin{itemize}
\item[{[E11]}] J.\ Eisert,
invited
review of ``M.J.\ Canty,
{\it Konfliktl{\"o}sungen mit Mathematica. Zwei\-per\-sonen\-spiele}\/
(Springer, Heidelberg, 2000)''
in {\it Phy\-si\-ka\-lische Bl{\"a}tter}\/, October issue (2000).

\bigskip

\item[{[E12]}] J.\ Eisert,
invited
review of ``R.\ Erd,
{\it OnlineRecht kompakt: Von der Domain zum Download.
Leitfaden f{\"u}r Internetnutzer}\/
(Fachhochschulverlag, Frankfurt/Main, 2000)''
in {\it Phy\-si\-ka\-lische Bl{\"a}tter}\/, scheduled
for March issue (2001).

\end{itemize}


\tableofcontents


\newpage

\lhead[\fancyplain{}{\bfseries\thepage}]%
   {\fancyplain{}{\bfseries Introduction}}
	
\rhead[\fancyplain{}{\bfseries Introduction }]%
   {\fancyplain{}{\bfseries\thepage}}%
\cfoot{}


\chapter*{Introduction}
\addcontentsline{toc}{chapter}{\protect\numberline{}Introduction}

\pagenumbering{arabic}

Any storage, transmission, and processing of 
information relies on a physical carrier \cite{Landauer}.
In a handwritten
note the sheet of paper serves as the carrier of 
information, in a desk top 
computer it is the random access memory
and the hard drive on which the relevant 
data are stored. 
Communication 
makes use of sound waves, radio waves, or 
light pulses.
The new field of {\it quantum information}
is based on the idea that single quantum systems 
can be used as the elementary carriers of information, 
such as single photons, atoms, and ions.
Quantum theory -- the theory that
describes physical systems on the atomic
scale -- opens up new possiblities for information
processing and communication
\cite{Bow,Ekert,MikesBook,Steane98,Ekert,Contemp}. 
Envisioned  applications
range from the factorization of large numbers
on a quantum computer to
communication protocols,
and key distribution in
quantum cryptography.

Quantum theory may become relevant to
technical development in
information processing mainly for two reasons.
On the one hand, 
the information processing and storage 
units in ordinary, ``classical'' 
computers are becoming smaller and smaller. 
The dimensions of
transistor elements in silicon-based 
microchips are decreasing to the extent that they
will be approaching 
scales in which quantum effects 
become relevant  in the near future
(see Fig. I.1). 
On the other hand, it has become 
possible to store and manipulate 
single quantum systems,
e.g., with sophisticated methods from quantum optics
and solid state physics \cite{Bow,MikesBook}.

The superior ``performance'' of quantum systems 
in computation and communication applications
is predominantly rooted in a property of
quantum mechanical states called {\it entanglement}\/.
Essentially, entanglement comes along with
new kinds of correlations.  
Entangled quantum states may show stronger
statistical correlations than those
attainable in a classical composite
system, where the correlation is produced by
a classical random generator.
%
%
\footnote{In 1935 A.\ Einstein, B.\ Podolsky, and N.\ Rosen (EPR)
published a seminal paper entitled
 ``Can Quantum-Mechanical Description of 
 Physical Reality Be Considered Complete?'', 
which started a long lasting
debate about the status 
of quantum theory \cite{EPR}. 
On the basis of the predicted 
outcomes of measurements on two spacelike separated 
quantum particles in an entangled state,
EPR came to the conclusion that quantum mechanics
could not be a complete
theory, suggesting the view that 
additional hidden 
variables should be appended
to a quantum state
in order to restore causality and locality.
N.\ Bohr, one of the spokesmen of the so-called
Copenhagen school of the interpretation of quantum mechanics,
argued against the assumption of a
more complete underlying local deterministic level \cite{BohrReply}.
It was not until 1964 that J. Bell presented
a way to empirically test the two competing hypotheses 
\cite{BellPaper,Bell}.
Bell's theorem
is not strictly about quantum mechanics. 
It is a statement concerning correlations of
measurement outcomes at distant events
that any physical theory may predict
under the assumption of an underlying local
classical model \cite{Mermin,Ballentine}. 
Starting from the 1980s many experiments 
were performed,
each in favor of quantum mechanics and against local
hidden variable models \cite{ExperimentAspect,Experiment,GisinNon}.\index{Bell's inequality}
}

The first prototol for quantum cryptography 
-- proposed in the early 1980s by
S.\ Wiesner, C.H.\ Bennett, and G.\ Brassard --
did not yet rely on entanglement. 
It made use of the fact that 
the state of a transmitted 
quantum system can under no circumstances 
be measured without introducing noise. As a consequence, 
a secure key can be established \cite{BenBra84}.
The general idea of the quantum computer was born at around the
same time. 
R.\ Feynman envisaged a quantum system the purpose of which
was not so much universal computing, but 
rather the simulation of the dynamics of other quantum systems 
\cite{Feynman}. Perhaps the first theoretical proposal
of a quantum computer in the ``modern'' sense 
was set forth in 1985
\index{Quantum computer} 
by D.\ Deutsch \cite{Deu85}.
Based on the work of C.H.\ Bennett
who had demonstrated 
that a universal
classical computing machine 
can be made reversible \cite{BenRev},
he introduced the concept of quantum
networks \cite{Deu89}, and showed that any unitary
operation can be generated by appropriately putting together
ingredients taken from a small set of operations, 
called quantum gates. The registers of such a 
quantum computer are not classical binary registers,
but two-level quantum systems with two orthogonal states. 
Quantum computers with a few registers can already
be built with present technology. However,
the experimental realization of a large-scale
quantum computer is extraordinarily difficult and is
hindered by challenging practical problems.
\index{Quantum gate}

\newfigure{9.5cm}{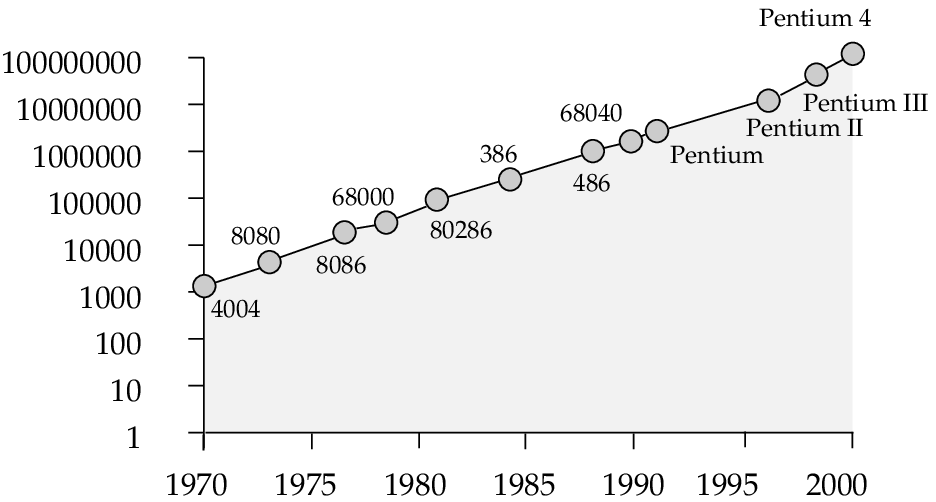}
{Figure I.1: This figure shows the number of transistors
of several CPUs from 1970 to 2000.
In 1965 G. Moore, later 
one of Intel's founders,
formulated a hypothesis
known as ``Moore's Law''.
In retrospect the hypothesis proved
to be in astonishing 
accordance with the technological development
between 1970 and 2000. It states that 
``the number of transistors integrated on 
leading edge circuits in silicon-based
chips would continue to double every 
18 months''. If the 
pace of development did not slow down 
(which does not seem very likely),
one would have to expect quantum effects to 
become predominant from 2016 on, as
the size of the information processing units
would be 
of the same order of magnitude as the size of atoms.}

Algorithms have been 
proposed in the framework of quantum computing
which are capable of solving particular problems much
more efficiently than any classical computer.
The most prominent example is the celebrated 
polynomial-time algorithm of P.\ Shor for finding 
the prime factors of large integers \cite{Shor,Ekert}, 
an algorithm that astonished 
the scientific community in 1995. To date,
no classical polynomial-time algorithm 
is  known for this problem. 
L.\ Grover's quantum algorithm 
for search in a database offers a square root
speedup compared to any classical algorithm \cite{Grover}. 
It has become obvious that in order to 
be able to perform quantum information
processing in the presence of noise, elaborate
protection
methods would be necessary. Such
methods were developed
under the names
quantum error correction, fault tolerant quantum
computing,  quantum error correcting codes, 
and stabilizer codes,
pioneered by the works of 
B.W. Schumacher, A.M.\ Steane, M.\ Nielsen,  D.P.\
DiVincenzo, P.W.\ Shor, R.\ Calderbank, A.\ Ekert, 
E.\ Knill, R.\ Laflamme, and
E.\ Rains.
An overview of recent developments is given in 
the excellent
comprehensive introductions in 
Refs.\ \cite{Bow,MikesBook,Steane98}.


The performance of Shor's algorithm on a quantum computer
can be traced back to properties of 
multi-particle entanglement, although entanglement 
enters  in a rather subtle way. 
Several schemes in quantum cryptography rely
on entanglement, and it plays a major role 
in considerations of 
quantum communication complexity
\cite{BuhrCom}.
It was only in recent years that the significance
of these correlations 
in quantum information theory was
fully appreciated. To put it in a catchy manner, one 
may say that it is entanglement that makes
quantum information theory
different from its classical counterpart.

\newfigure{13.0cm}{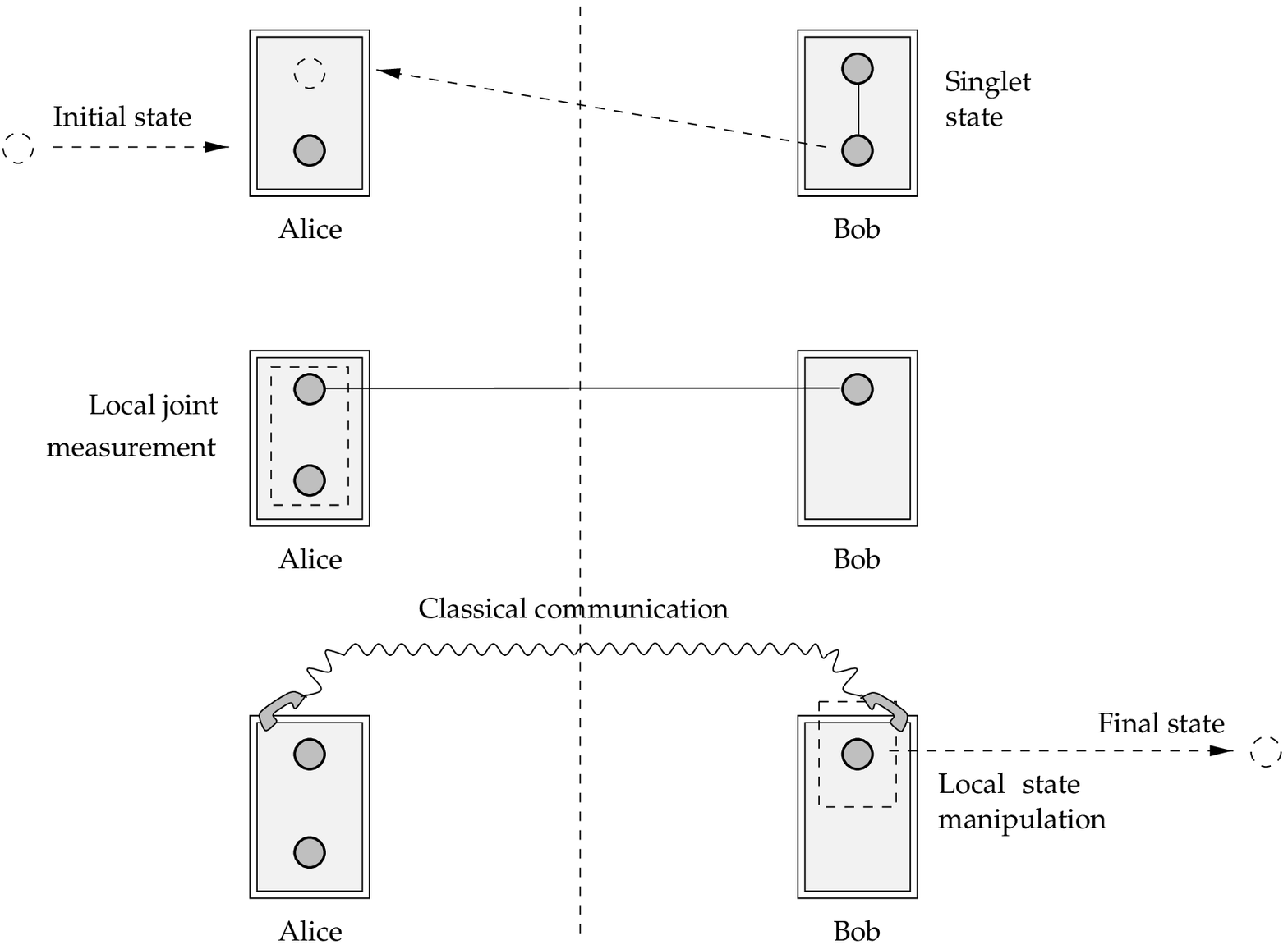}
{Figure I.2: The teleportation protocol 
as in Ref.\ \cite{BennettTele}.}

On a purely theoretical level 
it is important to
understand what kinds of tasks may be achieved
with entangled quantum systems.
It is impossible to 
transmit the particular 
``quantum information'' of a 
quantum system through a classical channel. 
This means that the statistical predictions
of quantum mechanics cannot fully be 
reproduced when trying to extract
information about the preparation procedure
from a state of one quantum system, 
transmitting the classical
information through the channel, 
and preparing another quantum system in a 
certain state on the basis of this information.
It is nevertheless possible to
transfer a quantum state to a different 
quantum system at a distant location
without physically transmitting the actual
quantum system in the particular state --
provided that
the parties initially share a pair of two-level
systems in a maximally entangled state.
This transfer of a quantum state was named
``teleportation'', a term
borrowed from science fiction literature.

The teleportation protocol, as illustrated in Fig. I.2, was 
proposed in 1993 by 
C.H.\ Bennett, G.\ Brassard, C. Cr{\'e}peau, R. Jozsa, A. Peres, 
and W.K.\ Wootters. It represented a major
breakthrough in the field \cite{BennettTele}. 
Assume that one experimenter, from now on referred
to as Alice, would like to send the unknown state of a given
quantum system to Bob, another experimenter 
at a distant location.
Bob prepares a bi-partite quantum system in a particular
entangled state and
gives one part of the system to Alice. In the next
step Alice performs a local joint 
quantum measurement on both her quantum system and
on the one she has received
from Bob. Then she phones Bob and tells him about the outcome.
Depending on the
outcome of Alice's measurement, Bob can finally
transform his part of the maximally entangled system
by use of a simple manipulation of the state.
The state of his system is eventually
identical to the state of Alice's original system: the state
has been ``sent'' from Alice to Bob.
\index{Teleportation}
\index{Dense coding}

The significance of the proposal 
is not predominantly derived from its obvious practical 
implication. Rather, it
has an immense paradigmatical value as it
is a scheme that necessarily relies on entanglement. 
Another important proposal of this type,
also by C.H.\ Bennett and S.\ Wiesner, is the dense
coding protocol \cite{BennettDense} 
concerning the transmission of classical information. 
A single quantum two-level system sent
from Alice to Bob can carry a single bit of classical 
information. Surprisingly, if
the two parties initially share a maximally
entangled state, 
two bits of classical information can be transmitted with
the same  two-level system.
Successful experimental quantum optical 
implementations of dense coding and
teleportation were performed by
A. Zeilinger \cite{Zeilinger} and by F. DeMartini \cite{Martini},
in addition to their respective co-workers. 
Many other applications of 
entanglement were suggested, the spectrum ranging 
from quantum cryptography using entangled
quantum systems by B.\ Huttner and A.\ Ekert \cite{Ekert2}
to improved frequency standards \cite{Huelga} and
clock synchronization \cite{Clock}.

In the wake of these developments, 
a more systematic theoretical 
exploration of entanglement has
begun. To give
a very fragmentary list of  pioneering work
in the field:
The resource character of entanglement was
first emphasized by C.H. Bennett, W.\ Wootters, and 
co-workers \cite{BennettTele,BennettLett,BennettMust,BiPart}. 
They also presented ways
of partly restoring the (extremely fragile)
entanglement 
degraded in purity 
due to decoherence.
These investigations have led to a more clearcut
conception of how to quantify entanglement.
The issue of quantification has also been systematically 
addressed by M.B.\ Plenio, V.\ Vedral, H.K. Lo, 
S.\ Popescu,
N.\ Linden, and G.\ Vidal, to name just a few 
\cite{Plenio,QuantLong,BennettMust,LMP,Multi1,VidalModOpt}. 
It has become apparent 
that the state space of a composite
quantum mechanical system contains
a wealth of structures, and 
that not all entangled states 
can be transformed
into maximally entangled states. 
For such states
the term ``bound entangled states'' was 
coined, a concept
introduced by the Horodecki family 
\cite{HoroBoundEnt,HoroSpin1/2}. 
The structure of the set of entangled states was further clarified by
A.\ Peres, A. Sanpera, M.\ Lewenstein, I.C.\ Cirac, B.\ Terhal,
and others (for a primer see Ref.\ \cite{Primer}).
M.\ Nielsen, 
G.\ Vidal, 
D.\ Jonathan, and 
M.B.\ Plenio investigated the
question which transformations can be accomplished
from one entangled state into another
\cite{Nielsen,Vidal,VidalJonathanNielsen,Jonathan}.
The role of symmetry was emphasized and explored by
R.F. Werner \cite{Wer89,VollWerner}. 
Possibly triggered by the recent work on
the theory of entanglement by physicists and mathematicians, 
there has also been a renewed interest in entanglement
in the philosophy of physics;
see, e.g.,  Refs.\ \cite{Berkovitz,Belnap} and the
references therein.

\newsl

This thesis deals with the resource character
of quantum entanglement, in particular, with
its quantification and its mathematical characterization.
%
The introductory 
chapter reviews some important 
ideas that will be built on in later chapters.  
In particular, the concepts of states and operations
in quantum mechanics will be explained, and
the notions of separability and distillability will
be introduced.

Chapter 2 deals with the quantification of
entanglement. It will be investigated how entangled
a given state of a bi-partite quantum system is. Related to
this is the question of
how well a particular task can be accomplished.
%
Several entanglement monotones -- proper measures of
entanglement -- will be proposed and their 
properties will be explored. 
The considerations will then be extended to multi-partite
quantum systems.

In a sense, Chapter 2 already addresses the
interconvertibility of the resource entanglement.
Local operations alone supplemented with communication
via a classical channel can only increase the classical
correlations, but not the entanglement. But given a
{\it single}\/ quantum system in an entangled state, is it
possible to transform the state
into every other entangled state, 
provided that the 
``amount of entanglement'' does not increase on average? 
The answer is no, and this fact motivates the quest for
criteria under what circumstances 
a particular transformation from one {\it known}\/ initial
state to a certain final state is possible with local 
quantum operations and classical communication. 
This is the topic of Chapter 3.
Particular emphasis will be put on transformations of 
mixed quantum states, and on so-called entanglement-assisted
local operations, where the two experimenters may borrow quantum
systems in an entangled state, but they must not use up 
the entanglement.

The ensuing question is
how non-local
quantum operations can be implemented, if the experimenters
{\it do not know} what state they share.
Chapter 4 investigates how certain elementary quantum gates 
may be implemented  in an optimal
way in 
a distributed quantum computer. 
Optimality will be measured by the minimal
amount of necessary resources.

The practically usable entanglement of a composite
quantum system is related to the classical knowledge
about the  state of the system. 
In Chapter 5 a relation between the 
amount of accessible  entanglement on the one hand
and the classical information about the order of
several entangled quantum systems on the other hand
will be established. 
A  physically relevant situation 
will be investigated in greater detail. Group theoretical
methods will help to 
clarify the connection between information and entanglement
in a more general set-up.
Finally,
Chapter 6 has a more visionary character. In this last
chapter the relationship between quantum information
theory and  game theory will be explored.


\chapter{Quantum States, Operations, and Correlations}
\chapterbreak

The questions that will be addressed in this chapter are:
What is a state? What is a quantum operation? 
What is, very roughly, the structure
of the state space of a bi-partite quantum system?
These questions tell a lot about 
the emphasis of  this field of research.
It is of interest to see what kind of manipulations
can be implemented in principle, and it turned
out to be convenient to abstract from 
the actual implementation of the 
manipulation.
Unitary operations,
for example, correspond to Schr{\"o}dinger dynamics,
and this dynamics is governed by
a  Hamiltonian 
associated with a physical set-up. 
In the context of the thesis it is however
more appropriate 
to speak about a unitary manipulation of the state,
without bothering to investigate the Hamiltonian
itself.

\renewcommand{\chaptermark}[1]{\markboth{#1}{}}
\renewcommand{\sectionmark}[1]{\markright{\thesection.\, #1}}
\lhead[\fancyplain{}{\bfseries\thepage}]%
   {\fancyplain{}{\bfseries\rightmark}}
	
\rhead[\fancyplain{}{\bfseries\chaptername\, \thechapter. \,\, \leftmark}]%
   {\fancyplain{}{\bfseries\thepage}}%
\cfoot{}

In the spirit of the recent development of quantum
information theory it is only natural to 
think of entanglement in operational terms:
Entanglement is conceived as a resource for
computational and communicational tasks, and
the emphasis is put on the usefulness of entanglement.
Notions like distillability and separability
of a state are defined in terms of 
local quantum operations.
These two concepts will also be  explained 
subsequently. This  chapter is an introductory chapter
which reviews important concepts that will
be frequently used in the later chapters.
It does not contain any original research 
material.

\section{States in Quantum Mechanics}

A state of a physical system collects 
the information 
\index{System}
which is available about the system 
that one has obtained in 
measurements. 
Associated with a quantum system
is a complex
Hilbert space
${\cal H}$  with scalar product
$\langle\cdot |\cdot\rangle$ and corresponding 
norm $\|\cdot\|$. 
If one has maximal information about the state of
a quantum system, in the sense that one has performed
a preparation such that the values of a complete
set of observables have been fixed, one can say that
the system is in a {\it pure state}\/ associated with a 
(state) vector
\index{Pure state}
$|\psi\rangle$
in Hilbert space satisfying $\| \psi \|=1$. 
This vector is 
\index{Unit ray}
uniquely determined
except for a phase factor, and the state itself corresponds
to the associated {\it unit ray} $\{ \exp(i\phi)|\psi\rangle\left|
\right.
\phi\in{\mathbbm{R}}\}$. 

The most elementary 
quantum system one can think of has
an only  two dimensional Hilbert space ${\mathbbm{C}}^2$:
the two level system. The basis elements of the
vector space can be labeled as
$ |0\rangle$ and $|1\rangle $. Any unit vector
in this Hilbert space is of the form
\begin{equation}\label{superpo}
        |\psi\rangle=\alpha|0\rangle+\beta |1\rangle,
\end{equation}
\index{Qubit}
where $\alpha$ and $\beta$ are complex numbers
satisfying $|\alpha|^2+|\beta|^2=1$. Such a 
quantum system may be in the state associated with
$|0\rangle$ or $|1\rangle$, but according to
Eq.\ (\ref{superpo}) it can also occur 
in a coherent superposition of 
$|0\rangle$ and $|1\rangle$. Simple as this system
is, it is of major importance in quantum information 
theory. 
It is the elementary physical carrier of information in
an information processing device using quantum systems.
Such a two-level
\index{Bit} 
system is called quantum bit in the context of
quantum information theory, or, in short,
{\it qubit}\/ \cite{Schumacher}. 
This term has been shaped in analogy to
the {\it bit}\/, 
which is the fundamental unit of classical 
information: The qubit is the physical system that 
can be in two different orthogonal pure states.
A possible implementation is a photon with
its polarization degree of freedom, or a two-level
system with a ground state and an excited state.

If one wants to include the
possibility of partial information to the state concept,
unit rays are not sufficient as a description of a state.
The concept of mixed states also incorporates 
ignorance about a quantum state.\footnote{This statement 
should of course not be understood in the sense 
that a mixed state corresponds to 
a mere classical probability distribution on the 
set of rays.}
For example, in a beam of unpolarized
spin-$1/2$ particles the state of
the quantum systems is given by the
classical mixture with uniform distribution
of particles in the
state corresponding to the unit ray of 
$|1\rangle$
and $|0\rangle$, respectively. 
Another example of a mixed state is one which can be 
prepared in the following way.
One takes a classical random number generator
which produces an output labeled $1$ with the classical
probability $p_1$, and the output labeled $2$
with probability $p_2$, where $p_1+p_2=1$.
If one gets the outcome $1$, the 
preparing procedure of the
pure state corresponding to $|\psi_1\rangle$ is 
implemented. In the other case  $|\psi_2\rangle$
is prepared. This procedure amounts
\index{Mixing}
to a preparation of the {\it mixed state}
\begin{equation}
	\rho=p_1 |\psi_1\rangle\langle\psi_1|+
	p_2 |\psi_2\rangle\langle\psi_2|.
\end{equation}

\newfigure{6cm}{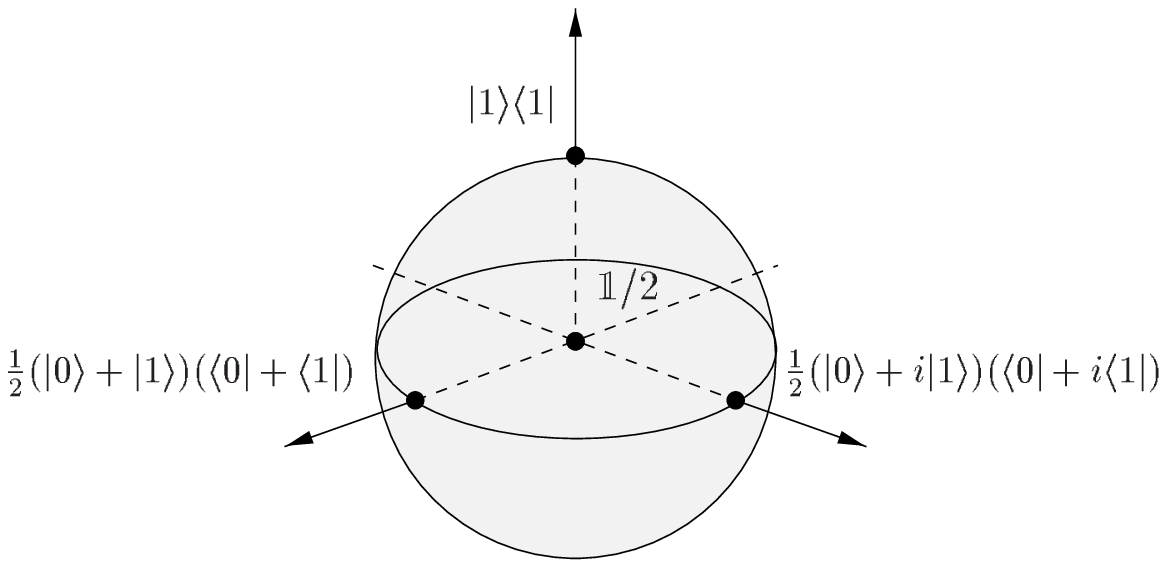}
{Figure 1.1: Schematic representation of  the state space 
of a single qubit. Each point in the ball of radius 1 
corresponds to a 
(mixed) state, pure states are represented by points
on the sphere, the {\it Bloch sphere}\/.
The coordinates of the point associated with a 
state $\rho$ are in this  representation given by
$x=\langle0|\rho|1\rangle+  \langle1|\rho|0\rangle$,
$y=\langle1|\rho|1\rangle-  \langle0|\rho|0\rangle$,
$z=i( \langle0|\rho|1\rangle - \langle1|\rho|0\rangle)$.
The {\it maximally mixed state}\/ 
${\mathbbm{1}}/2$ corresponds to the center of the sphere.}
\index{Bloch sphere}

A {\it state}\/ of a quantum mechanical
system with Hilbert space ${\cal H}$
will from now on be identified with a (bounded) operator $\rho$
fulfilling three requirements.
Firstly, $\rho$ and $\rho^\dagger$ are defined
on the entire Hilbert space and $\rho$ is self-adjoint,
        $\rho=\rho^\dagger$.
Secondly, it is a (semi)-positive operator
        $\rho\geq 0$,
and, thirdly, 
due to the conditions imposed on probabilities 
        ${\rm{tr}}[\rho]=1$.
\index{Pure state}
A state $\rho$ is a {\it pure state}\/ if $\rho^2=\rho$,
otherwise it is said to be a {\it mixed state}\/.
\index{Mixed state}
\index{State space}
For a given Hilbert space, the associated set of all states is 
denoted by ${\cal S}({\cal H})$ and referred to as
{\it state space}\/. The state space
is a convex set, that is, if $\rho_1$ 
is element of  ${\cal S}({\cal H})$
and the same is true for $\rho_2$, then also all states on the straight
line segment
$\lambda \rho_1+(1-\lambda) \rho_2$, $\lambda\in[0,1]$, 
are included in ${\cal S}({\cal H})$. The above example involving
the random number generator already points towards
this property: The procedure of the convex 
combination corresponds to a {\it mixing}\/ of two preparing 
procedures. One may prepare $\rho_1$ with probability $\lambda$
and the state $\rho_2$ with probability $1-\lambda$, and then
ignore the information about what preparation procedure has
actually been chosen. $\rho$ is then the state 
reflecting 
all the information which is finally available.
The extreme points in this convex set are just one dimensional
projectors $|\psi\rangle\langle\psi|$ with $\|\psi\|=1$.

Conversely, 
any mixed state $\rho$ of a quantum system admits a 
representation of the form
\begin{equation}\label{rep}
        \rho=\sum_{i=1}^n p_i |\psi_i\rangle\langle\psi_i |
\end{equation}
with a probability distribution $p_1,...,p_n$,
and   projections 
$|\psi_i\rangle\langle\psi_i|$, 
$i=1,...,n$.
The state space ${\cal S}({\cal H})$ of a quantum system
is no {\it Choquet simplex}\/, 
\index{Choquet simplex}
i.e., not all elements are unique mixtures of the extremal
boundary. Instead, for a given state there 
exists in general an
infinite number of such representations, and both
the weights 
and the projections 
may be different for two such decompositions.

In this thesis, all quantum systems that will be considered 
consist of at least two distinct parts. 
The Hilbert space of such a
composite quantum system with parts
$A$ and $B$ is appropriately
constructed as the {\it tensor product}\/ of the
Hilbert spaces of its constituents,
\begin{equation}
{\cal H}={\cal 
H}_A\otimes{\cal H}_B.
\end{equation} 
Almost all Hilbert spaces
will be finite dimensional. In this case,
if $ \{|1\rangle_A,...,|N\rangle_A \}$, $N={\text{dim}}[{\cal H}_A]$,
is a basis of ${\cal H}_A$ and 
$\{ |1\rangle_B,...,|M\rangle_B \}$, $M={\text{dim}}[{\cal H}_B]$,
is a basis of ${\cal H}_B$, then $\{|i\rangle_A\otimes |j\rangle_B|
i=1,...,N; j=1,...,M\}$ is a basis of ${\cal H}$. 
A {\it product state vector}\/ \index{Product state}
is a vector of the form 
\begin{equation}
|\psi\rangle=
|\phi\rangle_A\otimes |\varphi\rangle_B=
	\biggl(\sum_{i=1}^{N} 
	\alpha_i |i\rangle_A
	\biggr)\otimes
	\biggl(
	\sum_{j=1}^{M} 
	\beta_j |j\rangle_B\biggr)
\end{equation}
with complex coefficients $\alpha_i$ and $\beta_j$,
$\sum_{i=1}^{N} |\alpha_i|=1$ and $\sum_{j=1}^{M}
|\beta_j|=1$. 
The tensor product symbol will often
be omitted. The scalar product in ${\cal H}$ is induced
by the scalar products in ${\cal H}_A$ and
${\cal H}_B$ according to
$\langle\psi_1|\psi_2\rangle = \langle\phi_1|_A \phi_2\rangle_A
\langle\phi_2|_B \varphi_2\rangle_B$, where
$|\psi_1\rangle=|\phi_1\rangle_A\otimes |\varphi_1\rangle_B$
and 
$|\psi_2\rangle=|\phi_2\rangle_A\otimes |\varphi_2\rangle_B$,
and extended to ${\cal H}$ by linearity.

A very useful tool is the so-called 
{\it Schmidt decomposition}
\index{Schmidt decomposition}
of a pure state of a bi-partite system 
\cite{Schmidt,Everett}. 
Let ${\cal H}={\cal H}_A\otimes {\cal H}_B$, ${\cal H}_A={\cal H}_B=
{\mathbbm{C}}^N$, and
let $|\psi\rangle\in{\cal H} $ be a state vector.
Then there exists an orthonormal basis $\{|1\rangle_A,...,|N\rangle_A\}$
of ${\cal H}_A$ and an orthonormal basis
$\{|1\rangle_B,...,|N\rangle_B\}$ 
of ${\cal H}_B$
such that
\begin{equation}	
	|\psi\rangle=\sum_{i=1}^N \sqrt{\alpha_i}\; |i\rangle_A|i\rangle_B,
\end{equation}
where $\alpha_i$, $i=1,...,N$, 
are real positive numbers satisfying 
$\sum_{i=1}^N \alpha_i=1$. The numbers
 $\alpha_1,...,\alpha_N$
are called {\it Schmidt coefficients}\/;
the {\it Schmidt rank}\/ of the state is the number
of 
non-vanishing Schmidt coefficients.\index{Schmidt coefficient}\index{Schmidt rank}

\section{Operations}

\subsection{Elementary Quantum Operations}
The dynamics of  states of isolated
quantum systems is governed by
the {\it Schr{\"o}dinger equation}\/. 
\index{Schr{\"o}dinger equation}
Time evolution
of a state of a system with Hilbert space
${\cal H}$
\index{Dynamical map}
corresponds to the unitary 
{\it dynamical map}\/
\begin{equation}
        \rho\longmapsto \sigma=U \rho U^\dagger,
\end{equation}
where $U:{\cal H}\longrightarrow
{\cal H}$ is a time-dependent
unitary operator. States
at a later time are hence unitarily equivalent
to states at an earlier time.
In particular, pure states remain pure
throughout such a time evolution.

Also associated with an alteration of the state
is the process of {\it measurement}\/ in quantum mechanics
\footnote{ 
The process of measurement in quantum
mechanics is a subtle issue. While the basic
formalism of quantum theory is  
very well-understood,
and the  applications of quantum mechanics
are numerous, there are conceptual difficulties
in quantum theory with the reconciliation of the measurement
process and continuous Schr{\"o}dinger-type
time evolution. 
\index{Measurement problem}
The term ``measurement problem'' 
could be conceived as a collective
term for several issues
related to these difficulties:
In Ref.\ \cite{Espagnat}
five different aspects of this
problem are distinguished, among them
the famous and-or-problem or the
so-called pointer basis problem. The
latter problem is considerably weakened
by the insight that no quantum mechanical
system is fully isolated from its
\index{Decoherence}
environment, and the decoherence program
has dealt with this issue \cite{Dec,PT,Breuer}. 
Although one can learn much from
such an approach about the classical limit
of quantum mechanics, this viewpoint
does not ``solve'' the problems
with measurement, as the 
final mixture does correspond to an improper
mixture in the words of Ref.\ \cite{Espagnat}, 
unless one is in the position to accept some kind 
of Everett-type 
many-world interpretation \cite{Everett}.
Also, there have been attempts to 
alter the standard description of
time evolution in quantum mechanics, 
which try to give an explicit stochastic
dynamical of the collapse of the state
(see, e.g., Ref.\ \cite{GRW}).
For an overview about the topics 
related to quantum measurement see, e.g., Ref.\
\cite{Wheeler}. Refs.\ \cite{Sklar,Jammer}
describe the topic of measurement
from the perspective of
philosophy of physics, Ref. \cite{Zajonc}
with an emphasis on quantum optical
experiments.
This thesis -- pragmatic in its scope -- will not be 
concerned with these issues.
}.
%
Let $i=1,...,K$ be the labels of the possible 
outcomes in a   measurement. Each outcome of the measurement
is furnished with a projector $\pi_i$, 
\begin{equation}
	\pi_i \pi_j =\delta_{ij}\pi_{i},\,\,\,\,\,\,
	\sum_{i=1}^K \pi_i =\mathbbm{1}.
\end{equation}
If the quantum system is initially
in the state $\rho$, then the state 
immediately after the measurement is given by \index{Selective measurement}
\begin{equation}
        \rho\longmapsto \sigma_i=\frac{\pi_{i}\rho\pi_i}{{\text{tr}}[ \pi_i\rho\pi_i]}.
\end{equation}
The outcome with label $i$ is obtained with 
 probability   $p_i={\text{tr}}[\pi_i \rho]={\text{tr}}[\pi_i \rho \pi_i]$.
This type of measurement will be referred to as
{\it selective projective measurement}.
It is called {\it complete}\/ if all 
\index{Complete measurement}
projectors $\pi_i$ are one-dimensional. 
A {\it non-selective projective measurement}\/ corresponds to
a map
\begin{equation}
	\rho\longmapsto\sum_{i=1}^K\pi_i \rho\pi_i.
\end{equation}
\index{Non-selective measurement}

In addition to unitary transformations and projective
measurements
there are two more actions one can take: 
If $\rho_1$ is a state of a quantum system with 
a Hilbert space ${\cal H}$, one can append a
quantum system with Hilbert space ${\cal K}$
in a state $\omega$ such that 
\begin{equation}
	\rho \longmapsto \rho  \otimes \omega.
\end{equation}
These auxiliary
quantum systems are typically called 
{\it ancillae}.\index{Ancilla} Similarly, 
one may {\it dismiss a local part}\/ of the whole quantum system.
This is taken into account by the partial trace operation:
If a joint quantum system with Hilbert space
${\cal H} \otimes {\cal K} $ is in the state $\rho$, then
\begin{equation}
	\rho\longmapsto \sigma={\text{tr}}_{{\cal K}}[\rho].
\end{equation}
$\sigma$ is the final state 
of the first system with Hilbert 
space ${\cal H}$ alone.

Accordingly, 
one can apply any combination of these four basic 
ingredients. One may for example
 allow a coupling of the original 
quantum system to an auxiliary quantum system and let the two unitarily
interact. After performing a non-selective
projective measurement 
on the composite system one could eventually
 consider
the original system only again by taking a partial trace with respect
to the auxiliary part. In each step well-defined states are mapped on other
states, and the concatenation of these operations
clearly amounts to a positive linear map from the state space
onto itself.

\subsection{Generalized Measurements}
Alternatively, one can approach the issue 
of admissible quantum operations
in a quite different spirit. From an axiomatic point of
view one may
look for all maps ${\cal E}:{\cal}S({\cal H})
\longrightarrow {\cal}S({\cal H}')$
which are consistent with the statistical interpretation of
quantum theory. Certainly, ${\cal E}$ must be linear, such that
${\cal E}$ respects 
convex combinations of states as described above. Since
the semi-positivity of states has to be preserved, ${\cal E}$
is also required to be a positive map.

But ${\cal E}$ being positive is not a sufficient
criterion as will be argued subsequently.
Any quantum system
of interest can be conceived as being a part of a larger
quantum system. Bearing this in mind, 
one can always append an additional system; in the Hilbert space 
${{\cal K}}$ of the
additional system the map under consideration then 
simply acts as the identity operation, which 
leaves this part  of the enlarged system in the same state.
Hence, the map ${\cal E}\otimes \mathbbm{1}_N$ has to be a valid operation,
where $N$ stands for the dimension of ${{\cal K}}$, and
in particular, it must be positive.
 
This leads to the concept of complete positivity. 
A map ${\cal E}$ is called {\it completely positive}\/ if 
${\cal E}\otimes \mathbbm{1}_N$ is a positive map for all $N\in 
{\mathbbm{N}}$.
Quite surprisingly, this condition is stronger than mere 
positivity of ${\cal E}$. 
Indeed, there exists maps which are positive but fail to fulfil
complete positivity, such as the 
transposition operation. 
In the course of this thesis, 
{\it quantum operations}\/ 
are identified with linear completely positive maps on the state space.
%
%
%
\index{Completely positive map}

Unitary operations, non-selective
projective measurements, addition of uncorrelated
systems, and the dismissal of parts of a compound system: all
these operations
can be cast into the form
\begin{equation}\label{fullform}
        \rho\longmapsto {\cal E}(\rho) = 
	\sum_{i=1}^K E_i \rho E_i^\dagger,
\end{equation}
where the so-called  {\it Kraus operators}\/ 
$E_i:{\cal H}\longrightarrow{\cal H}$, 
$i=1,...,K$, 
\index{Kraus operator}
are not necessarily Hermitian operators \cite{Kraus}. 
Even more can be said about Eq.\ (\ref{fullform}):
Any linear completely positive map 
${\cal E}:{\cal}S({\cal H})\longrightarrow {\cal}S({\cal H})$
can be written in this form, and thus, 
Eq.\  (\ref{fullform})
gives the most general possible trace-preserving
operation in quantum mechanics.
The {\it trace-preserving property}\/ of ${\cal E}$ manifests as
\begin{equation}\label{trace}
        \sum_{i=1}^K E_i^\dagger E_i=\mathbbm{1},
\end{equation}
because it implies that 
$\sum_{i=1}^K {\text{tr}} [ E_i \rho E_i^\dagger]={\text{tr}} 
[\rho]=1$. 
As in general $E_i^\dagger$ and $E_i$ do not commute,
\begin{equation}\label{unital}
        \sum_{i=1}^K E_i E_i^\dagger=\mathbbm{1}
\end{equation}
is different from Eq.\ (\ref{trace}). The latter condition
is equivalent with the statement that the identity
is a fixed point of the map.
A completely positive \index{Unital map}
map satisfying Eq.\ (\ref{unital})
is called {\it unital}\/; if it is also trace-preserving it is
said to be a {\it doubly stochastic operation}. \index{Doubly stochastic map}
Many operations in quantum mechanics are
doubly stochastic -- in particular in the context
of quantum information theory -- but not all.
Doubly stochastic maps are   those
operations which increase the {\it von Neumann entropy}\/, that is,
operations ${\cal E}$ for which $S({\cal E}(\rho))\geq S(\rho)$ 
for all states $\rho$, where $S(\rho)=-{\text{tr}}[\rho \log_2 \rho]$.
\index{Von Neumann entropy}
For properties of the von Neumann entropy and the 
relative entropy functional see Appendix A.

Quantum operations ${\cal E}_i$, $i=1,...,K$, \index{Quantum operation}
may also be non-trace-preserving,  
\begin{equation}
	\rho\longmapsto {\cal E}_i (\rho)= 
	\sum_j E_{i,j} \rho E_{i,j}^\dagger,
\end{equation}
with $\sum_j E_{i,j}^\dagger E_{i,j}\leq \mathbbm{1}$. 
This corresponds to the fact
that the operation is partly classical in the sense that each label
$i$, $i=1,...,K$, 
belongs to a classical outcome in a measurement. 
The final state in an outcome associated with
label $i$ is then given by
\begin{equation}
        \rho\longmapsto
	\frac{\sum\limits_j E_{i,j} \rho E_{i,j}^\dagger}{
	{\rm{tr}} \bigl[\sum\limits_j E_{i,j} \rho 
	E_{i,j}^\dagger\bigr]},
\end{equation}
\index{Generalized measurement}
which occurs with probability
$
	p_i= {\rm{tr}} [\sum_j E_{i,j} \rho 
	E_{i,j}^\dagger]$.
Such an operation may be conceived as a 
generalization of 
a selective projective measurement. In order to guarantee that
the corresponding non-selective operation is a valid one,
the condition
\begin{equation}
	\sum\limits_{i=1}^K
	\sum\limits_{j} E_{i,j}^\dagger  E_{i,j}
	=\mathbbm{1}
\end{equation}
has to be satisfied.

\newfigure{7.5cm}{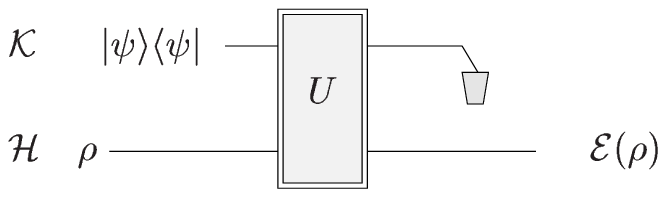}
{Figure 1.2: A general trace-preserving quantum operation ${\cal E}$.}

As has been pointed out before, all quantum operations
can be realized with unitary evolution, projective measurements,
addition of ancillae, and discarding parts of the system. 
Interestingly, such a procedure can be done with a single step only.
This is a direct consequence of the 
\index{Stinespring dilation}
{\it Stinespring dilation theorem}\/ \cite{Stine}, 
which has originally been formulated in the language of
C$^{\ast}$-algebras.
\index{C$^{\ast}$-algebra}
To be specific, 
let ${\cal H}$ be a Hilbert space with
dimension ${\text{dim}}[{\cal H}]=N$, and let
${\cal E}:{\cal S}({\cal H})\longrightarrow
{\cal S}({\cal H})$ be a trace-preserving quantum operation.
Then there exists a Hilbert space ${\cal K}$ with 
${\text{dim}}\tilde[{\cal K}]\leq N^2$ and, for any fixed 
$|\psi\rangle\in{\cal K}$, there exists a unitary
operator $U:{\cal H}\otimes {\cal K}
\longrightarrow
{\cal H}\otimes {\cal K}$ such that 
\begin{equation}
	{\cal E}(\rho)= {\text{tr}}_{{\cal K}}
	[U (\rho\otimes |\psi\rangle\langle\psi|) U^\dagger].
\end{equation}
That is, every trace-preserving 
quantum operation  may be
realized by appending an appropriate ancilla once,
applying a joint unitary operation on both
systems and finally dismissing the ancilla (see Fig.\ 1.2).

\subsection{Operations in Composite Quantum Systems}
Most of the quantum systems that will be
considered are
{\it bi-partite systems}\/ with two distinct parts $A$ and $B$.
Each part can be manipulated in an arbitrary way, but general 
joint quantum operations on both parts at the same time
are not possible. To illustrate this setup imagine two
experimenters named Alice and Bob who are separated
in space from each other, and 
they are each holding a part of a composite system.
%
Alice may take actions in system $A$ with associated Hilbert space 
${\cal H}_A$ while Bob is restricted to act in $B$ with ${\cal H}_B$,
where
$
{\cal H}={\cal H}_A\otimes {\cal H}_B
$.

\newfigure{10cm}{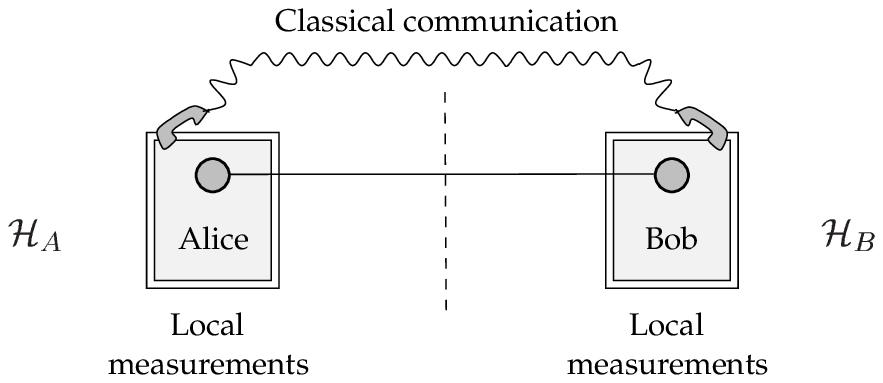}
{Figure 1.3: Schematic representation of a LOCC operation in a bi-partite setting.
Alice performs a local generalized measurement in which she obtains one of
the possible
outcomes with labels $i=1,...,K$. The measurement result is transmitted 
via the classical channel to Bob, who implements a generalized measurement 
depending on the outcomes $j=1,...,L$.
Again, Bob may transmit the classical information about the outcomes
to Alice, and so on. In a one-local operation 
it is only possible to transmit information 
from one party to the other.}

\index{Class of operations}
For bi-partite systems several natural classes of operations 
can be distinguished. A {\it class of operations}\/ is defined as 
a closed set which includes the identity: ``not to do anything at all''
is always a valid option \cite{Rains}.
A {\it local operation ${\cal E}_A$ 
on system $A$}\/ 
can be formulated by means of Kraus operators 
$A_1,...,A_K$ acting on ${\cal H}_A$. 
In an operation of this type Alice performs an operation on the part
of the state space belonging to her, 
while Bob remains inactive, i.e., he performs the identity
operation.
In the same way one can think
of a local operation ${\cal E}_B$ on Bob's side associated
with Kraus operators $B_1,...,B_L$.
A generic {\it local operation}\/ ${\cal E}$
acts as
\begin{equation}
        \rho\longmapsto {\cal E}(\rho) = \sum_{i=1}^K
	 \sum_{j=1}^L
	(A_i\otimes B_j) \rho 
	(A_i\otimes B_j)^\dagger,
\end{equation}
with operators 
$A_i:{\cal H}_A\longrightarrow {\cal H}_A$ and
$B_i:{\cal H}_B\longrightarrow {\cal H}_B$.
Local operations can also be generalized
measurements. \index{Locally distinguishable}
Two states $\sigma_1$ and $\sigma_2$
are said to be {\it locally distinguishable}\/, if ${\text{tr}}_A[\sigma_1]$
and ${\text{tr}}_A[\sigma_2]$ or ${\text{tr}}_B[\sigma_1]$
and ${\text{tr}}_B[\sigma_2]$ are orthogonal, such that there
exists a local projective measurement discriminating between
both states with unit probability. \index{Local operation}

\index{Classical communication}
A more general class of operations is one in which both Alice
and Bob can only act locally, but they may communicate by classical
means -- such as ordinary telephone lines -- 
to coordinate their actions. 
Such scenarios are ubiquitous in
applications of quantum information theory
\cite{BennettMust,BennettLett,Contemp,MikesBook}. 
In a situation with
one-way communication Alice performs a local operation 
including measurements 
and communicates the results of
her measurements to Bob. Bob can then implement an operation
depending on the values he has received from Alice
via the classical channel. Such operations are called 
{\it one-local operations}\/ \cite{Rains,BennettMust}. 
\index{One-local operation}
In terms of Kraus operators such an operation reads as
\begin{equation}\label{onelocal}
        \rho\longmapsto {\cal E}(\rho) = \sum_{j,k}
	\sum_{i=1}^K
	({\mathbbm{1}}_A\otimes B^{(i)}_k)
	(A_{i,j}\otimes {\mathbbm{1}}_B)
	\rho
	(A_{i,j}\otimes {\mathbbm{1}}_B)^\dagger
	({\mathbbm{1}}_A\otimes B^{(i)}_k)^\dagger.
\end{equation}
Eq.\ (\ref{onelocal}) gives the most general one-local
quantum operation, where the measurement results with
labels $1,...,K$ are transmitted classically from Alice
to Bob: the Kraus operators $B^{(i)}_k$ of Bob's quantum operation
may therefore depend on the actual outcome $i$.

In {\it two-local operations}\/ any sequence of local operations 
and transmission of classical data about measurement outcomes is
allowed (see Fig.\ 1.3). 
One assumes that communicating by classical means, that
is, using the telephone, is rather cheap, and that 
there are no limits for classical communication.
This class of operations is simply called 
{\it local quantum operations with classical communication},
typically abbreviated as {\it LOCC}\/ or {\it LQCC}\/. 
\index{LOCC operation}
Analogous to Eq.\ (\ref{onelocal}) 
LOCC operations can be represented with the help
of Kraus operators, but in general LOCC operations
any number of rounds of performing measurements and 
communicating the results from one party to the other
is possible.
LOCC operations are of paramount importance
to issues related to quantum entanglement,
and it will be the most important class
of completely positive maps throughout this thesis.

The set of LOCC operations is included in the set
of  {\it separable operations}\/ \cite{Rains,QuantLong,Plenio}. 
In these operations 
all Kraus operators
correspond to product operations, \index{Separable operation}
and the full operation amounts to a mixing of states
which have been manipulated in a classically 
correlated way. 
Thus, separable operations can be written in the 
general form
\begin{equation}
        \rho\longmapsto {\cal E}(\rho) = \sum_{i=1}^K (A_i\otimes B_i) \rho 
	(A_i\otimes B_i)^\dagger.
\end{equation}
Since $E_i=A_i\otimes B_i$ in this case, the map ${\cal E}$ is 
trace-preserving if both
\begin{equation}
        \sum_{i=1}^K A_i^\dagger  A_i=\mathbbm{1}_A\,\,\,\,{\text{and}}\,\,\,\,
        \sum_{i=1}^K B_i^\dagger  B_i=\mathbbm{1}_B.
\end{equation}
Note that not all trace-preserving
separable operations can be
implemented by means of LOCC operations \cite{Rains}, whereas
the converse statement is obviously true (but see Ref.\ \cite{SmallGate}).

The concept of separable operations can also be applied if
Alice and Bob each hold more than one quantum system. 
Local operations by Alice
still refer to operations in ${\cal H}_A$
only, but they may include joint operations involving
the different quantum systems on her side. Often 
an {\it asymptotic limit}\/ $n\longrightarrow \infty$
\index{Asymptotic limit}
is investigated of $n$ equal copies $\rho^{\otimes n}$ 
of the same state $\rho$.
These concepts can naturally also 
be extended to more than
three parties. LOCC operations is the set of
all local operations of all parties,
Alice, Bob, Claire, ..., together with an arbitrary amount
of two-way-classical communication. 

The theoretical study of operations is strongly linked to 
the theory of entanglement, and there is a deeply rooted
connection between
positive and completely positive maps and quantum entanglement
(see, e.g., Refs. \cite{TerhalPositive,HoroCrit,Primer}).
On the one hand several positive (but not completely positive) 
maps such as the reduction operation 
provide powerful tools to uncover nonseparability of mixed
quantum states. On the other hand separable operations
play an important role in the context of quantification 
of entanglement.
One of the fundamental properties of entanglement monotones
measuring the degree of entanglement in a given state
is, e.g., that separable operations can never increase the amount
of entanglement \cite{BennettMust,Contemp,QuantLong,Vidal,MikesBook}.

\section{Correlated Quantum States}

Assume that a bi-partite quantum system with parts $A$ and $B$
is in a pure product state. 
Then the probabilities associated with any local
measurement 
performed in systems $A$ and $B$ 
factorize: the measurement
outcomes are statistically independent.
This comes to no surprise, as
the preparation of such a pure product state
can be done locally by two independently acting
experimenters. \index{Bell's inequality}

This is not so if a pure state is no product state.
\index{Product state}
Any pure state that is no product state cannot be
prepared with local operations, and it will definitely
violate some kind of {\it Bell's inequality}\/. There
are several criteria for the non-locality
of a quantum state, but all classify 
pure states along the lines of product and
''correlated'' states. 
Such correlated states are called
{\it entangled states}\/, the singlet state
with state vector $(|01\rangle-|10\rangle)/\sqrt{2}$
of two qubits being the prototypical example.
The singlet is a particular 
maximally entangled state:
A pure state of a bi-partite system
is called {\it maximally entangled}\/
if the reduced states of both parties are
maximally mixed.

\subsection{Separability}

If the state of a bi-partite system is mixed, then 
new complications arise. It is no longer true that
any state which cannot be prepared locally necessarily
violates Bell-type inequalities. 
In mixed states both intrinsic quantum correlations and
classical correlations may be present. It is the
notion of separability that sharpens the distinction
between these two types of correlations.
A state of a bi-partite system
is called {\it classically correlated}\/ or {\it separable}\/ \cite{Wer89} 
if it is a convex combination of product states. 
In technical terms, a state $\rho$ is separable if
it can be written in the form \index{Separable state}
\begin{equation}\label{Separ}
	\rho=\sum_{i=1}^n p_i \rho_A^{(i)}\otimes \rho_B^{(i)},
\end{equation}
where $0\leq p_1,...,p_n\leq 1$ and $\sum_{i=1}^n p_i =1$.
The states 
$\rho_A^{(i)}$, $i=1,...,n$,
are taken from the state space
${\cal S}({\cal H}_A)$, $\rho_B^{(i)}$ are elements
of ${\cal S}({\cal H}_B)$. 
A state of the form of Eq.\ (\ref{Separ})
can be prepared with LOCC operations
by locally producing one of 
the product states $\rho_A^{(i)}\otimes \rho_B^{(i)}$
with probability $p_i$ and disregarding
the information which one of the product states
with label $i=1,...,n$  
has been prepared. 
States that cannot
be cast into the form  Eq.\ (\ref{Separ}) are
called {\it entangled}\/.
\index{Entangled state}

The set of separable states shall be denoted
as ${\cal D}({\cal H})$.
It is a convex set -- as
mixing of two separable states always produces 
again a separable state -- it is compact and
it includes the maximally mixed state.
In the full state space the set 
 ${\cal D}({\cal H})$ is not a
set of measure zero \cite{Volume}.
The product states on the right hand side of Eq.\ 
(\ref{Separ}) can always be chosen to be pure states,
according to 
\begin{equation}\label{puresep}
	\rho=\sum_{i=1}^n \sum_{j,k}
	p_i \lambda_{i,j}
	\mu_{i,k}
	\left(
	|\psi^{(i,j)}\rangle\langle\psi^{(i,j)}|_A\otimes
	|\phi^{(i,k)}\rangle\langle\phi^{(i,k)}|_B
	\right),
\end{equation}
where $\rho_A^{(i)}=\sum_j \lambda_{i,j}
|\psi^{(i,j}\rangle\langle\psi^{(i,j)}|_A$
and $\rho_B^{(i)}
=\sum_k \mu_{i,k}
|\phi^{(i,k}\rangle\langle\phi^{(i,k)}|_B$.
Due to the convexity property of 
the set of separable states
the number of product terms in Eq.\ (\ref{puresep})
can without loss of generality be restricted to
$(NM)^2$, 
where $N=\text{dim}[{\cal H}_A]$ and
$M=\text{dim}[{\cal H}_B]$, 
by virtue of a theorem by 
Caratheodory \cite{ConvexAna}.
\index{Caratheodory's theorem} 
\index{Caratheodory's theorem}
It is one of the major issues of quantum information
theory to judge whether a
given state $\rho$ of an $N\times M$ dimensional
bi-partite quantum system is separable or not
\cite{Primer,Peres,HoroCrit,CharacterWit,OptimizationWit};
it is a highly non-trivial task.

The set of separable states allows for a useful normal
form of quantum states of bi-partite systems. 
For any state $\rho$ there exists a unique \cite{Sini}
decomposition
\begin{equation}
	\rho=\lambda \rho_s+ (1-\lambda)
	\delta \rho,
\end{equation}
where $\rho_s$ is a separable state, $\delta\rho$
is a state with no product vector in its range,
and $\lambda$ is maximal.
This decomposition
is referred to as 
\index{Best separable approximation}
{\it best separable approximation}\/ \cite{Lewenstein}. 
This normal form can be constructed using
the method of maximally subtracting product vectors
from a state \cite{Lewenstein,Sanpera}, and has
proven to be a practical tool in investigations of
the structure of state space.

Included in the set of separable states ${\cal D}({\cal H})$
is another important convex 
subset of  ${\cal S}({\cal H})$,
the set ${\cal P}({\cal H})$ 
of so-called 
{\it positive-partial-transpose states}\/, in short
\index{PPT state}  PPT states, 
which will be explained successively.
The {\it partial transposition}\/
with respect to system $B$ is the transposition operation
in ${\cal H}_B$. If $\rho_{m\mu,n\nu}$ is the
matrix element of a state $\rho$ in some orthonormal
product basis, then the {\it partial transpose}\/ 
of $\rho$ with respect to $B$ can be written as
\begin{equation}
	\rho^{T_B}_{m\mu,n\nu}=\rho_{m\nu,n\mu}.
\end{equation}
While the partial transposition
is basis-dependent, this not true of the eigenvalues of
the partial transpose. 
Since the transposition operation \index{Partial transposition}
is no completely positive map, the partial transposition
does not necessarily map states on states. Indeed,
the partial transpose with respect to $B$, $\rho^{T_B}$,
is not always positive. However, $\rho^{T_B}$
is positive if and only if $\rho^{T_A}$
is positive. A state $\rho$
is called {\it PPT state}\/, if $\rho^{T_A}\geq 0$.
In Ref.\ \cite{Peres} it has been pointed out that
for all states $\rho$
\begin{equation}
	\rho \, \text{ is separable }
	\Longrightarrow
	\rho^{T_A}\geq 0.
\end{equation}
The converse has been proven in Ref. \cite{HoroCrit} 
for the case of bi-partite quantum systems\index{Peres-Horodecki-criterion}
of dimension $2\times 2$ and $2\times 3$:
let $\rho\in{\cal S}({\cal H}_A\otimes {\cal H}_B)$, where
$\dim[{\cal H}_A]+\dim[{\cal H}_B]\leq 5$, then
\begin{equation}\label{PHC}
	\rho  \, \text{ is separable }
	\Longleftrightarrow
	\rho^{T_A}\geq 0,
\end{equation}
such that ${\cal P}({\cal H})={\cal D}({\cal H})$, while
in general ${\cal D}({\cal H})\subset {\cal P}({\cal H})$ 
\cite{HoroPPT}.
The statement of Eq.\ (\ref{PHC})
implies an extraordinarily useful criterion
for separability in bi-partite quantum systems of 
small dimensions.
Another non-equivalent criterion is the so-called
{\it reduction criterion}\/ proposed in Ref.\ 
\cite{HoroReduction,OtherReduction}, \index{Reduction criterion}
\begin{equation}
	\rho \,
	 \text{ is separable }
	\Longrightarrow
	\text{ the map }\,
	\rho\longmapsto
	(\rho_A\otimes\mathbbm{1}_B)-\rho
	\,\text{ is positive.}
\end{equation}
where $\rho_A=\text{tr}_B[\rho]$ denotes Alice's
local state. 

Recent progress in the quest for criteria for
separability includes criteria for $2\times N$
systems \cite{2xNLong}.
For quantum states which have a low rank
(and are hence no generic states) surprisingly
strong necessary and sufficient 
criteria can be derived even for
$N\times M$-systems with $N,M>2$ \cite{LowRankCheck}.
It has turned out that the concept
of  {\it entanglement witnesses}\/
\cite{TerhalPositive,TerhalPositive2,OptimizationWit,CharacterWit}
is a powerful tool to study separability properties.

\subsection{Distillability}

Different from the 
question whether
a state is separable is the question if it is
{\it distillable}\/. 
\index{Distillation}
Assume that a 
source produces pairs of quantum systems in a certain
state $\rho$. This state $\rho$ 
is distillable if one can -- starting from a 
large number $n$ of copies $\rho^{\otimes n}$
of the state -- produce
a smaller number $k$ of approximately maximally
entangled states by just applying LOCC operations.
Such maximally entangled states are needed
for many protocols in quantum information theory,
and the issue is whether those desired maximally 
entangled states can be extracted from copies
of some mixed state. This mixed state
could be the state of an entangled composite 
quantum system that has in part been transmitted
through a 
\index{Noisy quantum channel}
{\it noisy quantum channel}\/: noise
cannot be avoided, and one tries to recover
as much as possible of the original entanglement.
The property to be distillable is in this
sense practically much more important than
separability.

In mathematical terms, a state $\rho\in{\cal S}({\cal H})$ 
is {\it distillable}\/ if there exists a $K\in{\mathbbm{N}}$
and a state vector $|\psi\rangle$
taken from a $2\times2$-dimensional subspace ${\cal C}\subset 
{\cal H}^{\otimes K}=({\cal H}_A\otimes {\cal H}_B)^{\otimes K}$ 
such that
\begin{equation}\label{distentdef}
	\langle\psi|(\rho ^{T_A})^{\otimes K}|\psi\rangle<0.
\end{equation}
It is not at all immediately obvious that this 
definition is compatible with
the above intuitive definition. To clarify this
connection two important results are necessary:
Firstly, it has been demonstrated in Ref.\  \cite{HoroBoundEnt}
that in order to study distillability, not the full
class of LOCC operations has to be considered, but
just those quantum operations 
who are a concatenation of a projection
on a $2\times 2$-dimensional subspace and further steps.
Secondly, according to the Peres-Horodecki criterion 
states of $2\times2$-systems are entangled if and only
if their partial transpose is not positive.
It is the central statement of Ref.\ \cite{HoroSpin1/2}
that by filtering and using the method proposed
in Ref.\ \cite{BennettLett} a large
number of copies of any entangled state
can be mapped onto a smaller number of approximate
singlets with LOCC operations. 
The fidelity of the output states can be 
made arbitrarily close to 1, if the number of input states
is increased.

In particular, it has been proved that if a state is
distillable, then it must have a non-positive 
partial transpose \cite{HoroBoundEnt},
\begin{equation}
	\rho^{T_A}\geq 0\Longrightarrow 
	\rho\,\,\,\text{ is not distillable}.
\end{equation}
Since there exist states which
are entangled but for which the partial transpose
is again a state, this fact implies that there
exist entangled states which cannot be distilled.
Motivated
by the metaphor of distilling entanglement 
the term {\it bound entangled states}\/ has been 
coined for these states
\cite{HoroBoundEnt,HoroBoundAct}
\index{Bound entangled state}
(the entanglement is bound and cannot be set
\index{Free entangled state}
free by LOCC operations) in contrast to
{\it free entangled states}\/.

It should however be clear that while
Eq.\ (\ref{distentdef}) specifies a well-defined
problem, it does not imply a constructive check for
distillability. Essentially, the general 
distillability problem
remains unsolved, although much progress was made in recent
months \cite{LewensteinNPTBE,Primer,BennettBound}. 
In Ref.\ \cite{LewensteinNPTBE} the number 
number $K$
\index{$K$-distillability}
in Eq.\ (\ref{distentdef}) is restricted 
to small numbers, and the so-defined {\it $K$-distillability}\/
for $K=1,2,...$
has been applied to the investigation of distillability
properties of states with high symmetry. 
This concept provides strong evidence that certain 
states 
may be bound entangled, although they have
a non-positive partial transpose 
\cite{LewensteinNPTBE,BennettBound}. In a sense this
is bad news, as this means that to check whether
the partial transpose is positive is not
sufficient to know whether a state is distillable.
\index{Non-PPT bound entangled state}

\chapter{Quantification of Quantum Entanglement}
\chapterbreak
\section{Introduction}

What is the degree of entanglement
of a given state of a composite quantum system? 
It is one of the crucial issues of
a theory of quantum entanglement to
find a general answer to this basic 
question,
and a lot of research effort has been devoted to this subject.
Since entanglement is conceived as a resource, 
there is a wide range of questions related to quantification:
On the one hand one could ask 
how much entanglement in 
pure, approximately 
maximally entangled form can be extracted
from quantum systems in entangled states,
on the other hand 
how much entanglement is actually
needed to prepare a system in a particular state.
Knowing about the amount of entanglement inherent in a 
quantum state 
also means knowing how well a certain
task can be accomplished.

For bi-partite systems in pure states the quantification
problem is essentially solved.
The origin of the difficulties in the quantification of
mixed-state 
entanglement of bi-partite systems is
rooted in the fact that classical and
quantum correlations are intertwined. 
A state may be classically
correlated without being entangled. Also, 
if a state is entangled, it is not clear 
whether many copies of it can be transformed
into copies of maximally entangled states
by means of 
local operations and classical communication.
A proper measure of entanglement should take
this interplay of classical and quantum correlations
into account. In particular, the value for
the entanglement as given by some proper
measure of entanglement 
should not increase on average, if two
experimenters implement local operations only. 

In multi-partite systems additional complications arise,
and even the pure-state case is not
well-understood yet. It is not obvious at all 
what ``standard unit'' 
to take in order to quantify the amount of entanglement
in a multi-partite setting, as it  became clear that
several inequivalent ``kinds of entanglement'' 
exist. It is definitely not sufficient to 
consider only two-party entanglement
in a quantum system consisting of more that two parts
in order to characterize the entanglement of a 
general composite quantum system.
This fact can probably most clearly be illustrated by 
investigating the
properties of the so-called GHZ-state of three qubits: it
is an entangled three-party state, but each two parties do not 
share any bi-partite entanglement at all. The reduced state
of any two qubits is a separable state.

This chapter is concerned with several issues related to the
quantification problem. Firstly,
it will be clarified what
it means to quantify the entanglement of a bi-partite system
in a meaningful way, and several good measures of entanglement
that can be found in the literature will be briefly reviewed.
From Subsection 2.2.3 on new material will be presented.
New entanglement measures will be proposed: this is done
in order to simplify the technical difficulties in actually
evaluating the degree of entanglement of a bi-partite system,
and in order to get useful
upper bounds for the amount of entanglement
that can be distilled from many copies of the same state.
Entropic and non-entropic quantities will be introduced, and 
their properties concerning continuity and additivity will be
investigated. Thirdly, 
the last section of the chapter deals with multi-partite
entanglement. After a short introduction to some issues of 
multi-particle entanglement a new measure will be
introduced and its properties will be studied.
Most of the material of this last section has 
been published in Ref.\ [E3].

\section{Entanglement Monotones}

\index{Entanglement monotone}
Entanglement monotones
are good measures of entanglement. They 
are functionals mapping states on positive
numbers that appropriately
quantify the amount of entanglement. 
It is the main feature of entanglement monotones, very
roughly speaking, that they are capable of distinguishing 
between quantum and classical correlations.
If a state
is separable, the entanglement monotone gives the value $0$,
and one could say that the more entangled the state,
the larger  the value of the entanglement monotone.
\index{Measure of entanglement}
By no means is an entanglement monotone a uniquely defined functional.
Instead, one would call any functional
$E:{\cal S}({\cal H})\longrightarrow \mathbbm{R}^+$
an entanglement monotone that satisfies a number of
natural criteria that are a manifestation
of physically motivated
properties which 
any good measure of entanglement should have.
They have been formulated 
in terms of the behavior under certain
local quantum
operations that can be implemented by remotely located parties.
Conditions for acceptable functionals go back to
Refs.\ \cite{BennettMust,BennettLett,Quant1,Plenio}.
In the important paper Ref.\ \cite{QuantLong}
the major condition has been clarified: 
the monotonicity
under local generalized measurements.
In Refs.\ \cite{VidalModOpt} and \cite{Horobound}
these criteria have been cast into the elegant form that 
will be explained below.

 

\subsection{General Properties}

An {\it entanglement monotone}\/ is a positive
functional and  it vanishes for separable states.
These statements are
put together in the first condition.
The second property is concerned with mixing. As stated in the
first chapter a convex combination of states corresponds
to the mixing of preparation procedures. If one has prepared
two particular states with certain values for the degree
of entanglement, then the postulate states that
the entanglement inherent in the state obtained from 
mixing can only be smaller than or equal 
to the weighted sum of
the previous degrees of entanglement. In other words,
mixing of preparing procedures alone never leads to an 
increased amount of entanglement. The
other postulate deals with local generalized
measurements. A local
operation performed on one part of a bi-partite
quantum system alone cannot   on average  increase
the entanglement of the composite system. ``On average'' 
means that a particular outcome of the generalized
measurement can well exhibit a larger amount of entanglement:
this is, e.g., the basis of entanglement distillation procedures.
However, the expected entanglement obtained from 
weighting the entanglement of the outcomes with
the respective probabilities must not grow.

\begin{itemize}
\item[(i)]
$E:{\cal S}({\cal H})\longrightarrow \mathbbm{R}$ is a positive functional,
and
$E(\sigma)=0$ for any separable state $\sigma\in{\cal D}({\cal H})$.
\item[(ii)]
$E$ is a convex functional, that is, 
\begin{equation}
	E(\sum_{i=1}^n p_i \sigma_i)\leq
	\sum_{i=1}^n p_i E(\sigma_i)
\end{equation}
for $p_i\in[0,1]$ and $\sigma_i\in{\cal S}({\cal H})$, $i=1,...,n$, with
$\sum_{i=1}^n p_i=1$.
\item[(iii)] 
$E$ is  monotone under local operations:
if one of the parties performs a local generalized measurement, then
the expected entanglement cannot increase. This means that if, say,
Alice implements a local operation leading to states
\begin{equation}
	\sigma_i = \frac{\sum_j (A_{i,j}\otimes \mathbbm{1}_B) \sigma (A_{i,j}\otimes 
	\mathbbm{1}_B)^\dagger}{p_i},\,\,\,\,i=1,...,K,
\end{equation}
with probability
$p_i= {\text{tr}}[\sum_j A_{i,j} \sigma A_{i,j}^\dagger ]$,
where 
$\sum_{i=1}^K \sum_j  A_{i,j}^\dagger  A_{i,j} =\mathbbm{1}_A$, then
\begin{equation}
	E(\sigma) \geq
	\sum_{i=1}^K p_i E(\sigma_i).
\end{equation}
\end{itemize}

As any local trace-preserving completely positive map ${\cal E}$
amounts to an operation as specified in (iii) together
with a mixing, that is, convex combination of different
\index{Mixing}
outcomes as in (ii), it follows that for such maps
\index{Mixing}
$
	E(\sigma)\geq E({\cal E}(\sigma)).
$
In earlier writings on the quantification of entanglement
the latter inequality 
was used as the starting point \cite{Plenio}. 
This 
implies that the degree of entanglement may not increase
under a local non-selective generalized measurement. 
Condition (iii) on its own leads to an
invariance under local unitary operations, that is,
$E(U\rho U^\dagger)=E(\rho)$ for all $\rho\in{\cal S}({\cal H})$
and all local unitary operators $U:{\cal H}\longrightarrow{\cal H}$.
\index{Local unitary operation}

Even for pure states this set of conditions 
-- together with an appropriate normalization --
does not
fully specify a measure for entanglement. However, if 
one appends two more conditions concerning the asymptotic
regime, then $E$ is completely determined for pure states
according to the so-called {\it uniqueness theorem}\/ 
\index{Uniqueness theorem for entanglement measures}
\cite{Horobound,VidalModOpt,Rohrlich}.
These additional conditions are \cite{Horobound}
\begin{itemize}
\item[(iv)] $E$ is
{weakly additive}\/, 
meaning that\index{Weakly additive}
$
	E(|\psi\rangle\langle\psi|^{\otimes n})=
	n E(|\psi\rangle\langle\psi|)
$
for all $|\psi\rangle\in {\cal H}$ 
and all $n\in\mathbbm{N}$.

\item[(v)] For a given $|\psi\rangle\in{\cal H}$ let
$(\sigma_n)_{n\in\mathbbm{N}}$ be a series
of states $\sigma_n\in{\cal S}({\cal H}^{\otimes n})$
with the property that
$
	\lim_{n\rightarrow\infty}\| |\psi\rangle\langle\psi|^{\otimes n}
	-\sigma_n\|= 0$, where $\|.\|$ denotes the trace norm.\footnote{The
	{\it trace norm}\/ of a matrix $A$ is defined
	as $\|A\|=\text{tr}|A|=\text{tr} [\sqrt{A^\dagger A}]$.}
Then $E$ satisfies
\index{Trace norm}
\begin{equation}
	\lim_{n\rightarrow\infty}
	\frac{1}{n}\left| E(|\psi\rangle\langle\psi|^{\otimes n})
	-
	E(\sigma_n)\right|
	= 0.\label{weaklycont}
\end{equation}
This property is called {\it weak continuity}\/. Essentially,
it is required that close to many 
products of pure states $E$ is sufficiently
continuous (see also Subsection 2.2.6).

\end{itemize}
\index{Weakly continuous}
The unique measure of entanglement for pure states of
bi-partite systems is given by 
\index{Unique measure of entanglement}
\begin{equation}\label{unimeasure}
E(|\psi\rangle\langle\psi|)=S(\text{tr}_A[|\psi\rangle\langle\psi|])=
S(\text{tr}_B[|\psi\rangle\langle\psi|]),
\end{equation}
where $S$ denotes the von Neumann entropy (see Appendix A).

\subsection{Distillable Entanglement and Entanglement of Formation}
\index{Distillation}

The {\it distillable entanglement}\/ grasps the resource
character of entanglement in mathematical
terms \cite{BennettMust,BennettLett}. 
It is the maximal number of maximally
entangled states per copy
that can be extracted from many copies of a given 
state $\sigma$ by means of local operations
and -- possibly --
classical communication. ``Many copies''
means that the 
\index{Asymptotic limit}
asymptotic limit $n\rightarrow\infty$ of $n$ identically
prepared systems in a state $\sigma$ is considered. As one
transforms a certain number of non-maximally entangled
states into a smaller number of approximately
maximally entangled
states with the use of LOCC operations, 
the metaphor
of ``distilling'' entanglement has been used. 
With the procedure of distilling entanglement it is
possible to partially reverse the process of decoherence:
imagine that the degree of entanglement of many bi-partite
quantum systems has degraded due to channel noise. One
can then extract a smaller number of approximately
maximally entangled states from the larger ensemble 
\cite{repeater,Privacy}.

Let $C$ be a class of operations \index{Class of operations},
e.g., the set of LOCC operations. 
The $C$-distillable entanglement of a state $\sigma$  on 
${\cal H}_A\otimes {\cal H}_B$ is then the optimal
rate of maximally entangled quantum systems that can
be achieved using generalized measurements from the set
$C$ of operations. As in Ref.\  \cite{HoroChannel}
 $D_{\leftrightarrow}$ denotes
the distillable entanglement with respect to LOCC
operations, also called 
\index{Two-way distillable entanglement}
{\it two-way distillable entanglement}. If not 
otherwise specified, distillable entanglement 
is meant to be $D_{\leftrightarrow}$. 
\footnote{
In accordance with Refs.\ \cite{Rains,Rains2}
a more precise definition of distillable entanglement
can be given.\index{Class of operations}
Let $C$ be a class of operations, e.g., the set of LOCC operations.
\index{Distillable entanglement}
The $C$-distillable entanglement of a state $\sigma$  on 
${\cal H}_A\otimes {\cal H}_B$ is
the maximum number $D_C(\sigma)$ such that there 
exists a sequence of generalized measurements from $C$
with
labels $j=1,2,...$ 
with the following properties: 
For each $j$ the generalized measurement takes the 
input state $\sigma^{\otimes n_j}$,
that is,
$n_j$ copies of $\sigma$, and maps it on
$\rho^{(j)}_{i}$ with probability $p^{(j)}_{i}$,
where $\rho^{(j)}_{i}\in{\cal S }({\cal H}^{(j)}_{i}\otimes{\cal H}^{(j)}_{i})$. 
In the limit $j\rightarrow\infty$, $n_j\to\infty$ and
\begin{eqnarray}
{1\over n_j} \sum_i p^{(j)}_{i} \log_2 (\dim [{\cal H}^{(j)}_{i}] )
&\to& D_C(\sigma),\\
{1\over n_j} \sum_i p^{(j)}_{i} (1-F^{(j)}_{i})\log_2(\dim [{\cal H}^{(j)}_{i}])
&\to& 0.
\end{eqnarray}
$F^{(j)}_{i}=\langle\psi^{(j)}_{i}|\rho^{(j)}_{i}|\psi^{(j)}_{i}\rangle$
denotes the fidelity of $\rho^{(j)}_{i}$ with respect to
a maximally entangled state on ${\cal H}^{(j)}_{i}\otimes {\cal 
H}^{(j)}_{i}$. This is the definition that will be used
in subsequent considerations.}


For pure states the distillable entanglement can be
evaluated: 
$S(\text{tr}_A[|\psi\rangle\langle\psi|])$
quantifies the amount of EPR entanglement contained
asymptotically in the state $|\psi\rangle\langle\psi|$. That is,
the optimal rate in a distillation that can be achieved
is given by
\begin{equation}
	D_{\leftrightarrow}(|\psi\rangle\langle\psi|)=
	S(\text{tr}_A[|\psi\rangle\langle\psi|])=
	S(\text{tr}_B[|\psi\rangle\langle\psi|]).
\end{equation}
It is not known how to evaluate $D_{\leftrightarrow}$
for general mixed states \cite{HoroIrrer}, 
and only a single example 
for $2\times 2$ systems is known, and some examples
in higher dimensional systems that will be presented
in Chapter 5. Bound entangled states are defined as 
entangled states
for which the  distillable entanglement 
$D_{\leftrightarrow}$ vanishes.

Given the fact that it is so extraordinarily difficult
to deal with this quantity, powerful upper bounds are
necessary. Such an upper bound is a functional introduced
in Refs.\ \cite{QuantLong,Plenio,Plenio,Contemp} 
and slightly modified in Ref.\ \cite{Rains2}.
This is the {\it relative entropy of entanglement}\/,
defined as
\index{Relative entropy of entanglement}
\begin{equation}
	E_R(\sigma)=\min_{\rho\in {\cal D}({\cal H})} S(\sigma||\rho)
\end{equation}
for states $\sigma$; it is an entanglement monotone \cite{QuantLong}.
In this expression the minimum is attained (and hence the expression does
not have to be formulated with an infimum) as
${\cal D}({\cal H})$ is a compact set and due to
the lower semi-continuity of the relative entropy functional. The properties
of the relative entropy functional will be explained in 
detail in Appendix A.

This bound can be made stronger -- and this is the
modification of Ref.\ \cite{Rains2} -- in that the variation
is not performed over
the set ${\cal D}({\cal H})$ of separable states, but over the
set ${\cal P}({\cal H})$ of PPT states,
\begin{equation}
	B_R(\sigma)=\min_{\rho\in {\cal P}({\cal H})} S(\sigma||\rho).
\end{equation}
In particular, 
this has the implication that bound entangled states
that are included in ${\cal P}({\cal H})$ are mapped on zero,
just as the distillable entanglement of such bound entangled
states vanishes. 

\index{Entanglement of formation}
The distillable entanglement specifies how much entanglement
can be extracted from copies of a certain mixed state
by means of LOCC operations. The entanglement of formation 
is in a sense  the dual measure to the distillable
entanglement $D_{\leftrightarrow}$. 
It gives an answer to the question:
how many maximally entangled states are needed in order
to prepare copies of a particular state, again meant
in the asymptotic limit.\footnote{This statement is only
correct if $E_F$ is weakly additive, in the
sense that $E_F(\sigma^{\otimes n})=n E_F(\sigma)$ for all states
$\sigma$. Then it can be  rigorously  shown that
$E_F$ quantifies the 
asymptotic entanglement cost of preparing a mixed state
of a bi-partite system \cite{Cost}.
So far, however, no proof of additivity is available for the
entanglement of formation. Without using the additivity conjecture
one can still identify the entanglement of formation with the
asymptotic entanglement cost in a preparation procedure, but one
has to replace $E_F(\sigma)$ by the regularized version of it, 
$E_F^\infty(\sigma)=\limsup_{n\rightarrow\infty} E_F(\sigma^{\otimes 
n})/n$, see Subsection 2.2.5.}
The {\it entanglement of formation}\/ is defined by \cite{Quant1}
\begin{equation}
E_F(|\psi\rangle\langle\psi|)=S(\text{tr}_A[|\psi\rangle\langle\psi|])
\end{equation}
for pure states $|\psi\rangle\langle\psi|$
and extended to mixed states according to
\begin{equation}
	E_F(\sigma)=\text{min} \sum_i \mu_i E(|\psi_i\rangle\langle\psi_i|),
\end{equation}
where the minimum is taken over all possible decompositions
\begin{equation}
\sigma=\sum_i \mu_i |\psi_i\rangle\langle\psi_i|
\end{equation} 
of $\sigma$ in terms of pure states
$ |\psi_1\rangle\langle\psi_1|$, $|\psi_2\rangle\langle\psi_2|$, 
with a probability distribution $\mu_1,\mu_2,...$.
An extension of this type to mixed states is also 
called {\it convex roof extension}\/ \index{Convex roof extension}
to mixed states \cite{UhlmannRoof,UhlmannRoof2}.
According to this construction the entanglement of formation is
the largest convex function that is consistent with 
the von Neumann entropy of the local state for pure states.
Therefore, it
can only be larger than the relative entropy of entanglement,
and
\begin{equation}
	D_{\leftrightarrow}(\sigma)
	\leq E_R(\sigma)\leq E_F(\sigma)
\end{equation}
holds for all $\sigma\in
{\cal S}({\cal H})$, implying that
the entanglement of formation is also an
upper bound for distillable entanglement, although a strictly
less tight one than the relative entropy of entanglement.
A very useful lower bound of $E_R(\sigma)$ is given by
$\max \{S(\text{tr}_A[\sigma]) - S(\sigma), S(\text{tr}_B[\sigma]) - S(\sigma)\}$
\cite{ShashLower}.

For pure states, $D_{\leftrightarrow}$, 
$E_R$, and $E_F$ all coincide \cite{BennettMust,QuantLong}. 
It is extremely hard to directly
evaluate any of these
quantities for general mixed states, 
the spectrum ranging from 
difficult to hopeless. 
For quantum systems consisting of two qubits
a general formula for the entanglement of formation
is known \cite{BiPart,Hill}.

\subsection{Non-Entropic Entanglement Monotones}


The use of the relative entropy functional
is motivated by an interpretation in terms of
statistical distinguishability \cite{QuantLong,Fuchs}.
Moreover, the relative entropy of entanglement provides a 
tight bound for distillable entanglement.
The entanglement of formation is important due to its
interpretation as a measure of the
cost of the preparation of a state.
The major drawback of these entanglement monotones is 
that a minimization over
a set of high dimension is always 
necessary, and analytically, this task can hardly 
be performed. 
In systems consisting of two qubits
the Bell basis with its 
particular properties is available,
which makes the task of evaluating
entanglement measures more accessible, the 
most spectacular example being the
analytical formula for the entanglement of
formation \cite{BiPart,Hill}.
In higher dimensional Hilbert spaces, however,
there is no equivalent of the 
Bell basis \cite{SpecialVoll}.

It would therefore 
be desirable  to have a  very simple quantity at hand which is
a good measure of entanglement in the sense
that is fulfils the above 
conditions (i), (ii), and
(iii). This quantity should not involve a 
minimization over a high dimensional set.\footnote{The main advantage of $E_N$
measure is the fact that no minimization is necessary
to evaluate the degree of entanglement of a given quantum state.
There exist, however, also non-entropic entanglement monotones
that are defined via a minimization of a high dimensional set.
Such an entanglement monotone will be investigated in Appendix C.}
It is the purpose of this subsection 
to show that a quantity first investigated in
Ref.\ \cite{Volume} 
fulfils those conditions. This quantity
has been given the name {\it negativity}\/ $E_N$, 
as it quantifies the ``negativity'' of the partial
transpose of a state. If a 
state is a PPT state, it is assigned a positive
value, otherwise the negativity vanishes.
As the entanglement measure  involves
just a trace norm of the partial transpose,
it may well be calculated with paper
and pencil, and a complicated minimization is
not necessary. \footnote{Note that in 
independent research an alternative proof of 
this statement has been found
by G.\ Vidal and R.F.\ Werner.
However, the results are unpublished and they 
were not available to the author of this thesis.}

The significance of this measure stems also from the
fact that the logarithm of the trace norm of the partial transpose
is known to be a useful  upper bound for distillable
entanglement $D_{\leftrightarrow}$ \cite{HoroIrrer,Horobound}, 
although certainly not a particularly tight one. It gives a
simple answer to the question how much entanglement
can at most be distilled from copies of a certain state.
As the logarithm is no convex 
function, the logarithm of the trace norm of the partial 
transpose cannot be guaranteed to be an entanglement 
monotone any more.\\

\medskip

\noindent
{\bf Proposition 2.1. --} 
{\it Let ${\cal H}={\mathbbm{C}}^N\otimes {\mathbbm{C}}^N$, and
let for $\sigma\in{\cal S}({\cal H})$
\begin{equation}\label{negdef}
	E_N(\sigma)=\|\sigma^{T_B}\|-1,
\end{equation}
where $\|.\|$ denotes the trace norm.
Then $E_N$ is an entanglement monotone.}
\index{Negativity}

\probreak

\proof
In Eq.\ (\ref{negdef}) the negativity is defined via
the trace norm of the partial transpose with respect to
system $B$. Equivalently, one could define it as
$E_N(\sigma)=\|\sigma^{T_A}\|-1$, since
$\|\sigma^{T_B}\|=\|\sigma^{T_A}\|$ for all $\sigma\in{\cal 
S}({\cal H})$. 
To see that the first condition is satisfied, 
note that
$\|\sigma^{T_B}\|\geq \text{tr}[\sigma^{T_B}]=1$ 
for all $\sigma\in{\cal S}({\cal H})$,
and hence, $E_N$ is a positive functional. For a separable state
$\rho\in{\cal D}({\cal H})$ the partial transpose $\rho^{T_B}$
is again a state, which results into $E_N(\rho)=0$. That is, condition (i)
is satisfied. The convexity of $E_N$ (ii) follows 
from the triangle inequality with respect to 
the trace norm. The remaining task
is to prove the validity of condition (iii). Assume that Alice
performs a local generalized measurement on a system prepared in the state
$\sigma$. Any final state $\sigma_i$, $i=1,...,K$, in a 
local generalized measurement can be 
represented with the help of
Kraus operators $A_{i,j}$, $i=1,...,K$, $j=1,2,...$,
acting in ${\cal H}_B$
as the identity as
\begin{equation}
        \sigma_i=\frac{\sum_j {A_{i,j}}\sigma A_{i,j}^\dagger}{p_i},
\end{equation}
where
\begin{equation}
	p_i={\text{tr}}\left[\sum_j {A_{i,j}}\sigma A_{i,j}^\dagger\right],
\end{equation}
$\sum_{ij}A_{i,j}^\dagger A_{i,j}={\mathbbm{1}}_A$.
Acting as the identity means that
the Kraus operators can be written in the
form $A_{i,j}\otimes \mathbbm{1}_B$.
Since the sum over $i$ corresponds to a 
\index{Mixing}
mixing which -- due to the convexity property --
can only reduce $E_N$, it 
suffices to consider final states of the form
\begin{equation}
	\sigma_i=\frac{{A_{i}}\sigma A_i^\dagger}{p_i}
\end{equation}
with   $p_i={\text{tr}}[{A_{i}}\sigma A_i^\dagger]$,
where the Kraus operators $A_1, ..., A_K$ 
satisfy
$\sum_{i=1}^K A_i^\dagger  {A_{i}}={\mathbbm{1}}_A$
according to the trace-preserving property of the quantum
operation.
Let 
\begin{equation}
	\sigma^{T_B}=T^+-T^-
\end{equation} 
be the 
\index{Jordan decomposition}
{\it Jordan decomposition}\/
\cite{Bhatia} of the partial transpose of a state
$\sigma$. Both $T^+$ and $T^-$ are positive and
Hermitian, and $|\sigma^{T_B}|=((\sigma^{T_B})^2)^{1/2}=
T^+ +T^-$. Then
\begin{equation}
	\sum_{i=1}^K
	p_i E_N(\sigma_i)=
	\sum_{i=1}^K
	p_i \left(\frac{\left\| 
	(A_i \sigma A_i^\dagger)^{T_B}\right\|}{p_i}
	-1\right) 
	=
	\sum_{i=1}^K
	\| A_i \sigma^{T_B} A_i^\dagger\| -1.
\end{equation}
For each $i=1,...,K$
\begin{equation}
	\| A_i (T^+ - T^-) A_i^\dagger\| \leq
	\| A_i T^+ A_i^\dagger\|+
	\|A_i  T^- A_i^\dagger\|
	=
	{\text{tr}}[A_i T^+ A_i^\dagger + A_i T^-  A_i^\dagger].
\end{equation}
As $\sum_{i=1}^K A_i^\dagger A_i ={\mathbbm{1}}_A$,
\begin{equation}
	\sum_{i=1}^K p_i E_N(\sigma_i)\leq {\text{tr}}[ T^+ +
	T^-]-1=\|\sigma^{T_B}\|-1=E_N(\sigma).
\end{equation}
Hence, $E_N$ is an entanglement monotone.

\proofend

\bigskip


It is an immediate consequence of 
Proposition 2.1 that a quantity  considered in Ref. \cite{HoroCrit} 
is also a good measure of entanglement. This measure is 
defined only for two-qubit-systems. It is proportional
to the absolute value of the smallest eigenvalue of
the partial transpose of a state. More precisely,
for a state $\sigma$ this measure of entanglement is
given by $2\max\{-\lambda_4,0\}$, where $\lambda_4$ is the
smallest eigenvalue of $\sigma^{T_B}$. It turns out that
this measure of entanglement and $E_N$ are simply identical
for systems with  ${\cal H}=
{\mathbbm{C}}^2\otimes {\mathbbm{C}}^2$.\\
\index{Negative eigenvalue measure}

\noindent
{\bf Remark 2.2. --} {\it
Let ${\cal H}={\mathbbm{C}}^2\otimes {\mathbbm{C}}^2$. Then
\begin{equation}
	E_N(\sigma)=2\max\{-\lambda_4,0\}
\end{equation}
for states $\sigma\in{\cal S}({\cal H})$,
where $\lambda_4$ is the smallest eigenvalue of 
$\sigma^{T_B}$. }

\probreak
\proof
Let $\lambda_1, ..., \lambda_4$ with
 $\lambda_1\geq ...\geq \lambda_4$ be the ordered
list of eigenvalues of $\sigma^{T_B}$. It has been
shown in Ref.\ \cite{Sanpera} that
$\lambda_1,\lambda_2,\lambda_3\geq0$ for all states
$\sigma$. If $\sigma$ is separable, $\lambda_4\geq 0$
according to the Peres-Horodecki-criterion \cite{Peres,HoroCrit}.
In the other case 
$E_N(\sigma)=
	\|\sigma^{T_B}\|-1=\lambda_1+\lambda_2+\lambda_3
	+|\lambda_4|-1 =2 |\lambda_4|$.
\proofend

\medskip

In $2\times2$-systems the negativity $E_N$ 
coincides with another useful quantity  for pure 
states and for {\it Werner states}\/
\cite{Wer89}, i.e.,
states of  the form\index{Werner state}
\begin{equation}
\rho_W 
        =\lambda |\psi^-\rangle \langle \psi^-|+(1-\lambda) 
        \mathbbm{1}/4,
\end{equation}
where $\lambda\in[0,1]$. This quantity is the so-called
{\it concurrence}\/ $E_C$, 
which is defined as follows:
let $\sigma$ be a state of a $2\times 2$-system, and let
$\mu_1,...,\mu_4$ with $\mu_1\geq ...\geq \mu_4$ be the
ordered list of eigenvalues of 
$(\sqrt{\sigma} \tilde\sigma 
\sqrt{\sigma})^{1/2}$, where $\tilde\sigma= (\sigma_y \otimes \sigma_y)
\sigma^*(\sigma_y \otimes \sigma_y) $, and $\sigma_y$ is 
one of the Pauli-matrices. The asterisk denotes
complex conjugation. Then
\begin{equation}
	E_C(\sigma)=\max\{ 0,\mu_1-\mu_2-\mu_3-\mu_4\}.
	\end{equation} 
The concurrence is -- according to
Wootters' formula \cite{BiPart} --  
related to the entanglement
of formation as $E_F(\sigma)=
- \nu_1 \log_2 (\nu_1)
- \nu_2 \log_2(\nu_2)$,
where $\nu_{1,2}=
(1\pm \sqrt{1-E_C(\sigma)^2})/2$.\\\index{Concurrence}

\noindent
{\bf Remark 2.3. --} {\it
Let ${\cal H}={\mathbbm{C}}^2\otimes {\mathbbm{C}}^2$, and
let $\sigma\in{\cal S}({\cal H})$ either be a pure state
or a Werner state. Then $
	E_N(\sigma)=E_C(\sigma)$.
}

\probreak
\proof 
The proof follows directly from the definition of 
the concurrence. 
For pure states one can make use of the fact
that any state vector can be
written in the Schmidt decomposition as
$|\psi\rangle=\sqrt{\alpha} |00\rangle+
\sqrt{\beta}  |11\rangle$, with $1\geq \alpha,\beta \geq 0$ 
and $\alpha+\beta=1$. 
\proofend

The quantity $\log_2(\|\sigma^{T_B}\|)=\log_2(E_N(\sigma)+1)$ 
is a useful
upper bound for distillable entanglement $D_{\leftrightarrow}(\sigma)$ 
 of a state $\sigma$ \cite{Horobound}. 
This fact will later be made use of when it comes
to a numerical investigation concerning the additivity
properties of the relative entropy of entanglement.
For some states this bound is tight, e.g., for
the singlet state on ${\mathbbm C}^2 \otimes {\mathbbm C}^2$
with state vector $|\psi^{-}\rangle=
(|01\rangle-|10\rangle)/\sqrt{2}$. The bound $\log_2(\|\sigma^{T_B}\|)$
will be referred to as {\it log negativity}\/.\index{Log negativity}

\subsection{Entropic Entanglement Monotones}

The relative entropy of entanglement with respect to the
set ${\cal P}({\cal H})$ 
of PPT states is a good upper bound for $D_{\leftrightarrow}$.
Unfortunately, for most states it is rather difficult to evaluate.
Say, for a quantum system consisting of two systems
with a Hilbert space of dimension $N$
one has to find a minimum in an $N^4-1$-dimensional set.
Fortunately, both the set of separable states and
the set of PPT states are convex sets
with non-empty interior, and as the relative
entropy functional is convex with respect to
both arguments one may use efficient 
algorithms to  find the minimum
in these sets numerically. 
However, in higher dimensional
Hilbert spaces even a numerical optimization is
an extraordinary expensive procedure.

One very fruitful way out is to consider
only particular states which exhibit a 
certain symmetry, and hope that
characteristic features of the general 
picture remain, an idea
going back to Ref.\ \cite{Wer89}.
Historically -- if one can speak
of historical events in a topic just a 
decade old -- such states with high
symmetry were the first for which 
entanglement measures could be 
computed \cite{Wer89,BennettMust,BennettLett,Plenio}. 
The problem of quantifying
the entanglement of {\it isotropic states}\/ and
so-called {\it Werner states}\/ is  much more feasible
than that of general states, and a general
strategy has been outlined 
in Ref.\ \cite{VollWerner}.

For general quantum states 
it seems appropriate to try to
restrict the set of reference states while keeping 
certain desired
features of the relative entropy of entanglement.
The aim is to lessen the dimension of the set
over which the variation has to be 
performed, but the restricted quantity should still be
an entanglement monotone. Let 
${\cal H}={\mathbbm{C}}^N\otimes {\mathbbm{C}}^N$,
and let for a given $\sigma\in {\cal S}({\cal H})$
\begin{equation}\label{dsigmadef}
	{\cal D}_\sigma({\cal H})=
	\left\{
	\rho\in {\cal D}({\cal H})\bigl|
	\text{tr}_A[\rho]=
	\text{tr}_A[\sigma],
	\text{tr}_B[\rho]=
	\text{tr}_B[\sigma]
	\right\}
\end{equation}
be the subset of separable states that 
are locally identical to $\sigma$. 
In the same way one can define 
\begin{equation}
	{\cal P}_\sigma({\cal H})=
	\left\{
	\rho\in {\cal P}({\cal H})\bigl|
	\text{tr}_A[\rho]=
	\text{tr}_A[\sigma],
	\text{tr}_B[\rho]=
	\text{tr}_B[\sigma]
	\right\}
\end{equation}\label{psigmadef}
as a subset of the set ${\cal P}({\cal H})$ of PPT states.
For each $\sigma$
both sets are convex and compact subsets of the state space
and both include the maximally mixed state
${\mathbbm{1}/{N^2}}$. \index{Maximally mixed state}
Now let
\begin{equation}\label{minsep}
	E_M(\sigma)=\min_{\rho\in{\cal D}_\sigma({\cal H})} 
	S(\sigma||\rho).
\end{equation}
This functional is  referred to as
{\it modified relative entropy of entanglement}\/.
\index{Upper bound for distillable entanglement}
\index{Modified relative entropy of entanglement}
Analogously, one can define\footnote{In Appendix C
a very similar measure of entanglement will be 
defined that is based on the trace norm distance.}
\begin{equation}
	B_M(\sigma)=\min\limits_{\rho\in{\cal P}_\sigma({\cal H})}
	S(\sigma||\rho).
\end{equation}
Quite surprisingly, $E_R$ and $B_R$, respectively, do
not lose their monotonicity
property by this restriction:\\

\noindent {\bf Proposition 2.4. --}
{\it $E_M$ and $B_M$ are entanglement monotones.}\\

\proof
The reasoning for $E_M$ and $B_M$ is fully analogous. 
For brevity,  only $E_M$ is considered in this proof.
Condition (i) is satisfied due to the nilpotence
property of the relative entropy functional. 
In order to show that $E_M$ is convex, let
$\sigma_1,\sigma_2\in{\cal S}({\cal H})$ and $\lambda\in[0,1]$.
Let 
\begin{equation}
	\rho_1 \in {\cal D}_{\sigma_1}({\cal H})\text { and }
	\rho_2 \in {\cal D}_{\sigma_2}({\cal H})
\end{equation}
be the separable states for which the respective 
minimum in Eq.\ (\ref{minsep}) is attained.
This minimum is assumed due to the compactness
of the sets ${\cal D}_{\sigma_1}$ and ${\cal D}_{\sigma_1}$
and since the relative entropy functional is 
lower semi-continuous. Then, as the relative entropy is joint convex,
\begin{eqnarray}
	\lambda E_M( \sigma_1)+(1-\lambda) E_M( \sigma_2)&=&
	\lambda
	S(\sigma_1||\rho_1)+(1-\lambda)
	S(\sigma_2||\rho_2)\nonumber \\
	&\geq&
	S(\lambda\sigma_1+(1-\lambda)\sigma_2||
	\lambda\rho_1+(1-\lambda)\rho_2  ).
\end{eqnarray}
The convex combination of $\rho_1$ and $\rho_2$
with weight $\lambda$ is locally identical to
the corresponding mixture of $\sigma_1$ and $\sigma_2$
with the same weight,
\begin{equation}
\lambda\rho_1+(1-\lambda)\rho_2\in
	{\cal D}_{\lambda\sigma_1+(1-\lambda)\sigma_2}({\cal H}),
\end{equation}
and hence,
\begin{equation}
	\lambda E_M( \sigma_1)+(1-\lambda) E_M( \sigma_2)\geq
	E_M(\lambda\sigma_1+(1-\lambda)\sigma_2)
\end{equation}
holds.

Let Alice perform a local generalized measurement on
${\cal H}_A$. As again, mixing can only
reduce the value of $E_M$ due to the convexity property, one may
without loss of generality assume that the posterior
\index{Posterior state}
states can be
written in the form 
$\eta_i=A_i\sigma A_i^\dagger/p_i$, where 
$p_i={\text{tr}}[A_i\sigma A_i^\dagger]$ for
$i=1,2,..., K$, $\sum_{i=1}^K A_i^\dagger A_i=\mathbbm{1}_A$.

Let $\sigma_i =A_i \sigma A_i^\dagger$
and $\rho_i =A_i \rho A_i^\dagger$ 
for $i=1,..., K$. Note that in general
$\text{tr}[\sigma_i]\leq 1$ and $\text{tr}[\rho_i]\leq 1$.
It follows from the monotonicity
of the relative entropy that
\begin{equation}
	\sum_i {\text{tr}}[\sigma_i] S\bigl(\sigma_i/{\text{tr}}[\sigma_i] \bigl|\bigr| 
	\rho_i/{\text{tr}}[\rho_i]\bigr) \leq S(\sigma||\rho),
\end{equation}
see Ref.\ \cite{QuantLong,Cover}.
Let $\omega$ 
be the state that attains the minimum in
Eq.\ (\ref{minsep}) with respect to $\sigma$.
The fact that
$\omega\in {\cal D}_{\sigma}({\cal H})$ implies that 
both $\text{tr}_A[A_i \sigma A_i^\dagger ]=
	\text{tr}_A[A_i \omega A_i^\dagger]$ and
	$\text{tr}_B[A_i \sigma A_i^\dagger]=
	\text{tr}_B[A_i \omega A_i^\dagger]$ hold, 
as the quantum
operation acts only
locally,	
	such that
\begin{equation}
	\omega_i/{\text{tr}}[\omega_i]\in{
	\cal D}_{\sigma_i/{\text{tr}}[\sigma_i]}({\cal H})
\end{equation}
for all $i=1,2,...,K$, where $\omega_i= A_i \omega A_i^\dagger$.		
Therefore,
\begin{eqnarray}
	E_M(\sigma)=S(\sigma||\omega)&\geq &
	\sum_i {\text{tr}}[\sigma_i] S\bigl(\sigma_i/{\text{tr}}[\sigma_i] \bigl|\bigr| 
	\omega_i/{\text{tr}}[\omega_i]\bigr) \nonumber\\
	&\geq& \sum_i {\text{tr}}[\sigma_i]
	E_M(\sigma_i/{\text{tr}}[\sigma_i]).
\end{eqnarray}
The same argument applies to the other party.
On average $E_M$ can only decrease when performing  a
generalized local measurement. 

\proofend

\medskip

By construction, $E_M\geq E_R\geq D_{\leftrightarrow}$,
meaning that the modified relative entropy of entanglement is a 
weaker bound than $E_R$ itself. 
Similarly, $B_M\geq B_R\geq D_{\leftrightarrow}$. 
For a large class
of states it can nevertheless be shown that 
$E_R$ and $E_M$ give the same value.\\

\noindent {\bf Proposition 2.5. --}
{\it  For pure states $|\psi\rangle\langle\psi|
\in{\cal S}({\mathbbm{C}}^N\otimes {\mathbbm{C}}^N)$
\begin{equation}
	E_M(|\psi\rangle\langle\psi|)=
	E_R(|\psi\rangle\langle\psi|)=D_{\leftrightarrow}
	(|\psi\rangle\langle\psi|).
\end{equation}
In ${\cal H}={\mathbbm{C}}^2\otimes {\mathbbm{C}}^2$
also $E_M(\sigma)=E_R(\sigma)$ for 
\begin{itemize}
\item[(i)] Bell diagonal
states,\footnote{
\index{Bell diagonal state}
{\it Bell diagonal states}\/ on 
${\cal H}={\mathbbm{C}}^2\otimes {\mathbbm{C}}^2$ 
are states which are diagonal in the basis 
consisting of the four state vectors
$|\psi^+\rangle =(|01\rangle +
        |10\rangle)/\sqrt{2}$,
        $|\psi^-\rangle =(|01\rangle -
        |10\rangle)/\sqrt{2}$,
        $|\phi^+\rangle =(|00\rangle +
        |11\rangle)/\sqrt{2}$,
        $|\phi^-\rangle =(|00\rangle -
        |11\rangle)/\sqrt{2}$. The corresponding states
are called {\it Bell states}\/, the basis is the 
{\it Bell basis}\/.\index{Bell basis}
\index{Bell state}
}
\item[(ii)] states of the form 
$\sigma= \lambda |\phi^+\rangle\langle\phi^+|+
(1-\lambda) |01\rangle\langle 01|$, 
\item[(iii)]
$\sigma=\lambda  |\phi^+\rangle\langle\phi^+|
+(1-\lambda)|00\rangle\langle 00|$,
\item[(iv)]
$\sigma=\lambda|00\rangle\langle 00 | + \mu|00\rangle\langle11|
+ \mu^\ast |11\rangle\langle 00 | +(1-\lambda)|11\rangle\langle11|$,
and 
\item[(v)] $\sigma=a|00\rangle\langle00|+b|00\rangle\langle11|
+b^\ast |11\rangle\langle 00| +(1-2a)|01\rangle\langle01|
+a|11\rangle\langle 11|$, 
\end{itemize}
where
$\lambda\in[0,1]$, $\mu\in{\mathbbm{C}}$ with $\lambda(1-\lambda) 
-|\mu|^2\geq 0$, and $a\in[0,1/2]$, $b\in{\mathbbm{C}}$
with $a^2 -|b|^2\geq 0$.}
\probreak

\proof In the Schmidt decomposition any $|\psi\rangle\in{\cal H}$
can be written as $|\psi\rangle=\sum_{i=1}^N \sqrt{\alpha_i} \,|ii\rangle$.
A state $\rho\in{\cal D}({\cal H})$ that minimizes the
relative entropy functional is given by $\rho=\sum_{i=1}^N \alpha_i$
$|ii\rangle\langle ii|$, which is included in ${\cal 
D}_{|\psi\rangle\langle\psi|}({\cal H})$. In Ref.\ \cite{QuantLong}
closest separable states of the classes of states 
(i) -- (v) are computed. For each $\sigma$ the
investigation of ${\text{tr}}_A[\sigma]$, 
${\text{tr}}_B[\sigma]$ and ${\text{tr}}_A[\rho]$, 
${\text{tr}}_B[\rho]$ of the respective
separable reference state $\rho$ 
shows that in all cases 
$\rho\in{\cal D}_\sigma( {\cal H})$.
\proofend

In this spirit one can restrict
the set of separable reference states 
also to other subsets of the separable states in order
to make the minimizing procedure
more accessible, or to simplify
symmetry arguments.
Let ${\cal C}_\sigma({\cal H})$ be a compact and 
convex subset of ${\cal P}({\cal H})$. Then one may
-- under certain circumstances --
minimize the relative entropy over the set
${\cal C}_\sigma({\cal H})$ without losing the
monotonicity property:\\

\noindent {\bf Proposition 2.6. --}
{\it Let ${\cal H}={\mathbbm C}^N\otimes {\mathbbm C}^N$ and
let $\Phi$ be a map 
\begin{equation}
        \sigma\longmapsto \Phi(\sigma)=
        {\cal C}_\sigma({\cal H})
\end{equation}
mapping states $\sigma\in{\cal S}({\cal H})$ on 
compact sets ${\cal C}_\sigma({\cal H})\subset {\cal S}({\cal H})$.
If $\Phi$ has the property that 
\begin{eqnarray}
        A\rho A^\dagger/\text{tr}[ A\rho A^\dagger]
        &\in&{\cal C}_{A \sigma A^\dagger/\text{tr}[ A\sigma A^\dagger]}
        ({\cal H})\label{restrictiona}\\
        \lambda \rho_1 +(1-\lambda) \rho_2 
        &\in& {\cal C}_{\lambda \sigma_1 +(1-\lambda) \sigma_2 }
        ({\cal H})\label{restriction}
\end{eqnarray}
for all $\sigma, \sigma_1, \sigma_2 \in{\cal S}({\cal H})$,
for all
$\rho \in{\cal C}_\sigma({\cal H})$,
$\rho_1 \in{\cal C}_{\sigma_1}({\cal H})$,
$\rho_2 \in{\cal C}_{\sigma_2}({\cal H})$,
and all 
$\lambda\in[0,1]$ and all 
$A:{\mathbbm C}^N\longrightarrow  {\mathbbm C}^N$,
then  
\begin{equation}\label{strangeinf}
        E_G(\sigma)=\min_{\rho\in{\cal C}_\sigma({\cal H})} S(\sigma||\rho)
\end{equation}
is an entanglement monotone.}\\

\probreak

\proof One may proceed as in Proposition 2.5. 
Eq.\ (\ref{restriction}) guarantees that 
if $\rho_1$ and $\rho_2$ are optimal
separable states achieving the respective minima 
in Eq.\ (\ref{strangeinf})
for two states $\sigma_1$ and $\sigma_2$,
then $\lambda\rho_1+(1-\lambda)\rho_2$
is a possible (but not necessarily optimal)
separable reference state for 
 $\lambda\sigma_1+(1-\lambda)\sigma_2$. Hence,
$E_G$ is convex.
A similar argument can be used to show that
on average, $E_G$ can only decrease
in the course of a local general quantum
measurement. Let $A_i$, $i=1,...,K$,
satisfying 
$\sum_{i=1}^K {A_i^\dagger A_i}=\mathbbm{1}_A$
be the Kraus operators associated with
a local generalized measurement performed by, say, 
Alice. If $\rho$ is the optimal separable
state corresponding to $\sigma$
taken from the set ${\cal C}_\sigma({\cal H})$,
then Eq.\  (\ref{restrictiona})
ensures that
$A_i\rho A_i^\dagger/
\text{tr}[ A_i\rho A_i^\dagger]$ is an
allowed separable reference state of 
the posteriori state 
$A_i\sigma A_i^\dagger/
\text{tr}[ A_i\sigma A_i^\dagger]$. Using the same
argument as in Proposition 2.5
one can conclude that $E_G$ 
is an entanglement monotone. 
\proofend


So far, the considered
entropic entanglement monotones have been defined
in such a way that 
the reference state $\sigma$ appeared 
in the
first argument of the relative entropy functional.
In the subsequent quantity $\sigma$ is 
part of the second argument. Depending on a parameter
$\mu\in[0,1]$ let
\begin{equation}\label{DivMon}
	E_\mu(\sigma)= 
	\min_{\omega\in{\cal D}_\sigma({\cal H}) }
	\min_{\rho\in{\cal D}_\sigma({\cal H})}
	S(\rho|| \mu \sigma + (1-\mu) \omega ).
\end{equation}
\medskip

\noindent {\bf Proposition 2.7. --}
{\it $E_\mu$ is an entanglement monotone
for all $\mu\in[0,1]$.}
\probreak

\proof
$E_\mu$ is clearly positive.
Also, if $\sigma\in{\cal D}({\cal H})$, then also
$\mu \sigma + (1-\mu) \omega\in {\cal D}({\cal H})$
for all $\mu\in[0,1]$ and all $\omega\in   {\cal D}_\sigma({\cal H})$.
Let $\sigma_1,\sigma_2\in {\cal S}({\cal H})$, 
$\lambda\in[0,1]$, and 
$\sigma=\lambda\sigma_1+(1-\lambda)\sigma_2$.
Then
\begin{eqnarray}
	\lambda E_\mu (\sigma_1)+(1-\lambda)E_\mu (\sigma_2)&=&
	\lambda S(\rho_1||\mu \sigma_1 + (1-\mu) \omega_1)+
	(1-\lambda) S(\rho_2||\mu \sigma_2 + (1-\mu) \omega_2)\nonumber\\
	&\geq&
	S(\lambda\rho_1+(1-\lambda)\rho_2||
	\mu \sigma +(1-\mu) (\lambda \omega_1+(1-\lambda)\omega_2))\nonumber\\
	&\geq& E_\mu (\sigma).
\end{eqnarray}
That is, $E_\mu$ is a convex functional for all $\mu\in[0,1]$.

Again, to show that on average $E_\mu$ does not
increase under local generalized measurements let
$A_i$, $i=1,...,K$, $\sum_{i=1}^K A_i^\dagger A_i=\mathbbm{1}_A$,
be the Kraus operators of 
the local generalized measurement implemented
by Alice. The first observation is that
\begin{equation}
	\sum_{i=1}^K \text{tr}[A_i \rho A_i^\dagger] 
	S\left( \frac{A_i \rho A_i^\dagger}{\text{tr}[A_i \rho A_i^\dagger]}
	\right|\left|
	\frac{\mu A_i \sigma A_i^\dagger +(1-\mu) A_i \omega A_i^\dagger}{
	\text{tr}[
		\mu A_i \sigma A_i^\dagger +(1-\mu) A_i \omega A_i^\dagger
	]}
	\right)\leq
	S(\rho||\mu\sigma+(1-\mu)\omega)\label{dunno}
\end{equation}
(see Proposition 2.4).
Let then $\rho,\omega\in{\cal D}_\sigma({\cal H})$. 
The Kraus operators act in the Hilbert space of one
party only and therefore,
\begin{equation}
	\text{tr}[A_i \rho A_i^\dagger] =\text{tr}[A_i \sigma A_i^\dagger]=
	\text{tr}[A_i \omega A_i^\dagger]
\end{equation}
for all $i=1,...,K$, 
because $\text{tr}[A_i \rho A_i^\dagger]=\text{tr}_A[A_i 
\text{tr}_B[\rho] A_i^\dagger]=\text{tr}_A[A_i 
\text{tr}_B[\sigma] A_i^\dagger]=\text{tr}[A_i \sigma A_i^\dagger]$ 
and similarly for $\omega$.
It follows that Eq.\ (\ref{dunno}) can be written as
\begin{equation}
	\sum_{i=1}^K \text{tr}[A_i \sigma A_i^\dagger] 
	S\left( \frac{A_i \rho A_i^\dagger}{\text{tr}[A_i \rho A_i^\dagger]}
	\right|\left|
	\mu \frac{A_i \sigma A_i^\dagger}{
	\text{tr}[
	A_i \sigma A_i^\dagger
	]
	} +(1-\mu) 
	\frac{A_i \omega A_i^\dagger}{
	\text{tr}[
	A_i \omega A_i^\dagger
	]
	}
	\right)\leq
	S(\rho||\mu\sigma+(1-\mu)\omega),
\end{equation}
such that
\begin{equation}
	\sum_{i=1}^K \text{tr}[A_i \sigma A_i^\dagger]
	E_\mu \left(\frac{A_i \sigma A_i^\dagger}{\text{tr}[A_i \sigma A_i^\dagger]}
	\right)\leq
	E_\mu (\sigma).
\end{equation}
This is the desired monotonicity property.\proofend

Practically, this measure is not very 
useful.
The interesting aspect is that for $\mu=1$
it is a monotone that is fully additive, see the subsequent
section, 
disproving the conjecture
that fully additive entanglement monotones do not 
exist.

\subsection{Additivity of Entanglement Measures}\label{Addit}
\index{Additivity of entanglement monotones}

Problems of additivity occur in different contexts 
in quantum information theory, most notably
in the characterization 
of quantum channels \cite{AddiWerner}
and in the theory of  
the quantification
of entanglement. In the latter
case it is the additivity
of entanglement measures 
that attracts major interest.
The {\it additivity property}\/ of a functional
quantifying the degree of entanglement is
-- in a sense -- a manifestation of the idea that 
entanglement is an extensive quantity: 
Assume two parties holding a pair of quantum systems
in a certain mixed state $\sigma$. The 
entanglement -- as quantified by an appropriate
measure of entanglement $E$ -- is given by $E(\sigma)$.
Later, the parties 
get another copy of the same state
which has been prepared by the same source. The question
that arises is whether they now share  
two times $E(\sigma)$ units of entanglement?
Note that the two states of the pairs of quantum systems do not
show any correlations. Intuitively, 
one might be tempted to assume that every
entanglement measure automatically has this property,
or that the structure of tensor products of Hilbert spaces
is simple enough such that the  
decision is trivial whether a given 
measure quantifies entanglement as 
extensive in this sense.

It turned out that the problem 
of additivity is among the most notorious
key problems of a theory of entanglement \cite{VollWerner,Benatti}.
Suppose the underlying Hilbert space has the structure
${\cal H}=({\cal H}_A^{(1)}\otimes{\cal H}_A^{(2)}) 
\otimes ({\cal H}_B^{(1)}\otimes{\cal H}_B^{(2)})$. 
Local operations of Alice act in ${\cal H}_A^{(1)}\otimes{\cal H}_A^{(2)}$,
whereas local operations of Bob
act in ${\cal H}_B^{(1)}\otimes{\cal H}_B^{(2)}$.
A measure of entanglement $E$ is called 
\index{Fully additive}
{\it fully additive}\/ \cite{Contemp,QuantLong}, 
if 
\index{Fully additive}
\begin{equation}\label{fully additive}
	E(\sigma\otimes \rho)=E(\sigma)+E(\rho),
\end{equation}
for a state $\sigma\otimes \rho$ with
$\sigma\in{\cal S}({\cal H}_A^{(1)}\otimes {\cal H}_B^{(1)})$
and
$\rho\in {\cal S}({\cal H}_A^{(2)}\otimes {\cal H}_B^{(2)})$ (see Fig.\ 2.1).
Subadditivity is equivalent with  
\index{Subadditivity}
$E(\sigma\otimes \rho)\leq E(\sigma)+E(\rho)$ for all such states
$\sigma$ and $\rho$.
It is said to be {\it weakly additive}\/ \cite{Horobound}, if 
\begin{equation}
	E(\sigma^{\otimes n})= n E(\sigma),
\end{equation}
for all $n\in{\mathbbm{N}}$, that is, if 
the state $\sigma^{\otimes n}$ is the state of $n$ quantum systems 
prepared in a state $\sigma$
by a stationary memoryless source. 
Strong additivity actually implies weak additivity.

\newfigure{6.6cm}{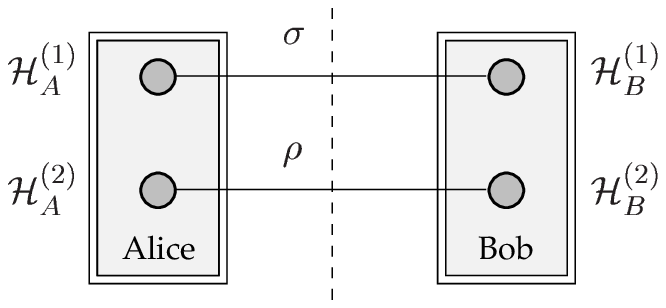}
{Figure 2.1: An entanglement monotone $E$ is fully additive, if
a state of this structure is assigned a value $E(\sigma\otimes\rho)=
E(\sigma)+E(\rho)$, where $\sigma\in{\cal S}({\cal H}_A^{(1)}\otimes {\cal H}_B^{(1)})$
and $\rho\in  {\cal S}({\cal H}_A^{(2)}\otimes {\cal H}_B^{(2)})$.}

The unique measure of entanglement for pure states
-- the von Neumann entropy of a local state -- is fully
additive on pure states. This is due to the fact that
the von Neumann entropy satisfies $S(\sigma^{(1)}\otimes \sigma^{(2)})=
S(\sigma^{(1)})+S(\sigma^{(2)})$ for all states $\sigma^{(1)}$ and 
$\sigma^{(2)}$.
The interpretation of the additivity
of the unique measure for pure states 
can be stated as follows:
If two parties share $n$ copies ($n$ very large)
of a system in a not maximally
entangled state $|\psi\rangle\langle\psi|$, then they
convert them into $n 
E_F(|\psi\rangle\langle\psi|)$ copies of systems in a 
maximally entangled state. Also, to prepare
$n$ copies of $|\psi\rangle\langle\psi|$, approximately
$n E_F(|\psi\rangle\langle\psi|)$ copies of systems in 
a maximally entangled state are needed.

For mixed states the situation is very different.
The distillable entanglement is by definition weakly additive.
$D_{\leftrightarrow}$ is nevertheless 
not known to be fully additive. 
However, for no entanglement monotone that is not
defined via a limit of infinitely many copies of
a state  a proof of weak additivity is known
(but see Proposition 2.13).
In order to achieve weak additivity for a generally
subadditive entanglement
monotone $E$ one may consider the
{\it regularized}\/ 
version of it. This is defined as 
\begin{equation}\label{regular}
	E^\infty(\sigma)={\limsup\limits_{n\rightarrow \infty}}
	\frac{E(\sigma^{\otimes n})}{n},
\end{equation}
which is the mean entanglement of several copies of the state, 
evaluated in the 
limit of infinitely many copies.
\index{Regularized entanglement monotone}
In practice it is hardly possible to calculate
this quantity 
for any entanglement monotone
for a given mixed state. Proposition 2.14,
however, will give a first 
example of the value of a regularized 
measure of entanglement.

It has been shown \cite{Horobound}
that all weakly additive
and weakly continuous entanglement
monotones are confined by
the distillable entanglement $D_{\leftrightarrow}$
and the regularized entanglement of formation,
\begin{equation}
	D_{\leftrightarrow}(\sigma)\leq
	E(\sigma)=\limsup\limits_{n\rightarrow \infty}
	\frac{E(\sigma^{\otimes n})}{n}
	\leq
	E_F^\infty(\sigma),
\end{equation}
As a corollary it follows that $E_M^\infty$ is a lower
bound for $E_F^\infty$, and, as the entanglement of
formation can only be subadditive, it can be 
concluded that
\begin{equation}
	E_M^\infty(\sigma)\leq E_F(\sigma)
\end{equation}
holds for any state $\sigma$.
$E_N$ is not additive in any sense; $\log_2(E_N+1)$ is
fully additive, as $\|\sigma^{T_A}\otimes\rho^{T_B}\|=
\|\sigma^{T_A}\| \, \|  \rho^{T_B}\|$ for
all $\rho\in{\cal H}^{(1)}$ and all 
$\sigma\in {\cal H}^{(2)}$,
but it is no entanglement monotone. Another fully 
additive quantity is $E_1$ defined in Eq.\ 
(\ref{DivMon}):\\

%

\noindent {\bf Proposition 2.8. --} {\it $E_1$ is fully additive.}

\probreak

\proof The additivity of $E_1$ follows from 
Lemma 2.9 
together with the additivity property of the 
relative entropy \cite{Ohya}, 
i.e.,
\begin{equation}\label{eq:6}
        S(\rho^{(1)}\otimes\rho^{(2)}||\sigma^{(1)}\otimes\sigma^{(2)})
        =
        S(\rho^{(1)}||\sigma^{(1)})+ S(\rho^{(2)}||\sigma^{(2)})
\end{equation}
for all $\sigma^{(1)},\rho^{(1)}\in {\cal S}({\cal H}^{(1)})$
and
$\sigma^{(2)},\rho^{(2)}\in {\cal S}({\cal H}^{(2)})$. 

\proofend

 \medskip

\noindent {\bf Lemma 2.9. --} 
{\it Let ${\cal H}={\mathbbm{C}}^N\otimes {\mathbbm{C}}^N$
and let ${\cal C}$ 
be the (compact and convex)
subset of ${\cal D}({\cal H})$
of all states which can be written in the form $\rho^{(1)}\otimes\rho^{(2)}$,
where $\rho^{(1)}\in {\cal D}({\cal H}^{(1)})$ and
$\rho^{(2)}\in {\cal D}({\cal H}^{(2)})$. 
Then 
\begin{eqnarray}\label{eq:5}
	\min\limits_{\rho\in {\cal D}({\cal H})} 
	S(\rho\|\sigma^{(1)}\otimes\sigma^{(2)})
 	=\min\limits_{\rho\in {\cal C}}  
	S(\rho\|\sigma^{(1)}\otimes\sigma^{(2)})
\end{eqnarray}
for all $\sigma^{(1)}\in {\cal S}({\cal H}^{(1)})$ and
$\sigma^{(2)}\in {\cal S}({\cal H}^{(2)})$.
}

\probreak

\proof\/ 
From the {\it conditional expectation property}\/ 
of the relative entropy 
\cite{Ohya}\index{Conditional expectation property}
with respect to the partial trace projection
\footnote{Originally, the 
\index{Conditional expectation property}
conditional expectation property has been 
\index{C$^{\ast}$-algebra}
formulated in terms of finite dimensional C$^{*}$-algebras \cite{Ohya}.
On a finite dimensional
C$^{*}$-algebra  ${\cal A}$ 
there exists a (unique) 
trace functional $\text{tr}$ with the property that $\text{tr}[A B]=
\text{tr}[B A]$ for all $A,B\in{\cal A}$ 
 (isomorphic algebras are not distinguished).
Associated with every functional
 $\sigma$ on the algebra ${\cal A}$
is a density operator
$D_\sigma\in{\cal A}$ via
$\sigma(A)=\text{tr}[D_\sigma A]$. That is, states
are identified with functionals on the algebra ${\cal A}$.
A conditional expectation is now defined as follows.
Let ${\cal B}\subset {\cal A}$ be a C$^{*}$-algebra.
A conditional expectation is a linear map 
$E:{\cal A}\longrightarrow{\cal B}$ with the properties that
(i) for all $B\in{\cal B}$ $E(B)=B$,
(ii) if $A\in{\cal A}^+$ (the positive part of
${\cal A}$) then $E(A)\in {\cal B}^+$,
(iii) $E(A B)=E(A) B$ for all $A\in{\cal A}$ and
$B\in{\cal B}$.

The conditional expectation property 
can be formulated with the help of such maps $E$.
Let ${\cal B}\subset{\cal A}$ be a subalgebra of
${\cal A}$ (also a C$^{*}$-algebra), 
and let $\rho$ be a state of ${\cal A}$
with an invertible density $D_\rho$.
If there exists a conditional expectation 
$E:{\cal A}\rightarrow{\cal B}$ with $\rho\circ E=\rho$,
then for any state $\sigma$ of ${\cal A}$
the equality
\begin{equation}
	S(\sigma||\rho)=S(\sigma|_{\cal B}||\rho|_{\cal B})
	+
	S(\sigma||\sigma\circ E)
\end{equation}
holds. In the words of Ref.\ \cite{Ohya}, 
the interpretation of this equality is that the 
''informational divergence'' of $\sigma$ from $\rho$
on the algebra ${\cal A}$ is given by the sum of the
corresponding divergence
on the subalgebra ${\cal B}$ and the divergence
of $\sigma$ from $\sigma\circ E$, which could be conceived
as the extension of $\sigma|_{\cal B}$ to the 
full algebra ${\cal A}$.} 
it follows that 
\begin{equation}
        S(\rho||  \sigma^{(1)}\otimes\sigma^{(2)})=
        S(\rho^{(1)}|| \sigma^{(1)})
	+S(\rho||\rho^{(1)}\otimes \sigma^{(2)})
\end{equation}
for all 
$\sigma^{(1)}\in {\cal S}({\cal H}^{(1)})$,
$\sigma^{(2)}\in {\cal S}({\cal H}^{(2)})$,
and
$\rho\in {\cal S}({\cal H})$, 
where $\rho^{(1)}=\text{tr}_2 [\rho]$ and $\rho^{(2)}=\text{tr}_1 [\rho]$
are the reduced density operators, 
such  that
\begin{equation}
        S(\rho||\sigma^{(1)}\otimes \sigma^{(2)})=
        S(\rho^{(1)}||\sigma^{(1)})+S(\rho^{(2)}||\sigma^{(2)})+
        S(\rho||\rho^{(1)}\otimes \rho^{(2)}),
\end{equation}
and hence
\begin{equation}\label{ineq:1}
	S(\rho||\sigma^{(1)}\otimes \sigma^{(2)})\geq
        S(\rho^{(1)}\otimes\rho^{(2)}||\sigma^{(1)}\otimes \sigma^{(2)}).
\end{equation}
This in turn implies that 
the state $\rho\in{\cal D}({\cal H})\subset {\cal S}({\cal H})$ 
which minimizes 
$S(\rho\| \sigma^{(1)}\otimes\sigma^{(2)})$
can always be taken out of the smaller subset 
${\cal C}$.

\proofend

\bigskip

This is the only known example
of an entanglement monotone that is (i)
additive and (ii) that is not defined as a 
regularized entanglement monotone in an asymptotic limit
as in Eq.\ (\ref{regular}). In particular, it is the
only fully additive entanglement monotone.
\index{Asymptotic limit}
However, from a practical point of view $E_1$
is useless: this is because 
pure states are not mapped on real numbers; it diverges 
on pure states. $E_1$ maps ${\cal S}({\cal H})$
on ${\mathbbm{R}}\cup \{\infty\}$ (see also Ref.\
\cite{QuantLong} in this context).\\

\noindent {\bf Conjecture 2.10. --} 
{\it Under the assumptions of Lemma 2.9 
\begin{eqnarray}\label{eq:5}
	S(\sigma^{(1)} \otimes \sigma^{(2)}||\rho)\geq
	S(\sigma^{(1)} \otimes \sigma^{(2)}|| \rho^{(1)}\otimes \rho^{(2)})
\end{eqnarray}
holds, where $\rho^{(1)}={\text{tr}}_2[\rho]$ and
$\rho^{(2)}={\text{tr}}_1[\rho]$.}

\probreak
\medskip

If this conjecture was
true, there would be wide-reaching 
consequences: It 
would imply that the relative entropy of entanglement 
was fully additive, and that the regularized relative
entropy of entanglement was identical to $E_R$
itself. Unfortunately, 
the statement of Conjecture 2.10 is wrong in general.
A random matrix test provides counterexamples
to this statement. \footnote{
Let ${\cal H}={\mathbbm{C}}^4 \otimes {\mathbbm{C}}^4 $. Every
state $\rho\in{\cal S}({\cal H})$ can be represented
according to
\begin{equation}
	\rho = U D U^\dagger,
\end{equation}
where $U$ is a unitary $16\times 16$-matrix
and $D$ is a diagonal $16\times 16$ matrix, $D_{ij}=p_i 
\delta_{ij}$. 
A random state may now be drawn as follows \cite{Volume}, [E7]:
A plausible (but by no means the only) choice for the ensemble
of random unitaries is the one with a uniform distribution on
unitaries of the above type (corresponding to the
Haar measure on the group $U(16)$), 
which is called {\it circular unitary
ensemble}\/ \cite{Kus}. For the diagonal matrix $D$
a uniform 
distribution is chosen on the manifold defined by $\sum_i p_i=1$.
If one checks the validity of Conjecture 2.10 with the help 
of random matrices drawn from this ensemble, one 
can easily find counterexamples. The relative frequency of a 
violation of Eq.\ (\ref{eq:5})
in a test with $N=1000$ runs yielded as an estimate
for the probability of a violation the (surprisingly small) value
\begin{equation}
	p=0.007\pm0.001.
\end{equation}
}

Quite recently, it has been shown that 
$E_R$ is  not  weakly additive, refuting a 
common belief \cite{WernerPriv,VollWerner}.
Instead, the relative entropy of entanglement
with respect to separable states can be shown
to be strictly subadditive.
In the remainder of this subsection a
numerical investigation of the asymptotic
limit of infinitely many copies
of a state will be presented
which goes along the lines of the
counterexample of
 Ref.\ \cite{VollWerner}. The 
considered state is a state with high 
symmetry: a Werner state \cite{Wer89,VollWerner}.
The result of the numerical investigation will be
summarized in Proposition 2.14.

The Hilbert space is taken to be 
${\cal H}={\cal H}_A\otimes {\cal H}_B$,
${\cal H}_A={\cal H}_B=
{\mathbbm{C}}^N $, and later restricted to 
${\mathbbm{C}}^3\otimes {\mathbbm{C}}^3$. Let $\pi$
be the {\it permutation operator}\/
\index{Permutation operator}
that interchanges the states of both parties.
In terms of the permutation operator
the projections on the {\it symmetric}\/ and the
\index{Symmetric subspace}
{\it antisymmetric subspaces}\/ of 
${\mathbbm{C}}^N \otimes {\mathbbm{C}}^N$ 
can be written  as
\index{Antisymmetric subspace}
\begin{equation}
	\pi_s=(\mathbbm{1}+\pi)/2,\,\,\,\,\,\,\,\,\,\,\,\,
	\pi_a=(\mathbbm{1}-\pi)/2,
\end{equation}
respectively. The trace of these operators is given by
$\text{tr}[\pi_s]= N(N+1)/2$ and $\text{tr}[\pi_a]= N(N-1)/2$.
Let
\begin{equation}
	\sigma_s=\pi_s/\text{tr}[\pi_s],
	\,\,\,\,\,\,\,\,\,\,\,\,
	\sigma_a=\pi_a/\text{tr}[\pi_a].
\end{equation}
It is obvious that $\sigma_a$ and $\sigma_s$ are
invariant under the map $\omega\longmapsto
(U\otimes U)\omega(U\otimes U)^\dagger$, 
where $U:{\cal H}_A\rightarrow {\cal H}_A$
is a unitary operator. In fact, all states
that are invariant under a random local unitary
operation with operators $U\otimes U$ are a convex
combination of these two states $\sigma_a$ and $\sigma_s$. More 
precisely, let the projection $\Pi:{\cal S}({\cal H})
\longrightarrow {\cal S}({\cal H})$ be
\begin{equation}
	\Pi(\rho)= \int
	d\mu_U 
	(U\otimes U)
	\rho
	(U\otimes U)^\dagger,
\end{equation}
where the integral is performed with respect to the normalized
invariant measure of the unitary group
\index{Haar measure}
({\it Haar measure}\/) \cite{Wer89,LewensteinNPTBE}. 
This quantum operation is typically referred to as
{\it twirling operation}\/; 
it is a trace-preserving completely positive unital map
mapping arbitrary states on \index{Werner state}
{\it Werner states}\/ \cite{Wer89}. Then 
operators $O$ that satisfy $\Pi(O)=O$ 
are just those operators 
for which $[U\otimes U, O]=0$:
they form the {\it commutant}\/ of the group $G$ having
unitaries of the type $U\otimes U$ as elements.
\index{Commutant}
It turns out that the commutant is a vector space 
spanned by 
${\mathbbm{1}}$ and $\pi$ \cite{Wer89,VollWerner},
and all states $\rho$ that are invariant under
$\Pi$ are a mixture of $\sigma_a$
and $\sigma_s$. The weights of the convex
combination are given by
	$\Pi(\rho) =
	\sigma_a
	\text{tr}[\rho \pi_a]+
	\sigma_s
	 \text{tr}[\rho \pi_s]$ \cite{LewensteinNPTBE}.

Of interest to the issue of this subsection
is a Hilbert space of the structure ${\cal H}^{\otimes n}$,
$n\geq 2$.
Suppose that the two parties share
$n$ copies of the state $\sigma=
\lambda \sigma_s+(1-\lambda)
\sigma_a$.
To be specific, let 
${\cal H}={\mathbbm{C}}^3 \otimes {\mathbbm{C}}^3$.
The state $\sigma^{\otimes n}$ is invariant
under unitary operations of the type
\begin{equation}\label{uuu}
	\omega\longmapsto
	\left(U^{(1)}\otimes U^{(1)}\right)
	\otimes ...\otimes
	\left(U^{(n)}\otimes U^{(n)}\right)
	\omega
	\left(U^{(1)}\otimes U^{(1)}\right)^\dagger
	\otimes ...
	\otimes
	\left(U^{(n)}\otimes U^{(n)}\right)^\dagger.
\end{equation}
For each copy of the state
the same unitary local operations are applied
on both Alice's part and Bob's part of the composite
quantum system. The group of local unitaries
is hence the group $G^{\otimes n}$, where $G$ is as
before the group with unitary operators $U\otimes U$
as elements.
By iterating the argument of Ref.\ \cite{VollWerner}
one finds that
states that are invariant under
this operation are necessarily
a convex combination of
states of the form $\sigma_{i_1}\otimes ...\otimes \sigma_{i_n}$,
where $i_1,...,i_n\in\{a,s\}$.

Due to the symmetry of the state the evaluation
of the relative entropy of entanglement is 
simplified by large: If $\sigma$ is invariant under
a group $G$, then the variation over the set
${\cal D}({\cal H})$ or ${\cal P}({\cal H})$
can also be restricted to the subset which 
is also invariant under the same group $G$
\cite{Rains2,PlenioMREGS,VollWerner}. For $\lambda=0$
and $n=2$ this reasoning 
leads to the counterexample
of Ref.\ \cite{VollWerner}:\\

\noindent
{\bf Example (Werner and Vollbrecht). --}
For a single copy
of the system in the state $\sigma_a$ 
the relative entropy of entanglement with
respect to separable states
can be evaluated as
\begin{equation}
	E_R(\sigma_a)=S(\sigma_a||(\sigma_a+\sigma_s)/2),
\end{equation}
leading to $E_R(\sigma_a)=1$.
For two copies of the same state $\sigma_a$ 
one finds that $E_R(\sigma_a\otimes \sigma_a)
\leq  S(\sigma_a\otimes \sigma_a||\rho)$,
where $\rho=(1/3)\sigma_a + (3/4) \sigma_s$, and hence,
\begin{equation}
	2=2 E_R(\sigma_a)> S(\sigma_a\otimes 
\sigma_a||\rho)= \log_2(3) 
\geq E_R( \sigma_a\otimes \sigma_a).
\end{equation}
The statement
that $\rho$ is a separable state can be proved
by designing a protocol preparing the state
locally starting from a product state. 
Therefore, for $\sigma_a$
the relative entropy of entanglement is subadditive.\\

\smallskip

A consequence is that $E_M$ has the same property. 
Let $\rho$ be defined 
as in the previous example, then the
subadditivity follows from the fact that
${\text{tr}}_A[\rho]={\text{tr}}_A[\sigma_a\otimes \sigma_a]=
{\mathbbm{1}}_B$,
${\text{tr}}_B[\rho]={\text{tr}}_B[\sigma_a\otimes \sigma_a]=
{\mathbbm{1}}_A$,
and 
${\text{tr}}_A[(\sigma_a+\sigma_s)/2]={\text{tr}}_A[\sigma_a]={\mathbbm{1}}_B$,
${\text{tr}}_B[(\sigma_a+\sigma_s)/2]={\text{tr}}_B[\sigma_a]={\mathbbm{1}}_A$:\\

\noindent
{\bf Corollary 2.11. --} {\it $E_M$ is truly subadditive.}\\

\probreak

From now on let $n=1,2,...$ be arbitrary.
The numerical investigation does not involve
the relative entropy of entanglement with
respect to the separable states $E_R$, but
the tighter bound $B_R$, that is, the
relative entropy with respect to PPT states.
It has already been pointed out that this
quantity is subadditive \cite{Rains2}. However,
in this subsection the case of many copies
will be addressed. Let $\sigma=\lambda\sigma_s + (1-\lambda) \sigma_a$.
The elements of series $(e_n)_{n\in{\mathbbm{N}}}$ with
\begin{equation}
	e_n(\lambda)=\frac{B_R(\sigma^{\otimes n})}{n}
\end{equation}
are the average amounts of entanglement of $n$ copies
of the state $\sigma$. The regularized quantity
$B_R^{\infty}(\sigma)$ can then be formulated as
\begin{equation}
	B_R^{\infty}(\sigma)= 
	\limsup_{n\rightarrow\infty} e_n(\lambda).
\end{equation}
The series $(e_n(\lambda))_{n\in{\mathbbm{N}}}$ 
is actually a convergent series: It has been proved
in Ref.\ 
\cite{NielsenOld} that if $(a_n)_{n\in{\mathbbm{N}}}$
is a series with the property 
$a_n+a_m\geq a_{n+m}$ for all $n,m\geq0$
and for which a $C\geq 0$ exists such that
$a_n\leq C n$ for all $n$, then $(a_n/n)_{n\in{\mathbbm{N}}}$
is a convergent series. 
\index{Convergent series}
If follows from the subadditivity of $E_R$ that the sequence 
with elements $a_n=n e_n(\lambda)$ satisfies these 
criteria.

In order to show that every known element of the series is also 
an upper bound for the limit,
assume that the value of $e_m(\lambda)$ for an $m\in{\mathbbm{N}}$ is known. 
Then the sequence $e_m(\lambda)$, $e_{2m}(\lambda)$, $e_{3m}(\lambda)$, ... 
is a subsequence satisfying 
$e_m(\lambda)\geq e_{km}(\lambda)$ for all $k\in{\mathbbm{N}}$,
and hence, 
\begin{equation}
	e_m (\lambda) \geq \liminf_{n\rightarrow\infty}  e_n(\lambda).
\end{equation}
Since $(e_n )_{n\in{\mathbbm{N}}}$ is convergent, also
$e_m(\lambda)\geq \lim_{n\rightarrow\infty} e_n(\lambda)$
holds.

Once it is clear that the limit is well defined,
a recipe for calculating the value of
$e_n(\lambda)=
B_R(\sigma^{\otimes n})/n$ has to be developed.
Again, the symmetry of the state simplifies the problem
considerably.
First, note that due to the invariance of $\sigma^{\otimes n}$
under $G^{\otimes n}$, the closest PPT state must be a
mixture of $\sigma_{i_1}\otimes ...\otimes \sigma_{i_n}$,
$i_1,...,i_n\in\{a,s\}$. Second, $\sigma^{\otimes n}$ 
is invariant
under the map 
\begin{equation}
	\omega\longmapsto (\pi\otimes \pi)
	\omega (\pi\otimes \pi),
\end{equation}
where $\pi\in S_n$. 
$S_n$ is the {\it symmetric group}\/ of
\index{Symmetric group}
degree $n$, whose elements are the {\it permutation operators of
degree $n$}\/. $\pi$ denotes both a permutation and
the associated unitary.
Note that $\pi\otimes \pi$ corresponds
to a local operation, as each permutation operator
acts in Alice's or Bob's system only. Hence,
in the variation over PPT states 
it suffices to consider states that are
invariant under both groups:\\

\noindent
{\bf Lemma 2.12. --} {\it Let
$e_n(\lambda)=
B_R(\sigma^{\otimes n})/n$ be defined as above.
Then
\begin{equation}
B_R(\sigma^{\otimes n})=S(\sigma^{\otimes n}||\rho_n),
\end{equation}
where the state $\rho_n\in{\cal P}({\cal H})$ 
is of the form
\begin{equation}
	\rho_n=\sum_{k=0}^{n} 
	\frac{p_{k}}{\left(
	\begin{array}{c}
		n\\
		k 
	\end{array}\right)}
	\sum_{\pi\in S_n} (\pi\otimes \pi) \left(
	\sigma_a^{\otimes k}
		\sigma_s^{\otimes(n-k)}\right)(\pi\otimes \pi).\label{nform}
\end{equation}
$p_0,...,p_n$ is a probability distribution.
The second sum is performed over all elements of the
symmetric group $S_n$ of degree $n$. }\\

\probreak

For simplicity of notation 
assume from now on that ${\cal H}={\cal H}_A\otimes{\cal H}_B$
with ${\cal H}_A={\cal H}_B={\mathbbm{C}^3}$.
Explicitly, if the basis elements of ${\cal H}_A$
and ${\cal H}_B$ are labeled $\{|1\rangle,|2\rangle,
|3\rangle\}$, the states $\sigma_a$ and $\sigma_s$ can
be written as
\begin{eqnarray}
\sigma_s&=& \frac{1}{6}\biggl(
\sum_{i=1}^3 |ii\rangle\langle ii|
+ \sum_{i,j=1 (i<j)}^3
(|ij\rangle+|ji\rangle)(\langle ij|+\langle ji|)
\biggr),\\
\sigma_a &=& \frac{1}{3}\biggl(
\sum_{i,j=1 (i<j)}^3
(|ij\rangle-|ji\rangle)(\langle ij|-\langle ji|)
\biggr).
\end{eqnarray}
The operators
$\sigma_a^{T_A}$ and $\sigma_s^{T_A}$ commute 
and can be simultaneously 
diagonalized. In an appropriate basis they can be represented
according to
\begin{eqnarray}
\sigma_s^{T_A} &=& 
\text{diag} (1/12,1/12,1/12,1/12,1/12,1/12,1/12,1/12,1/3), \\
\sigma_a^{T_A}&=&
\text{diag} (1/6,1/6,1/6,1/6,1/6,1/6,1/6,1/6,-1/3),
\end{eqnarray}
where ${\text{diag}}$ 
denotes a diagonal matrix with the vector entries
as main diagonal elements. When the task is
to find a criterion under what circumstances
a convex combination of products of these states
is a PPT state, the degeneracy of the largest
eigenvalue is of no relevance, and one can treat
$\sigma_a^{T_A}$ and $\sigma_s^{T_A}$ as if they
were operators $\omega_a$ and $\omega_s$
with a matrix representation $\text{diag} (1/6,-1/3)$
and $\text{diag} (1/12,1/3)$, respectively. 
That is, $\rho_n$ is a PPT state if and only
if 
\begin{equation}
	\sum_{k=0}^{n} 
	\frac{p_{k}}{
	\left(
	\begin{array}{c}
		n\\
		k 
	\end{array}
	\right)}\sum_{\pi\in S_n} \pi \left(
	\omega_a^{\otimes k}
		\omega_s^{\otimes(n-k)}\right)\pi\geq 0.\label{cold}
\end{equation}
For $n=1$ the state $\rho_n$ 
is a PPT state if and only if
$p_0+ 2  p_1\geq 0$
and $p_0 - p_1\geq 0$. For larger $n$ one
can explicitly  show  \cite{PlenioPriv}  that  
 Eq.\ (\ref{cold})
is equivalent with
\begin{equation}
	\sum_{k=0}^n
	\frac{p_k}{\left(
	\begin{array}{c}
	n\\ k
	\end{array}
	\right)}\sum_{l=0}^k
	\left(-
	\frac{1}{2}\right)^l
	\left(
		\begin{array}{c}
	n-s\\ k-l
	\end{array}
	\right)\left(
		\begin{array}{c}
	s\\ l
	\end{array}
	\right) 2^k\geq 0
\end{equation}\label{constr}
for all $s=0,...,n$. 
The sum can be evaluated as
\begin{eqnarray}
	&&\sum_{r=1}^n p_r 
	\frac{2^r (n-r)!}{n!(n-s)! (n-r-s)!}
	 {_2 F_1}\bigl(
	 -r,-s,1+n-r-s,-1/2
	 \bigr)+p_0
	\geq 0,
\end{eqnarray}
where $_2 F_1$ denotes the
\index{Hypergeometric function}
{\it hypergeometric function}\/. This condition, together
with the general form of the 
subset of PPT states that has to be considered of Lemma 2.12 
allows for a numerical evaluation of 
the regularized entanglement
measure $B^\infty_R(\sigma)$ with arbitrary
accuracy. The elements of the series
$(e_n)_{n\in{\mathbbm{N}}}$  
are given by
\begin{eqnarray}
	e_n(\lambda)=
	\frac{S(\sigma^{\otimes n}||\rho_n)}{n}
	&=&
	\frac{1}{n}
	\sum_{k=1}^n 
	\left(
	\begin{array}{c}
	n\\ k
	\end{array}
	\right)
	\lambda^{n-k}
	(1-\lambda)^{k}
	\log_2 
	\left(\left(
	\begin{array}{c}
	n\\ k
	\end{array}
	\right)
	\lambda^{n-k}
	(1-\lambda)^{k}
	/p_k\right)\nonumber\\
	&+& \frac{1}{n}\lambda^n
	\log_2\left(
		\lambda^n/ p_0
	\right).\label{nearly} 
\end{eqnarray}

\medskip

\noindent 
{\bf Example 2.13. --} 
For $\lambda=0$, that is, $\sigma=\sigma_a$, the remaining
minimization
can be performed numerically in an efficient way.
Then Eq.\ (\ref{nearly}) reduces to
\begin{equation}
	\frac{S(\sigma_a^{\otimes n}||\rho_n)}{n}=
	\frac{1}{n}\log_2(p_n). 
\end{equation}
In order to minimize the relative
entropy functional of $\sigma_a^{\otimes n}$
with respect to a state of the form given
by Eq.\ (\ref{nform}) one has to
maximize $p_n$ under the affine
constraints given by Eq.\ (\ref{constr}), since
the logarithm is a monotone increasing function.
Minimization problems in which the objective function
is linear and the constraint functions are affine
are \index{Linear programming}
{\it linear programming problems}\/. 
For such linear programming problems 
efficient and stable
numerical algorithms are available.\footnote{
The above problem can be cast into the so-called
{\it restricted normal form}\/. 
\index{Restricted normal form}
This means that
by introducing auxiliary variables 
the function that has to be maximized can be written
as
\begin{equation}
	z=a_{01} x_1 + ... + a_{0m} x_m
\end{equation}
with primary constraints
$
	x_1\geq 0,\,...\,, x_m\geq 0
$
and additional constraints of the form
$
	a_{k1} x_1+...+ a_{km} x_m = b_k,\,\,\,\,
	k=1,...,l,
$
where $(a_{ij})_{i=1,...,l;j=1,...,m}$ is a real
matrix and $b_1,..., b_l$ are real positive numbers.
The most prominent algorithm for solving such a
linear programming problem in restricted normal
form is the \index{Simplex method}
{\it Simplex method}\/ \cite{Rec,Stoer}.
The subsequent analysis has been carried out using
this method. 
}
The values of $(e_n(0))_{n\in{\mathbbm{N}}}$ can therefore
be evaluated with  high accuracy. 
The first 7 values
$e_1(0),...,e_7(0)$ are given by
\begin{eqnarray}
	e_1(0) &=&1,\\
	e_2(0)  &=&\log_2(3)/2,\\
	e_3(0)  &=&\log_2(5)/3,\\
	e_4(0)  &=&3/4,\\
	e_5 (0) &=&\log_2(13.25)/5,\\
	e_6(0)  &=&\log_2(21.75)/6,\\
	e_7(0)  &=&\log_2(36)/7,
\end{eqnarray}	
see Fig.\ 2.2. The corresponding closest PPT states will be presented in 
Appendix B. With not too much numerical effort the program can 
evaluate the average value of $B_R$ for $40$
copies of the state $\sigma_a$. The series
converges quickly, and the values for $20$ and
$30$ copies are identical to four significant
digits. The main result can be stated as follows.\\

\noindent 
{\bf Proposition 2.14. --} 
{\it Let ${\cal H}={\mathbbm{C}}^3\otimes {\mathbbm{C}}^3$, and let 
$\sigma_a=\pi_a/{\text{tr}}[\sigma_a]$ be the state that is
proportional to the projector on the antisymmetric subspace
of ${\cal H}$ as defined above. Then $B_R^\infty(\sigma_a)$ satisfies
$B_R^\infty(\sigma_a)\leq B_R(\sigma_a^{\otimes n})/n$
for all $n\in{\mathbbm{N}}$. In particular, 
\begin{equation}
	B_R^\infty(\sigma_a)\leq 
	B_{R}(\sigma_a^{\otimes 40})/40
	= 0.73697.
\end{equation}
}
\bigskip

Interestingly, this value strongly suggests
that $B^\infty_R(\sigma_a)$ is identical with
another upper bound for distillable entanglement 
$D_{\leftrightarrow}$ (see also
Ref.\ \cite{ShashPriv}), 
namely, the log negativity.\index{Negativity}\index{Log negativity}
For the state $\sigma_a$ the log negativity is given by
\begin{equation}
	\log_2 (E_N(\sigma_a)+1)=
	\log_2 (\|\sigma_a^{T_A}\|) = \log_2 (5/3)=0.73697.
\end{equation}
The log negativity is fully additive, and hence,
$
\lim_{n\rightarrow\infty}
\log_2 (\|(\sigma_a^{T_A})^{\otimes n}\|)/n= \log_2(\|\sigma_a^{T_A}\|)$.

\newfigure{7cm}{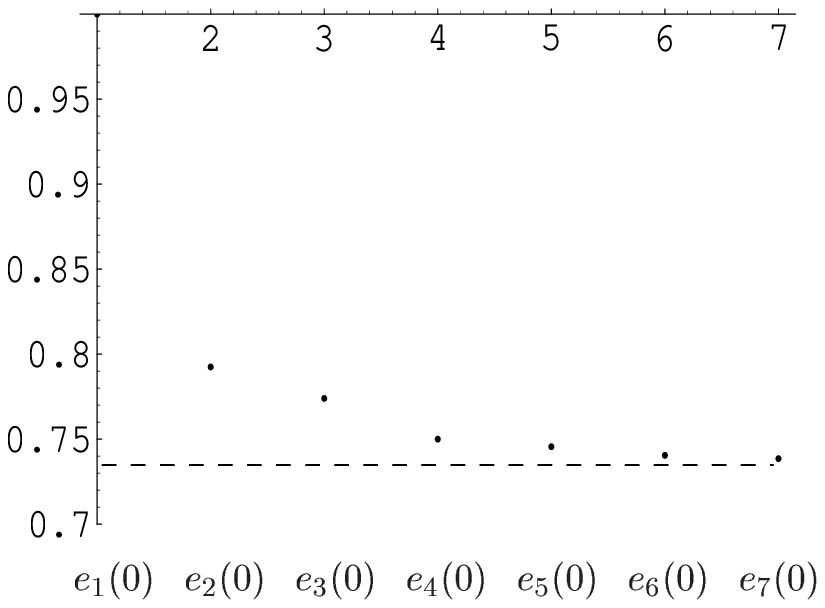}
{Figure 2.2: The average entanglement per copy of $\sigma_a$ for $n=1,...,7$
copies. The dashed line
corresponds to  $B_R(\sigma_a^{\otimes 40})/40$.}

This result has an implication on the issue of
``irreversibility'' of asymptotic
\index{Irreversibility of entanglement manipulations}
manipulation of entanglement. In Ref.\ \cite{HoroIrrer}
it is stated that the process
of distillation is truly
irreversible, meaning that there
exist states for which the distillable entanglement
$D_{\leftrightarrow}$ and the regularized entanglement
of formation $E_F^\infty$ are different from each other:
More resources are needed in a preparation procedure
even in the asymptotic limit than can be distilled
from many copies of the same state. 
The argument goes as follows: if one has two
quantities $E_1$ and $E_2$ satisfying
\begin{equation}
D_{\leftrightarrow}(\sigma)\leq E_1(\sigma),\,\,
E_1(\sigma)<E_2(\sigma),\,\,E_2(\sigma)\leq 
E_F^\infty(\sigma) 
\end{equation}
for a mixed state $\sigma$,
then existence of irreversibility is confirmed.
The two considered quantities are $E_1=B_R^\infty$
and $E_2=\log_2 (E_N+1)$, and the crucial ingredient 
is that $B_R^\infty(\sigma)< \log_2 (E_N(\sigma)+1)$, where $\sigma$
is taken from a small subset of Werner states.
Unfortunately, a theorem from Ref.\ \cite{Rains2}
has been used in Ref.\ \cite{HoroIrrer} that has turned
out to be wrong in general. For the case $\sigma=\sigma_a$
the above
numerical investigation even suggests that 
these quantities are equal, which leaves the
question again open whether this type of irreversibility
exists in quantum theory.
It is conceivable that $D_{\leftrightarrow}$ and
$E_F^\infty$ are identical for all states, 
meaning that one can extract the same 
amount of entanglement from
many copies of a mixed state as one has to invest in
order to ``form'' copies of this state.

It would also be very interesting to carry out a similar
analysis in the case that $\sigma=\lambda\sigma_s+(1-\lambda)\sigma_a$
with $0<\lambda<1$. Of particular interest
is the regime $\lambda\in[2/5, 1/2]$.  
It has been conjectured on indeed 
well-footed grounds \cite{LewensteinNPTBE}
that in this regime $\sigma$ is a bound entangled state, meaning that
$D_{\leftrightarrow}(\sigma)=0$, despite of the fact that 
$\sigma$ is by definition not a PPT state (see also \cite{BennettNPTBE}). 
This state would be a counterexample to 
the statement 
``all states with a non-positive partial 
transpose are distillable''.
As $B_R^\infty(\sigma)$ is a tight upper bound for 
$D_{\leftrightarrow}(\sigma)$, it might well be that a
similar analysis as above yields the value $B_R^\infty(\sigma)=0$.
The implication would be that testing whether the
partial transpose is positive is not enough to
find out whether copies of a state can be distilled
into a useful form.
\index{Non-PPT bound entangled state}

\subsection{Continuity Properties}

As has already been mentioned, there is yet another property
of an entanglement monotone that is important when
\index{Asymptotic limit}
considering the asymptotic limit: it
is the appropriate {\it continuity}\/ 
of the entanglement measure. 
Take, say, a
distillation process in which one tries
to distill a certain pure state $|\psi\rangle\langle\psi|$
from a large number of identically prepared quantum
systems in a mixed state $\sigma$. 
More precisely, one has a number $m$ of
copies of quantum systems in a state $\sigma$ at hand.
By applying LOCC operations, one maps the
state $\sigma^{\otimes m}$ with unit probability
on $\rho_n$. This state can be made
arbitrarily close to the desired state 
$|\psi\rangle\langle\psi|^{\otimes n}$,
and the rate $n/m$ approaches the optimal rate to which this
transformation is possible as $n\rightarrow \infty$.

Against the backdrop of these considerations,
a good measure of entanglement $E$ should fulfil
the following condition that equals the weak
continuity introduced in Eq.\ (\ref{weaklycont}):
the difference in the degree of entanglement per copy
$(E(\rho_n)-E(|\psi\rangle\langle\psi|^{\otimes 
n}))/n$ tends to zero as
$n\rightarrow \infty$. 
Roughly speaking, the requirement is that the measure of entanglement is 
sufficiently continuous  close to many copies of 
products of pure states. 

This form of 
continuity contrasts
with  the {\it strong continuity}\/ 
\index{Strong continuity}
of the unique measure of entanglement 
for pure states \cite{Cont1}
\index{Von Neumann entropy} 
via Fannes inequality \cite{Fannes}\footnote{{\it Fannes' 
inequality}\/ 
states that
\begin{equation}
	 |S(\sigma)-S(\rho)| 
	\leq   \|\sigma-\rho\|\log_2(d) -  
	\|\sigma-\rho\|\log_2\left(\|\sigma-\rho\|\right)
\end{equation}
for two states $\sigma$ and $\rho$ satisfying
$\|\sigma-\rho\|<1/3$, $d$ is the dimension of the underlying Hilbert 
space.}:
\index{Fannes' inequality}
For {\it pure states}\/ 
$\sigma_1$ and $\sigma_2$
with $\| \sigma_1 - \sigma_2\|<1/3$
\begin{equation}
	|E_F(\sigma_1)-E_F(\sigma_2)|
	\leq  \log_2(d) \| \sigma_1 - \sigma_2\|
	-
	\| \sigma_1 - \sigma_2\|
	\log_2(\| \sigma_1 - \sigma_2\|)\label{Purecont}
\end{equation}
holds, where $d$ is the dimension of the Hilbert space of the quantum system.
For the relative entropy of entanglement $E_R$ and $B_R$
\cite{Cont2} and the 
entanglement of formation $E_F$
\cite{Cont1} similar results to Eq.\ (\ref{Purecont}) hold
for mixed states.

In this subsection a single statement will be 
presented:
The subsequent proposition shows that the modified relative
entropy of entanglement $E_M$ is weakly continuous. The proof is technical, and
therefore, the major 
part of the proof is presented in Lemma 2.16. 
The result will not be necessary for the understanding of
the later considerations, so the proof may be skipped.\\

\index{Weakly continuous}
\noindent {\bf Proposition 2.15. --}
{\it  $E_M$ is weakly continuous:
Let ${\cal H}={\mathbbm{C}}^N\otimes {\mathbbm{C}}^N$,
and let $|\psi\rangle\in{\cal H}$. 
Let
$(\sigma_n)_{n\in\mathbbm{N}}$ be a series
of states $\sigma_n\in{\cal S}({\cal H}^{\otimes n})$
with the property 
$
	\lim_{n\rightarrow\infty}\| |\psi\rangle\langle\psi|^{\otimes n}
	-\sigma_n\|= 0$, where $\|.\|$ denotes the trace norm.
\index{Trace norm}
Then $E_M$ satisfies
\begin{equation}\label{Co2}
	\lim_{n\rightarrow\infty}
	\frac{1}{n}\left| E_M(|\psi\rangle\langle\psi|^{\otimes n})
	-
	E_M(\sigma_n)\right|
	= 0.
\end{equation}
}
\probreak

\index{Uniqueness theorem for entanglement measures}
\proof Let $|\psi\rangle\in{\cal H}$ and 
$(\sigma_n)_{n\in\mathbbm{N}}$ be a series
of states $\sigma_n\in{\cal S}({\cal H}^{\otimes n})$
as above. The first step is to introduce a certain 
appropriate  series of pure states:
There exists a series $(|\psi\rangle_n)_{n\in\mathbbm{N}}$
with $|\phi_n\rangle\in{\cal H}^{\otimes n}$
for $n=1,2,...$ with
the properties (i)
\begin{equation}
	\text{tr}_A[|\phi_n\rangle\langle\phi_n]
	=\text{tr}_A[\sigma_n],\,\,\,
	\text{tr}_B[|\phi_n\rangle\langle\phi_n]
	=\text{tr}_B[\sigma_n],
\end{equation}
and (ii)
\begin{equation}
	\lim_{n\rightarrow\infty}
	\left\| 
	|\phi_n\rangle\langle\phi_n|-
	|\psi\rangle\langle\psi|^{\otimes n}
	\right\|=0,\,\,\,
	\lim_{n\rightarrow\infty}
	\left\| 
	|\phi_n\rangle\langle\phi_n|-
	\sigma_n
	\right\|=0.
\end{equation}
That is, $|\phi_n\rangle\langle\phi_n|$ is locally
identical to $\sigma_n$.
Note that 
the trace distance of two states is non-increasing
under trace-preserving completely positive maps ${\cal E}$ 
\cite{Ruskai}, 
\begin{equation}\label{undercp}
\left\| \sigma-\rho\right\| \geq \|  {\cal E}(\sigma)-{\cal E}(\rho)  \|
\end{equation}
for all states $\sigma,\rho$. In particular, this statements holds
for the partial trace operation. Thus,
such a sequence always exists. As a consequence,
\begin{eqnarray}
	\frac{\left| E_M(|\psi\rangle\langle\psi|^{\otimes n})
	-
	E_M(\sigma_n)\right|}{n}&\leq&
	\frac{\left| E_M(|\psi\rangle\langle\psi|^{\otimes n})
	- E_M(|\phi_n\rangle\langle\phi_n|)\right|}{n}\nonumber\\
	&+&
	\frac{\left|E_M(|\phi_n\rangle\langle\phi_n|)
	-
	E_M(\sigma_n)\right|}{n}.\label{tozero}
\end{eqnarray}
The first term on the right hand side of Eq.\ (\ref{tozero})
will vanish in the limit $n\rightarrow\infty$.
According to Lemma 2.6 $E_M$ and 
the von Neumann entropy of a local state 
(see Eq.\ (\ref{unimeasure})) coincide for pure states. Therefore,
the result for pure states given by Eq.\ (\ref{Purecont})
can be applied. It follows that
\begin{eqnarray}
	\frac{\left| E_M(|\psi\rangle\langle\psi|^{\otimes n})
	-
	E_M(\sigma_n)\right|}{n}&=&
	\frac{|  
	S(\text{tr}_A[|\psi\rangle\langle\psi|^{\otimes n}])
	-
	S(\text{tr}_A[|\phi_n\rangle\langle\phi_n|])|}{n}\nonumber\\
	&\leq& 	x \log_2(n N^2)- x \log_2 (x)
\end{eqnarray}	
for all $n\geq n_0$, where $n_0\in\mathbbm{N}$ is
sufficiently large, as ${\text{dim}}[{\cal H}^{\otimes n}]=n N^2$.
In this step, both Eq.\ (\ref{undercp}) and
Fannes' inequality \cite{Fannes,Ohya} 
have been used. 

The technical part of the proof is to show that 
the second term on the right hand side of  Eq.\ (\ref{tozero})
tends to zero in the limit $n\rightarrow \infty$ as well.
It will be shown in Lemma 2.16  
that 
\begin{equation}
	\lim_{n\rightarrow\infty} 
	\frac{\left|E_M(|\phi_n\rangle\langle\phi_n|)
	-
	E_M(\sigma_n)\right|}{n}=0.
\end{equation}
That is, $E_M$ is weakly continuous.
\proofend

\probreak

Again, this observation is consistent with
the uniqueness theorem for entanglement measures
for pure states.
\index{Uniqueness theorem for entanglement measures}
$E_M$ satisfies all the criteria of the theorem, and
hence, 
\begin{equation}
	E_M(|\psi\rangle\langle\psi|)=D_{\leftrightarrow}(|\psi\rangle\langle\psi|)
\end{equation}	
for all $|\psi\rangle\in{\cal H}$.
The same argument can be applied to $B_M$.
The subsequent lemma provides 
part of the proof of Proposition 2.15.
The first part of
the proof of Lemma 2.16
is closely related to a proof of the strong
\index{Strong continuity}
continuity of the relative entropy of entanglement 
given in Ref.\ \cite{Cont2}. However, 
an important
assumption of the proof in Ref.\ \cite{Cont2}
is not available in this case. The 
rather elaborate detour in the proof
is necessary due to the
additional constraint in $E_M$ compared to the
relative entropy of entanglement.\\

\noindent {\bf Lemma 2.16. --} {\it 
Let ${\cal H}={\mathbbm{C}}^N\otimes {\mathbbm{C}}^N$,
and let $|\psi\rangle\in{\cal H}$. 
Let
$(\sigma_n)_{n\in\mathbbm{N}}$ be a series
of states $\sigma_n\in{\cal S}({\cal H}^{\otimes n})$
with the property that
$
	\lim_{n\rightarrow\infty}\| |\psi\rangle\langle\psi|^{\otimes n}
	-\sigma_n\|= 0$.
Let 
 $(|\phi_n\rangle\langle\phi_n|)_{n\in\mathbbm{N}}$,
 $|\phi_n\rangle\in{\cal H}^{\otimes n}$ for $n=1,2,...$,
be a series of pure states satisfying
\begin{equation}
	\text{tr}_A[|\phi_n\rangle\langle\phi_n]
	=\text{tr}_A[\sigma_n],\,\,\,
	\text{tr}_B[|\phi_n\rangle\langle\phi_n]
	=\text{tr}_B[\sigma_n],
\end{equation}
and 
\begin{equation}
	\lim_{n\rightarrow\infty}
	\left\| 
	|\phi_n\rangle\langle\phi_n|-
	|\psi\rangle\langle\psi|^{\otimes n}
	\right\|=0,\,\,\,
	\lim_{n\rightarrow\infty}
	\left\| 
	|\phi_n\rangle\langle\phi_n|-
	\sigma_n
	\right\|=0.
\end{equation}
Then the modified relative entropy of entanglement
$E_M$ satisfies
\begin{equation}
	\lim\limits_{n\rightarrow \infty}	
	\frac{\left|
	E_M(|\phi_n\rangle\langle\phi_n|)-E_M(\sigma_n)
	\right|}{n}=0.
\end{equation}
}
\probreak

\proof
For brevity, call $\eta_n= |\phi_n\rangle\langle\phi_n|$. 
Let $\eta_n^*\in {\cal D}_{\eta_n}({\cal H})$ 
and $\sigma_n^*\in {\cal D}_{\sigma_n}({\cal H})$ be 
states that satisfy
\begin{equation}
	E_M(\eta_n)=S(\eta_n||\eta_n^\ast),\,\,\,\,
	E_M(\sigma_n)=S(\sigma_n||\sigma_n^\ast), 
\end{equation}
respectively. Application 
of the triangle inequality yields
\begin{equation}\label{appenda}
	\left|E(\eta_n)-E(\sigma_n)\right|\leq
	|S(\eta_n)-S(\sigma_n)|+
	|{\text{tr}}[\eta_n \log_2(\eta_n^*)]-{\text{tr}}[ \sigma_n \log_2(\sigma_n^*)]|.
\end{equation}
The first step is to use
Eq.\  (\ref{Purecont}) to get
\begin{equation}
	|S(\eta_n)-S(\sigma_n)|\leq \|\eta_n - \sigma_n\| \log_2(n N^2)
	- \|\eta_n - \sigma_n\|\log_2(\|\eta_n - \sigma_n\|)
\end{equation}
for $n\geq n_0$, $n_0\in{\mathbbm{N}}$ sufficiently large,
as ${\text{dim}}[{\cal H}^{\otimes n}]= n N^2$.
It follows that 
\begin{equation}\label{resone}
	\lim_{n\rightarrow \infty}|S(\eta_n)-S(\sigma_n)|/n=0.
\end{equation}
The remaining task is to find an appropriate upper bound for the
right hand side in Eq.\ (\ref{appenda}). In order to construct
this upper bound an auxiliary quantity is helpful. Let
\begin{equation}
	M^n_\lambda(\sigma)=-\min\limits_{\omega\in{\cal D}_{\eta_n}(\cal H)}
	{\text{tr}}[\sigma  \log_2(\lambda \omega+ (1-\lambda) \tau_n)],
\end{equation}
which depends on $n$ and on an additional parameter $\lambda\in(0,1)$
that will later be fixed. $\tau_n$ stands in this definition for
\begin{equation}
	\tau_n= {\text{tr}}_B[\eta_n]\otimes
	{\text{tr}}_A[\eta_n].
\end{equation}
That is, $\tau_n$ is the mixed 
\index{Product state}
product state that is locally
identical to $\eta_n$ for each $n$.
The states taken from ${\cal D}_{\eta_n}(\cal H)$ 
which are optimal in $M^n_\lambda$ for
$\eta_n$ and $\sigma_n$  will be called
$\eta_n'$ and $\sigma_n'$, respectively. Then the right hand side
of  Eq.\ (\ref{appenda}) becomes
\begin{eqnarray}
	&&|-{\text{tr}}[\eta_n \log_2(\eta_n^*)]+{\text{tr}}[ \sigma_n 
	\log_2(\sigma_n^*)]|\\
	&\leq&
	|-{\text{tr}}[\eta_n \log_2(\eta_n^*)]-
	M_\lambda(\eta_n)|+
	|M_\lambda(\eta_n)- M_\lambda(\sigma_n)|
	+|M_\lambda(\sigma_n)+{\text{tr}}[ \sigma_n \log_2(\sigma_n^*)]|.
	\nonumber	
\end{eqnarray}
Now set $\lambda=1-\|\eta_n-\sigma_n\|$. 
Then
\begin{equation}
	|-{\text{tr}}[\eta_n \log_2(\eta_n^*)]-
	M_\lambda(\eta_n)|=
	|-{\text{tr}}[\eta_n (\log_2(\eta_n^*) + \log_2(\lambda\eta_n' + 
	(1-\lambda)\tau_n))]|.
\end{equation}
The logarithm is an operator
monotone function, that is, if $A$ and $B$ are Hermitian
matrices, then $A\leq B$ implies that $\log_2(A)\leq \log_2(B)$.
\begin{equation}\label{gettingweirder}
	-{\text{tr}}[\eta_n \log_2(\lambda\eta_n' + 
	(1-\lambda)\tau_n)]
	\leq 
	-{\text{tr}}[\eta_n \log_2(\lambda\eta_n^* + 
	(1-\lambda)\tau_n)]\leq {\text{tr}}[\eta_n \log_2(\eta_n^*)]
	-\log_2(\lambda).
\end{equation}
The first inequality in Eq.\ (\ref{gettingweirder})
holds since 
$\eta_n^*$ is included in ${\cal D}_{\eta_n}({\cal H})$, but
it is not necessarily optimal in $M^n_\lambda$ for
$\eta_n$.
Therefore,
\begin{equation}
	|-{\text{tr}}[\eta_n \log_2(\eta_n^*)]-
	M_\lambda(\eta_n)|\leq |\log_2(\lambda)|\leq 2 \|\eta_n-\sigma_n\|.
\end{equation}
The same argument applies to $|-{\text{tr}}[\sigma_n \log_2(\sigma_n^*)]-
	M_\lambda(\sigma_n)|$, and hence,
$|-{\text{tr}}[\sigma_n \log_2(\sigma_n^*)]-
	M_\lambda(\sigma_n)|\leq 2\|\eta_n-\sigma_n\|$. This means that
\begin{equation}\label{restwo}
	\lim_{n\rightarrow \infty}
	\frac{|-{\text{tr}}[\eta_n \log_2(\eta_n^*)]-
	M_\lambda(\eta_n)|	
	}{n}=0,\,\,\,\,
	\lim_{n\rightarrow \infty}
	\frac{|-{\text{tr}}[\sigma_n \log_2(\sigma_n^*)]-
	M_\lambda(\sigma_n)|	
	}{n}=0.
\end{equation}
The last step is to find an 
upper bound for $|M_\lambda(\eta_n)- M_\lambda(\sigma_n)|$.
In this last step it will be used that
$\lim_{n\rightarrow\infty}
	\left\| 
	\eta_n-
	|\psi\rangle\langle\psi|^{\otimes n}
	\right\|=0$, and employing
the fact that the trace norm can only decrease under
the application of the partial trace operation
\cite{Ruskai} it follows that also
\begin{equation}
\lim_{n\rightarrow\infty}
\left\| 
	|{\text{tr}}_A[\eta_n]-
	{\text{tr}}_A[|\psi\rangle\langle\psi|^{\otimes n}]
	\right\|=0=\lim_{n\rightarrow\infty}
	\left\| 
	|{\text{tr}}_A[\tau_n]-
	{\text{tr}}_A[|\psi\rangle\langle\psi|^{\otimes n}]
	\right\|.
\end{equation}
In particular, this means that there exists a constant $1\geq C>0$
independent of $n$ and an $n_1\in{\mathbbm{N}}$
such that
\begin{equation}
	\tau_n^\ast \geq C^n \mathbbm{1}
\end{equation}
for all $n\geq n_1$,
where $\tau_n^\ast$ is the state that is given by 
$\tau_n$ restricted 
to $\text{range}[\sigma_n] \cup \text{range}[\eta_n]$.
Using again the operator monotonicity of $\log_2$ one can
conclude that
\begin{eqnarray}\label{equip}
	\left(\log_2(1-\lambda)+n \log_2(C)\right){\mathbbm{1}}
	&\leq&
	\log_2(1-\lambda) {\mathbbm{1}} + \log_2(\tau_n^\ast)=
	\log_2\left((1-\lambda) \tau_n^\ast\right)\nonumber\\
	&\leq&\log_2(\lambda\sigma_n' +(1-\lambda)\tau_n^\ast )\leq 0.	
\end{eqnarray}
Equipped with Eq.\ (\ref{equip}) one can
find an appropriate 
upper bound for $|M_\lambda(\eta_n)- M_\lambda(\sigma_n)|$.
It is given by
\begin{eqnarray}\label{ui}
	|M_\lambda(\eta_n)- M_\lambda(\sigma_n)|&\leq& 
	|
	-{\text{tr}}[\eta_n \log_2(\lambda\sigma_n' + 
	(1-\lambda)\tau_n)]
	+{\text{tr}}[\sigma_n \log_2(\lambda\sigma_n' + 
	(1-\lambda)\tau_n)]
	|\nonumber\\
	&=&
	|
	{\text{tr}}[(\sigma_n-\eta_n) (\log_2(\lambda\sigma_n' + 
	(1-\lambda)\tau_n^\ast))]|\nonumber\\
	&\leq &\|\eta_n - \sigma_n\|
	\left(-\log_2(1-\lambda)-n \log_2(C)\right)
\end{eqnarray}
for $n\geq n_1$.
Thus
\begin{eqnarray}
	\lim_{n\rightarrow\infty}
	\frac{|M_\lambda(\eta_n)- M_\lambda(\sigma_n)|}{n}
	&\leq& 
	\lim_{n\rightarrow\infty}
	\left(
	-\frac{1}{n}
	\|\eta_n - \sigma_n\|
	\log_2\left(\|\eta_n - \sigma_n\|\right)
	-\log_2(C) \|\eta_n - \sigma_n\|
	\right)\nonumber\\
	&=&0.\label{resthree}
\end{eqnarray}
Combining Eqs.\ (\ref{resone}), (\ref{restwo}), and
(\ref{resthree}) it follows that
\begin{equation}
	\lim\limits_{n\rightarrow \infty}	
	\frac{\left|
	E_M(|\phi_n\rangle\langle\phi_n|)-E_M(\sigma_n)
	\right|}{n}=0.
\end{equation}
This is the statement of the lemma.\proofend

\section{Quantification of Multi-Partite Quantum Entanglement}
\index{Multi-particle entanglement}

In this subsection multi-partite entanglement
will be investigated.
As has been pointed out in the introduction of this chapter,
new complex structures emerge in the case when more
than two parties are present \cite{Multi2,Multi3,Kempe}. 
In a bi-partite setting
all pure-state entanglement is essentially equivalent to
the entanglement of a Bell state of two qubits:
A large number $n$ of 
copies of a particular pure state with a state vector
$|\psi\rangle$ 
can be transformed reversibly into a smaller number $m$ of Bell states
with local operations and classical communication. In the asymptotic
limit the ratio of the numbers $m/n$ is given by
$S(\text{tr}_A[|\psi\rangle\langle\psi|])=
D_{\leftrightarrow}(|\psi\rangle\langle\psi|)=
E_F(|\psi\rangle\langle\psi|)$ \cite{BennettMust,Multi3}.
Such a reasoning is not possible in the multi-partite domain.
There is no single ``unit'' of entanglement 
like the entanglement of the singlet in the bi-partite case
\cite{VidalModOpt,Wootters,Brun,MultiPuri}.
Accordingly, if more than two parties are present, a single
number indicating  the amount of entanglement of a pure state
is not sufficient: 
it has been shown that several inequivalent
``kinds of entanglement'' have to
be distinguished \cite{Duer,Classes2,Classes}.

The system which will be considered in this section
is a general $N$-partite quantum system
with parties $A_1,...,A_N$
holding quantum systems with dimension $d_1,...,d_N$,
that is, the state space
of the composite system is given by ${\cal S}({\cal H})$, 
where
\begin{equation}
{\cal H}={\cal H}_{A_1}\otimes ...\otimes {\cal H}_{A_N}
,\,\,\,\,\,\,\,\,\,\,
{\cal H}_{A_i}={\mathbbm{C}}^{d_i},
\,\,\,\,\,\,\,\,\,\,
i=1,...,N.
\end{equation}

\newfigure{5.6cm}{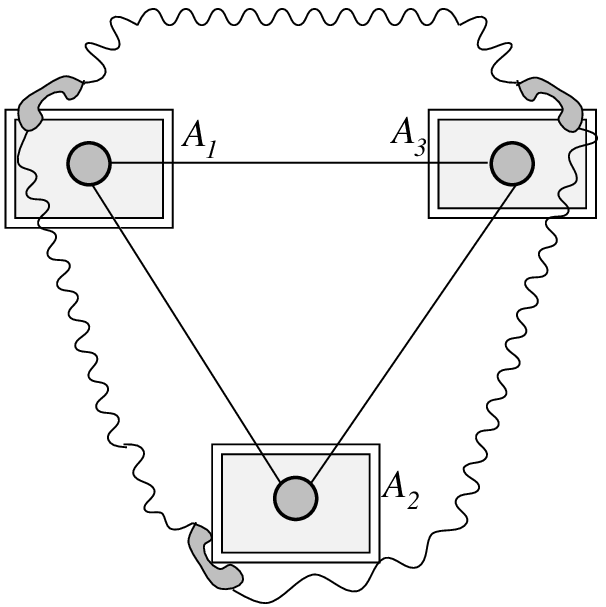}
{Figure 2.3: LOCC operations of three parties $A_1$, $A_2$, and $A_3$.}

\noindent 
If $N=2$, that is, 
for bi-partite systems,
\index{Schmidt decomposition} the entanglement
of pure states can be described by considering
the eigenvalues of the two
reduced states. Many of the particular
results about  entanglement manipulation 
in the bi-partite case can in some way be traced
back to the high symmetry of the Schmidt decomposition.
In the case of $N\geq 2$ a general Schmidt decomposition is 
not available, and the pure states which are Schmidt
decomposable form a small subset of all pure states.
Recently, considerable effort has been devoted
to a better understanding of equivalence classes
of pure states, where two states are
called equivalent if they can be transformed
into each other by local unitary operations
\cite{Multi1,Sudbery,Sudbery2,Acin,Grassl}.
Canonical forms of representants of these
{\it equivalence classes}\/ which may be 
conceived as generalizations 
of Schmidt decompositions to
multi-partite settings
can be found in
Refs.\ \cite{Sudbery}  and \cite{Acin}.

The structure of multi-particle  entanglement
cannot most appropriately be described by studying 
only the full composite system. Instead,
several properties are revealed only  when
one conceives the full system
as a composite system of several 
parts, taking into account that
each part may well include several 
quantum systems held by different parties. 
These parts are
taken to be systems on their own when 
examining entanglement properties.
The following investigations
involve arbitrary partitions
of the $N$-partite system
with parties $A_1,...,A_N$ 
into $k$ parts, $k=2,...,N$. In accordance with
Ref.\ \cite{Classes}
a division of the original system
into two parts will be called
a 2-split, and a division
into $k$ parts a \index{$k$-split} {\it $k$-split}\/. 
%
%
%
For a three-party system with parts
$A_1$, $A_2$, and $A_3$
the 3-split $A_1 A_2 A_3$ and the
three 2-splits $(A_1 A_2)A_3$, $(A_2 A_3)A_1$, and $(A_3 A_1)A_2$
are possible.

These splits reveal the \index{Separability structure} 
{\it separability structure}\/
of a state.
Let $\rho\in{\cal S}({\cal H}_{A_1}\otimes 
{\cal H}_{A_2}\otimes{\cal H}_{A_3})$
be an arbitrary state of such a tri-partite system. 
Following Ref.\ \cite{Classes2,Classes},
several classes of separable
states can be distinguished
for a tri-partite system.
\begin{eqnarray}
	\rho&=&\sum_i p_i 
	|\psi^{(i)}\rangle
	\langle \psi^{(i)}|_{A_1}
	\otimes 
	|\phi^{(i)}\rangle
	\langle \phi^{(i)}|_{A_2}
	\otimes
	 |\varphi^{(i)}\rangle
	\langle \varphi^{(i)}|_{A_3}	
	\label{FullySep}
	\\
	\rho&=&\sum_i p_i 
	|\psi^{(i)}\rangle
	\langle \psi^{(i)}|_{A_1}
	\otimes 
	|\phi^{(i)}\rangle
	\langle \phi^{(i)}|_{A_2 A_3}\label{S1}\\
	\rho&=&\sum_i p_i |\psi^{(i)}\rangle
	\langle \psi^{(i)}|_{A_2}
	\otimes 
	|\phi^{(i)}\rangle
	\langle \phi^{(i)}|_{A_1 A_3}\label{S2}\\
	\rho&=&\sum_i p_i |\psi^{(i)}\rangle
	\langle \psi^{(i)}|_{A_3}
	\otimes 
	|\phi^{(i)}\rangle
	\langle \phi^{(i)}|_{A_1 A_2},\label{S3}
\end{eqnarray}
where $p_1,p_2,...$ is a probability distribution.
If the state $\rho$
can be put into the form of Eq.\ (\ref{FullySep})
then it is called {\it fully separable}\/. For party
$A_1$ this state is \index{One-system bi-separable state}
{\it one-system  bi-separable}\/
if it is of the form of Eq.\ (\ref{S1}), but
not of the form of Eq.\ (\ref{S2}) and Eq.\ (\ref{S3}).
It is a {\it two-system bi-separable state} \index{Two-system bi-separable state} if
both Eq.\ (\ref{S1}) and Eq.\ (\ref{S2}) may
be satisfied, but not Eq.\ (\ref{S3}). 
{\it Three-system
bi-separable states}\/ \index{Three-system
bi-separable state} are states of the
form of Eq.\ (\ref{S1}), Eq.\ (\ref{S2}), and  Eq.\ (\ref{S3}),
but which fail to fulfil Eq.\ (\ref{FullySep}).
Finally, {\it fully inseparable states}\/ 
\index{Fully inseparable state} can be
formulated in any of these forms. Examples for each class
can be found in Ref.\ \cite{Classes}.
Such a terminology leads to a 
classification of states with respect to 
separability properties. 

\subsection{The Schmidt Measure \index{Schmidt measure}}

%
%
%
%

In Ref.\ \cite{Lattice} a certain class of 
multi-qubit states has been introduced, the
so-called 
$N$-party {\it cluster states}\/ 
$|\phi_N\rangle\langle\phi_N|$.
These states are of practical relevance in the
context of quantum computing:
It has turned out that such cluster states of a quantum 
Ising model may provide the ``carrier'' of a new type of quantum 
computing, as has been shown in Ref.\ \cite{Lattice2}. 
A quantum Ising system in a cluster state is a resource with 
which quantum computations can be performed by 1-qubit measurements 
only. Any computation can be 
realized by a proper sequence of
1-qubit measurements on the cluster. It is generally hoped
that this approach might amount to a new way of
lessening the difficulties 
in the realization of a large-scale universal quantum computer.
It has been demonstrated that
the minimal number 
of product terms for such a cluster state $|\phi_N\rangle\langle\phi_N|$
is given by $2^{\lfloor N/2 \rfloor}$, if
one expands $|\phi_N\rangle$ in
product state vectors of $N$ qubits. 
The observation concerning the
minimal number of product states
is related to the findings of
Ref.\ \cite{Duer}, in which
such numbers have first been
considered: it has been shown that
there are two classes of tripartite entangled pure 
states of three qubits which cannot be transformed
into each other with nonvanishing probability,
the so-called {\it W-state}\/ \cite{Duer}
and the  {\it GHZ-state}\/ \cite{Greenberger} 
being representatives. 
One has three and the other one has two product 
states in the minimal decomposition  
in terms of product states. 
This statement is also made stronger 
in that it is pointed out that this minimal number
of product terms can never be increased by means of
invertible local operations. Building upon these observations
one can define an entanglement
monotone on the entire state space, containing the mixed states,
of an arbitrary 
multi-partite system: 
\index{W-state}\index{GHZ-state}

Any $|\psi\rangle
\in{\cal H}= {\mathbbm{C}}^{d_1}\otimes ...\otimes {\mathbbm{C}}^{d_N}$ 
can be written in the form 
\begin{equation}\label{Dec}
	|\psi\rangle=\sum_{i=1}^R
	\alpha_i
	|\psi^{(i)}\rangle_{A_1} \otimes ...\otimes 
	|\phi^{(i)}\rangle_{A_N},
\end{equation}
where 
$\alpha_i\in{\mathbbm{C}}$, $i=1,...,R$,
with some $R\in{\mathbbm{N}}$. Let $r$ be the minimal 
number of product terms $R$ in such a decomposition
of $|\psi\rangle$.
The {\it Schmidt measure}\/ 
is then defined as
\begin{equation}
E_S(|\psi\rangle\langle\psi|)=\log_2 (r).
\end{equation} 
In the case of a bi-partite system with parties $A_1$ and $A_2$ 
the minimal number of product terms $r$ is 
given by the Schmidt rank of the state, which is in
turn identical to the rank of the local state
\index{Local state} of the
respective party. \index{Schmidt rank}

\index{Convex roof extension}
The definition of $E_S$ can be 
extended to the full state space
in a natural way.
This is done by using 
a convex roof construction \cite{Quant1,UhlmannRoof}
as in the entanglement of formation.
For a $\sigma\in {\cal S}({\cal H})$
let
\begin{equation}\label{Definition}
        E_S(\sigma)=\min \sum_i \lambda_i E_S(|\psi_i\rangle\langle\psi_i|),
\end{equation}
where the minimum is taken over all possible
convex combinations of the form 
\begin{equation}
	\sigma=\sum_i \lambda_i 
|\psi_i\rangle\langle\psi_i|
\end{equation} 
in terms of pure states
$|\psi_1\rangle\langle\psi_1|,|\psi_2\rangle\langle\psi_2|,...\,$,
with
$0\leq \lambda_i\leq 1$ for all $i$.

The Schmidt measure serves as
an entanglement
monotone \index{Entanglement
monotone} in the sense
of Ref.\ \cite{VidalModOpt} (see also Ref.\ \cite{Horobound})
and is hence a proper measure of mixed state entanglement.
In particular, it can neither increase on average
under
LOCC operations nor under mixing. 
It vanishes for \index{Fully separable state}
fully separable states, that is,
for states  $\sigma\in{\cal S}({\cal H})$ that
can be cast into the form 
	\begin{equation}
	\sigma=\sum_{i=1}^n p_i
	|\psi^{(i)}\rangle\langle \psi^{(i)}|_{A_1}\otimes
	... \otimes
	|\phi^{(i)}\rangle\langle \phi^{(i)}|_{A_N}
\end{equation}
where 
$p_1,...,p_n$ is a probability distribution.
In this multi-partite setting the Schmidt
measure 
is said to be an entanglement monotone
because the following three conditions are satisfied.
%
%
\begin{itemize}
\item[(i)]   $E_S\geq 0$, and   $E_S(\sigma)=0$ if $\sigma$
is {\it fully}\/ separable.
\item[(ii)]   $E_S$ is a convex functional.
\item[(iii)]   $E_S$ is monotone under local generalized
measurements: Let $\sigma$ be the initial state, and let
{\it one} of the parties $A_1,...,A_N$ perform a 
(partly selective)
local generalized measurement leading to the final 
states $\sigma_1,...,\sigma_K$ with respective probabilities
  $p_1,...,p_K$. Then 
\begin{equation}\label{iii}
        E_S(\sigma)\geq \sum_{i=1}^K p_i E_S(\sigma_i).
\end{equation}
\end{itemize}

\bigskip

\noindent 
{\bf Proposition 2.17. --} {\it The
Schmidt measure $E_S$ is an entanglement
monotone.} \\

\proof Condition (i) follows immediately 
from
the definition. Due to the convex roof construction
  $E_S$ is also a convex functional (condition (ii)): 
let $\sigma_1$ and 
$\sigma_2$ be states from ${\cal S}({\cal H})$, and let
$\sigma_1=\sum_j \mu_j |\phi_j\rangle\langle\phi_j|$
and $\sigma_2=\sum_j \eta_k |\varphi_k\rangle\langle\varphi_k|$
be the two decompositions
for which the respective minima
in Eq.\ (\ref{Definition}) are attained.
Then 
\begin{equation}
	\sum_j \lambda \mu_j |\phi_j\rangle\langle\phi_j|
	+ \sum_k (1-\lambda) 
	\eta_k |\varphi_k\rangle\langle\varphi_k|
\end{equation}
is a valid decomposition of $\sigma=
\lambda \sigma_1 + (1-\lambda) \sigma_2$, but it is 
not necessarily the optimal one. Hence,
  $E_S(\lambda \sigma_1 + (1-\lambda) \sigma_2)\leq
        \lambda E_S(\sigma_1)+(1-\lambda)E_S(\sigma_2)$.

The local measurement of condition (iii)
can be assumed to be performed by party $A_1$. 
Again, it suffices to consider the
posterior state in
a local generalized measurement
that does not involve mixing. That is,
it is sufficient to 
consider final states of the form
\begin{equation}
	\sigma_i=\frac{{E_{i}}\sigma {E_i}^\dagger}{p_i},
\end{equation}
where   
$
	p_i={\text{tr}}[{E_i}\sigma {E_i}^\dagger]$,
	$\sum_{i=1}^K E_i^\dagger  {E_i}={\mathbbm{1}}$.
The Kraus operators $E_1,...,E_K$ act in 	
${\cal H}_{A_2}\otimes ...\otimes {\cal H}_{A_N}$
as the identity.	
For any pure state  
$|\psi\rangle\langle\psi|\in{\cal S}({\cal H})$
\begin{equation}\label{Crucial}
        E_S\left(
		  \frac{{E_i}|\psi\rangle\langle\psi| {E_i}^\dagger}
        {{\text{tr}}[
        {E_i}|\psi\rangle\langle\psi| {E_i}^\dagger
        ]}\right)\leq E_S(|\psi\rangle\langle\psi|)
\end{equation}
for all $i=1,...,K$. This can be seen as follows.
Let $|\psi\rangle=\sum_{i=1}^r \alpha_i
|\psi^{(i)}\rangle_{A_1} \otimes ...\otimes 
|\psi^{(i)}\rangle_{A_N}$ be the decomposition of $|\psi\rangle$
into products as in Eq.\ (\ref{Dec}) 
with the minimal number of terms $r$. Then $E_i$ is
either invertible, and then $E_i |\psi\rangle$ has the same
minimal number of product terms $r'=r$, or it is not invertible, such 
that $r'\leq r$.
Moreover, 
if Eq. (\ref{Crucial}) holds for pure states $|\psi\rangle
\langle\psi|$,
it is also valid for arbitrary states $\sigma\in {\cal S}({\cal 
H})$: 

Let $\sigma=\sum_k \lambda_k |\psi_k\rangle\langle\psi_k|$ be the
optimal decomposition of $\sigma$ belonging to the minimum in Eq.\ 
(\ref{Definition}), then 
\begin{eqnarray}
      E_S(\sigma)&=&\sum_k \lambda_k E_S(|\psi_k\rangle\langle\psi_k|)
        \nonumber\\
        &\geq& \sum_k \lambda_k E_S \bigl(
        E_i|\psi_k\rangle\langle\psi_k|E_i^\dagger/
        {\text{tr}}[
        E_i|\psi_k\rangle\langle\psi_k|
	E_i^\dagger]\bigr)\nonumber\\
        &\geq& E_S\bigl(E_i\sigma E_i^\dagger/
        {\text{tr}}[
        E_i \sigma E_i]\bigr)
\end{eqnarray}
for all $i=1,...,K$.
The statement of condition (iii) then follows from the fact that
\begin{equation}
       \sum_{i=1}^K p_i E_S( E_i\sigma  E_i^\dagger/{\text{tr}}[
         E_{i}\sigma  E_{i}
        ])
        \leq \sum_{i=1}^K p_i E_S(\sigma)=E_S(\sigma).
\end{equation} 
\proofend

The Schmidt measure cannot be increased on average
under LOCC.
It can be used as a functional appropriately
quantifying the entanglement of a given state of a 
$N$-partite quantum system. 
A number of noteworthy properties are listed below.
The normalization and the additivity on pure states
follow immediately from the definition of the
Schmidt measure for pure states. For mixed states
$E_S$ is subadditive.\\

\noindent 
{\bf Lemma 2.18. --} {\it 
The Schmidt measure 
is normalized, 
\begin{equation}
	E_S(|\psi\rangle\langle\psi|)=1 
\end{equation}
for all $N$-party GHZ-states, i.e.,
states with state vectors of the form}
\begin{equation}
	|\psi\rangle=(|0^{\otimes N}\rangle+|1^{\otimes N}\rangle)/\sqrt{2}.
\end{equation}
\index{N-party GHZ state}

\bigskip

\noindent 
{\bf Lemma 2.19. --} {\it $E_S$ is
fully
additive on pure states:
If the parties $A_1,...,A_N$ share $n$ $N$-partite
quantum systems in the state $|\psi_1\rangle\langle\psi_1|$
and $m$ $N$-partite systems in the state $|\psi_2\rangle\langle\psi_2|$, 
then it follows that
  \begin{equation}
  E_S(|\psi_1\rangle\langle\psi_1|^{\otimes n}\otimes
|\psi_2\rangle\langle\psi_2|^{\otimes m})=n E_S(|\psi_1\rangle\langle\psi_1|)
+ m E_S(|\psi_2\rangle\langle\psi_2|).
\end{equation} }\\

\noindent 
{\bf Lemma 2.20. --} {\it 
Let
$\sigma\in {\cal S}({\cal H})$ be a fully 
separable state, then 
  $E_S( \sigma^{\otimes n}\otimes
|\psi\rangle\langle\psi|^{\otimes m})=
m E_S(|\psi\rangle\langle\psi|)$ holds
for all pure states $|\psi\rangle\langle\psi|\in{\cal S}({\cal H})$
and all $m=1,2,...\,$.}\\

\noindent 
{\bf Lemma 2.21. --} {\it $E_S$ is
subadditive,
          $E_S(\sigma\otimes \rho )\leq E_S(\sigma)+E_S(\rho)$ 
        for all $\sigma,\rho
       \in{\cal S}({\cal H})$, where $\sigma$ and $\rho$
        are states of quantum systems
	held by the same parties. }\\
	
\proof Let $\sigma=\sum_j \mu_j |\phi_j\rangle\langle\phi_j|$
and $\rho=\sum_k \eta_k |\varphi_k\rangle\langle\varphi_k|$
be the optimal decompositions of $\sigma$ and $\rho$ respectively
for which the minima in Eq.\ (\ref{Definition}) are attained.
Then 
\begin{equation}
\sum_{j,k}  
{\eta_k \mu_j}  
{|\varphi_k\rangle\langle\varphi_k|\otimes
|\phi_j\rangle\langle\phi_j|}
\end{equation} 
is a decomposition of 
$\sigma\otimes\rho$, but not necessarily the
one for which Eq. (\ref{Definition}) becomes minimal.
\proofend

It should be noted that 
out of the conditions 
for an entanglement measure
in the strict sense of Ref.\ \cite{Horobound}
        $E_S$ 
        does not satisfy a continuity
        \index{Continuity} criterion. In particular, it
		  is not weakly continuous in the sense of Eq.\ (\ref{weaklycont}).
This is why
the uniqueness theorem for 
entanglement measures
\index{Uniqueness theorem for entanglement measures} 
of pure states (see Ref.\ \cite{Horobound})
cannot be applied. Hence,
	$E_S$ does not
	have to coincide with the von-Neumann
	entropy of
	one subsystem $S({\text{tr}_A[|\psi\rangle\langle\psi|]})$
	for a bi-partite system in a pure state $|\psi\rangle\langle\psi|$,
	and indeed, it does not coincide in general.

\subsection{Classification of Multi-Particle Entanglement}

Although the
Schmidt measure of a mixed state 
is defined via a minimization over all possible
realizations of the state,
it can be calculated exactly for a rather 
large class of states. This is mainly due to 
the fact that it is a coarse grained measure.
All terms that appear in Eq.\ (\ref{Definition})
are logarithms of 
natural numbers weighted with respective
probabilities. This quality of the Schmidt measure 
is both the strength and
the weakness of this quantity.
It allows, however, for a
detailed classification of multi-particle
entangled states.  The subsequent investigations
will be restricted to the multi-qubit case, where
${\cal H}=({{\mathbbm{C}}^2})^{\otimes N}$.
In this classification the 
Schmidt measure will also be furnished with 
an index indicating the respective split.
For example, in a system consisting of
parts $A_1$, $A_2$, and $A_3$
the Schmidt measure 
associated with the 3-split $A_1 A_2 A_3$
is written as 
  $E_S^{A_1 A_2 A_3}$, and the one belonging to
$(A_1 A_2) A_3$ is denoted by $E_S^{(A_1 A_2) A_3}$ (see Fig.\ 2.4).

For each $k$-split the Schmidt measure cannot increase 
on average in the course of
a LOCC operation, as the Schmidt measure
is an entanglement monotone with respect to every possible
split of the system. This means in turn that once
one knows that the Schmidt measure would become larger 
on average, one can be sure that the corresponding 
transformation
cannot be implemented under LOCC.\\

\newfigure{9cm}{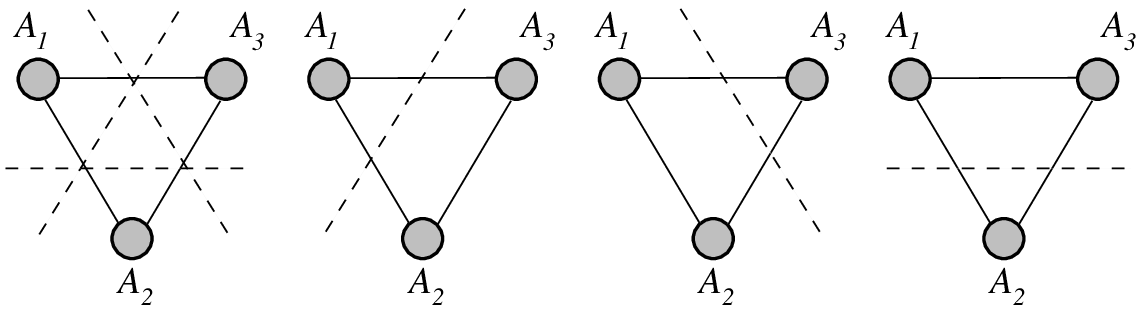}
{Figure 2.4: The 3-split $A_1 A_2 A_3$ and
the splits $(A_2 A_3)A_1$, $(A_1 A_2)A_3$,  and $(A_3 A_1)A_2$.}
 
 \index{W-state}\index{GHZ-state}
 
 \noindent
 {\bf Example 2.22. --}
 Consider 
a three party {\it W-state}\/ 
with
\begin{equation}
|\text{W}\rangle=
(|100\rangle+|010\rangle+|001\rangle)/\sqrt{3}
\end{equation}
that has been used in Ref.\  \cite{Duer}. 
$E_S^{A_1 A_2 A_3}(|\text{W}\rangle\langle \text{W}|)=\log_2 (3)$, 
while the three party {\it GHZ-state}\/
\begin{equation}
|\text{GHZ}\rangle=(|000\rangle+|111\rangle)/\sqrt{2}
\end{equation}
obtains 
the value   $E_S^{A_1 A_2 A_3}(|\text{GHZ}\rangle\langle 
\text{GHZ}|)=1$. 
Hence, in the 3-split
  $E_S$ discriminates between 
the GHZ-state, the
W-state and product states (value 0).
The 3-tangle, proposed in Ref.\ \cite{Wootters},
   is also an entanglement monotone (see Ref.\ \cite{Duer}),
    and it can distinguish the W from the GHZ
    state. However, it is defined only for
    three qubits, and it gives the same value for
    the W-state and for product states.\\
  
Other splits with $k=2$ reveal further information and give rise to
the full classification.\\

\noindent 
{\bf Example 2.23. --}
A two-party Bell state with state vector
 $|\phi^{+}\rangle_{A_1 A_2}=(|00\rangle_{A_1 A_2}+|11\rangle_{A_1 A_2})/\sqrt{2}$
held by $A_1$ and
$A_2$  with party $A_3$ in the state
$|0\rangle\langle 0|_{A_3}$
cannot be transformed into $|\text{GHZ}\rangle\langle 
\text{GHZ}|$ of $A_1$, $A_2$, and $A_3$
under local operations and classical communication. This
has not yet become obvious 
from the values of the Schmidt measure
corresponding to the split $A_1 A_2 A_3$, as
both states yield
\begin{equation}
E_S^{A_1 A_2 A_3}(|\phi^{+}\rangle
\langle \phi^{+}|_{A_1 A_2}
\otimes|0\rangle
\langle0|_{A_3})=
E_S^{A_1 A_2 A_3}(|\text{GHZ}\rangle\langle 
\text{GHZ}|)=1. 
\end{equation}
However, 
$
E_S^{(A_1 A_2) A_3}(|\phi^{+}\rangle\langle \phi^{+}|_{A_1 A_2}
\otimes|0\rangle\langle 0|_{A_3})=0$, and
\begin{equation}
E_S^{(A_1 A_2) A_3}(|\text{GHZ}\rangle\langle \text{GHZ}|)=1.
\end{equation} 
And it is for this reason that 
the split $(A_1 A_2) A_3$ indicates that
\begin{equation}
|\phi^{+}\rangle\langle \phi^{+}|_{A_1 A_2}
\otimes|0\rangle\langle 0|_{A_3}\not\longrightarrow
|\text{GHZ}\rangle\langle 
\text{GHZ}|
\end{equation}
under LOCC.\\

Further examples can be found in Table \ref{Table1}. It shows
the values of the Schmidt measure with respect to
all possible splits for some pure states of a
four-partite system.
\begin{table}
\caption{{\small Values of the Schmidt measure $E_S$ for 
some four qubit pure states
(the
four party GHZ-state
$(|0000\rangle+|1111\rangle)/\sqrt{2}$, the
generalized W-state [9]
$(|0001\rangle +|0010\rangle +|0100\rangle+|1000\rangle
)/2$, the cluster state \cite{Lattice,Lattice2} 
$|\phi_4\rangle=
(|0000\rangle+|0011\rangle+|1100\rangle-|1111\rangle)/2$,
and the state
$|\phi^+\rangle |\phi^+\rangle=
(|00\rangle+|11\rangle)\otimes (|00\rangle+|11\rangle)/2$).}}

\begin{center}
\begin{tabular}{|c||c|c|c|c|}
\hline\hline 
 & GHZ & W &  $|\phi_4\rangle$ & $|\phi^+\rangle |\phi^+\rangle$ \\
\hline
 & & & & \\
$A_1 A_2 A_3 A_4$     & 1 & 2 & $2$ & $2$\\
$A_1 A_2 (A_3 A_4)$   & 1 & $\log_2 3$ & $1$ & 1\\
$(A_1 A_2) A_3 A_4$   & 1 & $\log_2 3$ & $1$ & 1\\
$(A_1 A_2) (A_3 A_4)$ & 1 & $1$ & $1$ & 0\\
$(A_1 A_3) (A_2 A_4)$ & 1 & $1$ & $2$ & $2$\\
$(A_1 A_4) (A_2 A_3)$ & 1 & $1$ & $2$ & $2$\\
$(A_1 A_2 A_3) A_4$   & 1 & 1 & 1 & 1\\
 & & & &  \\
\hline\hline 
\end{tabular}
\end{center}
\label{Table1}
\end{table}
%
In order to calculate the Schmidt measure of a mixed state
a minimization over all possible
decompositions of the state
is required. The convex roof construction 
provides 
upper bounds of $E_S$: If
$\sigma=\sum\eta_i |\psi_i\rangle\langle\psi_i|$ is
any  not necessarily optimal decomposition of
a state $\sigma \in{\cal S}({\cal H})$, then 
$\sum_i\eta_i E_S(|\psi_i\rangle\langle\psi_i|)$ is
an upper bound of   $E_S(\sigma)$.
For many states   $E_S$ can however be fully
evaluated. \\

\noindent
{\bf Example 2.24. --}
Consider two parties $A_1$ and $A_2$
sharing two qubits in the 
\index{Werner state}
Werner state \cite{Wer89}
\begin{equation}
        \rho_W(\lambda)
        =\lambda |\psi^-\rangle \langle \psi^-|+(1-\lambda) 
        \mathbbm{1}/4,
\end{equation}
with $|\psi^-\rangle=
(|01\rangle-|10\rangle)/\sqrt{2}$, $0\leq \lambda\leq 1$.
As all pure states in the range of $\rho_W(\lambda)$ 
have Schmidt measure 0 or 1, one has to identify in any
decomposition $\rho_W(\lambda)=
\sum_i
\eta_i |\psi_i\rangle\langle\psi_i|$
the terms with Schmidt measure 0 (product states) or 1
(entangled states). Hence, the 
Schmidt measure is given by   $E_S(\rho_W(\lambda))=1-s$,
where $s$ is the weight of the separable state that
can maximally be subtracted from $\rho_W(\lambda)$
while maintaining the semi-positivity of the state.
As shown in  \cite{Lewenstein} it follows that
\begin{equation}
        E_S(|\psi\rangle\langle\psi|)=\left\{
        \begin{array}{ll}
        \frac{3}{2}\lambda -\frac{1}{2}, & {\text{
        for $1/3<\lambda\leq 1$,}}\\
        0,& {\text{for $0\leq \lambda\leq 1/3$.}}\\
        \end{array}
        \right.
\end{equation}
In other words, $E_S$ is given by
the weight of the inseparable state in the best
separable approximation in the sense of Ref.\ \cite{Lewenstein}.
The Schmidt rank of all states in the range of any state from 
${\cal S}({\mathbbm{C}}^2\otimes {\mathbbm{C}}^2)$ is smaller
or equal to 2, and hence,
this statement holds for all mixed states of systems consisting 
of two qubits. \\

\noindent
{\bf Corollary 2.25. --} {\it Let 
${\cal H}={\mathbbm{C}}^2\otimes {\mathbbm{C}}^2$, and
let $\lambda(\sigma)$ be the
weight of the
separable contribution of
the best separable approximation $\sigma_s$ in
\index{Best separable approximation}
$\sigma=\lambda(\sigma)\sigma_s+(1-\lambda(\sigma))\delta \sigma$ of
a state $\sigma\in{\cal S(\cal H)}$. Then
$\lambda$ is
an entanglement monotone.}\\

For more than two parties different entanglement
classes can be distinguished. 
The Schmidt measure is
again defined for all possible $k$-splits, $k=2,...,N$.
If the $N$-partite system is separable with respect
to a particular $k$-split, the value of the
corresponding Schmidt measure is 0:
$E_S(\sigma)=0$ if and only if
$\sigma$ is fully separable. Hence, the Schmidt measure
with respect to a certain split gives an account of 
the separability of the state.

For three systems, e.g., the classes of 
one-system bi-separable states,
\index{One-system bi-separable state}
two-system  bi-separable states,
\index{Two-system bi-separable state}
three-system  bi-separable states,
\index{Three-system bi-separable state} 
and
fully separable states
\index{Fully separable state} (see above) 
can be distinguished. For a 
state $\sigma$ which is taken to be a 
one-qubit bi-separable state with respect to
party $A_1$, $E_S$ takes the values
\begin{equation}
E_S^{A_1 (A_2 A_3)}(\sigma)=0, \text{ but }
E_S^{A_3 (A_1 A_2)}(\sigma)>0\text{  and }E_S^{A_2 (A_3 A_1)}(\sigma)>0.
\end{equation}
The Schmidt measure, however, can reveal 
more structure, since the entanglement is 
also quantified. \\

\noindent 
\begin{table}
\caption{{\small The Schmidt measure   $E_S$
for some mixed quantum states
($\rho_{\text{G}}(\lambda)$,
$\rho_M$, and 
$\rho(\lambda,\mu)$).
}}
\begin{center}
\begin{tabular}{|c||c|c|c|}
\hline\hline
 & $\rho_{\text{G}}(\lambda)$ & $\rho_M$ & $\rho(\lambda,\mu)$ \\
\hline
& & & \\
$A_1 A_2 A_3$      & $\lambda$ & 1 & 1 \\
$(A_1 A_2 ) A_3 $  & $\lambda$  & $2/3$ & $1-\lambda$ \\
$(A_1 A_3) A_2 $   & $\lambda$  & $2/3$ & $\lambda+\mu$ \\
$(A_2 A_3) A_1 $   & $\lambda$  & $2/3$ & $1-\mu$ \\
& & & \\
\hline\hline
\end{tabular}
\end{center}
\label{Table2}
\end{table}

\noindent
{\bf Example 2.26. --}
By way of an example, let
\begin{eqnarray}
        \rho(\lambda,\mu)&=&
        \lambda
        |\phi^+\rangle
        \langle \phi^+|_{A_1 A_2 }
	\otimes |0\rangle \langle 0|_{A_3}+
        \mu|\phi^+\rangle
        \langle \phi^+|_{A_2 A_3 }
	\otimes |0\rangle\langle 0|_{A_1}\nonumber \\
        &+&
        (1-\lambda-\mu)
        |\phi^+\rangle
        \langle \phi^+|_{A_3 A_1 }\otimes |0\rangle
	 \langle 0|_{A_2},
\end{eqnarray}
$0\leq \lambda,\mu\leq1$. For $\lambda=\mu=1/3$ this state
reduces to the three-party \index{Molecule state} molecule state
\begin{eqnarray}
\rho_M&=&\frac{1}{3}\bigl(
	|\phi^+\rangle
        \langle \phi^+|_{A_1 A_2 }
	\otimes |0\rangle \langle 0|_{A_3}+
        |\phi^+\rangle
        \langle \phi^+|_{A_2 A_3}\otimes 
	|0\rangle \langle 0|_{A_1}\nonumber \\
        &+&
        |\phi^+\rangle
        \langle \phi^+|_{A_3 A_1}
	\otimes |0\rangle \langle 0|_{A_2}
	\bigr)
\end{eqnarray}
studied in Ref.\
\cite{Molecules}. The Schmidt measure
  $E_S^{A_1 A_2 A_3}(\rho_M)=1$
is equal to the Schmidt measure 
of a state where $A_1$ and $A_2$
hold a $|\phi^+\rangle\langle\phi^+|$ 
state and $A_3$ is in the state $|0\rangle\langle0|$,
as the mere classical ignorance of which parties 
are actually holding 
the Bell state cannot increase the amount of entanglement.
In Table \ref{Table2} the values of the Schmidt measures
of all splits of the states $\rho(\lambda,\mu)$,
$\rho_M$, and
\begin{eqnarray}
        \rho_{\text{G}}(\lambda)=\lambda
        |\text{GHZ}\rangle\langle \text{GHZ}|
        +(1-\lambda) |000\rangle \langle 000|,
\end{eqnarray}
$0\leq \lambda\leq1$, are shown.\\

\subsection{Remarks on the Asymptotic Limit}
\index{Asymptotic limit}

The Schmidt measure is the basis for a rather detailed
classification of mixed state entanglement and it 
is hoped to be
giving useful information about the possibility to
transform one state into another using LOCC operations.
However, its relation to a number of other concepts
of multi-particle entanglement is not yet fully investigated. 
For cluster states \cite{Lattice}
\index{Cluster state}
the Schmidt measure coincides
with the {\it persistency of entanglement}\/ \cite{Lattice},
and it would be interesting to establish 
the exact connection between the definition of the
Schmidt measure and a -- possibly
\index{Persistency of entanglement}
refined -- definition of persistency. 

Another  potentially rewarding 
direction for further research is
the notorious problem of transformations of 
multi-partite
pure state entanglement in the asymptotic limit.  
Much is known about {\it stochastic transformations}\/
of single copies of three-qubit systems \cite{Duer}.
However, the question is not resolved whether
every pure state of three qubits can be prepared
in an asymptotically reversible way starting from copies of states
taken from a finite set of pure states.
To shed light on this issue, entanglement monotones
like the Schmidt measure would be desirable, which are
-- unlike $E_S$ -- continuous functionals. 

Before being able to state this problem in 
a more formal way
\cite{Multi3,MREGS,PlenioMREGS},
the concept of asymptotical reducibility \cite{Multi3}
needs to be introduced. 
A pure state
$|\psi\rangle\langle\psi|$ is said to be
{\it asymptotically reducible}\/
to $|\phi\rangle\langle\phi|$ if the following condition holds:
For all $\delta >0$ and all $\varepsilon>0$
there exist $n,m\in\mathbbm{N}$ and
an LOCC operation
${\cal E}$ taking inputs from ${\cal S}({\cal H}^{\otimes m})$
and mapping them on ${\cal S}
({\cal H}^{\otimes n})$
with the property
\begin{equation}
	\left|\frac{n}{m}-1\right|<\delta\,\,\,\,\,\text{ and }
	\,\,\,\,\,
	F({\cal E}(|\psi\rangle\langle\psi|^{\otimes m}),
	|\phi\rangle\langle\phi|^{\otimes n})
	\geq1-\varepsilon.
\end{equation}
The statement that $|\psi\rangle\langle\psi|$ 
\index{Asymptotically reducible}
is asymptotically reducible to
$|\phi\rangle\langle\phi|$ is abbreviated as
\begin{equation}
|\psi\rangle\langle\psi|\longrightarrow|\phi\rangle\langle\phi|\,\,\,\,
\text{ under ALOCC}.
\end{equation}
\index{ALOCC operation}
The yield in asymptotic reducibilities can also be 
non-integer. One writes in short
\begin{equation}
|\psi\rangle\langle\psi|^{\otimes x}\longrightarrow|\phi\rangle\langle\phi|^{\otimes y}
\,\,\,\,\text{ under ALOCC}
\end{equation}
with $x,y\geq 0$ if
for all $\delta >0$ and all $\varepsilon>0$
there exist $n,m\in\mathbbm{N}$ and
an LOCC operation
${\cal E}$ with the property
\begin{equation}
	\left|\frac{n}{m}-\frac{x}{y}\right|<\delta\,\,\,\,\,\text{ and }
	\,\,\,\,\,
	F({\cal E}(|\psi\rangle\langle\psi|^{\otimes m}),
	|\phi\rangle\langle\phi|^{\otimes n})
	\geq1-\varepsilon.
\end{equation}

With this concept at hand,  a 
\index{Reversible entanglement generating set}
{\it reversible entanglement
generating set}\/ (REGS) can be defined precisely:
Let ${\cal H}$ be the Hilbert space of a 
multi-partite quantum system. A REGS is a set
$\{|\psi_1\rangle ,...,|\psi_n\rangle\}$, $|\psi_i\rangle
\in{\cal H}$, $i=1,...,n$,
with the property that there are coefficients $x_i\geq 0$ such that
\begin{eqnarray}
	\bigotimes_{i=1}^n |\psi_i\rangle\langle\psi_i|^{\otimes x_i}
		&\longrightarrow& |\psi\rangle\langle\psi|
		\,\,\,\,\text{ under ALOCC and }\\
		|\psi\rangle\langle\psi|&
		\longrightarrow&
		 \bigotimes_{i=1}^n |\psi_i\rangle\langle\psi_i|^{\otimes x_i}
		\,\,\,\,\text{ under ALOCC}.\nonumber
\end{eqnarray}
An {\it MREGS}\/ 
\cite{Multi3,MREGS,PlenioMREGS}
is a minimal REGS in the sense that it is a 
reversible entanglement
\index{MREGS}
generating set with the minimal number of elements.

As is well known, the set $\{ |\phi^+\rangle \}$ is an 
MREGS for all systems
with ${\cal H}=\mathbbm{C}^2\otimes \mathbbm{C}^2$, that is, for
bi-partite qubit systems. Every state of a two-qubit
system can be prepared in an asymptotically reversible way from 
Bell states. Unfortunately,
it is not known yet what the MREGS is for 
systems consisting of three qubits $A_1$, $A_2$, and
$A_3$?
There is a strong indication that 
$
\{|\phi^+\rangle_{A_1 A_2}, 
|\phi^+\rangle_{A_2 A_3},
|\phi^+\rangle_{A_1 A_3},
|\text{GHZ}\rangle\}
$
is not sufficient 
(see Refs.\ \cite{PlenioMREGS} and \cite{PlenioPriv}). In order
to see whether particular states can be prepared in a reversible way
from certain ingredients in the asymptotic limit, it would be very
useful to dispose of continuous entanglement monotones. 
The Schmidt measure
alone is not appropriate to tackle this problem.

It can however be used to address the question of
the minimal amount of
resources needed to prepare a particular
multi-partite pure state. To be more specific, 
one may ask
how many Bell states are needed on average asymptotically
to generate a certain final three qubit state.
This procedure is typically expected to be irreversible. 
Three qubit GHZ-states can by no means 
be prepared irreversibly
from two qubit Bell states, not even in the asymptotic limit \cite{Multi3}. 
Nevertheless, 
if one allows the protocol to be irreversible 
all pure states of three qubits can be achieved by using only
Bell states 
as a resource. The Schmidt measure provides a
lower bound for this minimal average number of
Bell states. The subsequent proposition is concerned with
tri-partite systems, but an analogous statement also holds
for arbitrary $N$-party systems.\\
\index{GHZ-state}

\noindent
{\bf Proposition 2.27. --} {\it Consider a tripartite system
with parts $A_1$, $A_2$, and $A_3$, with
${\cal H}=\mathbbm{C}^2 \otimes \mathbbm{C}^2 \otimes
\mathbbm{C}^2$.
Let $|\phi\rangle
\in{\cal H}$. There exist numbers $x_1,x_2,x_3\geq 0$ such that
\begin{equation}
	|\phi^+\rangle\langle\phi^+|_{A_1 A_2}^{\otimes x_1}
	\otimes
	|\phi^+\rangle\langle\phi^+|_{A_1 A_3}^{\otimes x_2}
	\otimes
	|\phi^+\rangle\langle\phi^+|_{A_2 A_3}^{\otimes x_3}
	\longrightarrow |\phi\rangle\langle\phi|\,\,\,\,
	\text{ under ALOCC}
\end{equation}
and such that the sum $E_X(|\phi\rangle\langle\phi|)=x_1+x_2+x_3$ attains its infimum. Then
\begin{equation}
	E_X(|\phi\rangle\langle\phi|)\geq E_S^{A_1 A_2 A_3 }(|\phi\rangle\langle\phi|).
\end{equation}
}
\probreak

\proof This statement is a consequence of 
the fact that $E_S^{A_1 A_2 A_3}$
is an entanglement monotone and of its additivity property with
respect to pure states. For all $\varepsilon>0$ and all $\delta>0$
there exist $n,n_1,n_2,n_3\in{\mathbbm N}$ and an LOCC operation
${\cal E}$ 
such that
$|n_i/n-x_i|<\delta$ for $i=1,2,3$ and
\begin{equation}
	F( {\cal E}( |\phi^+\rangle\langle\phi^+|_{A_1 A_2}^{\otimes n_1}
	\otimes
	|\phi^+\rangle\langle\phi^+|_{A_1 A_3}^{\otimes n_2}
	\otimes
	|\phi^+\rangle\langle\phi^+|_{A_2 A_3}^{\otimes n_3}),
	|\phi\rangle\langle\phi|^{\otimes n})\geq 1-\varepsilon.
\end{equation}
The statement of the proposition follows from the fact that
\begin{equation}
n_1+n_2+n_3 \geq E_S^{A_1 A_2 A_3 } (|\phi\rangle\langle\phi|^{\otimes n})
=n E_S^{A_1 A_2 A_3 } (|\phi\rangle\langle\phi| ).
\end{equation}
\proofend

\noindent
{\bf Example 2.28. --} How many Bell states 
are needed to prepare copies of 
W-states in the asymptotic limit? A lower bound can be deduced from Proposition 
2.27,
but it seems  hard to tell whether this bound can be
achieved. To approach this problem one can at least 
produce copies of W-states in two steps as shown below.\\
\index{W-state}

\jbox{
\begin{enumerate}
\item Two parties prepare copies of 
\begin{equation}
	\frac{1}{\sqrt{3}}
	\biggl(
		|0\rangle_{A_1}
		|0\rangle_{A_2}
		|1\rangle_{A_2'}+
		|0\rangle_{A_1}
		|1\rangle_{A_2}
		|0\rangle_{A_2'}+
		|1\rangle_{A_1}
		|0\rangle_{A_2}
		|0\rangle_{A_2'}
	\biggr),
\end{equation}
where $A_2$ and $A_2'$ are held by the same party.

\item The state of $A_2'$ is teleported to $A_3$.

\item The total amount of entanglement used  per copy
of the final state is given by
	$1/3+\log_2(3)=1.918$.
\end{enumerate}
}

\medskip

\noindent While this procedure is far from being optimal, it is possible to 
state
that in this example
\begin{equation}
	1.585=\log_2 (3)=E_S^{A_1 A_2 A_3} (|W\rangle\langle W|)
	\leq E_X(|W\rangle\langle W|)\leq 1/3+\log_2(3)=1.918.
\end{equation}

\bigskip

This example concludes the considerations of the 
quantification
of quantum entanglement. To summarize,
several good measures of entanglement have been proposed,
and their properties have been studied. After all, 
the degree of entanglement  gives information about
the possibility to manipulate the resource entanglement.
Entanglement cannot be created on average with
local operations and classical communication. However, it
is in general not sufficient to know the degree of entanglement
present in the initial and in the final situation to find
out whether a certain state transformation can -- in principle --
be implemented. For single copies of quantum systems more information
is needed. This points towards the topic of the next chapter,
in which criteria for the transformation of quantum states
will be presented.


\chapter{Entanglement Transformations}
\chapterbreak

\section{Introduction}

Given a machine that is designed in such a way that
it can implement any LOCC operation on 
a bi-partite quantum system, what tasks may
this machine accomplish? 
If one inserts a {\it single}\/ 
copy of a bi-partite quantum system 
in an entangled state into this machine,
one can be sure that no output state with a larger
amount of entanglement can be produced with probability one: 
Local operations and classical communication
alone cannot increase the amount of entanglement in a 
such a deterministic
transformation, as quantified by any entanglement
monotone. But what particular states {\it can}\/ be prepared
with LOCC operations starting from a given state?

The theory of entanglement transformations aims at 
answering questions of this type. Essentially, 
the issue is what state transformations can be
done with a single copy of a bi-partite quantum 
system using LOCC operations. 
On the one hand  this analysis aims
at a classification of states via the
class of LOCC operations, on the other hand
it provides answers to practical questions
concerning the 
manipulation of the resource entanglement.

\newfigure{12.2cm}{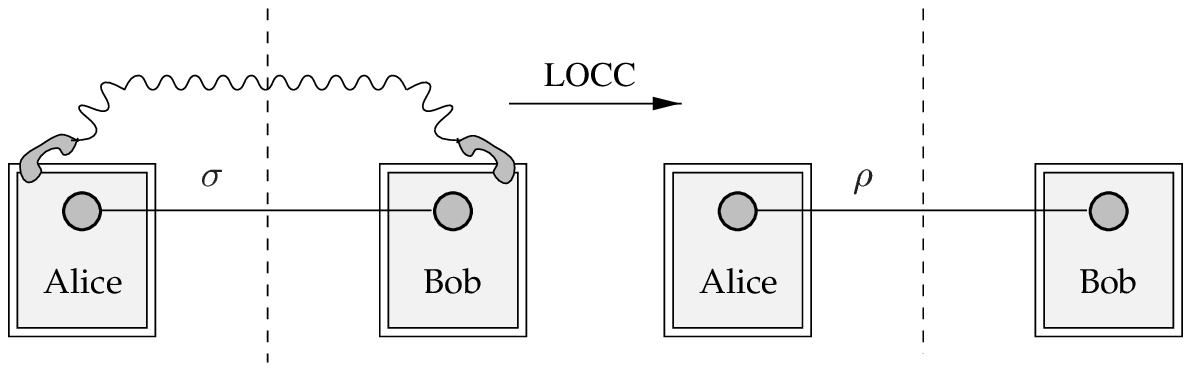}
{Figure 3.1: 
A schematic representation of a transformation of an initial state
$\sigma$ into the state $\rho$ with the use of 
local quantum operations and classical communication (LOCC).}



More formally, 
 let $\sigma$ and $\rho$ be states taken from
the state space ${\cal S}({\cal H})$,
where ${\cal H}={\cal H}_A\otimes {\cal H}_B$ is the 
Hilbert space associated with a bipartite quantum system
consisting of parts $A$ and $B$. 
If there exists a trace-preserving 
LOCC operation ${\cal E}$ 
mapping $\sigma$ on ${\cal E}(\sigma)=\rho$,
one writes \cite{Nielsen}\index{Majorization}
\begin{equation}
	\sigma\longrightarrow\rho\,\,\,{\text{\it under LOCC}}.
\end{equation} 
The statement that such a transformation is not
possible within this class of operations is abbreviated
as 
\begin{equation}
	\sigma\not\longrightarrow\rho\,\,\,{\text{\it under LOCC}}.
\end{equation}
States $\rho$ and $\sigma$ are called
{\it incommensurate}\/ if both $\sigma\not\longrightarrow\rho$
and $\rho\not\longrightarrow\sigma$ under LOCC. 


The question whether an initial state $\sigma$ can be
transformed into a particular final state $\rho$ 
is different from the question of asymptotic reducibility.
Under LOCC operations,
it may well be possible to transform a large
number of copies of $\sigma$ into a large number of 
copies of $\rho$ in a reversible way, 
while with a single copy at hand
the corresponding transformation $\sigma\longrightarrow \rho$ under
LOCC is not possible.

\index{Probabilistic transformation} 
Much work has been done on entanglement
transformations of pure states.
In Ref.\ \cite{Nielsen} a necessary and sufficient 
criterion for entanglement transformation has been 
presented. Ref.\ \cite{Vidal} concentrates on 
{\it probabilistic
transformations}\/, and the maximal
probability is calculated of the successful 
transformation from an initial
pure entangled state into another final pure state. 
In Ref.\ \cite{Jonathan} and \cite{Hardy}
these results are generalized to the case
where a pure state is mapped on an ensemble of pure states.
{\it Approximate transformations}\/ are considered in 
Refs.\  \cite{VidalJonathanNielsen} and \cite{Barnum}. 
\index{Approximate transformation}
In aggregate, pure-state entanglement 
transformations are rather well-understood.

The structure of the chapter is as follows:
First of all, two important
results \cite{Nielsen,Vidal,Jonathan}
for the transformation of pure states will be mentioned.
This chapter however deals mainly with 
transformations of mixed states,
about which much less is known. In the subsequent section
transformation criteria for mixed states will be derived.
The main emphasis of the chapter is on a new class
of entanglement transformations proposed in Ref.\ \cite{JonPle}
and developed for mixed states in Ref.\ [E1]:
the class of so-called {\it entanglement-assisted local quantum operations}\/.
Most of the results presented in this chapter
have been published in 
Ref.\ [E1].

\section{Entanglement Manipulation for Pure States}

The theorem proved in Ref.\ \cite{Nielsen}
provides a necessary and sufficient condition
for entanglement transformations.
It links the problem of manipulating quantum systems
in entangled states to the theory of {\it majorization}\/,
a topic from linear algebra and matrix analysis.
The criterion itself is given by a
{\it majorization relation}\/. In this chapter 
it will be
assumed that 
the dimensions 
$N=\dim[{\cal H}_A]=\dim[{\cal H}_B]$
of the Hilbert spaces ${\cal H}_A$ and
${\cal H}_B$ of the bi-partite system with Hilbert space
${\cal H}={\cal H}_A\otimes{\cal H}_B$ are identical.\\

\noindent{\bf Theorem (Nielsen). --} {\it 
Let ${\cal H}={\cal H}_A\otimes {\cal H}_B$ and 
let $\sigma, \rho\in{\cal S}({\cal H})$ be pure states.
Then $\sigma\longrightarrow\rho$ under
LOCC if and only if
\begin{equation}
	\sum_{i=1}^k
	\alpha_i\leq \sum_{i=1}^k \beta_i\,\,\,
	{\text{ for all }}
	k=1,...,N,\label{maj}
\end{equation}
where $\alpha_1, ..., \alpha_N$
and $\beta_1,...,\beta_N$ with
$\alpha_1\geq  ... \geq \alpha_N$ and 
$\beta_1\geq  ... \geq \beta_N$ are
the eigenvalues of ${\text{tr}}_A [\sigma]$
and ${\text{tr}}_A [\rho]$, respectively.}\footnote{It should
be noted that if an entanglement transformation is
possible, it suffices to use one-local operations
instead of the full class of LOCC operations.
This fact has already been demonstrated in Ref.\ \cite{LoBeyond}.}\\

\probreak

A list of the latter type \index{Ordered list}
will also be  referred to as {\it ordered list}\/.
The content of the $N$ conditions 
on the eigenvalues of ${\text{tr}}_A [\sigma]$ and ${\text{tr}}_A [\rho]$
in Eq.\ (\ref{maj}) \index{Majorization}
is typically abbreviated as 
\begin{equation}\label{MajEq}
	{\text{tr}}_A [\sigma]\prec {\text{tr}}_A [\rho].
\end{equation}
The symbol $\prec$ in Eq.\ (\ref{MajEq})
is the {\it (operator) majorization relation}\/ 
\cite{Majo,Uhlmann,Bhatia}
and means that ``${\text{tr}}_A [\rho]$
is {\it more mixed than}\/  ${\text{tr}}_A [\sigma]$''.
Such a  majorization relation is defined for all Hermitian
matrices. However, it does not lead directly 
to a criterion for entanglement transformations for mixed states.
For states the $N$-th condition given by
Eq.\ (\ref{maj}) is trivially satisfied due to
the normalization constraint.\\

\noindent{\bf Example 3.1. --} Take ${\cal H}={\cal H}_A\otimes
{\cal H}_B$,
${\cal H}_A= 
{\cal H}_B={\mathbbm{C}}^3$, and label the basis
vectors 
of ${\cal H}_A$ as $|1\rangle,|2\rangle,|3\rangle$.
Consider the states $|\psi\rangle\langle\psi|$ and 
$|\phi\rangle\langle\phi|$ with
\begin{eqnarray}
	|\psi\rangle&=& \sqrt{0.6}|11\rangle+\sqrt{0.3}|22\rangle+
	\sqrt{0.1}|33\rangle
	,\\
	|\phi\rangle&=& \sqrt{0.15}|11\rangle+\sqrt{0.7}|22\rangle
	+
	\sqrt{0.15}|33\rangle
	.
\end{eqnarray}
The transformation $|\psi\rangle\langle\psi|$ to
$|\phi\rangle\langle\phi|$ is not possible with 
local operations and classical communication, that is,
$|\psi\rangle\langle\psi|\not\longrightarrow 
|\phi\rangle\langle\phi|$
under LOCC,
by virtue of the above theorem. The ordered list of eigenvalues
of $\text{tr}_A[|\psi\rangle\langle\psi|]=0.6|1\rangle\langle 1|+
0.3|2\rangle\langle 2|+0.1 |3\rangle\langle 3|$ is given
by $0.6,0.3,0.1$, and the corresponding ordered
list for $\text{tr}_A[|\phi\rangle\langle\phi|]$ 
is $0.7,0.15, 0.15$. As
\begin{equation}
	0.6\leq 0.7,\,\,\,\text{ but } \,
	0.6+0.3 > 0.7+0.15,
\end{equation}
the majorization criterion of Nielsen's theorem
is not satisfied.\\

The second important result is related to 
transformations from a pure state to a mixed state.
In Ref.\ \cite{Vidal} an abstract criterion is
given for the probabilistic transformation from one pure state
to a set of other pure states. This result can be used to
give necessary and sufficient conditions for the deterministic
transformation from a pure to a mixed state, as the desired
mixed state can always be generated by mapping the initial state
to a known set of
certain final pure states and mixing of the results. 
%
That 
is, the classical information about the outcomes is discarded.
The explicit criterion is given in Ref.\ \cite{Jonathan}
and reads as follows.\\

\noindent{\bf Theorem (Plenio, Jonathan, Vidal) 
\cite{Vidal,Jonathan}. --} 
{\it Let $|\psi\rangle\in{\cal H}$ 
and let $\rho\in {\cal S}({\cal H}) $ be a mixed state.
For any $|\phi\rangle\in{\cal H}$
let 
\begin{equation}\label{Monoto}
	E_k(|\phi\rangle)=\sum_{i=k}^N \alpha_i,\,\,\,\,\,\,
	k=1,2, ...,N,
\end{equation}
where $\alpha_1,  ... ,\alpha_N$ is the
ordered list of
eigenvalues of ${\text{tr}}_A [|\phi\rangle\langle\phi|]$.
Then 
\begin{equation}
	|\psi\rangle\langle\psi|\longrightarrow \rho \,
	\text{ under LOCC }
\end{equation}
if and only if there exists a decomposition
$\rho=\sum_i \mu_i |\phi_i\rangle\langle\phi_i|$ of $\rho$
in terms of pure states  such that
\begin{equation}
	 \sum_i \mu_i E_k(|\phi_i\rangle)\leq E_k(|\psi\rangle)
\end{equation}
for all $k=1,2,..., N$.
}

\section{State Transformations for Mixed States}


In any practical application, however, one
would expect to always deal with entangled mixed states 
rather than with pure states. Unfortunately, much
less is known about entanglement transformations
of mixed states, and such a powerful tool as the 
criterion of Ref.\ \cite{Nielsen}
is missing in this case.
The question whether a particular entanglement transformation
from one mixed state into another mixed state is possible seems 
to be much more involved. As has been pointed out
in the previous chapter,
in quantum mechanical states both classical correlations and
intrinsic quantum mechanical correlations may be present, which
makes the structure of mixed-state entanglement a much more
complex matter. A different aspect of the same problem is the
well known fact that a representation of a mixed state in terms
of pure states is not uniquely defined, and
it is essentially 
this ambiguity that prohibits a straightforward
application of the techniques of the pure-state case.
Due to these difficulties 
one can hardly hope
for as convenient tools as are available for pure states.

In this subsection transformation criteria
for mixed states will be developed.
A still rather simple setting is one where the projections 
in the spectral decompositions of the initial state 
\index{Locally distinguishable}
can be locally distinguished. 
Examples of such states will be given in Chapter 5.
In this case Alice may design 
a projective measurement discriminating between the
respective components. A particular strategy
for manipulating such a  quantum state can easily be identified:
she performs a projective measurement yielding a pure state,
transforms this pure state to an appropriate other pure state,
and finally discards the classical information of the outcome of
the measurement in order to realize a certain mixed state.
Based on this strategy, one can introduce the
following lemma providing both necessary and sufficient
conditions for the transformation of states of this type
into other such states.\\

\noindent{\bf Lemma 3.2. --} {\it
Consider the states of rank two
\begin{eqnarray}
	\sigma&=&\lambda |\psi\rangle\langle\psi|+(1-\lambda)
	|\eta\rangle \langle\eta|_A\otimes |\xi\rangle \langle\xi|_B
	,\label{1a}\\
	\rho&=&\lambda |\phi\rangle\langle\phi|+(1-\lambda)
	|\varphi\rangle \langle\varphi|_A\otimes 
	|\nu\rangle\langle\nu|_B,\label{1b}
\end{eqnarray}
$\lambda\in(0,1)$,
where $|\psi\rangle\langle\psi|$ and $|\phi\rangle\langle\phi|$ 
are pure entangled states satisfying
$\langle \eta|_A (\text{tr}_B[|\psi\rangle\langle\psi|]) |\eta\rangle_A=0$,
$\langle \varphi|_A (\text{tr}_B[|\phi\rangle\langle\phi|]) |\varphi\rangle_A=
0$, 
$\langle \xi|_B (\text{tr}_B[|\psi\rangle\langle\psi|]) |\xi\rangle_B=0$,
$\langle \nu|_B (\text{tr}_B[|\phi\rangle\langle\phi|]) |\nu\rangle_B=
0$.
Then
$\sigma\longrightarrow\rho$
under LOCC if and only if
$|\psi\rangle\langle\psi|\longrightarrow
|\phi\rangle\langle\phi|$ under LOCC.\\
}

\probreak

\proof\/
$\sigma$ can 
be transformed into $\rho$ by applying LOCC
if $|\psi\rangle\langle\psi|\longrightarrow|\phi\rangle\langle\phi|$ 
under LOCC:
Alice performs a local selective and projective 
measurement
in ${\cal H}_A$
distinguishing between
$|\psi\rangle\langle\psi|$ and 
$|\eta\rangle \langle\eta|_A \otimes |\xi\rangle\langle\xi|_B$. 
If the final state of 
this selective measurement is $|\psi\rangle\langle\psi|$, she applies
an appropriate LOCC operation
to obtain the state $|\phi\rangle\langle\phi|$. 
In the other case she transforms
$
|\eta\rangle\langle\eta|_A\otimes |\xi\rangle\langle\xi|_B
$ into $|\varphi\rangle\langle\varphi|_A\otimes 
	|\nu\rangle\langle\nu|_B$
by using LOCC operations. In a last step
she disregards the classical
information about
the outcomes of the projective  measurement 
to achieve $\rho$.
%
%
Conversely, let ${\cal E}$ be the 
LOCC operation 
realizing ${\cal E}(\sigma)=\rho $.
Then ${\cal E}(|\eta\rangle \langle\eta|_A\otimes 
|\xi\rangle \langle\xi|_B)=
|\varphi\rangle\langle\varphi|_A\otimes 
	|\nu\rangle\langle\nu|_B$, as 
$|\varphi\rangle_A\otimes 
	|\nu\rangle _B$ 
is the only product
vector in $\text{range}[\rho]$. By linearity of
${\cal E}$ is follows that
${\cal E}(|\psi\rangle\langle\psi|)=
|\phi\rangle\langle\phi|$. \proofend

\probreak

In a sense the states in Lemma 3.2 can be regarded
as essentially pure states: the parties can go
back and forth from the initial mixed state to one of
the pure states in the spectral decomposition by
implementing measurements and discarding information.
Finding  criteria for 
the transformation of states for which this procedure
is not possible is a more challenging task.
Such states for which the projections
of the spectral decomposition cannot be locally
distinguished will be called {\it genuinely mixed states}.
\index{Genuinely mixed state}
The subsequent lemma
\index{Genuinely mixed state}
gives a necessary condition in order
for an entanglement transformation
to be possible. The class of mixed states considered in Lemma 3.3 is
actually rather artificial, and it presents a particular case
in which an explicit necessary condition can be given. Propositions 
3.4 -- 3.6 
will be more general statements; however, in the context of later 
considerations Lemma 3.3 will turn out to be very useful.\\

\noindent{\bf Lemma 3.3. --} {\it
Let $\sigma$ and $\rho$ 
be mixed states of rank two of the form 
\begin{eqnarray}
\sigma&=&\lambda |\psi\rangle\langle\psi|
+(1-\lambda)|\eta\rangle \langle\eta|_A\otimes |\xi\rangle \langle\xi|_B,\\
\rho&=&\mu 
|\phi\rangle\langle\phi|
+(1-\mu)|\eta\rangle \langle\eta|_A\otimes |\xi\rangle \langle\xi|_B,\label{class}
\end{eqnarray}
where $\langle \eta|_A (\text{tr}_B[|\phi\rangle\langle\phi|]) |\eta\rangle_A=
\langle \eta|_B (\text{tr}_A[|\phi\rangle\langle\phi|]) |\eta\rangle_B=0$, and
$\mu=\lambda \,{\text{tr}}[\chi]$, 
$\chi=\Pi|\psi\rangle\langle\psi|\Pi$.
The projector $\Pi$ is given by
\begin{equation}
\Pi=\mathbbm{1}-|\eta\rangle \langle\eta|_A\otimes |\xi\rangle \langle\xi|_B.
\end{equation} 
$|\psi\rangle\langle\psi|$ and $|\phi\rangle\langle\phi|$ are 
entangled pure states. 
Then 
\begin{equation}
	\sigma\longrightarrow\rho\,
	\text{ under LOCC }\,
	\Longrightarrow
	\frac{{\text{tr}}_A [\chi]}{{\text{tr}} [\chi]}
	 \prec
	{\text{tr}}_A [|\phi\rangle\langle\phi|].
\end{equation}\\

}

\proof 
Let $\sigma\longrightarrow\rho$ under
LOCC. The set of LOCC operations is a subset of 
the set of separable operations. Therefore, one
may consider a separable operation ${\cal E}$,
\begin{equation}
{\cal E}(\sigma)=
\sum_{i=1}^K 
(A_i\otimes B_i)\sigma (A_i\otimes B_i)^\dagger,
\end{equation}
such that ${\cal E}(\sigma)=\rho$. The
Kraus-operators $A_i$, $B_i$, $i=1,...,K$, act
in ${\cal H}_A$ and ${\cal H}_B$, respectively, and
satisfy
$
\sum_{i=1}^K A_i^\dagger A_i={\mathbbm{1}}_A$, $\sum_{i=1}^K 
B_i^\dagger B_i={\mathbbm{1}}_B$.
For each $i$ the image of $\sigma$ under this map
must be element
of the range of $\rho$, that is
\begin{equation} 
(A_i\otimes B_i)\sigma (A_i\otimes B_i)^\dagger\in {\text{range}}[\rho].
\end{equation} 
There is only a single product
vector included in the range of $\rho$, which 
then amounts to a best separable approximation in the
sense of \cite{Lewenstein}. As 
${\cal E}(|\eta\rangle \langle\eta|_A\otimes |\xi\rangle \langle\xi|_B)=
|\eta\rangle \langle\eta|_A\otimes |\xi\rangle \langle\xi|_B$,
it follows that
\begin{equation}
	{\cal E}(\sigma)=\lambda {\cal E}(|\psi\rangle\langle\psi|)
	+(1-\lambda) |\eta\rangle \langle\eta|_A\otimes |\xi\rangle \langle\xi|_B,
\end{equation}
and therefore, the state
$|\psi\rangle\langle\psi|$ must be mapped on
\begin{equation}
	{\cal E}(|\psi\rangle\langle\psi|)=\nu |\phi\rangle\langle\phi|+(1-\nu)
|\eta\rangle \langle\eta|_A\otimes |\xi\rangle \langle\xi|_B,
\end{equation}
where $\nu=\mu/\lambda$. Thus,
\begin{equation}
\Pi (A_i\otimes B_i) |\psi\rangle=
 \Pi (A_i\otimes B_i) \Pi |\psi\rangle
\end{equation}
for $i=1,...,K$. Hence, the following chain holds:
\begin{eqnarray}
\nu&=&{\text{tr}}\bigl[
\Pi\sum_{i=1}^K (A_i\otimes B_i) |\psi\rangle\langle\psi|(A_i\otimes B_i)^\dagger
\Pi\bigr]\nonumber\\
&=&{\text{tr}}\bigl[ \sum_{i=1}^K \Pi (A_i\otimes B_i) \chi (A_i\otimes B_i)^\dagger
\Pi\bigr]\\
&\leq&{\text{tr}}[\chi].
\end{eqnarray}
But ${\text{tr}}[\chi]=\nu$ by definition, and therefore,
\begin{equation}
\chi/{\text{tr}}[\chi]\longrightarrow |\phi\rangle\langle\phi|
\end{equation}
under LOCC, which in turn implies,
according to Nielsen's theorem,
that 
\begin{equation}
{\text{tr}}_A [\chi]/{\text{tr}} [\chi]
	 \prec
	{\text{tr}}_A [|\phi\rangle\langle\phi|].
	\end{equation}
This is the statement of the lemma.\proofend

\probreak

This lemma provides a convenient tool for  
the considerations of the next section. In particular,
it will be used to show that also in the mixed state
domain entangle\-ment-assisted operations
are more powerful than ordinary LOCC operations.
The general problem of finding necessary and sufficient
criteria for the transformation of general mixed states
of full rank into other mixed states is extremely difficult.
Ref.\ \cite{Schmidtnumber} partially 
addresses  this problem and gives some hints,
related to the observation that
the Schmidt number should not increase on average
under LOCC operations. The {\it Schmidt number}\/ -- 
which is 
different from the Schmidt measure proposed in the 
previous chapter -- can be conceived
as a useful
generalization of the Schmidt rank of pure states
to the mixed-state domain. \index{Schmidt number}
It seems likely that any 
simple generalization of Nielsen's theorem will 
not be appropriate. Even the connection of this problem
to majorization theory is not obvious at all. 
The following three statements are
necessary criteria, but unfortunately, their
practical implications are rather limited, as it is not
known how to find the optimal decompositions of the
involved mixed states in terms of pure states.\\

\noindent {\bf Proposition 3.4. --} {\it
Let $\sigma,\rho\in{\cal S}({\cal H})$ be states
satisfying
$\sigma\longrightarrow \rho$ under LOCC. Then for
any LOCC operation ${\cal E}$ for which
${\cal E}(\sigma)=\rho$ and any
decomposition $\sigma=\sum_{i=1}^n 
\mu_i |\psi_i\rangle\langle\psi_i|$
of $\sigma$ in terms of pure states
\begin{equation}
	\text{range}[{\cal E}(|\psi_i\rangle\langle\psi_i|)]\subset \text{range}[\rho]
\end{equation}
holds for all $i=1,...,n$\, .}

\bigskip

\noindent {\bf Proposition 3.5. --} {\it
Let $\sigma,\rho\in{\cal S}({\cal H})$
be states satisfying
	$\sigma\longrightarrow \rho$ under LOCC. Then
for any decomposition  
$\sigma=\sum_{i=1}^n \mu_i |\psi_i\rangle\langle\psi_i|$
of $\sigma$ in terms of pure states there exists a
decomposition $\rho=
\sum_{i=1}^n \mu_i \rho_i$ of $\rho$
and for each $i=1,2,...,n$ a
decomposition $\rho_i =\sum_j \eta_{i,j} 
|\phi_{i,j}\rangle\langle\phi_{i,j}|$ 
of $\rho_i$
in terms of pure states 
such that
\begin{equation}
	\sum_j \eta_{i,j}  
	E_k( |\phi_{i,j}\rangle)\leq E_k(|\psi_i\rangle ) 
	\,\,\,\,	
	\text{ for all }
	k=1,2,...,N\, \text{and all }i=1,2,...,n \,.
\end{equation}
}

\medskip
\proof Let ${\cal E}$
be the LOCC operation realizing ${\cal E}(\sigma)=\rho$,
then $\rho_i={\cal E}(|\psi_i\rangle\langle\psi_i|)$
satisfy $\sum_i \mu_i \rho_i =\rho$. The proof of the statement
then follows from the theorem 
of Ref.\ \cite{Jonathan}. \proofend

\noindent {\bf Proposition 3.6. --} {\it
Let $\sigma,\rho\in{\cal S}({\cal H})$
be states satisfying
	$\sigma\longrightarrow \rho$ under LOCC. Let $|\psi\rangle\in{\cal H}$
and let $\sigma=\sum_{i=1}^n \mu_i |\psi_i\rangle\langle\psi_i|$
be a decomposition of $\sigma$ in terms of pure states satisfying
\begin{equation}
	 \sum_{i=1}^n \mu_i E_k(|\psi_i\rangle)\leq E_k(|\psi\rangle)
\end{equation}
for all $k=1,2, ..., N$. Then there exists a decomposition
$\rho=\sum_{j=1}^m \eta_j |\phi_j \rangle\langle\phi_j|$
of $\rho$ in terms of pure states such that
\begin{equation}
	\sum_{j=1}^m \eta_j E_k(|\phi_j\rangle)\leq E_k(|\psi\rangle)
\end{equation}
for $k=1,...,N$.\\}

\medskip

\newfigure{6.5cm}{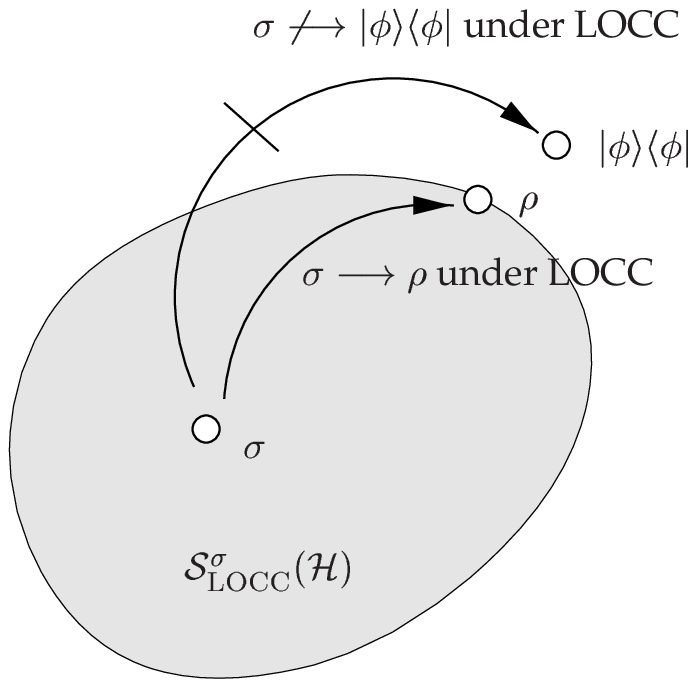}{Figure 3.2: Although the state $\sigma$
cannot be transformed into $|\phi\rangle\langle\phi|$  
by use of LOCC operations, the similar state $\rho$
can be reached. $\rho$ maximizes the fidelity
with respect to $|\phi\rangle\langle\phi|$  in the
set ${\cal S}_{\rm LOCC}^\sigma ({\cal H})$.}

\index{Approximate transformation}
Sometimes it is not of primary interest to find
out whether a particular
final state can exactly be reached when performing a LOCC
operation. Instead, if a transformation from a certain initial state
$\sigma$ to a final state $|\phi\rangle\langle\phi|$ is
not possible, the objective is to prepare a
mixed state $\rho$ that is closest 
to the desired pure state as measured by the fidelity.
Such transformations are called {\it approximate
transformations}\/ \cite{VidalJonathanNielsen}. 
In general 
the task is to increase the fidelity with respect to 
$|\phi\rangle\langle\phi|$,
\begin{equation}
F(\sigma,|\phi\rangle\langle\phi|)=\langle\phi|\sigma|\phi\rangle,
\end{equation}
to the maximal value which can be attained by applying LOCC
operations on the initial mixed state 
$\sigma$. This maximal value will hereafter be 
denoted as
\begin{equation}
	F_{\rm LOCC}(\sigma,|\phi\rangle\langle\phi|)
	= \max_{\rho\in{\cal S}_{\rm LOCC}^\sigma ({\cal H})}
	\langle \phi|\rho|\phi\rangle.\label{Max2}
\end{equation}
${\cal S}_{\rm LOCC}^\sigma ({\cal H})\subset{\cal S}({\cal H})$ 
is in this equation the set of states  
$\rho$ for which $\sigma\longrightarrow\rho$ under
LOCC. This set is a convex subset of
${\cal S}({\cal H})$:
If $\rho_1$ is accessible from $\sigma$ under LOCC and
the same is true for another state $\rho_2$, then
also the convex combination
$\lambda \rho_1+(1-\lambda)\rho_2$ can be 
achieved with the use of such transformations for
all $\lambda\in[0,1]$. This set is also a compact set.
As the fidelity with respect to
a fixed pure state is a linear functional, the
max in Eq.\ (\ref{Max2}) is justified because
the corresponding maximum is actually attained.

\section{Entanglement-Assisted Transformations}
\index{Catalysis}

As has been 
pointed out in the introduction, entanglement is 
often viewed as the essential resource for many 
\index{Entanglement-assisted transformation}
tasks of quantum information processing. 
Typically, the entanglement 
is used up in the course of the implementation
of a certain protocol. So it comes as a surprise
that ``the mere presence of entanglement'' can be
an advantage when one intends to transform an
initial state into a particular final state with
the use of local quantum operations with
classical communication. It has been demonstrated
in  Ref.\ \cite{Jonathan} that 
there are indeed target states which cannot be 
reached by LOCC starting from a particular
initial state, but with the assistance of a
distributed pair of auxiliary quantum systems in a 
particular known state, 
%
%
even though these auxiliary quantum systems
are left in exactly the same state and remain finally
completely uncorrelated to the quantum system of interest.
This phenomenon is quite remarkable as the entanglement
which serves as a ``catalyst'' for the otherwise
forbidden ``reaction'' is not consumed.

The basis of the example given in Ref.\ \cite{Jonathan}
is the above majorization criterion of Ref.\ 
\cite{Nielsen}.
%
Let $\sigma$ and $\rho$ be states of a bi-partite
quantum system with Hilbert space ${\cal H}$. 
The notation
\begin{equation}
	\sigma\longrightarrow\rho\,\,\,\,\,\,
	\text{{\it  under ELOCC}\/ }
\end{equation}
will be used if there exists a Hilbert space
${\cal K}$ of another bi-partite system
and a pure state $\omega\in{\cal S}({\cal H})$ such that
\begin{equation}
	\sigma\otimes\omega
	\longrightarrow
	\rho\otimes\omega\,\,\,\,\,\,
	\text{ under LOCC}.
\end{equation}
${\cal H}$ and ${\cal K}$ 
are Hilbert spaces corresponding to bi-partite
quantum systems, 
${\cal H}={\cal H}_A\otimes {\cal H}_B$
and
${\cal K}={\cal K}_A\otimes {\cal K}_B$,
respectively, such that local operations of Alice, say, 
are operations in ${\cal H}_A\otimes{{\cal K}}_A$ (see Figs. 3.3 and 
3.4).
The phenomenon of ``catalysis'' of entanglement
transformations might serve
as a basis for applications in cryptography. For
first steps in this direction see Refs.\ \cite{Barnum}
and \cite{SchackCata}.

\newfigure{12.2cm}{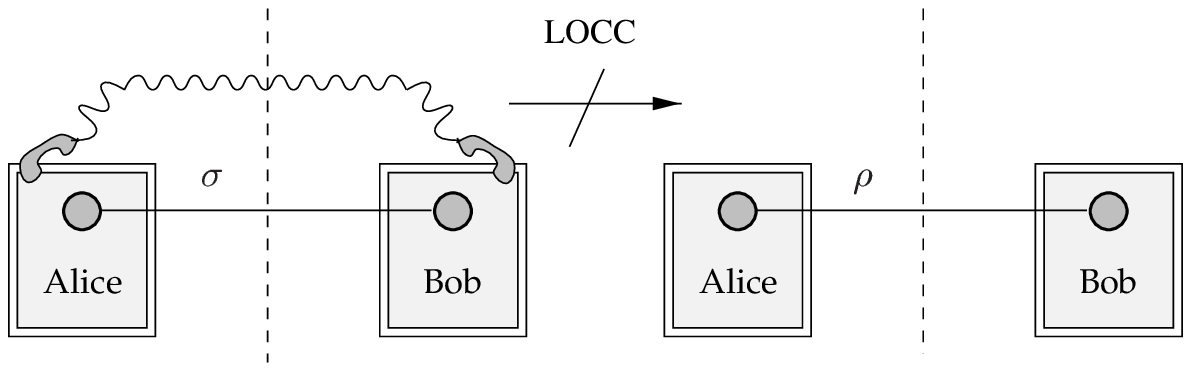}{Figure 3.3: 
If $\sigma$ and $\rho$ are entangled states of a bi-partite
quantum system, it may well be that $\sigma\not\longrightarrow\rho$
under LOCC, even if $\rho$ is less entangled than $\sigma$
with respect to some  entanglement monotone.}

\vspace*{-0.6cm}

\newfigure{12.2cm}{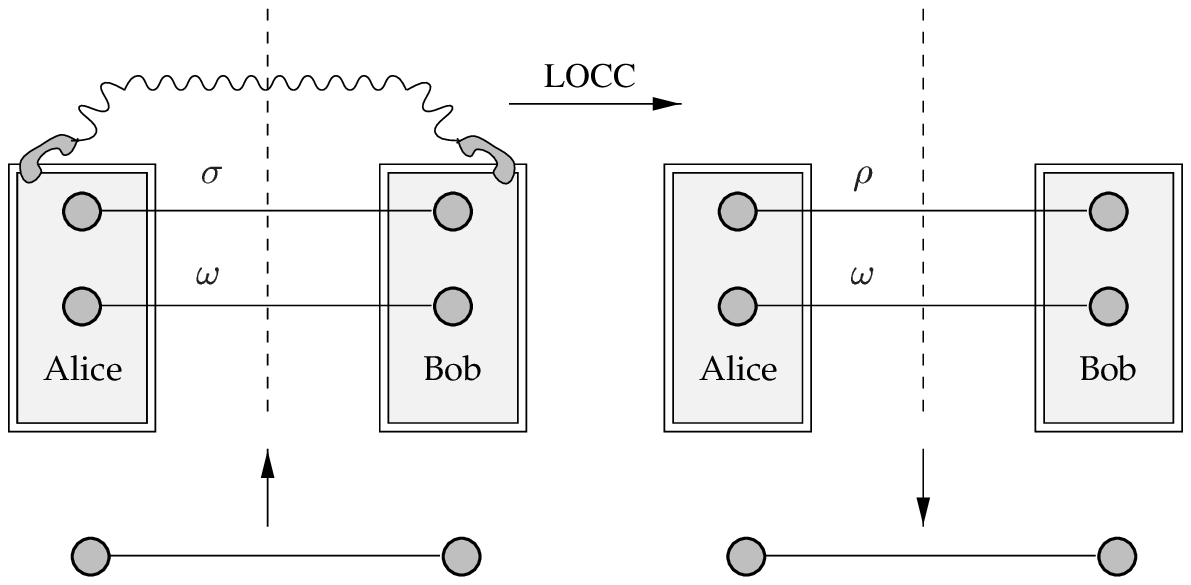}
{Figure 3.4: 
There are, however, cases for which  
$\sigma\not\longrightarrow\rho$
under LOCC, but $\sigma\otimes\omega\longrightarrow\rho
\otimes\omega$
under LOCC for an appropriately chosen state $\omega$
of another bi-partite quantum system. 
That is, the
auxiliary system in the state $\omega$ is borrowed
by Alice and Bob, such that they hold a composite system
in the state $\sigma\otimes\omega$. This state can then be
transformed into $\rho\otimes\omega$ under LOCC
operations. The auxiliary system can in a last step 
be returned.
The state $\omega$ remains unchanged in the course
of the protocol, and the auxiliary system is finally
fully uncorrelated to the original 
quantum system.}

\subsection{Pure State Case}

For pure states one can use the strong tool that is provided
by Nielsen's theorem. On applying the majorization criterion 
one can look numerically
for examples for catalysis of entanglement
manipulation. \\

\noindent{\bf Example  (Jonathan and Plenio). --} 
In this example \cite{JonPle} 
the initial pure state $|\psi\rangle\langle\psi|$
and the final state $|\phi\rangle\langle\phi|$ are
taken from the state space corresponding to
${\cal H}={\cal H}_A\otimes
{\cal H}_B$,
${\cal H}_A= 
{\cal H}_B={\mathbbm{C}}^4$. The basis is denoted as
$\{|1\rangle,|2\rangle,|3\rangle,|4\rangle\}$.
The state vectors 
can be represented in the Schmidt decomposition as
\begin{eqnarray}
	|\psi\rangle&=& \sqrt{0.4}|00\rangle
		+\sqrt{0.4}|11\rangle
		+\sqrt{0.1}|22\rangle
		+\sqrt{0.1}|33\rangle,\\
		|\phi\rangle&=& \sqrt{0.5}|00\rangle
		+\sqrt{0.25}|11\rangle
		+\sqrt{0.25}|22\rangle.
\end{eqnarray}
It follows from Nielsen's theorem \cite{Nielsen}
that $|\psi\rangle\langle\psi|
\not\longrightarrow |\phi\rangle\langle\phi|$,
as the majorization relation is not satisfied, 
${\text{tr}}_A[|\psi\rangle\langle\psi|]\not\prec
	{\text{tr}}_A[|\phi\rangle\langle\phi|]$.
However, surprisingly, if one appends the catalyst
state 
\begin{equation}
	\omega=\left(\sqrt{0.6}|44\rangle+\sqrt{0.4}|55\rangle\right)
	\left(\sqrt{0.6}\langle44 |+\sqrt{0.4}\langle 55|\right),
\end{equation}
then 
\begin{equation}
	{\text{tr}}_A[|\psi\rangle\langle\psi|\otimes \omega]\prec
	{\text{tr}}_A[|\phi\rangle\langle\phi|\otimes \omega].
\end{equation}
In other words, $|\psi\rangle\langle\psi|$ can be
transformed into  $|\phi\rangle\langle\phi|$  by 
means of ELOCC operations.\\

The general problem of {\it catalysis for pure states}\/
remains  unresolved. 
The issue is basically a topic from 
matrix analysis and majorization theory. 
Let $M_1$ and $M_2$ be Hermitian matrices with $M_1^2=M_1$,
$M_2^2=M_2$, satisfying
$M_1\not\prec M_2$. 
What are the
(necessary and sufficient) conditions on $M_1$ and
$M_2$ for the existence
of a Hermitian matrix $C$, $C^2=C$, such that 
\begin{equation}
M_1\otimes C \prec M_1\otimes C.
\end{equation}
Translated into 
quantum mechanical terms this question reads:
Let $\sigma$ and $\rho$ be pure states on a Hilbert 
space ${\cal H}={\cal H}_A\otimes {\cal H}_B$.
What are the (necessary
and sufficient) conditions on $\sigma$ and $\rho$
for the existence of a 
Hilbert space ${\cal K}={\cal K}_A\otimes {\cal K}_B$
and a pure state $\omega\in {\cal S}({\cal K} )$ such that 
\begin{equation}
	\text{tr}_A[\sigma]\otimes \text{tr}_A[\omega]
	\prec \text{tr}_A[\rho]\otimes \text{tr}_A[\omega]?
\end{equation}
In the absence of a complete answer to this question,
practically relevant necessary conditions are already
helpful. A step in this direction could be the subsequent 
statement\footnote{After completion of the work on this proposition
the author found out that in unpublished lecture notes 
on majorization theory 
statements can be found  which 
are similar in spirit \cite{Notes}.}:\\

\noindent 
{\bf Proposition 3.7. --} {\it
Let $|\psi\rangle,|\phi\rangle\in {\cal H}$. 
Then
\begin{equation}\label{FState}
	 |\psi\rangle\langle\psi|\longrightarrow|\phi\rangle\langle\phi| \,\,
	\text{ under ELOCC } 
	\Longrightarrow 
	\sum_{i=1}^N 
	\alpha_i^\xi \leq
	\sum_{i=1}^N 
	\beta_i^\xi
\end{equation}
for all $\xi>0$,
where $\alpha_1,...,\alpha_N$ and $\beta_1,...,\beta_N$
are the eigenvalues of ${\text{tr}}_A[ |\psi\rangle\langle\psi|]$ and 
${\text{tr}}_A[|\phi\rangle\langle\phi| ]$, respectively.\\}

\probreak

\proof Assume that $|\psi\rangle\langle\psi| 
\longrightarrow |\phi\rangle\langle\phi|$ under ELOCC.
Then there exists a ${\cal K}={\cal K}_A\otimes {\cal K}_B$ 
and a pure state
$\omega\in {\cal S}({\cal K})$
such that $|\psi\rangle\langle\psi| \otimes\omega
\longrightarrow |\phi\rangle\langle\phi|\otimes\omega$ under LOCC 
and hence,
\begin{equation}
{\text{tr}}_A[|\psi\rangle\langle\psi| \otimes\omega]\prec
 {\text{tr}}_A[|\phi\rangle\langle\phi|\otimes\omega].
 \end{equation} 
 Let $\gamma_1,...,\gamma_M$, $M={\rm dim}[{\cal K}_B]$,
be the eigenvalues of ${\text{tr}}_A[\omega]$, such that
$\alpha_i\gamma_j$ and $\beta_i\gamma_j$, $i=1,...,N$, $j=1,...,M$
are the respective
eigenvalues of ${\text{tr}}_A[|\psi\rangle\langle\psi| \otimes\omega]$ and
${\text{tr}}_A[|\phi\rangle\langle\phi| \otimes\omega]$. So,
\begin{equation}
	\sum_{i=1}^N 
	\sum_{j=1}^M  
	g(\alpha_i\gamma_j) \leq
	\sum_{i=1}^N  \sum_{j=1}^M  
	g(\beta_i\gamma_j)
\end{equation} 
for any convex function $g:\mathbbm{R}\longrightarrow \mathbbm{R}$
\cite{Majo,Bhatia}. In particular, take
the function $f$ defined as $f(x)=x^\xi$, $\xi>0$. As
\begin{equation}
f(\beta_i\gamma_j)=f(\beta_i) f(\gamma_j) 
\end{equation}
for all $i=1,...,N$, $j=1,...,M$, the 
statement of Eq.\ (\ref{FState}) holds.
\proofend 

\medskip

This proposition provides pairs 
of states for which one can tell that the corresponding
en\-tangle\-ment-assisted transformation is not possible.
Accordingly, one can rule out a rather large class
of pairs of pure states which are otherwise
eligible for
enhancement of entanglement manipulations through 
additional catalyst quantum systems.\\

\noindent 
{\bf Example 3.8. --} 
Consider the initial state
$|\psi\rangle\langle\psi|$ and the final 
state $|\phi\rangle\langle\phi|$
with
\begin{eqnarray}
	|\psi\rangle&=&\sqrt{0.5}
	|11\rangle
	+\sqrt{0.2-\varepsilon}
	|22\rangle
	+
	\sqrt{0.2+\varepsilon}
	|33\rangle
	+
	\sqrt{0.1} |44\rangle,
	\\
	|\phi\rangle&=&\sqrt{0.5}
	|11\rangle
	+\sqrt{0.2}
	|22\rangle
	+
	\sqrt{0.2}
	|33\rangle
	+
	\sqrt{0.1} |44\rangle,
\end{eqnarray}
$\varepsilon\in(0,0.1)$.
As follows from Nielsen's theorem $|\psi\rangle\langle\psi|\not\longrightarrow
|\phi\rangle\langle\phi|$
under LOCC for all values of $\varepsilon$. In addition, one cannot gain 
from the possibility of using 
ELOCC operations,
as can be inferred from Proposition 3.7: take as an appropriate 
function 
$f:\mathbbm{R}\longrightarrow \mathbbm{R}$
with
$f(x)=x^2$. 
Since 
\begin{equation}
0.5^2 +(0.2-\varepsilon)^2
+(0.2+\varepsilon)^2 + 0.1^2 >0.5^2 +2\times 0.2^2+0.1^2
\end{equation} 
for all $\varepsilon\in(0,0.1)$, also $|\psi\rangle\langle\psi|\not\longrightarrow
|\phi\rangle\langle\phi|$ under ELOCC
for all $\varepsilon\in(0,0.1)$.\\

\noindent
{\bf Example 3.9. --} It has been shown 
that $|\psi\rangle\langle\psi|\not\longrightarrow
|\phi\rangle\langle\phi|$ under LOCC, but
$|\psi\rangle\langle\psi|\longrightarrow
|\phi\rangle\langle\phi|$ under ELOCC, where 
\begin{eqnarray}
	|\psi\rangle&=&\sqrt{0.4}
	|11\rangle
	+\sqrt{0.4}
	|22\rangle
	+
	\sqrt{0.1}
	|33\rangle
	+
	\sqrt{0.1} |44\rangle,
	\\
	|\phi\rangle&=&\sqrt{0.5}
	|11\rangle
	+\sqrt{0.25}
	|22\rangle
	+
	\sqrt{0.25}
	|33\rangle.
\end{eqnarray}
In order to investigate to what degree one can change
the coefficients without changing the structure of the example, 
take the function 
$f:\mathbbm{R}\longrightarrow \mathbbm{R}$ 
with $f(x)=x^2$. 
Then the criterion of Proposition 3.7 is satisfied, since
\begin{equation}
	2\times 0.4^2 + 2\times 0.1^2 =0.34< 0.5^2+ 2\times 0.25^2 = 0.375.
\end{equation}
The slightly modified states
\begin{eqnarray}
	|\psi\rangle&=&\sqrt{0.43}
	|11\rangle
	+\sqrt{0.43}
	|22\rangle
	+
	\sqrt{0.07}
	|33\rangle
	+
	\sqrt{0.07} |44\rangle,
	\\
	|\phi\rangle&=&\sqrt{0.5}
	|11\rangle
	+\sqrt{0.25}
	|22\rangle
	+
	\sqrt{0.25}
	|33\rangle.
\end{eqnarray}
are nevertheless not appropriate for an entanglement-assisted
transformation. From Proposition 3.7 one can infer  that the transformation
from $|\psi\rangle\langle\psi|$ to  $|\phi\rangle\langle\phi|$ under 
ELOCC is certainly not possible.\\

\subsection{Mixed-State Catalysis of Entanglement Manipulation}

It is not clear a priori whether in the mixed state domain the
set of tasks that can be accomplished with 
entanglement-assisted local operations is strictly larger than the set
of tasks which may be performed with mere LOCC.
The possibilities of manipulating entanglement
are often much different for pure and for mixed states.
For example, 
while (pure) singlets can be produced from non-maximally
entangled pure states by individual measurements, this
is not true for mixed states, not even if they are sufficiently
close to being singlets \cite{Kent}.

In an abstract form, the class of ELOCC operations
can be easily explored in terms of separable operations.
The class of LOCC operations is contained in the set of 
separable operations, which can be represented using
Kraus operators as explained in the first chapter. 
Certainly,
the statement that a certain operation is a ELOCC
operation implies particular restrictions on the Kraus
operators:\\

\noindent {\bf Proposition 3.10. --} {\it 
Let  $\sigma,\rho\in{\cal S}({\cal H})$
be states for which
\begin{equation}
	\sigma\longrightarrow \rho\,\,\text{ under ELOCC}. 
\end{equation}
Then there exist a Hilbert space ${\cal K}$, a state
vector $|\tilde\psi
\rangle\in{\cal K}$ and
Kraus operators $A_1,...,A_K:{\cal H}_A\rightarrow {\cal H}_A$ and 
$B_1,...,B_K:{\cal H}_B\rightarrow {\cal H}_B$ 
satisfying $\sum_{k=1}^K A_k^\dagger A_k={\mathbbm{1}}_A$ and
$\sum_{k=1}^K B_k^\dagger B_k={\mathbbm{1}}_B$,
such that for every decomposition
$\sigma={\sum}_{i=1}^n \mu_i |\psi_i\rangle\langle\psi_i|$ of 
$\sigma$ in terms of pure states 
\begin{equation}\label{complkraus}
	(A_k\otimes B_k) 
	|\psi_i\rangle \otimes
	|\tilde\psi\rangle\propto 
	|\phi^{(k)}_{i}\rangle\otimes |\tilde
\psi\rangle
\end{equation}
holds for all $k=1,...,K$, $i=1,...,n$, where $|\phi^{(k)}_{i}\rangle\in
{\cal H}_A$
are appropriate state vectors.\\
}

\probreak
\proof 
Let ${\cal E}$ be the separable operation realizing
${\cal E}(\sigma\otimes |\tilde\psi\rangle\langle\tilde\psi|)
=\sum_{k=1}^K (A_k\otimes B_k) 
(\rho\otimes |\tilde\psi\rangle\langle\tilde\psi|) (A_k\otimes 
B_k)^\dagger =\rho$, and let $\sigma=\sum_{i=1}^n \mu_i 
|\psi_i\rangle\langle\psi_i|$ be a decomposition of $\sigma$
in terms of pure states. Then 
$(A_k\otimes B_k)|\psi_i\rangle\otimes |\tilde\psi\rangle$ 
must be element of the range of 
$\rho\otimes  |\tilde\psi\rangle\langle\tilde\psi|$ for 
every $k=1,...,K$ and every $i=1,...,n$. This is condition
Eq.\ (\ref{complkraus}).
\proofend

These conditions are rather restrictive, but from this
statement alone it cannot be judged whether entanglement-assisted
operations are possible for mixed states. 
This is because Proposition 3.10 does not present 
a constructive way of actually 
finding the Kraus operators which satisfy Eq.\
(\ref{complkraus}).
Nevertheless, it can be shown that
examples of mixed-state catalysis of entanglement
transformations can indeed be found. 
The major result of this subsection is 
an example of incommensurate genuinely mixed states such
that with the use of some appropriately chosen
catalyst state, the initial state can be converted into 
the final state while fully retaining the
catalyst state.  That is, there exist mixed 
states $\sigma,\rho$ 
such that
        $\sigma\longrightarrow\rho$ under ELOCC
but not
        $\sigma\longrightarrow\rho$ under LOCC.
To show that the transformation $\sigma\longrightarrow  \rho$
is actually not possible
by means of LOCC operations, 
Lemma 3.3 will be used.\\

\noindent{\bf Example 3.11. --}
Consider the 
two one-parameter sets  of states of rank two, 
which are density operators on the Hilbert space 
${\cal H} ={\cal H}_A\otimes {\cal H}_B $, where
${\cal H}_A={\cal H}_B ={\mathbbm{C}}^5$ with
basis $\{|1\rangle,|2\rangle,|3\rangle,|4\rangle,|5\rangle \}$.
For a $\lambda\in(0,1)$ let
\begin{eqnarray}\label{example}
\sigma&=&\lambda|\psi\rangle\langle\psi|+(1-\lambda)|55\rangle\langle55|,\\
\rho&=&\mu |\phi\rangle\langle\phi|
+(1-\mu)|55\rangle\langle55|,
\end{eqnarray}
with $\mu=0.95\lambda$ and 
\begin{eqnarray}
	|\psi\rangle&=&
	\sqrt{0.38}|11\rangle+
	\sqrt{0.38}|22\rangle+
	\sqrt{0.095}|33\rangle\nonumber \\
	&+&
	\sqrt{0.095}|44\rangle+
	\sqrt{0.05} |55\rangle,\label{psi}\\
	|\phi\rangle&=&
	\sqrt{0.5}|11\rangle+
	\sqrt{0.25}|22\rangle+
	\sqrt{0.25}|33\rangle,\label{phi}
\end{eqnarray}
%
The initial state $\sigma$ is indeed
genuinely mixed for all $\lambda\in(0,1)$, as
the components of the spectral
decomposition cannot be locally distinguished.
These states are included in the sets of 
states considered
in Lemma 3.3, and this lemma can be applied.
$\chi/{\text{tr}}[\chi]=|\eta\rangle\langle\eta|$,
where 
\begin{equation}\label{tild}
|\varphi\rangle=\sqrt{0.4}|11\rangle+
\sqrt{0.4}|22\rangle+
\sqrt{0.1}|33\rangle+
\sqrt{0.1}|44\rangle
\end{equation}
as in Ref. \cite{Jonathan}. Hence, 
\begin{equation}
	\frac{{\text{tr}}_A [\chi]}{{\text{tr}} [\chi]}
	 \not\prec
	{\text{tr}}_A [|\phi\rangle\langle\phi|],
\end{equation}
and therefore, 
\begin{equation}
\sigma\not\longrightarrow\rho\,\,\text{ under LOCC}
\end{equation}
for all values of $\lambda\in(0,1)$.
However, it can be shown that
$\sigma\longrightarrow\rho$ under ELOCC by
designing a protocol that accomplishes this task:\\

\noindent
\jbox{
\begin{enumerate}
\item
Alice performs a local projective   measurement
in system $A$
associated with Kraus operators $E_1=A_1\otimes {\mathbbm{1}}_B$,
$E_2=A_1\otimes {\mathbbm{1}}_B$, where
\begin{equation}\label{FunnyPOVM}
A_1=\sum_{i=1}^4 |ii\rangle\langle ii|,\,\,\,
A_2=|55\rangle\langle55|
\end{equation}
satisfying 
$
A_1^\dagger A_1+
A_2^\dagger A_2={\mathbbm{1}}_A$. (
A similar decomposition will be used in Chapter 
5 in Eq. (\ref{FunnyPOVM2}).)
\item
If the outcome
corresponds to $A_2$, she does not implement
any further 
operations. 
\item
In the alternative 
case the final state is the pure state 
$
|\varphi\rangle\langle\varphi|$ given by
Eq.\ (\ref{tild}). 
As in Ref.\ \cite{Jonathan} this state can be
transformed into $|\phi\rangle\langle\phi|$
by the help of the pure catalyst state 
\begin{equation}
\omega=\bigl(\sqrt{0.4}|66\rangle+
\sqrt{0.6}|77\rangle\bigr)\bigl(\sqrt{0.4}\langle66|+
\sqrt{0.6}\langle77|\bigr).
\end{equation}
This follows from the fact that
\begin{equation}
{\text{tr}}_A [|\varphi\rangle\langle\varphi|
\otimes \omega]\prec 
{\text{tr}}_A [|\phi\rangle\langle\phi|
\otimes \omega],
\end{equation} 
as the criterion 
of Nielsen's theorem is satisfied.
\item 
Finally, 
the classical information about the outcomes in the
first step
is discarded
in order to achieve $\rho$.
 \end{enumerate}
}

\noindent 
Hence, it turns out that the initial state $\sigma$ 
cannot be transformed into $\rho$ if 
-- except for the classical communication to coordinate
the steps -- only 
local quantum operations on the quantum system itself
can be performed. In the presence of an additional quantum system
in the pure state $\omega$ this task
may be feasible.
 %

\subsection{Increasing the Proportion of a Pure State in a Mixture}

In the mixed-state regime
entanglement-assisted LOCC operations are
more powerful than mere local operations with
classical communication. This 
has an implication 
on the efficiency of attempts to 
increase the quota of an entangled 
state $|\phi\rangle\langle\phi|$ in a mixed state $\sigma$
by applying a trace-preserving operation. 
When the task is to maximize the fidelity 
\begin{equation}
F(\sigma,|\phi\rangle\langle\phi|)=\langle\phi|\sigma|\phi\rangle
\end{equation}
of an initial state with respect to a particular 
entangled pure state, protocols 
which make appropriate use of catalyst states
can indeed be more efficient than
protocols in which such additional quantum systems are
not available. This result is quite amazing since
in this case a ``number'' is increased (the efficiency 
of the protocol) without consuming the additional resource
at all (the entanglement of the catalyst state).\\

\noindent{\bf Proposition 3.12. --} {\it 
There exist mixed states $\sigma$ and
pure states $|\phi\rangle\langle\phi |$ with the property that 
the maximal average attainable value of the fidelity 
under ELOCC
\begin{equation}
	F_{\rm ELOCC}(\sigma,|\phi\rangle\langle\phi|)= 
	\max_{\rho\in{\cal S}_{\rm ELOCC}^\sigma
	({\cal H})}
	\langle \phi|\rho|\phi\rangle\label{Sup}
\end{equation}
is strictly larger than the 
maximal attainable fidelity under LOCC,
\begin{equation}
	F_{\rm LOCC}(\sigma,|\phi\rangle\langle\phi|)
	= \max_{\rho\in{\cal S}_{\rm LOCC}^\sigma ({\cal H})}
	\langle \phi|\rho|\phi\rangle.\label{Max}
\end{equation}
The set ${\cal S}_{\rm ELOCC}^\sigma ({\cal H})
\subset{\cal S}({\cal H})$
consists of the states $\rho$ for which
$\sigma\longrightarrow \rho$ under ELOCC.\\
}

\probreak

\proof 
This statement can be proven by taking
an initial state $\sigma$ 
of the form specified in
Eq.\ (\ref{example}), 
$\sigma=\lambda|\psi\rangle\langle\psi|+(1-\lambda)|55\rangle\langle55|$
, $\lambda\in[0,1]$, with
\begin{eqnarray}
	|\psi\rangle&=&
	\varepsilon(\sqrt{0.4}|11\rangle+
	\sqrt{0.4}|22\rangle+
	\sqrt{0.1}|33\rangle+
	\sqrt{0.1}|44\rangle)\nonumber\\
	&+&
	\sqrt{1-\varepsilon^2}|55\rangle,\label{fidexample}
\end{eqnarray}
and one may choose 
\begin{equation}
|\phi\rangle=
	\sqrt{0.5}|11\rangle+
	\sqrt{0.25}|22\rangle+
	\sqrt{0.25}|33\rangle
\end{equation} 
as in Eq.\ (\ref{phi}).
As the components of the initial state $\sigma$ are
not locally distinguishable for $\varepsilon<1$
 and since the achievable
fidelity can be no better than the sum of 
both largest possible fidelities of each contribution,
\begin{eqnarray}
	F_{\rm LOCC}(\sigma,|\phi\rangle\langle\phi|)&\leq&
	(1-\lambda)
	F_{\rm LOCC}(|55\rangle\langle55|,|\phi\rangle\langle\phi|)
	\nonumber\\
	&+&
	\lambda
	F_{\rm LOCC}(|\psi\rangle\langle\psi|,|\phi\rangle\langle\phi|).
\end{eqnarray}
The not necessarily pure separable state $\rho$ which 
maximizes the fidelity
	$F(\rho,|\phi\rangle\langle\phi|)=
	\langle\phi|\rho|\phi\rangle$
is given by $|11\rangle\langle11|$ as follows from Lemma 3.13, 
and for any mixed separable state
this fidelity takes a smaller value. Therefore, 
\begin{equation}
	F_{\rm LOCC}(|55\rangle\langle55|, |\phi\rangle\langle\phi|)=
	1/2.
\end{equation}
Finally, from 
\begin{equation}
F_{\rm ELOCC}(\sigma,|\phi\rangle\langle\phi|)\geq
\lambda\varepsilon^2+(1-\lambda\varepsilon^2)/2
\end{equation} 
and due to the continuity of 
$F_{\rm LOCC}(|\psi\rangle\langle\psi|,|\phi\rangle\langle\phi|)$
taken as a function of $\varepsilon$ 
one can conclude that there exists a
$\tilde\varepsilon\in (0,1)$ such that
the inequality
\begin{equation}
	F_{\rm ELOCC}(\sigma,|\phi\rangle\langle\phi|)>
	F_{\rm LOCC}(\sigma,|\phi\rangle\langle\phi|)
\end{equation}
certainly holds for all $\varepsilon\in(\tilde\varepsilon,1]$.
For all $\varepsilon<1$ the initial state is also genuinely mixed.
\proofend

\newfigure{8.0cm}{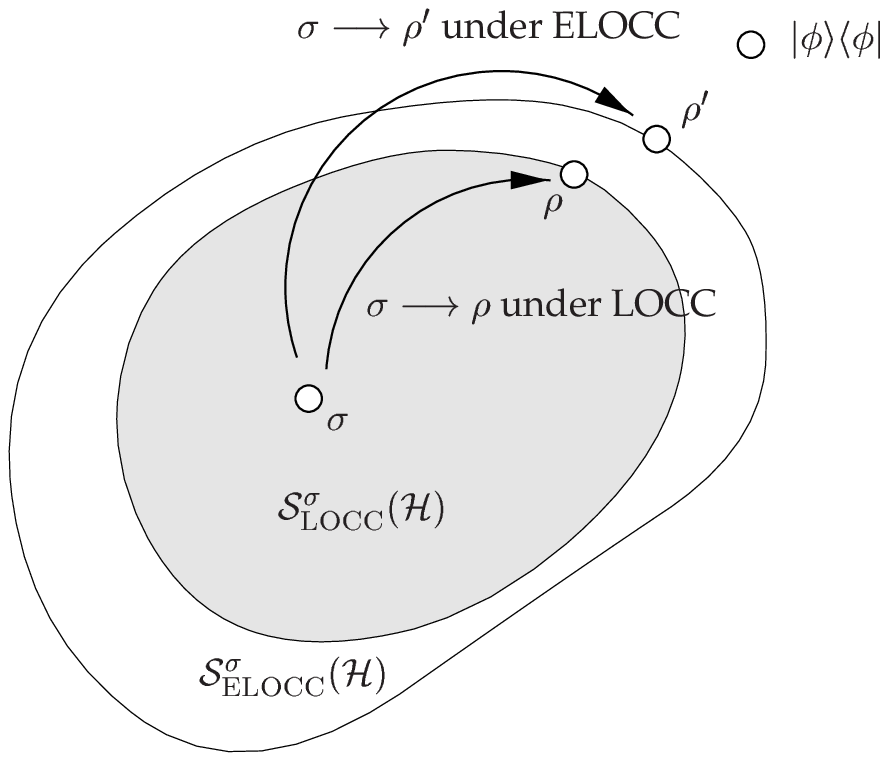}{Figure 3.5:
As in  Fig.\ 3.2
the state $\sigma$ cannot be transformed into
the desired final state $|\phi\rangle\langle\phi|$
under LOCC operations. 
The closest state to $|\phi\rangle\langle\phi|$
in the set ${\cal S}_{\rm LOCC}^\sigma({\cal H})$ as
measured by the fidelity is given by $\rho$.
In the set  ${\cal S}_{\rm ELOCC}^\sigma({\cal H})$ 
even the state $\rho'$ is accessible which is 
closer to $|\phi\rangle\langle\phi|$. Consequently,
the approximate transformation is more efficient
under ELOCC operations than under LOCC operations.}

\noindent{\bf Lemma 3.13. --} 
{\it Let $|\phi\rangle\in{\cal H}$, and let $|\psi\rangle\in{\cal H}$,
$|\psi\rangle=
|\psi\rangle_A\otimes |\psi\rangle_B$, 
be a product state vector.  Let
$\beta_1,...,\beta_N$   be the ordered
list of eigenvalues of ${\text{tr}}_A[|\phi\rangle\langle\phi|]$.
Then $F_{\rm LOCC}( |\psi\rangle\langle\psi| ,
|\phi\rangle\langle\phi|)=\beta_1$
holds.}

\probreak

\proof
Under LOCC operations
any separable state is accessible starting from
the product state $|\psi\rangle\langle\psi|$.
All separable states can be written in the form
$\rho=\sum_i p_i |\psi_i\rangle\langle\psi_i|$, 
where $|\psi_i\rangle\langle\psi_i|$ 
are pure product states and $p_1,p_2,...$ is a probability
distribution.  Under
these states the fidelity 
$\langle\phi|\rho| \phi\rangle$
is maximal if
$\rho=|11\rangle\langle 11 |$, and  
$F_{\rm LOCC}(|\psi\rangle\langle\psi|,|\phi\rangle\langle\phi|)=\beta_1$ holds.\proofend

\probreak

It should be noted that the set 
${\cal S}_{\rm ELOCC}^\sigma ({\cal H})$ is not automatically
a convex set. If both $\sigma\longrightarrow \rho_1$ and
$\sigma\longrightarrow \rho_2$ under ELOCC, then of course any
convex combination $\lambda \rho_1 + (1-\lambda) \rho_2$, 
$\lambda\in[0,1]$, can
be achieved through mixing. But it may not
be possible to assume
the additional states $\omega_1$
and $\omega_2$ that are needed to render the respective
entanglement manipulation possible to be identical. 
In general no state $\omega$ can be 
found such that
\begin{equation}
	\sigma\otimes \omega\longrightarrow
	\left(
	\lambda\rho_1 + (1-\lambda) \rho_2
	\right)\otimes \omega
\end{equation}
under LOCC. ${\cal S}_{\rm ELOCC}^\sigma ({\cal H})$
is nevertheless compact.\\

\noindent{\bf Example 3.14. --} The transformation 
investigated in Proposition 3.12  can be 
explored further, 
as $F_{\rm LOCC}(|\psi\rangle\langle\psi|,|\phi\rangle\langle\phi|)$
can be calculated exactly. 
Following Ref.\ \cite{VidalJonathanNielsen}
the quantities defined in  Eq.\ (\ref{Monoto})
are needed in order to evaluate the optimal achievable fidelity; 
they are given by
\begin{eqnarray}
E_1(|\psi\rangle)&=&1,\,\,\, 
E_2(|\psi\rangle)=1-0.4\varepsilon^2,\,\,\, 
E_3(|\psi\rangle)=1-0.8\varepsilon^2, \,\,\, 
E_4(|\psi\rangle)=1-0.9\varepsilon^2,\nonumber\\
E_5(|\psi\rangle)&=&1-\varepsilon^2;\nonumber\\
E_1(|\phi\rangle)&=&1,\,\,\, 
E_2(|\phi\rangle)=0.5,\,\,\, 
E_3(|\phi\rangle)=0.25, \,\,\, 
E_4(|\phi\rangle)=E_5(|\phi\rangle)=0.
\end{eqnarray}
According to Ref.\ \cite{VidalJonathanNielsen}
the state which maximizes the fidelity with respect to
$|\phi\rangle\langle\phi|$ among all states $\rho$
for which $|\psi\rangle\langle\psi|\longrightarrow \rho$ 
under LOCC is a pure state. This optimal pure state will be
referred to as 
$|\xi\rangle\langle\xi|$.
Now let $\varepsilon>\sqrt{15}/4$,
otherwise $F_{\rm LOCC}(|\psi\rangle\langle\psi|,|\phi\rangle\langle\phi|)=1$.
In this case $|\xi\rangle\langle\xi|$
can be evaluated as
\begin{equation}
	|\xi\rangle=
	\varepsilon\sqrt{\frac{8}{15} }|11\rangle+
	\varepsilon\sqrt{\frac{4}{15}  }|22\rangle+
	\sqrt{1-\frac{4}{5} \varepsilon^2 }|33\rangle,
\end{equation}
such that
\begin{equation}
	F_{\rm LOCC}(|\psi\rangle\langle\psi|,|\phi\rangle\langle\phi|)=
		|\langle\xi|\phi\rangle|^2=
		\left(
		5+8 \varepsilon^2 +4 \sqrt{3}\sqrt{5-4 \varepsilon^2} \varepsilon
		\right)/20.
\end{equation}
Fig.\ 3.6 shows an upper bound on
the maximally attainable fidelity under LOCC operations,
$F_{\rm LOCC}(\sigma,|\phi\rangle\langle\phi|)$, and
a lower bound on $F_{\rm ELOCC}(\sigma,|\phi\rangle\langle\phi|)$ 
for $\lambda=1/2$. 

%

\newfigure{7.5cm}{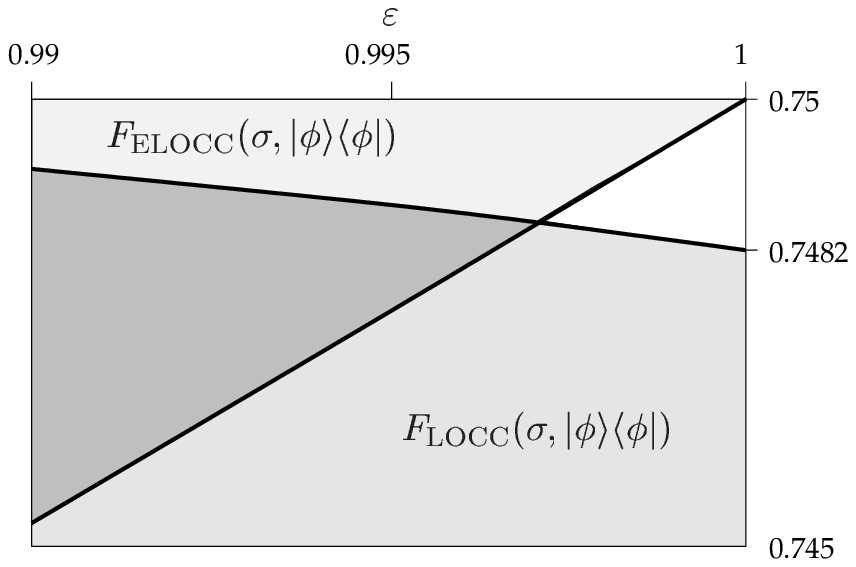}
{Figure 3.6: This diagram depicts an upper bound
on $F_{\rm LOCC}(\sigma,|\phi\rangle\langle\phi|)$ and
a lower bound on $F_{\rm ELOCC}(\sigma,|\phi\rangle\langle\phi|)$
as functions of $\varepsilon$. The parameter $\lambda$ is taken to 
be $1/2$. If $\varepsilon=1$,
$F_{\rm ELOCC}(\sigma,|\phi\rangle\langle\phi|)\geq3/4$
and $F_{\rm LOCC}(\sigma,|\phi\rangle\langle\phi|)\leq(5+8+4\sqrt{3})/20$.}

\subsection{Purification Procedures} 
The results of the two previous subsections
indicate that
the class of ELOCC operations is more powerful than
LOCC operations 
not only on the subset of
the boundary of ${\cal S}({\cal H})$ comprising 
the pure states, but 
also in the interior of the set 
${\cal S}({\cal H})$. This fact
suggests that the use of supplementary catalyst states opens up
possibilities to enhance general purification procedures.
However, ELOCC operations 
cannot improve the efficiency in the subsequent 
practically motivated problem:

\index{Probabilistic transformation}\index{SLOCC operation}
Clearly, any mixed state cannot be mapped on a maximally
entangled state in a deterministic transformation. However,
it may be transformed into a maximally
entangled state with a certain probability $p>0$.
Transformations in which the target state has to 
be reached with a probability $p>0$ are
called {\it probabilistic
transformations}\/ \cite{Vidal,Multi3} or
stochastic transformations, abbreviated as
SLOCC operations. The process of transforming
mixed states into maximally
entangled states under SLOCC operations is referred
to as {\it purification}\/ -- but the reader should be warned
that the term ``purification'' has at least four
different meanings\footnote{A) A SLOCC transformation
from a mixed state to a maximally entangled state, see, e.g., Refs.\
\cite{Kent,KentLinden}. B) Entanglement 
distillation \cite{BennettLett,BennettMust,Rains}
as described in Chapter 2.
C) The reconstruction of a pure state from many copies of 
mixed states which are a convex combination of the same pure state
and the maximally mixed state (this kind of purification aims
at  ``reversing'' the process of decoherence), see Refs.\ \cite{Cir,KeylRate}.
D) A pure state
on a larger Hilbert space
which is consistent 
with a given mixed state 
on a smaller Hilbert space. More general,
the purification of a
positive linear form is a 
lift to a pure linear form
of a larger algebra.}.\index{Purification}

This subsection concentrates on a 
particular physically
relevant class of mixed states, and a 
slightly more general 
problem than that of purification  is addressed.
This considered class of states
includes the set of states consisting
of a convex combination 
of some pure state 
and the maximally mixed state 
in the corresponding state space. 
It is of major interest in connection to
practical applications.\\

\noindent{\bf Proposition 3.15. --} {\it
Consider the class of states
\begin{equation}\label{combination}
	\sigma=\lambda |\psi\rangle\langle\psi|+(1-\lambda)\chi
\end{equation}
with the property that there exists a 
$\lambda_0\in(0,1)$ such that $\sigma$
is a separable state and so that every state with
a larger weight of $|\psi\rangle\langle\psi|$ is
entangled.  Furthermore, it is assumed that
$\langle\psi|\chi|\psi\rangle=0$.
Then the fidelity $F(\sigma,|\psi\rangle\langle\psi|)$
of $\sigma$ with respect
to $|\psi\rangle\langle\psi|$ cannot be
increased under ELOCC with a fixed operation
on the convex set $\{\sigma |
 \sigma=\lambda |\psi\rangle\langle\psi|+(1-\lambda)\chi
 \text{ for all }\lambda\in[0,1]\}$ 
with a non-vanishing probability.\\}

\probreak

The same class of states as in Proposition 3.15 has been
investigated in Ref.\ \cite{Kent} in the 
case that no catalyst state 
is available. It has been shown that 
for this class of states, the proportion of
$|\psi\rangle\langle\psi|$ cannot be increased
with SLOCC operations.
Proposition 3.15 shows that an analogous statement is
also true of probabilistic 
ELOCC operations: any catalyst
state cannot increase the efficiency of this
procedure.

Before continuing with the proof, the last statement
of Proposition 3.15 
might need some further explanation:
for the task of purification one specifies a 
certain LOCC operation, which is optimized with
respect to the states
$|\psi\rangle\langle\psi|$ and $\chi$
of the convex combination Eq.\ (\ref{combination}).
Under such conditions
it is then not possible 
to increase the proportion of $|\psi\rangle\langle\psi|$,
not even if one performs a selective generalized measurement
which corresponds to a non-trace preserving 
completely positive map. 
In a more general treatment, one could, however, 
include the possibility of modifying 
the operation on the particular initial state from the above set, 
such that the operation itself becomes dependent on the parameter $\lambda$.\\
\probreak

\proof\/ 
Let $\sigma\in{\cal S}({\cal H})$ be such a state,
and let $\omega\in {\cal S}({\cal K})$
be an appropriate catalyst state. The above
transformation then amounts to a map
\begin{equation}\label{trans}
	\sigma\otimes \omega\longmapsto
	\rho\otimes\omega=
	\frac{\sum\limits_{i=1}^K
	(A_i\otimes B_i)(\sigma\otimes \omega)(A_i\otimes B_i)^\dagger}
	{{\text{tr}} \left[\sum\limits_{i=1}^K
	(A_i\otimes B_i)(\sigma\otimes \omega)(A_i\otimes B_i)^\dagger
	\right]}\,\,.
\end{equation}
Since the map does not have to be trace-preserving,
$
	A_i$ and
$B_i$
corresponding to this separable operation
satisfy 
\begin{equation}
	\sum_{i=1}^K A_i^\dagger A_i\leq \mathbbm{1}_A,\,\,\,
	\sum_{i=1}^K B_i^\dagger B_i\leq \mathbbm{1}_B,
\end{equation}
but not necessarily $\sum_{i=1}^K A_i^\dagger A_i=\mathbbm{1}_A$ and
$\sum_{i=1}^K B_i^\dagger B_i=\mathbbm{1}_B$.
The operation defined by Eq.\ (\ref{trans})
is separable in the sense that 
$A_i$ and $B_i$ act only in
 ${\cal S}({\cal H}_A\otimes {\cal K}_A)$ and 
${\cal S}({\cal H}_B\otimes {\cal K}_B)$, 
respectively. 
The quantity to be considered is the fidelity 
\begin{equation}\label{fide}
	F(\rho, |\psi\rangle\langle\psi|)
	=\langle\psi|\rho|\psi\rangle
\end{equation}
of the normalized final state 
$\rho$ with respect to $|\psi\rangle\langle\psi|$.
This fidelity decreases under convex combination
of states, and therefore
it suffices to consider the smaller 
class of transformations which includes a single Kraus
operator $A_i\otimes B_i$ for some $i=1,...,K$.
Hence, it is no restriction on generality to 
investigate the map
\begin{equation}
\sigma\otimes \omega\longmapsto
	\rho\otimes\omega=
	\frac{
	(A_i\otimes B_i)(\sigma\otimes \omega)(A_i\otimes B_i)^\dagger}
	{{\text{tr}} \left[
	(A_i\otimes B_i)(\sigma\otimes \omega)(A_i\otimes B_i)^\dagger
	\right]}\,\,.
\end{equation}	
Regarded as a function of $\lambda$ the fidelity in Eq.\ (\ref{fide})
is given by
\begin{eqnarray}
	F(\lambda)&=&
	{\text{tr}}_{{\cal K}} \biggl(
	\lambda\,\langle\psi|
	\left[(A_i\otimes B_i) (
	|\psi\rangle\langle\psi|
	\otimes\omega)
	(A_i\otimes B_i)^\dagger
	\right] |\psi\rangle
	\nonumber\\
	&+&
	(1-\lambda)\langle\psi|\left[
	(A_i\otimes B_i) (\chi\otimes\omega)
	(A_i\otimes B_i)^\dagger\right] |\psi\rangle\biggr)/\xi(\lambda)\,\,,
	\nonumber
\end{eqnarray}
where the normalization constant $\xi(\lambda)$ is
\begin{eqnarray}
	{\xi}(\lambda)={\text{tr}}\left[
	(A_i\otimes B_i)
	((\lambda|\psi\rangle\langle\psi|
	+(1-\lambda)
	\tilde\sigma)	
	\otimes\omega)
	(A_i\otimes B_i)^\dagger
	\right]\,\,.\nonumber
\end{eqnarray}
%
The second derivative of $F$ with respect to $\lambda$ can
be calculated as
\begin{equation}
	\frac{dF^2(\lambda)}{d^2 \lambda} =
	\frac{
	{\text{tr}}_{{\cal K}}
	\langle\psi|
	\left[(A_i\otimes B_i) ((
	\chi-
	|\psi\rangle\langle\psi|)
	\otimes \omega)
	(A_i\otimes B_i)^\dagger
	\right] |\psi\rangle
        \,\xi'(\lambda)
	}{\xi(\lambda)^3}.
\end{equation}
In particular, it is of the form
\begin{equation}
	\frac{dF^2(\lambda)}{d^2 \lambda} = \frac{\Xi}{{\xi(\lambda)}^3}
\end{equation} 
with
a number $\Xi$ independent of $\lambda$, and it is
safe to argue that
the sign of the second derivative of the
function 
\begin{equation}
	f(\lambda)=
	F(\lambda)-\lambda
\end{equation} 
is constant for all $\lambda\in(0,1)$ 
(as in the case of local operations without
a catalyst state \cite{Kent}).
Therefore, the function $f$ must be either convex or concave or
linear over the whole interval. 
At $\lambda=0$, $f(0)\geq0$ as
	$f(\lambda)\geq-\lambda$ for all 
	$\lambda\in(0,1)$,
and $f(1)\leq0$. $f(\lambda_0)\leq0$ follows from the
fact that the map Eq.\  (\ref{trans}) cannot transform
the state pertaining to $\lambda_0$
to an entangled state. Hence, 
\begin{equation}
	f(\lambda)\leq0\,\text{ for all }\,
	\lambda\in[\lambda_0,1),
\end{equation}
i.e., the proportion
of $|\psi\rangle\langle\psi|$ can only decrease.\proofend

\subsection{Entanglement-Assisted Small Transformations}
A transformation from one pure state $|\psi\rangle\langle\psi|$
to another pure state $|\phi\rangle\langle\phi|$ is obviously
not always possible under LOCC operations,
even if the overlap $|\langle\psi|\phi\rangle|^2$
is arbitrarily close to 1 and the two states are therefore
arbitrarily similar as measured by the fidelity.
Instead, for a given  $|\psi\rangle\langle\psi|$ 
and for any $\delta$ there exist pure states $|\phi\rangle\langle\phi|$ 
with $|\langle\psi|\phi\rangle|^2>1-\delta$ 
which cannot be 
accessed by these operations.
One might suspect that if such a ``small transformation'' is 
not possible, entanglement-assisted operations might 
under certain circumstances 
provide the tools to perform the task, because 
the resources required would be small in any case.

However, this intuitive approach does not generate
a valid statement as will be shown in
Proposition 3.16:
For a $|\psi\rangle\langle\psi|$
there is a $\delta>0$
such that for all pure states $|\phi\rangle\langle\phi|$ 
with $|\langle\psi|\phi\rangle|^2>1-\delta$ the presence of
catalyst states does not imply an advantage at all,
and a transformation under ELOCC is only
possible if it can already be realized with LOCC operations. 
In the mixed state regime things are again different, see Fig.\ 3.7.
Quite surprisingly, it has become apparent that
pure states behave differently from mixed states as
far as small transformations are concerned.\\

\newfigure{7.0cm}{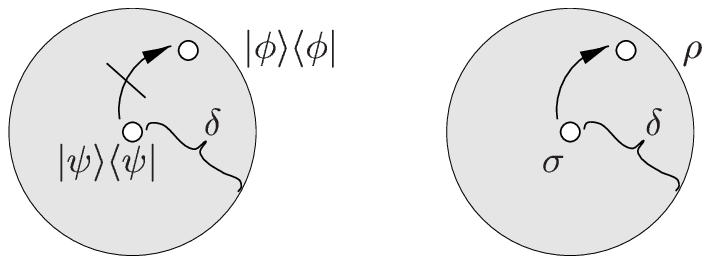}{Figure 3.7: A 
schematic representation of 
entanglement-assisted small transformations
in the pure state case and for mixed states.
For a given pure state $\sigma=|\psi\rangle\langle\psi|$
there exists a $\delta>0$
such that for all pure states 
$\rho=|\phi\rangle\langle\phi|$ with
$|\langle\psi|\phi\rangle|^2>1-\delta$
no transformation $\sigma\longrightarrow\rho$  
is possible under
ELOCC. This is represented in the
picture on the left hand side. The other 
figure depicts a possible transformation
in the mixed state case 
where there are states $\sigma$ such that
for every $\delta>0$ there exist other mixed states
with $F(\sigma,\rho)>1-\delta$ and
$\sigma\longrightarrow\rho$ under ELOCC. }

\noindent {\bf Proposition 3.16. --} {\it
For all
 $|\psi\rangle\in{\cal H}$ 
 and all $|\tilde\psi\rangle\in{\cal K}$
there exists a $\delta>0$ such that 
\begin{eqnarray}
|\psi\rangle\langle\psi|\not\longrightarrow
|\phi\rangle\langle\phi|{\text{ under LOCC}}
\,\,\,
\Longrightarrow
\,\,\,
|\psi\rangle\langle\psi|
\otimes
|\tilde\psi\rangle\langle \tilde\psi|
\not\longrightarrow |\phi\rangle
\langle\phi|\otimes
|\tilde\psi\rangle\langle \tilde\psi|
{\text{ under LOCC}}
\nonumber
\end{eqnarray}
for all $|\phi\rangle\in {\cal H}$
with $|\langle\psi|\phi\rangle|^2>1-\delta$. \\
}

\probreak

\proof Let 
\begin{equation}
\alpha_1, ..., \alpha_N,\,\,\,\,\,\, N=\dim[{\cal H}_A]
\end{equation}
be the ordered lists of eigenvalues of 
${\text{tr}}_A[|\psi\rangle\langle\psi|]$,
and let 
$\gamma_1, ..., \gamma_M$, $M=\dim[{\cal K}_A]$,
be the corresponding list of the pure catalyst state.
Let $\varepsilon>0$ 
and call an $\varepsilon$-list a list
\begin{equation}
\beta_1, ...,\beta_N,\,\,\,\,\,
1\geq \beta_1\geq ...\geq \beta_N\geq0
\end{equation} 
that has the
property 
$
|\beta_i-\alpha_i|<\varepsilon
$
for all $i=1,...,N$.
There exists an $\varepsilon_0>0$ such that
for all $\varepsilon$-lists
$\beta_1$, ..., $\beta_N$
the statement that 
$\alpha_i\gamma_j>\alpha_k\gamma_l$ for some
$i,k=1,...,N$ and  $j,l=1,...,M$ implies that
\begin{equation}
\beta_i\gamma_j>\beta_k\gamma_l.
\end{equation} 
Additionally, 
there exists a $\delta>0$ such that 
for each $|\phi\rangle\in {\cal H}$
with 
\begin{equation}
|\langle\psi|\phi\rangle|^2>1-\delta,
\end{equation}
the ordered
eigenvalues of ${\text{tr}}_A[|\phi\rangle\langle\phi|]$
form an $ \varepsilon_0$-list  -- and
hence, for such states it is not possible that
$\beta_i\gamma_j<\beta_k\gamma_l$ and
$\alpha_i\gamma_j>\alpha_k\gamma_l$. 
It follows that for all
such $|\phi\rangle\in {\cal H}$
with $|\langle\psi|\phi\rangle|^2>1-\delta$ 
the majorization relation
\begin{equation}
	{\text{tr}}_A[|\psi\rangle\langle\psi|\otimes\omega]\not\prec
{\text{tr}}_A[|\phi\rangle\langle\phi|\otimes\omega]
\end{equation} 
holds if 
\begin{equation}
{\text{tr}}_A[|\psi\rangle\langle\psi|]\not\prec
{\text{tr}}_A[|\phi\rangle\langle\phi|].
\end{equation} 
Finally, this  implies
the statement of Proposition 3.16.\proofend

A corresponding statement does not hold for mixed states.
The generalized fidelity
of two states $\sigma$ and $\rho$ can, e.g., be
taken to be \cite{Uhlmann,Fuchs}
\begin{equation}
F(\sigma,\rho)= \left({\text{tr}} \left[(\sqrt{\sigma}\rho\sqrt{\sigma})^{1/2}\right]\right)^2.
\end{equation}
The statement for mixed states can then be formulated
as follows.\\

\noindent {\bf Proposition 3.17. --} {\it
There exist 
states $\sigma\in{\cal S}({\cal H})$ such that
for every $\delta>0$ there exist states $\rho\in{\cal S}({\cal H})$
with the property that 
\begin{equation}
F(\sigma,\rho)>1-\delta
\end{equation}
and
$
	\sigma\not\longrightarrow\rho $
	 under LOCC,
but
$
	\sigma\longrightarrow\rho $
	under ELOCC.}\\
	
Such states can, e.g., be constructed using the class of states
defined in Eq.\ (\ref{example}), Eq.\ (\ref{psi}), and
Eq.\ (\ref{phi}). For any  given $\delta>0$
there is a sufficiently small $\lambda>0$ such that 
the fidelity satisfies 
$F(\sigma,\rho)>1-\delta$.
Hence, quite surprisingly, in the case of 
entanglement manipulations
from an initial pure state to a close pure state
entanglement-assisted operations do not add any power to LOCC 
operations. Put differently, there
is no catalysis for sufficiently close pure states. 
Yet, for mixed states there can be catalysis for 
such close states.


\section{Concluding Remarks}

In this chapter entanglement transformations
have been investigated. In particular,
it has been the emphasis of this chapter
to explore the power of entanglement-assisted 
manipulation of entangled quantum systems in 
mixed states. It became apparent that the
class of ELOCC 
operations, despite their counterintuitive character, 
is superior to mere LOCC operations
also in the interior of the state space, 
for which such strong tools as the 
majorization criterion of Ref.\ \cite{Nielsen}
are not available. 
It is the hope that these findings 
contribute significantly to
the quest for a 
better understanding of mixed-state entanglement 
transformations.

There remain, however, numerous open questions.
A general criterion for entanglement transformations to
be possible in the mixed state domain 
would be very useful. The isomorphism 
\index{Isomorphism}
between quantum operations and quantum states
of bi-partite systems as presented in Ref.\ \cite{SmallGate}
(going back to Ref.\ \cite{Jamiolkowski})
could possibly point towards a resolution of the general
problem. Also, the characterization of
mixing \index{Mixing} and measurement
in terms of majorization relations for
the eigenvalues of the involved matrices could
be useful \cite{NewNielsen} when approaching this 
problem.

There might also be a connection between entanglement-assisted
local operations and bound entanglement. \index{Bound entanglement}
As has been 
mentioned in the first chapter bound entangled states
cannot be transformed into free entangled states using
LOCC operations. The main implication is as follows:
Let $\sigma\in{\cal S}({\cal H})$
with $\sigma^{T_A}\geq 0$
on a Hilbert space ${\cal H}=\mathbbm{C}^N \otimes \mathbbm{C}^N$
of a bi-partite $N\times N$-system.
One can safely say that
any state $\rho$ 
which satisfies 
$\sigma\longrightarrow \rho$ under LOCC
is not distillable \cite{Primer}.
However, for stochastic
ELOCC operations one may
conjecture the validity of the following statement.\\

\noindent {\bf Conjecture 3.18. --} 
{\it
There exist states $\sigma\in {\cal S}({\cal H})$ 
and $\omega\in {\cal S}({\cal K})$ (${\cal H}$ and
${\cal K}$ are appropriate
Hilbert spaces of bi-partite systems)
satisfying}
\begin{enumerate}
\item[(i)] $\sigma^{T_A}\geq 0$,
\item[(ii)] $\sigma\otimes \omega \longrightarrow \rho\otimes \omega$
{\it under SLOCC, such that
$\rho$ is distillable, that is, such that
there exists a $K\in{\mathbbm{N}}$
and a state vector $|\psi\rangle$
taken from a $2\times2$-dimensional subspace ${\cal C}\subset 
{\cal H}^{\otimes K}$ 
such that
\begin{equation}\label{distentdef}
	\langle\psi|(\rho ^{T_A})^{\otimes K}|\psi\rangle<0.
\end{equation}
}
\end{enumerate}

\probreak

This is not the same as
activating bound entanglement in the sense
of Ref.\ \cite{HoroBoundAct}, as here
the entanglement of the free entangled 
catalyst state is not used up in the course of
the protocol.
A priori it is not clear  why 
such a transformation should not be possible under ELOCC
operations. 
This would imply that
bound entanglement could be 
\index{Unlocking bound entanglement}
''unlocked'' without
using up any free entanglement, that is, it could be 
transformed
into a form of entanglement which can be distilled to 
singlet form. \\


\chapter{Non-Local Implementation of Joint Unitary Operations}

\section{Introduction}

In order to better understand 
the properties and structure of quantum
entanglement the typical class of
quantum operations on bi-partite quantum
systems has been studied in the previous chapter.
In entanglement transformations both the initial state and
the desired final state are known, and it 
is a matter of investigation
to see if or to what degree the task of transforming the states
into each other can be accomplished with a given physical 
instrument.
But what if the initial and the final state
are not known in advance? 
In particular, joint unitary operations have major
practical implications. How can
{\it joint unitary operations}\/
\index{Joint unitary operations}
at different nodes be implemented
without knowing the involved states? And what resources
are needed to perform such a task?
In the words of Ref.\ \cite{Linden}, the problem is to 
characterize the non-local character of unitary
operations. This non-locality 
\index{Non-locality}
is measured in terms of
the used resources, namely, shared quantum systems
in entangled states and the
amount of necessary classical communication.

In mathematical terms, the problem of 
{\it non-local
implementation}\/ 
of joint unitary operations
can be formulated as follows.
Let ${\cal H}$ be the Hilbert space
of a multi-partite quantum system (at least
bi-partite). For any
unitary operator $U:{\cal H}\longrightarrow {\cal H}$ 
there exists a Hilbert space
${\cal K}$ of another multi-partite quantum system, 
a state $\omega\in{\cal S}({\cal K})$
and a trace-preserving
LOCC operation ${\cal E}$
such that \index{LOCC operation}
\begin{equation}
	 {\cal E}(\sigma\otimes\omega)
	= U\sigma U^\dagger\otimes \omega'
\end{equation}
for all $\sigma\in{\cal S}({\cal H}) $,
with $\omega'\in {\cal S}({\cal K})$. 
Local quantum operation here means that
the quantum operations are performed
on those parts of ${\cal H}$ and ${\cal K}$
that correspond to quantum systems which are
physically held by the same parties.
The question now is: for a given $U$, 
what is the minimal amount of classical communication
needed in order to implement ${\cal E}$, 
and what is the minimal entanglement
of the state $\omega$?

This matter is not only of academic interest. 
From a practical point of view,
the problem is essentially how to implement elementary
{\it quantum gates}\/ in a quantum computer which consists
of several parts at remotely located 
positions. Such a so-called 
\index{Distributed quantum computer}
\index{Quantum computer}
{\it distributed quantum computer}\/ 
may have -- once realized --
several advantages compared to a quantum computer
located at a single site.
The actual construction of a full large scale quantum computer
is hindered by
daunting problems. 
The main obstacle
to the experimental realization of such a 
quantum computer is the process of 
\index{Decoherence}
decoherence \cite{PT,Dec}. Quantum systems in
pure states get entangled with their environment
and are degraded in purity, which makes the   
manipulation and storing of sufficiently
many quantum systems 
\cite{Steane98,Wineland}
extraordinarily difficult.
Leaving the fundamental problems aside, imperfections in
applying 
the appropriate quantum operations cannot
be avoided. In a quantum optical implementation, say,
one would have to expect fluctuations in timing, length, and intensity
of the applied laser pulses.
It has been shown that under non-ideal conditions
a distributed quantum computer may be
superior to a non-distributed quantum 
computer for certain problems \cite{Grover,Huelga}.

In this chapter several protocols implementing gates
that effect qubits at different nodes will be presented
using only LOCC operations and previously shared 
entanglement. Optimality is
measured by 
the consumption of the basic experimental
resources of entanglement and classical communication between
nodes. As the emphasis of the chapter is on 
rather practical issues, 
all considerations are restricted to the case
where the quantum systems at different nodes are
qubits.
The results 
which are presented in this chapter
can be found in published
form in 
[E4]
\footnote{Independently, 
similar results were published 
on the eprint server 
of the Los Alamos National Lab on the same day 
by D.\ Collins, N.\ Linden, and
S.\ Popescu \cite{Linden}. Also, the results presented
in Ref.\ \cite{Gottesman} -- an 
expanded version of a plenary 
speech given at the 1998 International Conference on 
Group Theoretic Methods in Physics --
imply a scheme for the non-local 
implementation of a CNOT gate the authors of Ref.\ 
[E4] were not aware of.}.

\section{Quantum Gates}

Computation can be made reversible, both logically and 
thermodynamically \cite{BenRev}.
\index{Quantum gate}
Quantum computation amounts to the implementation of one
or more than one unitary quantum operation. 
In essentially all proposals for quantum computation
these unitary operations are 
built up of elementary quantum operations
-- or {\it quantum gates}\/ \cite{gates,Deu85,Deu89,MikesBook} --
which are applied on one, two or more separated
quantum systems (for an exception see Ref.\ \cite{Lattice2}).
{\it Single qubit gates}\/ are unitary operations on ${\mathbbm{C}}^2$
with basis $\{|0\rangle, |1\rangle\}$
and can be represented by unitary $2\times 2$ matrices.
\index{Hadamard gate}
Among the single qubit gates the {\it Hadamard gate}\/ and the
{\it $\sigma_z$ operation}\/ 
will be used subsequently.
The Hadamard gate is defined by the map
\begin{equation}
	|0\rangle\longmapsto (|0\rangle+|1\rangle)/\sqrt{2},\,\,\,\,\,\,\,\,\,\,
	|1\rangle\longmapsto (|0\rangle-|1\rangle)/\sqrt{2},
\end{equation}
a $\sigma_z$ operation is specified by
\begin{equation}
	|0\rangle\longmapsto |0\rangle,\,\,\,\,\,\,\,\,\,\,
	|1\rangle\longmapsto -|1\rangle.
\end{equation}
The quantum {\it CNOT gate}\/ \index{CNOT gate}
is an operation on ${\mathbbm{C}}^2\otimes
{\mathbbm{C}}^2$ and acts as
\begin{equation}
	|e \rangle_A|f\rangle_B\longmapsto |e \rangle_A |e\oplus f\rangle_B,
\end{equation}
where $e ,f\in\{0,1\}$; $\oplus$ denotes addition modulo two. 
In accordance with their roles
in the transformation the first qubit is called {\it control qubit}\/, 
the second
{\it target qubit}\/. 
For $N$ qubits acting as nodes, $N=2,3,...$, let 
\begin{equation}
	U=
	\left(
	\begin{array}{cc}
		u_{00}&u_{01}\\
		u_{10}&u_{11} 
	\end{array}
	\right)
\end{equation}
be the matrix representation of a unitary operator
and let
\begin{eqnarray}
	&&\Lambda_n(U) \, |x_1,...,x_{N-1},y\rangle\\
	&=&
	\left\{
	\begin{array}{ll}
	u_{y0}|x_1,...,x_{N-1},0\rangle
	+
	u_{y1}|x_1,...,x_{N-1},1\rangle, & 
	\text{if $x_1=...=x_{N-1}=1$},\\
	|x_1,...,x_{N-1},y\rangle & \text{otherwise.}
	\end{array}
	\right.\nonumber
\end{eqnarray}
$\Lambda_N(U) $ is an $N$ node {\it control-U gate}\/ \cite{gates}.
In matrix form the gate $\Lambda_N(U)$ can be represented
as 
\begin{equation}
\left(
\begin{array}{ccccc}
1& & & & \\
 &1& & & \\
 & &\ldots & & \\
 & & & u_{00}& u_{01}\\
 & & & u_{10}& u_{11}
\end{array}
\right),
\end{equation}
where again, the basis states are lexicographically ordered.
If
\begin{equation}
	U=
	\left(
	\begin{array}{cc}
		0&1\\
		1&0 
	\end{array}
	\right)
\end{equation}
and $N=2$, then such a control-U gate reduces
to the quantum CNOT gate. For $N=3$ this particular kind of
control-U gate is called {\it Toffoli gate}\/. \index{Toffoli gate}

\newfigure{10.0cm}{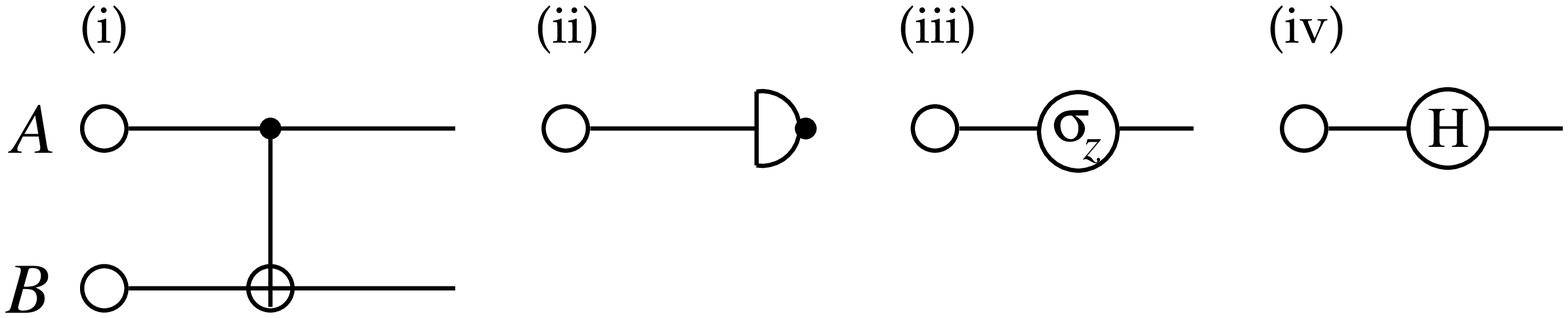}{Figure 4.1:
A schematic 
representation of (i) a quantum CNOT gate 
where $A$ is the control qubit and $B$ is the target.
The CNOT maps 
$|00\rangle\mapsto |00\rangle$,
$|01\rangle\mapsto |01\rangle$,
$|10\rangle\mapsto |11\rangle$, and
$|11\rangle\mapsto |10\rangle$.
(ii) is the symbol of 
a projective selective
measurement in the 
computational basis
(with Kraus operators $E_1=|0\rangle\langle0|$
and $E_2=|1\rangle\langle1|$),
(iii) stands for a $\sigma_z$  gate, and (iv) 
depicts a Hadamard transformation.}

\section{Implementation of Two-Qubit Gates in Distributed Quantum
Computation}

\newfigure{8.5cm}{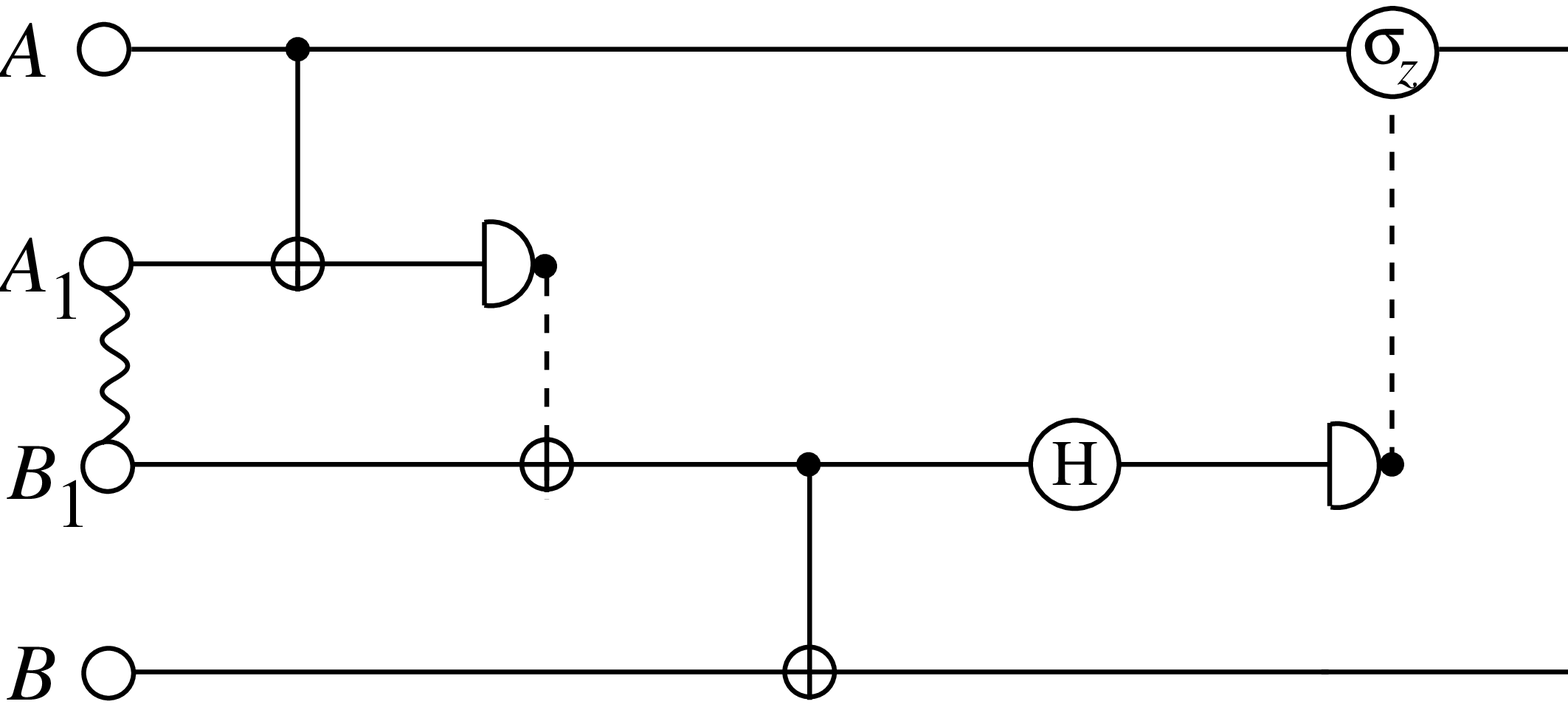}
{Figure 4.2: A representation of a quantum
circuit to perform a quantum CNOT gate with
the minimal use of resources. The qubits
$A$ and $B$ are the qubits on which the
CNOT gate is applied. In addition to $A$ and
$B$ Alice and Bob hold the qubits $A_1$ and
$B_1$ respectively, which are initially
in a maximally entangled state. Each dashed
line corresponds to a single bit of classical
communication.
}

\index{Two-qubit gate}
The first result is concerned with the optimal
non-local implementation of a quantum CNOT gate.
Optimal means in this context that the least
possible resources are used up, both as far as
shared entanglement and classical communication 
are concerned. \\
\index{Classical communication}
\index{Quantum circuit}

\noindent {\bf Proposition 4.1. --} {\it
A single shared ebit and one bit of classical communication in each direction 
are necessary and sufficient for the non-local
implementation of a quantum CNOT gate.}\\

\proof (i) 
 {\it Sufficiency}\/: This statement can be proved
 by constructing an explicit circuit
 which performs the CNOT using these resources, see
 Fig. 4.2. The CNOT is
performed between the qubits $A$ and $B$. Alice holds the qubits $A$
and $A_1$, and Bob holds the qubits $B$ and $B_1$. 
The state of $A_1$ and $B_1$ will be taken to be
a maximally entangled state corresponding to
$(|00\rangle_{A_1 B1} + |1 1 \rangle_{A_1 B1})/\sqrt{2}$.
The initial state of $A$ is 
necessarily arbitrary, it can, however, 
without loss of generality be taken to be a pure state
$\alpha |0\rangle_A +
\beta |1\rangle_A$.
The initial state of $B$ is also arbitrary and
is given by $\gamma |0\rangle_B + \delta |1\rangle_B$. 
First a local CNOT is
performed with $A$ as the control and $A_1$ as the target. After this
the combined state\footnote{
In this chapter both Hermitian, positive
trace-class operators with trace one and
state vectors 
are called states.} of $A$, $A_1$ and $B_1$ is
\begin{equation}
    \frac{1}{\sqrt{2}}(\alpha |000\rangle_{A A_1 B_1} + \alpha |011\rangle_{A A_1 B_1} +
    \beta |110\rangle_{A A_1 B_1} + \beta |111\rangle_{A A_1 B_1}).
\end{equation}
Alice then performs a measurement on $A_1$ in the computational basis,
and the line corresponding to this qubit terminates.
The result of the measurement is one bit of information, which is
communicated to Bob, and this communication is denoted by the
dashed line. If the result is 0 Bob does nothing, and if the
result is 1 Bob performs the not operation. At this point the
combined state of $A$ and $B_1$ is $\alpha |0 0\rangle_{A B_1} + \beta
|1 1\rangle_{A B_1}$. That is, this procedure amounts
effectively to the implementation of a CNOT
between $A$ and $B_1$.
Now particle $B_1$ contains the necessary information about the
state of $A$. One may 
now perform a CNOT between $B_1$ and $B$. The
combined state of $A$, $B_1$ and $B$ is 
\begin{equation}
  \frac{1}{\sqrt{2}}(\alpha\gamma |0 0 0 \rangle_{A B_1 B} + 
\alpha\delta |0 0 1 \rangle_{A B_1 B} +
  \beta \delta|1 1 0 \rangle_{A B_1 B} + 
\beta\gamma |1 1 1 \rangle_{A B_1 B})
\end{equation}
All that needs to be done is to remove $B_1$ from the state. This is
done by performing a Hadamard transformation on $B_1$, and then
measuring $B_1$ in the computational basis, at which point the line
denoting $B_1$ terminates. The result of the measurement (one bit) is
communicated to Alice. If the result is 0 Alice does nothing,
and if the result is 1 she performs a (state-independent)
$\sigma_z$ operation on particle $A$. This completes the non-local
CNOT. 

(ii) {\it Necessity}\/: 
The procedure consists of LOCC operations, and hence, 
all information which has been
sent at the end of the operation must have been sent classically.
If the target qubit is
initialized in the state $|0\rangle$, then its final state will be
$|0\rangle$ or $|1\rangle$ depending on the initial state of the
control qubit being $|0\rangle$ or $|1\rangle$
respectively. Therefore, the final result of the gate in this case
is the communication of one bit of information from
Alice (holding the control qubit) to Bob (holding the target
qubit). Consequently, in the non-local implementation, one bit of
classical information must have been sent classically from Alice
to Bob. 
In order to understand that one bit must also have been sent from Bob to
Alice, note that in the basis $|\pm\rangle =
(|0\rangle \pm |1\rangle)/\sqrt{2}$ the role of control and target
in a CNOT gate are reversed, see Fig.\ 4.3. 
Assume that Alice's particle is
prepared in the standard state $|+\rangle$ and Bob chooses to
prepare his particle either in state $|+\rangle$ or $|-\rangle$.
After the application of the CNOT gate,
Alice will hold a quantum system
which is either in the state $|+\rangle$ or $|-\rangle$ depending on
the state Bob's particle has been prepared in. Consequently, 
one bit of information has been transmitted from Bob to Alice. As the
implementation of the CNOT must be independent of the initial
state, the procedure has to  allow for one bit of communication in
each direction. Thus 
the non-local implementation
must involve, as a minimum, one bit of communication in both
directions.
A CNOT gate acting on the initial state
$(|0\rangle_A+|1\rangle_A)|0\rangle_B$ leads to a maximally entangled
state $(|0 0 \rangle_{AB}+|1 1 \rangle_{AB})/\sqrt{2}$, which is 
why at least one ebit is required.
\proofend

\probreak

\newfigure{8cm}{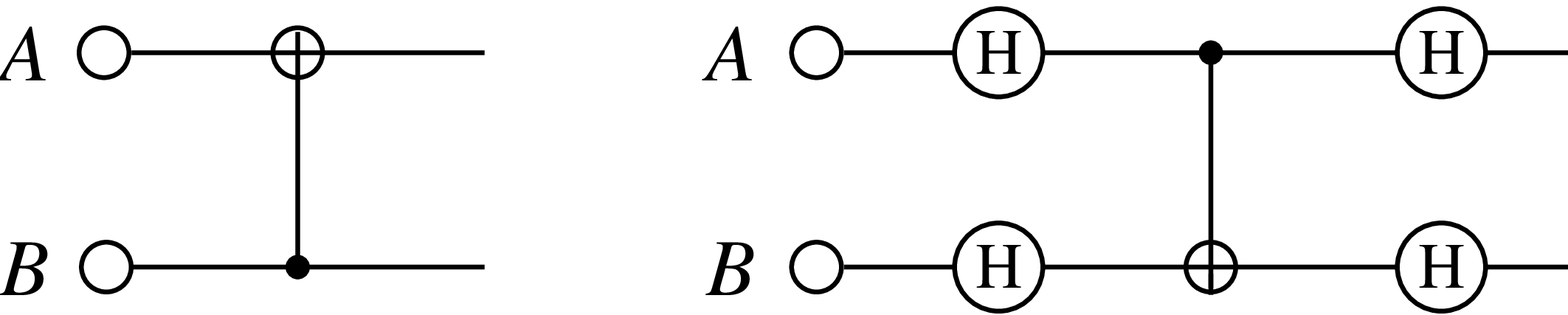}
{Figure 4.3: Two equivalent formulations of a quantum CNOT gate
with qubit $A$ as target and $B$ as control qubit.}

\index{Control-U gate}
A control-U gate can be implemented using one shared
ebit and one bit of classical communication in each direction.
A control-U gate is defined as a gate that applies
the identity on the target qubit if the control bit is in state
$|0\rangle$ and it applies the unitary operator $U$ to the target
if the control qubit is in state $|1\rangle$. It is possible to 
use the same quantum
circuit as in Fig. 4.2 except that the CNOT gate on Bob's
side is replaced by a control-U gate.

In general a single application of a control-U gate cannot be
employed to create one ebit from an initial product state of
two qubits. Also, the amount of classical information
that can be sent from Alice to Bob via a general control-U gate
is less than or equal to one bit. In fact, it has been shown very
recently 
that one can implement certain phase gates with
a very small amount of initial entanglement \cite{SmallGate}.
In particular, this proposal 
provides the first
example of a non-local implementation which makes use
of a non-integer number of ebits. The basis for this approach
is an isomorphism between certain quantum operations and
quantum states  going back to Ref.\ \cite{Jamiolkowski}.\\

\noindent {\bf Lemma 4.2. --} {\it 
Two bits of classical communication in both directions
and two shared ebits are 
sufficient for the non-local implementation
of a general two-bit gate. This upper bound cannot be made
smaller.}\\

\proof 
Any operation may be performed by teleporting
Alice's state to Bob \cite{BennettTele}, 
at which point Bob may locally perform the
operation and then teleport the resulting state back to Alice.
This procedure requires two bits of communication in each
direction and two shared ebits.
\index{State swapper}
This bound may be saturated: 
The {\it state swapper}\/ mapping unitarily
$|\psi\rangle_{A} \otimes |\phi\rangle_{B}$
on
$|\phi\rangle_{A}\otimes |\psi\rangle_{B}$
-- that is, applying a permutation operator --
requires two bits of classical communication
and two ebits of entanglement (see Fig. 4.4): 
By single
use of a state swapper on two qubits two ebits
of entanglement can be created. Accordingly, two
initial ebits of entanglement are necessary
in the local implementation. Using a 
state swapper of the above type enables
Alice to send two bits of classical information
to Bob (and vice versa). This can be done 
using a 
\index{Dense coding}
{\it dense coding protocol}\/ \cite{BennettDense}.
It thus becomes apparent that 
two bits of classical information
need to be sent in both directions. \proofend

\newfigure{5.5cm}{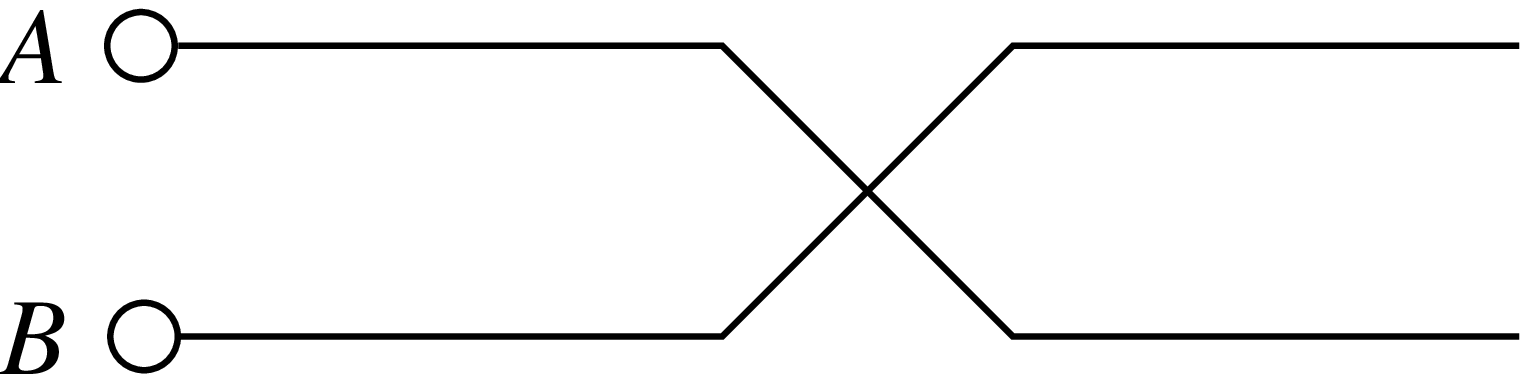}
{Figure 4.4: Schematic representation of the state swapper.}

As any quantum gate array can be constructed
from quantum CNOT gates and one-qubit
gates only \cite{gates}, there is a realization of a state swapper
on two qubits using only those gates. 
In can be shown that
in such an implementation three CNOT gates
would be necessary \cite{PlenioPriv} (see Fig.\ 4.5).
This procedure would, however, make use of
three ebits of entanglement instead of two.

\newfigure{4.4cm}{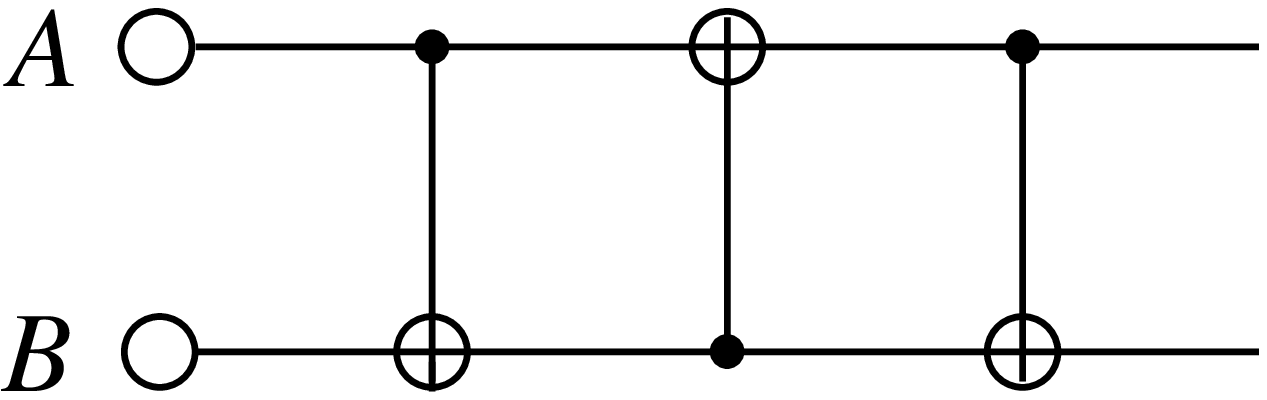}
{Figure 4.5: A state swapper represented in terms of CNOT gates.}

\section{Implementation of Multi-Qubit Gates}
 \index{Toffoli gate}
In this subsection, 
it will be presented how the implementation
of gates can be generalized to 
certain multi-qubit
gates, that is, to
the realization of 
gates where more than two parties are involved.
At first, 
the implementation of the Toffoli gate 
will be investigated. The generalization to multi-party
gates will then become evident.\\

\noindent {\bf Proposition 4.3. -- } {\it
Two shared ebits and four bits of classical communication
are necessary and sufficient for the local
implementation of a non-local three-party quantum Toffoli gate.}\\

\proof 
(i) {\it Sufficiency}\/: 
Again, a certain circuit will be constructed to
perform the task, see Fig.\ 4.6. 
Alice, Bob, and Charles 
initially hold the qubits labeled $A$, $B$, and
$C$ respectively such that
${\cal H}={\mathbbm{C}}^2
\otimes{\mathbbm{C}}^2\otimes {\mathbbm{C}}^2$.
In addition to this, let
Alice and Charles share a pair $A_1$ and $C_1$
of qubits in a maximally entangled state $|\phi^+\rangle=
(|0 0 \rangle_{A_1 C_1} + |1 1  \rangle_{A_1 C_1})/\sqrt{2}$,
and let Bob and Charles share another pair $B_1$
and $C_2$ in the same maximally entangled state. The initial
state $|\psi\rangle$ 
of $A$, $B$, $C$, $A_1$, $B_1$, $C_1$ and $C_2$
is then given by
\begin{equation}
	|\Psi\rangle=
	|\psi\rangle_A\otimes
	|\phi\rangle_B\otimes
	|\varphi\rangle_C\otimes
	|\phi^+\rangle_{A_1 C_1}
	\otimes
	|\phi^+\rangle_{B_1 C_2},
\end{equation}
where
\begin{equation}
	|\psi\rangle_A=\alpha|0\rangle_A+\beta|1\rangle_A,\,\,\,
	|\psi\rangle_B=\gamma|0\rangle_B+\delta|1\rangle_B,\,\,\,
	|\psi\rangle_C=\eta|0\rangle_C+\xi |1\rangle_C.
\end{equation}
The first step is a local quantum CNOT gate on
$A$ and $A_1$ with $A$ as control. 
Then Alice measures
particle $A_1$ and Charles performs a NOT operation
on his particle $C_1$ if Alice finds $1$ in the measurement
and the identity if Alice finds $0$. 
Qubit $A_1$ is then discarded, and
Bob applies a local CNOT with $B$ being the control and $B_1$
being the target. Then Bob measures particle 
$B_1$ and Charles 
performs
a NOT operation on his 
particle $C_2$ if Bob finds $1$ and
the identity if Bob finds $0$. Qubit $B_1$ is then
discarded.
Now the state of the remaining qubits $A$, $B$, $C$, $C_1$
and $C_2$ is given by
\begin{eqnarray}\label{stage}
    (\alpha|00\rangle_{A C_1}+\beta|11\rangle_{A C_1})\otimes
    (\gamma|0 0 \rangle_{B C_2} + \delta |1 1 \rangle_{B C_2}) \otimes
    |\psi\rangle_C 
\end{eqnarray}
In a further step 
Charles locally
applies a Toffoli with $C_1$ and $C_2$ being the
control qubits. Charles then
applies Hadamard gates to
both $C_1$ and $C_2$. He 
measures $C_2$ in the computational basis
and applies
$\sigma_z$ or the identity $\mathbbm{1}$ 
to $B$ if his result is $1$
or $0$, respectively. Finally he measures $C_1$ and applies
$\sigma_z$ or the identity to $A$ if his result is 1 or 0.
This completes the Toffoli gate.
The total number of classical bits which have to be communicated
is four, and two shared ebits of entanglement
are consumed.

(ii) {\it Necessity}\/: 
Two ebits of entanglement being necessary is taken to mean that
the pure initial state of the auxiliary quantum systems
with Hilbert space ${\cal K}$ 
is a product of two maximally
entangled states in 
${\mathbbm{C}}^2 \otimes {\mathbbm{C}}^2$ each.
Assume first that $|\psi\rangle_A=|1\rangle_A$.
Then the initial state
\begin{equation}
    |\psi\rangle = |1\rangle_A (\alpha |0\rangle_B + \beta
    |1\rangle_B) (\gamma |0\rangle_C + \delta |1\rangle_C) 
\end{equation}
is mapped on
\begin{equation}
    |\phi\rangle = |1\rangle_A (\alpha\gamma |00\rangle_{B C} + \alpha\delta|01\rangle_{B C}
    + \beta\gamma |11\rangle_{B C} + \beta\delta |10\rangle_{B C})
\end{equation}
by application of the three-party quantum
Toffoli gate. Hence, in this case, the quantum
Toffoli gate amounts to a quantum CNOT between the qubits
$B$ and $C$. According to Proposition 4.1 it follows that
a single classical bit of information has to be
exchanged in both directions between Alice
and Charles. Also, 
Alice and Charles have to effectively
share one ebit of entanglement. That is,
the initial state
of the auxiliary qubits labeled 
$A_1$ and $C_1$ has to have the
property that with local operations and classical
communication between Alice, Bob, and Charles
this state can be transformed into a maximally
entangled state between Alice and Charles
with unit probability. The same argument
can be applied when initially 
$|\psi\rangle_B=|1\rangle_B$. Hence,
four bits of classical information and two
ebits are the necessary resources. 
\footnote{Note that
while Alice and Charles on the one hand  and
Bob and Charles  on
the other hand  have to share
effectively a maximally entangled state
before the appropriate LOCC operation is 
performed, the parties
Alice, Bob, and Charles
do not necessarily have to share
the qubits $A_1$, $B_1$, $C_1$, and $C_2$ in
a state of the form
$|\phi^+\rangle_{A_1 C_1}\otimes
|\phi^+\rangle_{B_1 C_2}$ or unitarily
equivalent via local unitary operations. 
It might well be that
Bob has an additional qubit $B_2$
at hand; the initial state of 
$A_1$, $B_1$, $B_2$, and $C_2$
could then be 
$|\phi^+\rangle_{A_1 B_1}\otimes
|\phi^+\rangle_{B_2 C_1}$, as by
\index{Entanglement swapping}
{\it entanglement swapping}\/ \cite{Swapping1,Swapping2}
at Bob's site, Alice and Charles
may obtain a maximally entangled state
of two qubits.}
\proofend

Again, these results can be generalized to three-party
control-U operations.
The local Toffoli gate merely needs to be replaced 
by a local three-party
control-U. Also, on the basis of these findings, 
a statement about
control-U operations involving $N$ parties
can be made with $N\geq 3$.\\

\noindent {\bf Proposition 4.4. -- } {\it
Two shared ebits and four 
bits of classical communication
are sufficient for the local
implementation of a non-local 
control-U gate of three parties.}\\
\index{Three-party control-U gate}

\newfigure{8.035cm}{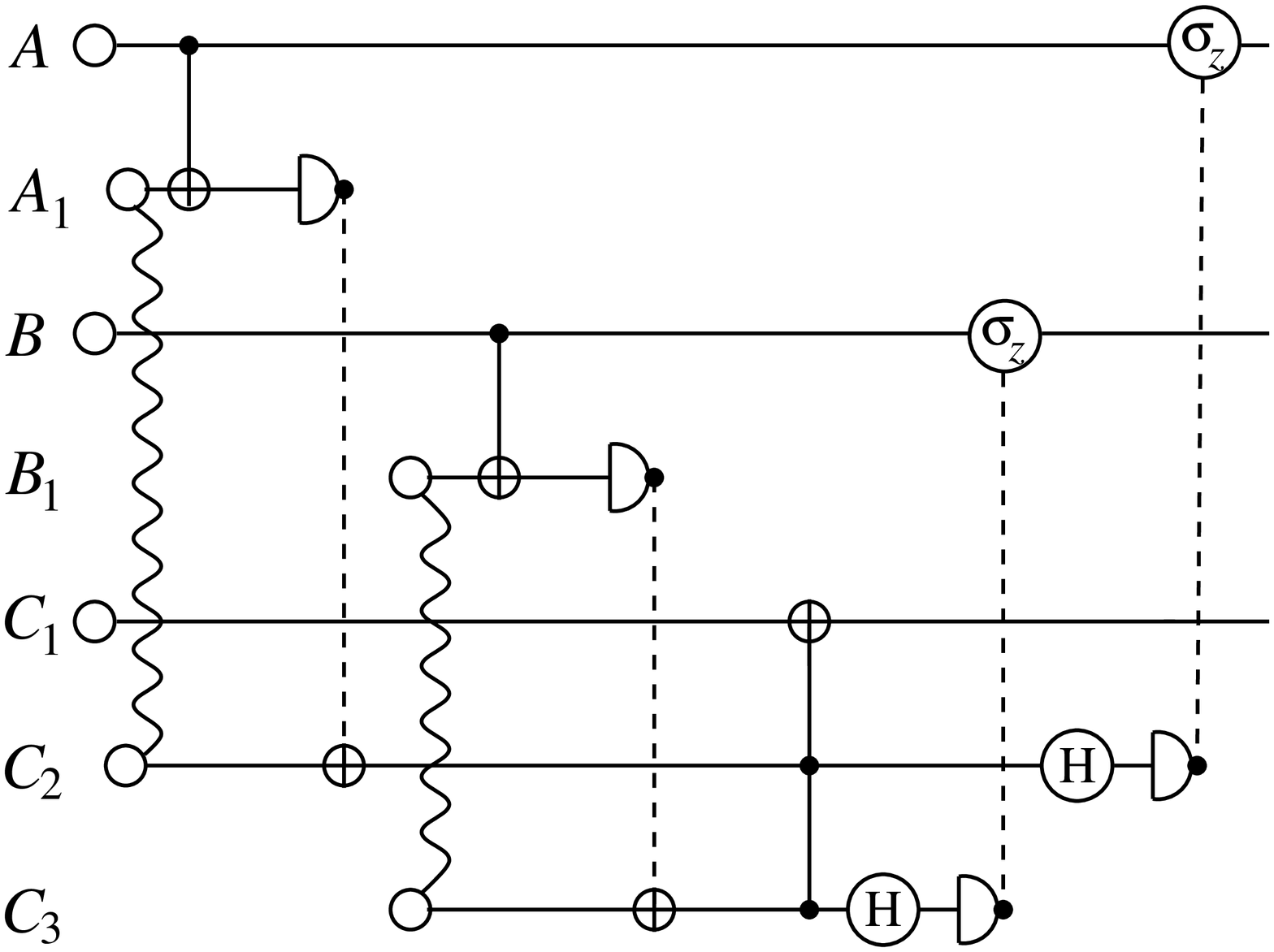}
{Figure 4.6: A representation of the optimal non-local implementation of a quantum
Toffoli gate.}

\noindent 
{\bf Proposition 4.5. --} {\it 
An $N$ party control-U gate with $N\geq 3$
can be implemented using $2(N-1)$ bits of 
classical communication and $N-1$ shared ebits.}\\

\proof The $N-1$ control qubits are
denoted by 
$A_1, ..., A_{N-1}$, the control qubit
is labeled $T$, 
${\cal H}=({{\mathbbm{C}}^2})^{\otimes N}$.
Initially there are $N-1$ shared ebits:
In addition to $A_i$, $i=1,...,N-1$,
the $i$-th party has 
the qubit $P_i$ at hand which is
maximally entangled with a qubit
$Q_i$ held by the $N$-th party, 
see Fig. 4.7.

The first $N-1$ steps of the protocol are essentially 
analogous to the previous protocol.
In the $i$-th step, $i=1,...,N-1$,
a local quantum CNOT gate is applied on $A_i$ and
$P_i$ with $A_i$ as control. 
Then party $i$ measures
particle $P_i$ (again a projective
measurement in the computational basis). The
target party, the $N$-th, 
performs a NOT operation on his 
ancillary qubit $Q_i$ if the $i$-th party 
finds $1$ and the identity
if the $i$-th party finds $0$ in the
measurement. Qubit $P_i$ is afterwards 
discarded.
In the $N$-th step an
$N$-party control-U gate is performed 
by the $N$-th party on
$Q_1,...,Q_{N-1}$ and $T$, where $T$ is the target
qubit.
Eventually, 
the target party performs Hadamard gates on each of
the qubits $Q_1,...,Q_{N-1}$.

The last $N-1$ steps again involve measurements,
each of them being of the following type:
$Q_i$, $i=1,...,N-1$ is measured
in the computational basis.
If the outcome is $1$, then $\sigma_z$ is applied
to $A_i$, if the outcome is $0$ then no action is
taken. 
Qubit $Q_i$ may finally be discarded.
Hence, the total required resources are
$2(N-1)$ bits of classical information and $N-1$
initially shared ebits.
\proofend

\newfigure{8cm}{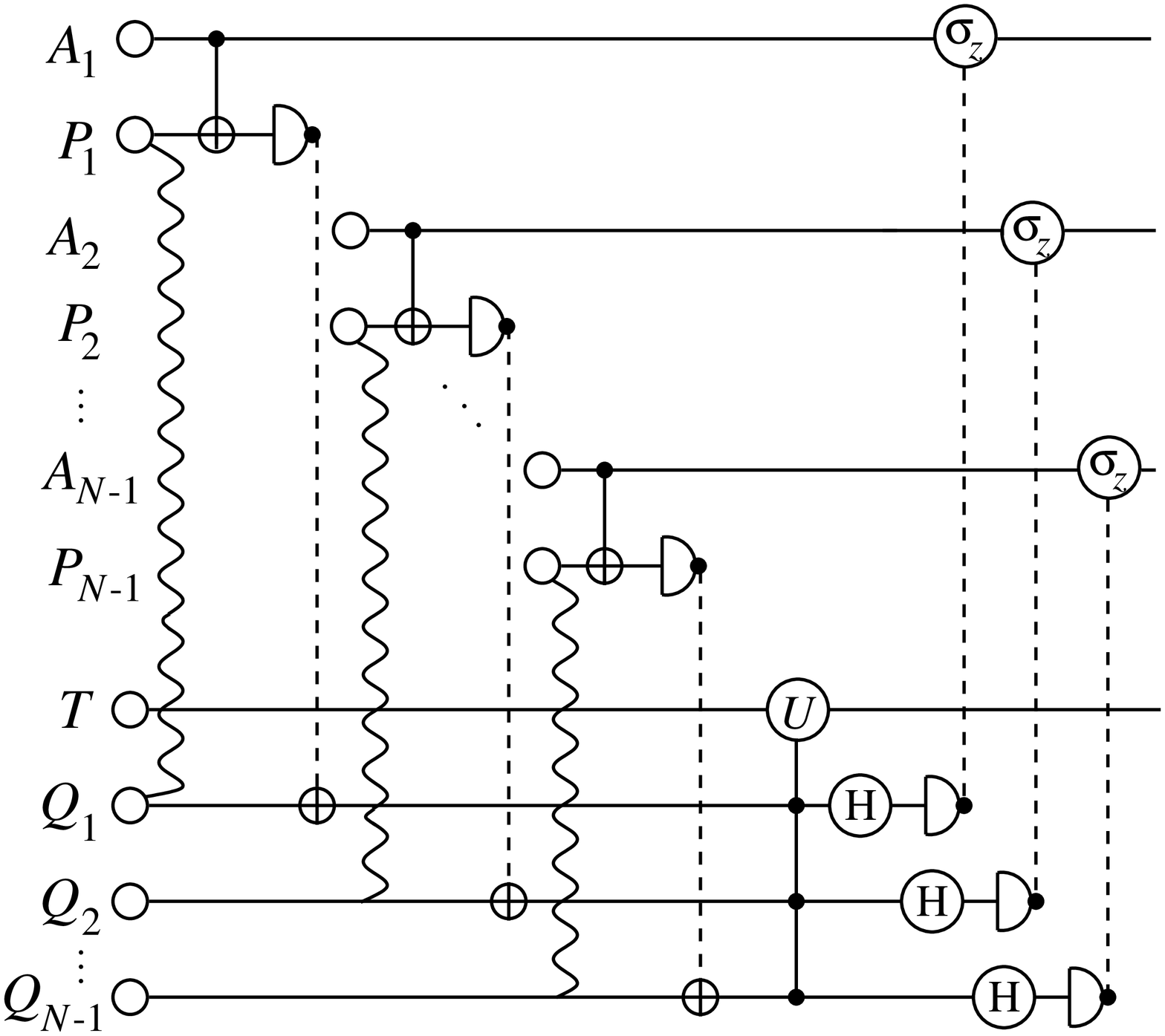}
{Figure 4.7: Representation of the protocol
described in Proposition 4.5.}

Quite surprisingly,  a mere 
$N-1$ ebits are required in this
protocol. Obviously, 
one could also construct a 
quantum gate array
\index{Multi-party control-U}
performing an $N$-party control-U
gate by using two-party control-U
gates and CNOT gates only, as
described in Ref.\ \cite{gates}.
In each step, the two-party gates
would then have to 
be realized non-locally.
In such a protocol
a supply of  $3\times 2^{N-1}-4$ ebits would be necessary.
A more efficient 
\index{Teleportation}
teleportation-based protocol
\cite{Chefles} in which the respective states
of the qubits at different nodes are twice
teleported would still use $2(N-1)$ ebits and
$4(N-1)$ bits of classical information.

\section{Concluding Remarks}

In all of the above prototols the
amount of classical bits that needed to be sent from Alice to Bob
and from Bob to Alice was the same. In this sense
there was a symmetry in the  classical communication cost
in 
the non-local implementation of a joint unitary operation. 
The same is true for the protocol 
presented in Ref.\ \cite{SmallGate}.
One may wonder 
whether this symmetry is a general property of such 
protocols.
For a large class of gates it can be proved that this 
symmetry holds [E4] -- in the sense that an implementation
can  be found for which the minimal amount of transmitted 
classical information is the same in both directions --
but not for all possible unitary operations. 
Another open question is whether the achievable
minimal (average) amount of entanglement
that is necessary to implement 
a gate is always identical 
to the entanglement that can be maximally created by the
gate. In all the above implementations this is the case,
but it is not obvious whether 
it is always possible 
to find a non-local implementation with this property.

%
%


\chapter{Entanglement and Classical Information}
\chapterbreak

\section{Introduction}

The amount of entanglement two parties share is
related to the knowledge of the parties about the
state of the quantum system. If a composite
quantum system is in one of a set of possible
entangled states, but the parties do not know in
which one, it may  happen
that the entanglement is
of no use. In principle, the parties could
transform the initial state into an appropriate
form, but in order to do so, they would have to
know what particular protocol 
they should implement.

This chapter deals with a relationship 
between the loss
of classical information about a quantum system and
the concomitant loss of entanglement.
For reasons of simplicity, a particularly clear
way of losing classical information will be
considered: the classical record of the identity of the quantum systems
will be discarded.
\index{Classical record}
\index{Erasure of classical information}
%
%
In the first section 
simple examples will be presented. 
In the second section 
the reflections will be generalized
using group theoretical methods. In the last section 
a number of remarks
on bounds for the loss of entanglement 
due to loss of classical information will be
added, and the
connection to the hashing
inequality will be pointed out.

In order to explain the structure of the problem 
in a more formal way,
assume that two parties, Alice and Bob, prepare 
$n$ copies of pairs of qubits in a particular
pure state. This
state is taken to be an entangled state which may
be used to implement a protocol for
quantum communication. 
Let us also assume 
that the following misfortune happens: 
the parties forget which system is which, such that
they do not know any more which qubit
is entangled with which, without having 
manipulated the systems. They may have, for instance,
written the relevant classical information on a sheet
of paper that has been taken away. After all,
there is no way
of finding out what pairs of qubits 
are in the original state.

The important question is whether a system of this 
type could still be used for the original purpose.
The two parties would like to recover as much
of the entanglement as possible by operational
means, and hence, the appropriate concept to quantify
the entanglement is the entanglement of distillation 
with respect to LOCC operations $D_{\leftrightarrow}$
\cite{BennettLett,Rains,Rains2}.
The loss of classical information
may essentially be regarded as a mixing
procedure; this is why the
degree of distillable entanglement 
\index{Distillable entanglement}
will decrease.
But 
{\it how much}\/ 
distillable entanglement will be lost in the 
course of the loss of
classical information? Is there a general upper bound
for $\Delta D_{\leftrightarrow}$, the difference
in distillable entanglement in the initial and the
final situation?
And is there a connection to the amount of lost
classical information $\Delta I$, as quantified by an
appropriate functional?

The relevance of these considerations is
twofold:  Firstly, 
from the perspective of a theory of 
entanglement this analysis is useful 
as the distillable entanglement 
$D_{\leftrightarrow}$ after
the permutation of the qubits 
can actually be evaluated. This means that
these investigations
provide 
one of the very few classes of mixed states 
for which this quantity can be computed. 
Secondly, 
the relevance in the context of channel capacities
arises from the fact that
the process of losing information can be regarded
as the result of the transmission of qubits through
a particular noisy quantum channel. 
The statement of the proposition
can then be interpreted as a statement
about the quantum capacity of the
corresponding
\index{Quantum channel}
quantum channel.
Most of the results of this chapter have been published in
a shorter form in 
[E2].

\section{Examples}

\newfigure{12.5cm}{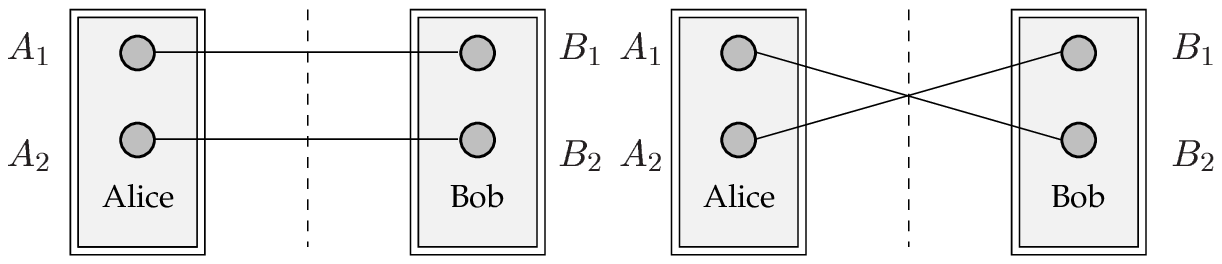}
{Figure 5.1: The situation in Example 5.1.}

\noindent {\bf Example 5.1. --} 
Consider the situation where Alice and Bob initially
share two pairs  of qubits each in a maximally 
entangled state, so that they share two ebits of entanglement.
The parties then lose the information about the
order of the quantum systems. 
This means that they do not know whether 
the two pairs are 
in the original order or have been 
permuted (see Fig.\ 5.1). 

Let 
\begin{equation}
|\psi\rangle=(|00\rangle_{A_1 B_1}+ |11\rangle_{A_1 B_1}) \otimes 
(|00\rangle_{A_2 B_2}+ |11\rangle_{A_2 B_2})/2
\end{equation}
be the state vector of the qubits 
labeled $A_1$, $B_1$, $A_2$, and $B_2$ in the original situation (see Fig.\ 6.1),
belonging to Alice and Bob, respectively;
\begin{equation}
{\cal H}={\cal H}_A\otimes {\cal H}_B\,\,\,\,\text{ with }\,\,\,\,
{\cal H}_A={\cal H}_{A_1}\otimes {\cal H}_{A_2}, \,\,\,\,
{\cal H}_B={\cal H}_{B_1}\otimes {\cal H}_{B_2},
\end{equation} 
${\cal H}_{A_1}={\cal H}_{A_2}={\cal H}_{B_1}={\cal H}_{B_2}={\mathbbm{C}^2}$.
In the computational basis $|\psi\rangle$ is given by
\begin{eqnarray}
	|\psi\rangle=(|0000\rangle_{A_1 B_1 A_2 B_2}+
	|0011\rangle_{A_1 B_1 A_2 B_2}+
	|1100\rangle_{A_1 B_1 A_2 B_2}+
	|1111\rangle_{A_1 B_1 A_2 B_2})/2  ,
\end{eqnarray}
while in the permuted case the role of $B_1$ and $B_2$
is interchanged, 
\begin{eqnarray}
	|\phi\rangle=
	(|0000\rangle_{A_1 B_1 A_2 B_2}+
	|0110\rangle_{A_1 B_1 A_2 B_2}+
	|1001\rangle_{A_1 B_1 A_2 B_2}+
	|1111\rangle_{A_1 B_1 A_2 B_2})/2  .
\end{eqnarray}
The index indicating which entry corresponds to which
quantum system will be omitted in the remainder of this example.
As a result of the loss of the record about the
order of the particles, 
the composite quantum system is 
now described by the
mixed state
\begin{equation}\label{sigma}
	\sigma=\frac{1}{2}\left(|\psi\rangle\langle\psi|+|\phi\rangle\langle\phi|\right).
\end{equation}
The question that now arises is 
 how much distillable entanglement the
state $\sigma$ holds, i.e.,
how much entanglement is still accessible
to Alice and Bob. In order to approach this problem 
consider the spectral 
decomposition of $\sigma$,
\begin{equation}
	\sigma= \frac{1}{4}|\phi_1\rangle\langle\phi_1| +
	\frac{3}{4}	|\phi_2\rangle\langle\phi_2|  ,
\end{equation}
where 
 \begin{eqnarray}
	|\phi_1\rangle&=&\frac{1}{2}
	\left(|0011\rangle-
	|0110\rangle-
	|1001\rangle+
	|1100\rangle
	\right),\\
	|\phi_2\rangle&=&	
	\frac{1}{\sqrt{12}}
	\bigl(2|0000\rangle+
	|0011\rangle+
	|0110\rangle\nonumber\\
	&+&
	|1001\rangle+
	|1100\rangle+
	2|1111\rangle\bigr).
\end{eqnarray}
The structure of the problem will become more transparent
if one represents the states in terms of the
angular momentum eigenstates.
In the basis of angular momentum eigenstates $|j,m\rangle$ 
with $j=0,1$, $m=-1,0,1$,
\index{Angular momentum}
\begin{eqnarray}
	|1,-1\rangle=|00\rangle,\,\,\,\,&&
	|1,1\rangle=|11\rangle,\\
	|1,0\rangle=(|01\rangle+|10\rangle)/\sqrt{2},\,\,\,\,&&
	|0,0\rangle=(|01\rangle-|10\rangle)/\sqrt{2}  ,
\end{eqnarray}
the eigenstates $|\phi_1\rangle$ and $|\phi_2\rangle$ read
\begin{eqnarray}
	|\phi_1\rangle&=&
	|0,0\rangle|0,0\rangle,\\
	|\phi_2\rangle&=&
	\frac{1}{\sqrt{3}}
	\left(
	|1,-1\rangle|1,-1\rangle
	+
	|1,0\rangle|1,0\rangle
	+
	|1,1\rangle|1,1\rangle
	\right) .
\end{eqnarray}

An upper bound  for the distillable 
entanglement is given by the relative entropy of entanglement 
$E_R(\sigma)$ of $\sigma$ which in turn 
is smaller or equal to the relative entropy with respect
to any separable state $\rho\in{\cal D}({\cal H})$. Hence, the distillable
entanglement $D_{\leftrightarrow}(\sigma)$ of $\sigma$ is bounded by
\begin{equation}
	D_{\leftrightarrow}(\sigma) \le S(\sigma||\rho)= \frac{3}{4}\log_2 3,
	\label{bound1}
\end{equation}
where the separable state $\rho$ 
is taken to be
\begin{eqnarray}\label{frho}
	\rho = \frac{1}{4} \sum_{j=0}^{1} \sum_{m=-j}^{j} |j,m\rangle|j,m\rangle\langle j,m|\langle j,m| \, .
\end{eqnarray}
Here,
the relative entropy of $\sigma$ with respect to
$\rho$ can be evaluated
using the direct sum property of the functional (see Appendix A).
Surprisingly, 
the upper bound given in Eq. (\ref{bound1}) can be 
achieved.  This task may be accomplished by means of the
optimal distillation protocol. This distillation protocol
involves one-way classical communication only.\\

\jbox{
\begin{enumerate}
\item
Alice performs a projective measurement
with the two projections 
\begin{equation}\label{FunnyPOVM2}
A_1=|0,0\rangle\langle0,0|,\,\,\, 
A_2=	\sum_{m=-1}^{1} |j=1,m\rangle\langle j=1,m|,
\end{equation}
while Bob remains inactive, i.e., $E_1=A_1\otimes \mathbbm{1}_B$,
$E_2=A_2\otimes \mathbbm{1}_B$. In
this measurement, two locally distinguishable subspaces 
are discriminated (compare also Eq.\ (\ref{FunnyPOVM})).

\item
With probability $p_1=1/4$ they obtain the normalized 
output state $|\phi_1\rangle\langle\phi_1|$, which is a product
state that has no further use 
in the distillation protocol.
\item
The other final pure state of the selective measurement
they get with probability $p_2=3/4$ 
is $|\phi_2\rangle\langle\phi_2|$.
The entanglement of this state amounts to  $\log_2 (3)$ 
ebits. 
\item
The average number of maximally entangled 
states that can be distilled from $\sigma$ is given by 
\begin{equation}
 	D_{\rightarrow}(\sigma)
	=D_{\leftrightarrow}(\sigma)=\frac{3}{4}\log_2 (3) \approx 1.189 .\label{sp1}
\end{equation}
\end{enumerate}
}

As this realizes the bound Eq. (\ref{bound1}) it is the maximally 
possible value. It is worth noting that this value 
is greater than one. This means that 
less than one ebit of entanglement is 
erased due to the loss of the classical information about 
the order. 
The classical information can be taken to be 
the von Neumann entropy of the mixed state that 
Alice and Bob share afterwards, i.e., 
\begin{equation}
\Delta I = S(\sigma)=
-(1/4)\log_2(1/4)-(3/4)\log_2(3/4)=\frac{3}{4}\log_2 (3).\label{sp2}
\end{equation} 
This choice
is justified as -- according to Schumacher's
noiseless coding theorem \cite{SchumacherCoding} -- the amount
of classical information that can be 
encoded in the two non-orthogonal states with state vectors
$|\psi\rangle$ and $|\phi\rangle$ associated with weights
$p=1/2$ and $1-p=1/2$
is given by $S(\sigma)=
S((|\psi\rangle\langle\psi|+|\phi\rangle\langle\phi|)/2)$.
This quantifies the classical uncertainty about the
order of the particles. The change in distillable
entanglement is from now on denoted by
\begin{equation}
\Delta D_{\leftrightarrow}=D_{\leftrightarrow}(|\psi\rangle\langle\psi|)-
D_{\leftrightarrow}(\sigma).
\end{equation}
From  Eq.\ (\ref{sp1}) and Eq.\ (\ref{sp2}) it follows that
\begin{equation}
	\frac{\Delta D_{\leftrightarrow}}{\Delta I} = 1  . \label{equal}
\end{equation}
This means that in this case the amount of lost classical
information and the loss of distillable entanglement are
identical. This is the desired result for this particular case
connecting the loss of distillable with the loss of classical information.\\

\medskip

\noindent {\bf Example 5.2. --} 
The above scenario can be generalized
to a situation where Alice and Bob initially do not hold 
quantum systems in maximally
entangled states but in arbitrary pure states
with a given degree of entanglement. This case is suitable
for investigation because
it leads to an operationally defined one-parameter 
class of states for which the distillable entanglement
can be analytically computed.
This class could provide a useful tool for investigating
distillable entanglement.
The class of states to be considered is according
to the Schmidt decomposition given by
$
\sigma=(|\psi\rangle\langle\psi|+|\phi\rangle\langle\phi|)/2
$
with
\begin{eqnarray}
	|\psi\rangle&=&
	(\sqrt\alpha |00\rangle_{A_1 B_1}+\sqrt\beta |11\rangle_{A_1 B_1})
	\otimes
	(\sqrt\alpha |00\rangle_{A_2 B_2}+\sqrt\beta  |11\rangle_{A_2 B_2})/2
	\nonumber\\
	&=&\alpha |0000\rangle_{A_1 B_1 A_2 B_2}+
	\sqrt{\alpha\beta} |0011\rangle_{A_1 B_1 A_2 B_2}\nonumber\\
	&+&
	\sqrt{\alpha\beta}|1100\rangle_{A_1 B_1 A_2 B_2}+
	\beta|1111\rangle_{A_1 B_1 A_2 B_2},\\
	|\phi\rangle&=&
	\alpha|0000\rangle_{A_1 B_1 A_2 B_2}+
	\sqrt{\alpha\beta}|0110\rangle_{A_1 B_1 A_2 B_2}\nonumber\\
	&+&
	\sqrt{\alpha\beta}|1001\rangle_{A_1 B_1 A_2 B_2}+
	\beta|1111\rangle_{A_1 B_1 A_2 B_2},
\end{eqnarray}
where $\alpha\in[0,1]$, $\beta=1-\alpha$. Again, the indices will be omitted
subsequently.
The spectral decomposition of $\sigma$ is given by
\begin{eqnarray}
	\sigma= \alpha \beta
	|\phi_1\rangle\langle\phi_1|
	+
	(\alpha^2+\alpha \beta+\beta^2)
	|\phi_2\rangle\langle\phi_2|,
\end{eqnarray}
where,  written in the same basis as above,
\begin{eqnarray}
	|\phi_1\rangle&=&
	|0,0\rangle|0,0\rangle,\\	
	|\phi_2\rangle&=&
	\frac{
	\alpha
	|1,-1\rangle|1,-1\rangle
	+
	\beta
	|1,1\rangle|1,1\rangle
	+
	\sqrt{\alpha\beta}
	|1,0\rangle|1,0\rangle
	}{\sqrt{\alpha^2+\beta^2+\alpha\beta}}.
\end{eqnarray}

\newfigure{7.0cm}{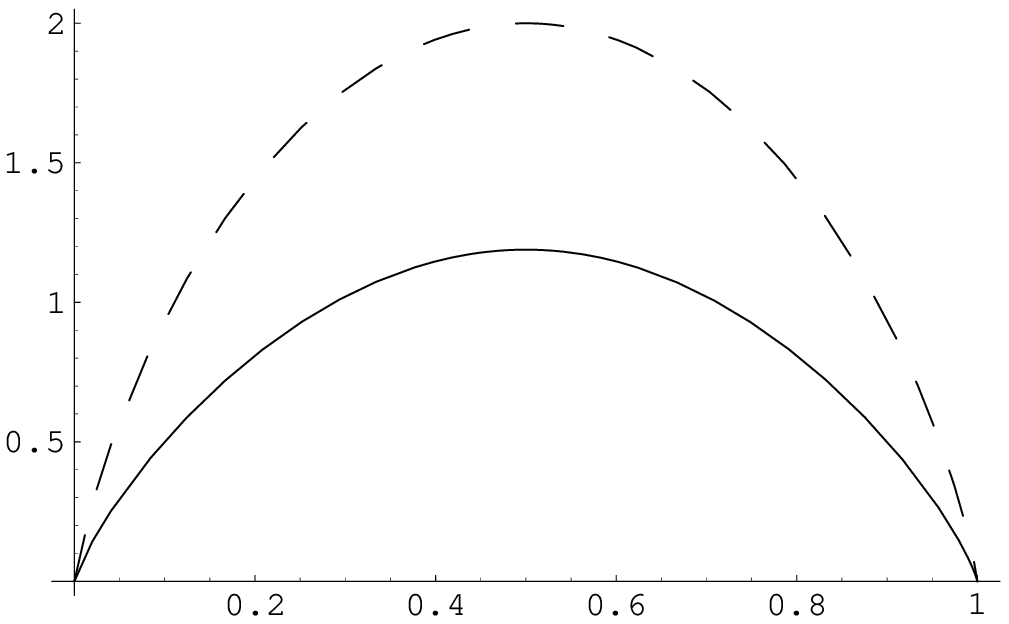}
{Figure 5.2:
The distillable entanglement $D_{\leftrightarrow}(\sigma)$ 
of $\sigma$ (solid line) and the
entanglement of the initial pure state
$D_{\leftrightarrow}(|\psi\rangle\langle\psi|)$ (dashed line)
as functions of $\alpha$ in Example 5.2.}

Now one can proceed in a similar fashion as before. 
The upper bound for distillable entanglement, given by the
relative entropy with respect to the separable state
\begin{eqnarray}
	\rho&=&
	\alpha \beta 
	\left(|0,0\rangle| 0,0\rangle\langle 0,0 |\langle 0,0|
	+
	|1,01,0\rangle\langle1,0|\langle1,0|
	\right)\\
	&+&
	\beta^2 |1,1\rangle|1,1\rangle\langle1,1|\langle1,1|
	+
	\alpha^2 |1,-1\rangle|1,-1\rangle\langle1,-1|\langle1,-1|,\nonumber
\end{eqnarray}
can again 
be reached using the protocol of Example 5.1.
It follows that the distillable entanglement is in this case
given by
\begin{eqnarray}
	D_{\leftrightarrow}(\sigma)=E_R(\sigma)&=&
	(1-\alpha \beta)
	\log( 1-\alpha \beta)\label{Full}\\
	&-&
	(\alpha^2\log(\alpha^2)+
	\beta^2 \log_2 (\beta^2)
	+
	\alpha \beta \log_2 (\alpha \beta)),\nonumber
\end{eqnarray}
see Fig.\ 5.2. 
The entanglement of the initial pure state
was given by the entropy of the reduced states of Alice
or Bob, that is, by 
\begin{equation}
D_{\leftrightarrow}(|\psi_1\rangle\langle\psi_1|=
-2(\alpha \log_2 (\alpha)+ \beta \log_2 (\beta)),
\end{equation}
and since $S(\sigma)=-2(\alpha \log_2 (\alpha)+ \beta \log_2 
(\beta))$, 
again, 
\begin{equation}
	\frac{\Delta D_{\leftrightarrow}}{\Delta I}=1
\end{equation} 
for all $\alpha\in[0,1]$.
Of course, Eq.\ (\ref{Full}) reduces to 
$
	D_{\leftrightarrow}(\sigma)=
	(3/4) \log_2 (3)
$
for $\alpha=\beta=1/2$.\\

\noindent {\bf Example 5.3. --} 
Assume that
Alice has merely 
a single qubit $A$ at hand, while Bob
possesses two qubits $B_1$ and $B_2$;
${\cal H}={\cal H}_A\otimes 
{\cal H}_{B_1}\otimes {\cal H}_{B_2}$, ${\cal H}_A={\cal H}_{B_1}={\cal H}_{B_2}=
\mathbbm{C}^2$.
Let initially, Alice and Bob share one pair
of qubits in a maximally entangled state, Bob's
second qubit is in the state with state vector $|0\rangle_{B_2}$.
The state of $A$, $B_1$, and $B_2$
after permutation can then be taken to be
$\sigma=(|\psi_1\rangle\langle\psi_1|+|\psi_2\rangle\langle\psi_2|)/2$,
with
\begin{eqnarray}
		|\psi_1\rangle&=&
		(|000\rangle_{A B_1 B_2}+|110\rangle_{A B_1 B_2})/\sqrt{2},\\
		|\psi_2\rangle&=&
		(|000\rangle_{A B_1 B_2}+|101\rangle_{A B_1 B_2})/\sqrt{2}.
\end{eqnarray}
Similarly as before, 
one may 
obtain a constructive method of distillation by
investigating 
the spectral decomposition of $\sigma$ which reads
\begin{equation}
	\sigma=\frac{1}{4}
	|\phi_1\rangle\langle\phi_1|+
	\frac{3}{4}|\phi_2\rangle\langle\phi_2|,
\end{equation}
where
\begin{eqnarray}
	|\phi_1\rangle&=&\label{late1}
	(|110\rangle-|111\rangle)/\sqrt{2},\\
	|\phi_2\rangle&=&
	(2|000\rangle+ |110\rangle+|111\rangle)/\sqrt{6}.\label{late2}
\end{eqnarray}
In the basis
\begin{equation}
|\hat{0}\rangle=|00\rangle,\,\,\, 
|\hat{1}\rangle=|01\rangle,\,\,\, 
|\hat{2}\rangle=(|10\rangle-|11\rangle)/\sqrt{2},\,\,\, 
|\hat{3}\rangle=(|10\rangle+|11\rangle)/\sqrt{2},
\end{equation}
satisfying $|\langle\hat{i}|\hat{j}\rangle|^2=\delta_{ij}$
for $i,j=0,1,2,3$, Eqs.\ (\ref{late1}) and (\ref{late2}) become
\begin{eqnarray}
	|\phi_1\rangle&=&
	|1\hat{2}\rangle,\\
	|\phi_2\rangle&=&
	(\sqrt{2}|0\hat{0}\rangle+ |1\hat{3}\rangle)/\sqrt{3}.
\end{eqnarray}
Hence, in this scenario
\begin{equation}
	\Delta I= 2-\frac{3}{4}\log_2 (3),
	\,\,\,\,\,\,\,\,
	\Delta D_{\leftrightarrow}= \frac{3}{4}\log_2 (3) 
	-\frac{1}{2}
\end{equation}
and therefore $\Delta D_{\leftrightarrow}/\Delta I< 1$.
%
%
%
%
%
This means that the loss of classical information and the
loss of distillable entanglement are not the same any
more, but $\Delta D_{\leftrightarrow}$ is still bounded
by above by the loss of classical information $\Delta I$.

\section{A General Result}

In this section a generalized result will be proved.
Alice and Bob hold $n$
pairs of qubits in any pure state. Then 
the classical record
about the order of the particles is lost. It will be
shown that the distillable
entanglement can be evaluated  exactly 
for any state and any 
number of copies. 
The value of $D_{\leftrightarrow}$ does
not vanish in the limit of infinitely many copies.
For any number of copies some entanglement can be recovered,
and the distillable entanglement grows monotonically with
the number of copies.

The permutation of the quantum systems is incorporated
by applying elements of the 
\index{Symmetric group}
symmetric group $S_n$ on the state. A
$\pi\in S_n$ acts as
\begin{equation}
	\pi|\psi \rangle_{1} \otimes...\otimes |\psi\rangle_{n}=
	|\psi\rangle_{\pi(1)}\otimes...\otimes |\psi\rangle_{\pi(n)}.
\end{equation}
Again, $\pi$ denotes a permutation of degree $n$ and
at the same time the associated unitary. 
The subsequent proposition 
is the general statement in case Alice and Bob are
initially sharing pairs of
qubits in a maximally entangled state. 
First, the case will be considered where
Alice and Bob hold an even number $n$ of pairs
of qubits, such that the 
Hilbert space of the composite system is given by
\begin{equation}
{\cal H}={\cal H}_A\otimes {\cal H}_B,\,\,\,\,\,
{\cal H}_A={\cal H}_B=({\mathbbm{C}}^2)^{\otimes 2J},
\end{equation}
with an appropriate $J$. 
In Proposition 5.6 this restriction will be given up.\\

\noindent {\bf Proposition 5.4. --} {\it 
Let Alice and Bob share 
$n=2J$ pairs of qubits each in the same maximally entangled
state, such that the initial state of the composite
system is given by $|\psi\rangle\langle\psi|^{\otimes n}$,
$|\psi\rangle=(|00\rangle+|11\rangle)/\sqrt{2}$.
The associated Hilbert space is ${\cal H}={\cal H}_A\otimes {\cal H}_B$, where
${\cal H}_A={\cal H}_B=({\mathbbm{C}}^2)^{\otimes n}$,
$J=1,2,...$. Both parties then lose all classical
information about the order of their $n$ particles, 
such that
the state of the composite system becomes}
\begin{equation}\label{AfterPer}
	\sigma = 
	\sum_{\pi_A,\pi_B \in S_n}
	(\pi_A\otimes \pi_B)
	|\psi\rangle\langle\psi|^{\otimes n}
	(\pi_A\otimes \pi_B).
\end{equation}
{\it The distillable entanglement of the state $\sigma$
can then be calculated as}
\begin{equation}\label{PerResult}
	D_{\leftrightarrow}(\sigma)=\sum_{j=0}^J
		d_j^2
		p_j \log(2j+1),
\end{equation}
{\it 
where
\begin{equation}
	d_j=\frac{2j+1}{2J+1} \left( {2J+1 \atop J-j } \right)
\end{equation}
and 
\begin{equation}
	p_j=(2j+1)/(2^{2J} d_j).
\end{equation}	
The ratio between the change of distillable entanglement 
$\Delta D_{\leftrightarrow}$ and the
amount of erased information $\Delta I=S(\sigma)$ for any $J=1,2,...$ 
obeys the inequality
\begin{equation}\label{ineq}
	\frac{\Delta D_{\leftrightarrow}}{\Delta I}\le 1  ,
\end{equation}
with equality for $J=1$.}
\probreak

The following considerations 
will prepare the proof of this proposition.
The Hilbert space corresponding to Alice's system, ${\cal H}_A=(\mathbbm{C}^2)^{\otimes n}$,
can be decomposed into a 
\index{Direct sum decomposition}
 direct sum  according to
\begin{equation}
{\cal H}_A=\bigoplus_{j}{\cal H}_A^{(j)}=
		\bigoplus_{j}{\cal M}_A^{(j)}\otimes {\cal K}_A^{(j)}.
\end{equation}
(see also  Ref.\ \cite{KeylRate}).
Here, ${\cal K}_A^{(j)}$ is a 
\index{Multiplicity space}
{\it multiplicity space}\/. It carries a 
{\it representation}\/ of the 
\index{Symmetric group}
symmetric group $S_n$. Let $\pi_A\in S_n$, and
denote the corresponding unitary on ${\cal H}_A$
also by $\pi_A$.
Then 
\begin{equation}\label{permrep}
	\pi_A= \bigoplus_{j} \mathbbm{1}\otimes \lambda^{(j)}_A(\pi_A),
\end{equation}
where $\lambda^{(j)}_A$ is a suitable 
\index{Unitary irreducible representation}
{\it unitary irreducible representation}\/ of $S_n$.
${\cal M}_A^{(j)}$ carries the representation of $\text{SU}(2)$. Let
$U_A\in \text{SU}(2)$, then the $n$-fold
tensor product $U_A^{\otimes n}$ reads as
\begin{equation}
	U_A^{\otimes n}= \bigoplus_{j} D_A^{(j)}(U_A)\otimes \mathbbm{1},
\end{equation}
where $D_A^{(j)}$ is a spin-$j$ 
\index{Spin coupling}
irreducible representation of
the group $\text{SU}(2)$. For the relationship 
between the 
symmetric group $S_n$ and $\text{SU}(2)$ see, e.g., Refs.\ 
\cite{Weyl,Sternberg}.
The Hilbert space ${\cal H}_B$ of Bob's quantum systems can
be decomposed in a fully analogous fashion. \\

To be more specific, it is helpful to specify a particular basis for the sets
${\cal H}_A^{(j)}=
{\cal M}_A^{(j)}\otimes {\cal K}_A^{(j)}.
$
Let -- as in Ref.\ \cite{Cir} --
\begin{equation}
	|j,m,1\rangle=
	|j,m\rangle\otimes \left(\frac{|01\rangle-|10\rangle}{\sqrt{2}}\right)^{\otimes(J-j)}
\end{equation}
for $j=1,...,J$,
where $|j,m\rangle$ is the state of $2j$ qubits with a fixed value of 
$j$ and $m$ with $j-m$ qubits in $|0\rangle$.
$|j,m,\alpha_j\rangle$ with $\alpha_j=1,...,d_j$ and $m=-j,...,j$
are then constructed using 
permutation operators as
\begin{equation}
	|j,m,\alpha_j\rangle=\sum_i \eta_i \pi_A^{(i)}|j,m,1\rangle
\end{equation}
(compare Eq.\ (\ref{permrep})),
where $\eta_j\in[0,1]$, $\pi_A^{(i)}\in S_n$, $i=1,2,...$, such that
the set 
\begin{equation}
\{|j,m,1\rangle,...,
|j,m,\alpha_j\rangle\}
\end{equation} 
forms an orthonormal set.
The degeneracy (the dimension of the multiplicity space) is
given by
\begin{equation}
	d_j=\frac{2j+1}{2J+1} \left( {2J+1 \atop J-j } \right).
\end{equation}

The state given by Eq.\ (\ref{AfterPer})
can be further specified according to Lemma 5.5.
It makes use of
\index{Schur's second lemma}
{\it Schur's second lemma}\/ \cite{Weyl,Sternberg}.
If the matrices $D(g)$ are the irreducible
representations of a group $G$, and if
\begin{equation}
	[A,D(g)]=0
\end{equation}
for all $g\in G$, then $A=\text{const}\times \mathbbm{1}$.\\

\noindent {\bf Lemma 5.5. --} {\it 
Let ${\cal H}={\cal H}_A\otimes {\cal H}_B$, where
${\cal H}_A={\cal H}_B=({\mathbbm{C}}^2)^{\otimes 2J}$, $n=2J$.
$\sigma=|\phi\rangle\langle\phi|$, where
\begin{equation}
	|\phi\rangle=
	\sum_{j=0}^J \sum_{m=-j}^{j}\sum_{\alpha_j=1}^{d_j}
		\frac{|j,m,\alpha_j\rangle|j,m,\alpha_j\rangle}{\sqrt{2^{n}}}.
\end{equation}
Then the state
\begin{equation}
	\sigma=
	\sum_{\pi_A,\pi_B\in S_n}
	(\pi_A\otimes \pi_B)
	|\phi\rangle\langle\phi|
	(\pi_A\otimes \pi_B).
\end{equation}
can be written as
\begin{equation}\label{FinalSigma}
	\sigma=\sum_{j=0}
	\sum_{\alpha_j,\beta_j=1}^{d_j}
	p_j |\psi_j(\alpha_j,\beta_j)\rangle\langle\psi_j(\alpha_j,\beta_j)|,
\end{equation}
where 
\begin{equation}\label{FinalSigma2}
	|\psi_j (\alpha_j, \beta_j)\rangle=
	\sum_{m=-j}^j 
	\frac{
	|j,m,\alpha_j\rangle
	|j,m,\beta_j\rangle}
	{\sqrt{2j+1}}
\end{equation}
and 
\begin{equation}\label{FinalSigma3}
p_j=\frac{2j+1}{2^{2J} d_j}.
\end{equation}
}
\probreak

\proof Before applying the permutation operators, 
the reduced state of
Alice is given by
\begin{equation}
	\text{tr}_B[|\phi\rangle\langle\phi|]=
		\sum_{j=0}^J \sum_{j'=0}^J
		\sum_{m=-j}^j
		\sum_{m'=-j'}^{j'}
		\sum_{\alpha_j=1}^{d_j}
		\sum_{\alpha'_{j'}=1}^{d_{j'}}
		\frac{|j,m,\alpha_j\rangle\langle j',m',\alpha'_{j'}|}{2^n}.
\end{equation}
After applying of the
permutation operators
the reduced state of Alice,
\begin{equation}
	\text{tr}_B[\sigma]=\sum_{\pi_A\in S_n}
	\pi_A \text{tr}_B[|\phi\rangle\langle\phi|]
	\pi_A,
\end{equation}
commutes with all $\pi_A\in S_n$ by construction. Hence,
Schur's Lemma can be applied with respect to 
all spin$-j$ irreducible
representations, $j=0,...,J$. 
In accordance with
Schur's Lemma $\text{tr}_B[|\phi\rangle\langle\phi|]$ 
can be written in the form
\begin{equation}
	\text{tr}_B[\sigma]=
		\sum_{j=0}^J 
		\sum_{m=-j}^j
		\sum_{m'=-j'}^{j'}
		\sum_{\alpha_j=1}^{d_j}
		p_j|j,m,\alpha_j\rangle\langle j,m',\alpha_{j}|,
\end{equation}
with $p_j=((2j+1))2^{2J})/d_j$. Bob's reduced state
$\text{tr}_A[\sigma]$ is identical to 
$\text{tr}_B[\sigma]$. The full final state $\sigma$
can be identified by investigating the structure
of $|\phi\rangle\langle\phi|$. Since
\begin{equation}
	|\phi\rangle=
	\sum_{j=0}^J \sum_{m=-j}^{j}\sum_{\alpha_j=1}^{d_j}
	\frac{|j,m,\alpha_j\rangle|j,m,\alpha_j\rangle}{\sqrt{2^{n}}},
\end{equation}
the final state
$\sigma$ is given by
Eq.\ (\ref{FinalSigma}).\proofend

\medskip

This partial result can be used to tackle the statement
of the proposition:\\

\noindent 
\proof (Proposition 5.4)
It may initially be observed that
$|\psi\rangle^{\otimes n}$
can be written in the form 
\begin{equation}
	|\psi\rangle^{\otimes n}=
	|\phi\rangle=
	\sum_{j=0}^J \sum_{m=-j}^{j}\sum_{\alpha_j=1}^{d_j}
		\frac{|j,m,\alpha_j\rangle|j,m,\alpha_j\rangle}{\sqrt{2^{n}}}
\end{equation}
as in Lemma 5.5. It is thus possible to apply Lemma 5.5 in order
to find out the particular form of the state after applying
the permutation operators: the form is given by 
Eqs. (\ref{FinalSigma}), (\ref{FinalSigma2}), and  
(\ref{FinalSigma3}).
The next step is
constructing the optimal  
entanglement distillation protocol.
As before, the following distillation protocol is based on
the fact that the subspaces of the state space corresponding
to the above components of the underlying Hilbert space
are locally distinguishable. Interestingly, this
protocol is related to the algorithm proposed in Ref.\
\cite{Cir}
for the \index{Optimal purification}
optimal purification of qubits.\\

\jbox{
\begin{enumerate}
\item 
Alice starts the protocol by implementing a local
projective measurement. This measurement is designed 
in order to project 
the reduced state belonging to her system $A$ 
on one of the orthogonal subspaces with fixed
$j$ and $\alpha_j$ for 
some $j=0,...,J$ and $\alpha_j=1,...,d_j$.

\item  In this measurement she will obtain a 
value of  $\alpha_j=1,...,d_j$.
If $\alpha_j\neq1$, she further
applies a local unitary operation 
$U^A_{j,\alpha_j}$ such that her reduced state is 
included in the subspace of the state space 
belonging to $\alpha_j=1$.
%
In general,
$|j,m,\alpha_j\rangle$ can be written in the form
\begin{equation}
	|j,m,\alpha_j\rangle=\sum_i \eta_i (\pi_A^{(i)}\otimes \mathbbm{1}_B)
	|j,m,1\rangle,
\end{equation}
where $\eta_i\in[0,1]$.
$\pi_A^{(i)}\in S_n$, $i=1,2,...$, are
appropriate
permutation operators acting in ${\cal H}_A$ only.
That is, 
$|j,m,\alpha_j\rangle$ is a linear superposition 
of $\pi_A^{(i)}|j,m,1\rangle$. It is for this reason that 
the task of this step can always be performed.

\item  The reduced state $\sigma_A$ of Alice is at this stage
of the structure 
\begin{equation}
\sigma_A=\omega_A\otimes 
\left(\frac{(|01\rangle-|10\rangle)(\langle01|-\langle 10|)}{2}\right)^{\otimes(J-j)}.
\end{equation}
The last $J-j$ pairs of qubits in the singlet state are neither entangled
with the other qubits on her side nor entangled with any of Bob's qubits. 
They will thus be of no further use in the remainder of the 
protocol.

\item[4.] Bob performs a local measurement projecting his reduced state on 
one of the subspaces associated with ${\cal H}^B_{k,\beta_k}$ 
for a $k=0,1,...,J$ and a
$\beta_k=1,2,...,d_k$. Due to the particular form of the initial state
he will obtain the value
$k=j$, but he may get a $\beta_j$ different from $\alpha_j$.

%

\item[5.] Equal to the precedent protocol 
Bob applies a 
local unitary operation $U^B_{j,\beta_j}$ such that his
reduced state is 
included in the subspace of the state space 
belonging to $\beta_j=1$. The structure of the state is
depicted in Fig.\ 5.3.

\item[6.] Alice and Bob attain  with
probability
\begin{equation}
	d_j^2 p_j=\frac{(2j+1) d_j}{2^{2J}}
\end{equation} 
one of the pure states 
$|\psi_j\rangle\langle\psi_j|$, where
\begin{equation}
|\psi_j\rangle = \frac{1}{\sqrt{2j+1}}\sum_{m=-j}^j |j,m,1\rangle|j,m,1\rangle.
\end{equation} 
This state contains $\log_2 (2j+1)$ ebits of entanglement. 

\item[7.]
The
total average number of ebits achieved in this protocol is 
\begin{equation}
	\sum_{j=0}^J d_j^2 p_j S({\rm{tr}}_A[|\psi_j\rangle\langle\psi_j|])=
\sum_j d_j^2 p_j \log(2j+1).
\end{equation}
\end{enumerate}
}

In order 
to show that the above protocol is actually optimal, the
relative entropy functional of the state $\sigma$ after permutation
with respect to an appropriate separable state $\rho$ will be 
calculated. The separable state $\rho$ is taken to be 
$
	\rho=
	\sum_{j=0}^J 
	p_j \rho_j , 
$
where 
\begin{equation}
\rho_j=
	\sum_{\alpha_j,\beta_j=1}^{d_j}
	\sum_{m=-j}^j
	\frac{
	|j,m,\alpha_j\rangle\langle j,m,\alpha_j|\otimes
	|j,m,\beta_j\rangle\langle j,m,\beta_j|
	}
	{2j+1}.
\end{equation}
All subspaces associated with different values of $j$, $m$,
$\alpha_j$, and $\beta_j$ are orthogonal. Therefore, one may 
use the direct sum property of the relative entropy to 
achieve
\begin{equation}\label{edfull}
	S(\sigma||\rho)=
	\sum_{j=0}^J d_j^2 p_j \log(2j+1).
\end{equation}
This is identical to 
the value given for the average number of maximally
entangled states obtained when applying the above
procedure. It is therefore 
also identical to the
distillable entanglement $D_{\leftrightarrow}(\sigma)$ with respect to LOCC 
operations. As $\Delta I= S(\sigma)=
-\sum_{j=0}^J 
p_j \log_2(p_j)$,
it follows that $\Delta D_{\leftrightarrow}/\Delta I\leq1$ for all $n$ for this 
particular initial state. 
\proofend

\newfigure{8.0cm}{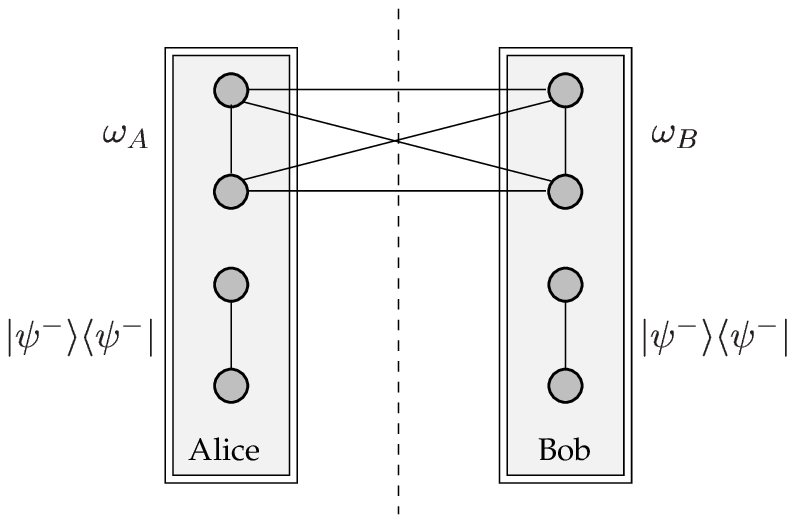}{Figure 5.3:
The structure of the state after step 5. in the above protocol.
This figure corresponds to $n=4$, $J=2$, and $j=1$. Both
Alice and Bob hold at this stage a pair of qubits in the
singlet state  $|\psi^-\rangle\langle\psi^-|$
with state vector $|\psi^-\rangle=(|01\rangle-|10\rangle)/\sqrt{2}$,
which can be discarded.
$\omega_A$ and $\omega_B$ denote the reduced state of Alice and Bob, 
respectively, of the first two pairs of qubits.}

\medskip

\noindent {\bf Proposition 5.6. --} {\it 
Let Alice and Bob share 
$n$ pairs of qubits each in the same pure
state with a state vector $|\psi\rangle$.
The associated Hilbert space is ${\cal H}={\cal H}_A\otimes {\cal H}_B$, where
${\cal H}_A={\cal H}_B=({\mathbbm{C}}^2)^{\otimes n}$.
Both parties then lose all classical
information about the order of their $n$ particles, 
such that the state of the composite system becomes}
\begin{equation}
	\sigma = 
	\sum_{\pi_A,\pi_B \in S_n}
	(\pi_A\otimes \pi_B)
	|\psi\rangle\langle\psi|^{\otimes n}
	(\pi_A\otimes \pi_B).
\end{equation}
{\it The distillable entanglement of the state $\sigma$
can then be calculated exactly, and
the ratio between the change of distillable 
entanglement $\Delta D_{\leftrightarrow}$ and the
amount of erased information $\Delta I=S(\sigma)$  for any $n=1,2,...$ 
obeys the inequality
\begin{equation}\label{ineq}
	\frac{\Delta D_{\leftrightarrow}}{\Delta I}\le 1  ,
\end{equation}
with equality for $n=2$.}

\probreak

\proof
One can set according to the Schmidt decomposition
\begin{equation}
	|\psi\rangle=\sqrt{\alpha}|00\rangle+\sqrt{\beta}|11\rangle,
\end{equation}
with $\alpha\in[0,1]$, $\beta=1-\alpha$. Assume
that $n=2J$ with a $J=1,2,...$. 
It is then possible to pursue the same argument as in 
the preceding case, and the
same protocol is optimal for distillation with
respect to separable operations. 
The state $\sigma$ after permutation will be 
found to be
\begin{equation}
	\sigma=\sum_{j=0}^J p_j 
	|\psi_j(\alpha_j,\beta_j)\rangle
	\langle  \psi_j(\alpha_j,\beta_j)|
\end{equation}
with 
\begin{equation}
	p_j=\sum_{m=-j}^j \frac{\alpha^{(J-m)}\beta^{(J+m)}}{d_j}.
\end{equation}
The unnormalized state vectors 
$
|\psi_j(\alpha_j,\beta_j)\rangle\langle \psi_j(\alpha_j,\beta_j)|
$
are defined as 
\begin{equation}
|\psi_j(\alpha_j,\beta_j)\rangle=\sum_{m=-j}^j
	\alpha^{j-m}\beta^{j+m}
	|j,m,\alpha_j\rangle |j,m,\beta_j\rangle.
	\end{equation}
The distillable entanglement is given by
\begin{equation}
D_{\leftrightarrow}(\sigma)=
\sum_{j=0}^J d_j^2 p_j S(\text{tr}_A
	[|\psi_j(1,1)\rangle \langle \psi_j(1,1)|])
	\end{equation}
and $\Delta I=S(\sigma)$. In addition to this, there are initially 
\begin{equation}
	D_{\leftrightarrow}(|\phi\rangle\langle\phi|^{\otimes n})=
-n(\alpha \log_2 (\alpha)+ \beta \log_2 (\beta))
\end{equation}
ebits of entanglement, so that it may be concluded that 
$\Delta D_{\leftrightarrow}/\Delta I\leq 1$ for all $n$ holds for this
case as well. The case that $n=2J+1$ with a $J=1,2,...$ 
can be analyzed in an analogous fashion, and again, one obtains
$\Delta D_{\leftrightarrow}/\Delta I\leq 1$. In the particular
case that $n=2$ it follows
that $\Delta D_{\leftrightarrow}=\Delta I$.

\proofend

This proposition provides a general statement in the case
that the lost classical information is the information about
the order of several qubits. If a certain number of bits
of classical information disappears, not more than the
same amount of distillable entanglement is lost -- as measured
in ebits. Again, it should be emphasized that this setting
can be interpreted in terms of channel capacities \cite{HoroChannel,NielsenOld}.

\section{Concluding Remarks}

Motivated by these findings one
may conjecture that the inequality 
$
	\Delta D_{\leftrightarrow}/ \Delta I\leq 1
$
holds in general, also if the loss of classical information
is not due to the loss of a classical record about the
identity of a number of qubits. \\

\noindent {\bf Conjecture 5.7. --} {\it 
Let $\sigma_1,...,\sigma_n\in{\cal S}({\cal H})$ be pure states,
each one of which is assigned a classical 
probability $p_1,...,p_n$. Let
\begin{eqnarray}
	\Delta D_{\leftrightarrow}&=&\sum_{i=1}^n p_i D_{\leftrightarrow}(\sigma_i)-
		D_{\leftrightarrow}\bigl(\sum_{i=1}^n p_i \sigma_i\bigr),\\
		\Delta I&=&S\bigl(\sum_{i=1}^n p_i \sigma_i\bigr).
\end{eqnarray}
Then the change in distillable entanglement
$\Delta D_{\leftrightarrow}$ and the loss of classical information $\Delta I$
obey the inequality
\begin{equation}
\frac{\Delta D_{\leftrightarrow}}{\Delta I}\leq 1.
\end{equation}}

\probreak

The problem can be considerably simplified if
not the distillable entanglement $D_{\leftrightarrow}$
but the relative entropy of entanglement $E_R$
is taken to be the figure of merit instead of the
distillable entanglement (see also Ref. \cite{VedralInfo} and 
Ref.\ \cite{PlenioPriv}).\\

\noindent {\bf Proposition 5.8. --} {\it 
Let $\sigma_1,...,\sigma_n\in{\cal S}({\cal H})$ be pure states,
each one of which is assigned a classical 
probability $p_1,...,p_n$. Let
\begin{eqnarray}
	\Delta E_R&=&\sum_{i=1}^n p_i E_R(\sigma_i)-
		E_R\bigl(\sum_{i=1}^n p_i \sigma_i\bigr),\\
		\Delta I&=&S\bigl(\sum_{i=1}^n p_i \sigma_i\bigr).
\end{eqnarray}
Then the change in relative entropy of entanglement
$\Delta E_R$ and the loss of classical information $\Delta I$
obey the inequality
\begin{equation}
	\frac{\Delta E_R}{\Delta I}\leq 1.
\end{equation}
}

\probreak

\proof The functional $f:{\cal S}({\cal H})\times
{\cal S}({\cal H})\longrightarrow \mathbbm{R}^{+}$
defined as $f(\sigma||\rho)=S(\sigma||\rho)+S(\sigma)$
is linear in its first argument. Let
$\omega\in{\cal D}({\cal H})$ 
be a state satisfying
\begin{equation}
	S(\sum_{i=1}^n p_i \sigma_i||\omega)=\min_{\rho\in{\cal D}({\cal H})}
	S(\sum_{i=1}^n p_i  \sigma_i||\rho),
\end{equation}
then
\begin{eqnarray}
	f(\sum_{i=1}^n p_i \sigma_i||\omega)&=&
	E_R(\sum_{i=1}^n p_i \sigma_i)+S(\sum_{i=1}^n p_i\sigma_i)=
	\sum_{i=1}^n p_i S(\sigma_i||\omega)\nonumber \\&\geq&
	\sum_{i=1}^n p_i E_R(\sigma_i),
\end{eqnarray}
and therefore
$\sum_{i=1}^n p_i E_R(\sigma_i)- E_R(\sum_{i=1}^n p_i \sigma_i)
\leq S(\sum_{i=1}^n p_i \sigma_i)$.
\proofend

The statement of Conjecture 5.7 is intimately related to
another inequality which is highly relevant in the
context of the 
\index{Quantum capacity}
{\it quantum capacity of a quantum channel}\/.
In Ref.\ \cite{HoroChannel} Conjecture 5.7 
has been discussed
with reference to Ref.\ [E2]. It has 
therein been demonstrated that
if the {\it hashing inequality}\/
\begin{equation}\label{hashing}
	D_{\rightarrow}(\rho)
	\geq
	I_B(\rho)
\end{equation}
holds for all states $\rho\in{\cal S}({\cal H})$, 
then Conjecture 5.7 is true as well. 
In inequality (\ref{hashing}),
$D_{\rightarrow}(\rho)$ denotes the distillable entanglement
of $\rho$ with respect to one-local operations with
classical communication from Alice to Bob, and 
$I_B(\rho)$ is the {\it coherent information}\/ of $\rho$
with respect to Bob, that is,\index{Coherent information}
\begin{equation}
I_B(\rho)=S(\text{tr}_B[\rho])-S(\rho).
\end{equation}
It is in fact one of the major issues of quantum information theory
to prove the validity of the hashing inequality.
%
\index{Hashing inequality}
With this inequality at hand, the quantum channel capacity
problem would essentially be solved. This astonishing fact
has been pointed out
in Ref.\ \cite{HoroChannel}, which presents
an elegant unifying approach to 
several quantum channel capacity problems.
However, the hashing inequality is already known to be 
true for a rather large class of states. 
These are the 
pure states, mixtures of two Bell states, and all states
for which the distillable entanglement is equal to
the entanglement of formation \cite{HoroChannel,PlenioPriv}.




\chapter{Quantum Information and Game Theory}
\chapterbreak

\section{Introduction}
Game theory is the theory of conflict between rationally
acting opponents. 
The subject of game theory is the
mathematical
analysis of situations in which several
parties
are interested only in winning and
make decisions according to their personal
interest, situations that involve 
contest, rivalry, or struggle.
%
%
The fundamentals of game theory were laid by 
{\' E}mile Borel, and, first and foremost, 
by the mathematician
John von Neumann and the economist Oskar
Morgenstern, who realized that several
problems in economic behavior resemble
in structure the
mathematical notions of ``games of strategy''. 
Since the appearance of their seminal work 
(Ref. \cite{NeuMor47}) the theory has developed
rapidly and has found manifold
applications
in the social sciences, biology, or economics.
Of particular interest to the theory
are
\index{Games of incomplete information}
{\it games of incomplete information}\/ in which  
the parties may choose their plan of action 
with complete knowledge of the situation on rational 
grounds, but without knowing what decision the other parties
would actually take\footnote{Jacob Brownowski, 
a later co-worker of
von Neumann, recalls a conversation with
von Neumann during a taxi ride in which he explained
his understanding of a game:
''I naturally said to him, since I am an enthusiastic
chess player, 'You mean, the theory of games like chess.'
'No, no,' he (von Neumann) 
said. 'Chess is not a game. Chess is a well-defined
form of computation. You might not be able to work out the
answers, but in theory there must be a solution, a right
procedure in any position. Now real games,' he said, 'are
not like that at all. Real life is not like that. Real life
consists of bluffing, of little tactics of deception,
of
asking yourself what is the other man going to think I
mean to do. And this is what games are about in my theory'.''
Cited after Ref.\ \cite{Pou92}.}.
This chapter is devoted to the
idea of identifying 
strategic moves in the sense of game theory 
with quantum operations as introduced in Refs. [E6] and
\cite{Mey99}. 
This approach
appears to be fruitful in several ways:

Firstly, some recently proposed
applications of quantum information theory
can already be conceived as competitive 
situations in which 
several parties -- or players -- interact
with more or less opposed
motives.  Quantum cloning has been
formulated as a game between two players \cite{Wer98}. 
Eavesdropping in quantum cryptography \cite{BenBra84,GisHut97}
can equally 
be regarded as a game between the eavesdropper
and the sender, and there are similarities
of the extended form of quantum versions of games
and quantum algorithms \cite{Shor,Ekert}.

Secondly, a generalization of
the theory of decisions into the domain of
quantum probabilities seems interesting,
because the roots of game theory are partly in 
classical
probability theory. In this context 
the question to be addressed would be what
solutions could be attained if superpositions
of strategies are allowed for.

Thirdly, while game theory
rarely deals with the 
transmission of information explicitly,
it should nevertheless
be noted that the practical implementation 
of any (classical) game inevitably involves the
exchange of information by classical means. 
Bearing this in mind, it
becomes legitimate to ask what happens if
the carriers of information are taken to be
quantum systems.
\footnote{Game theoretical settings in the quantum
domain have first been
considered in Refs. [E6] and \cite{Mey99}. 
In Ref. \cite{Mey99} an elegant quantum analogue of 
the Penny Flip game has been introduced, and the 
connection to quantum algorithms has been sketched.
Some popular articles 
about the issues raised 
in Refs.\
[E6] and \cite{Mey99} 
can be found in Refs.\ 
\cite{Game1,Game2,Game3,Game4,Game5}.
In the meantime 
a series of other articles has appeared on 
investigations of such quantum games: 
in Ref.\ \cite{WeberGames} (see also
\cite{Comment2}) an alternative  quantization scheme has been
proposed. Multi-player extensions have been 
introduced and investigated in Refs. \cite{Hayden,MultiPlayer2,JohnsonGames,MultiPlayer3}.
Ref. \cite{Hall} deals with the so-called Monty Hall problem. Evolutionary
games have been considered in Ref.\ \cite{Iqbal}.
The connection between quantum algorithms and game theory
has been further developed in Ref.\ \cite{Meyer2}, in which has
been shown that a quantum algorithm 
for an oracle problem may be 
understood as a quantum strategy for a player 
in a certain game.
This chapter deals with the results presented
in Refs.\ [E5] and [E6].}

\section{Game Theory}

To start off, a few concepts will be
summarized which are essential for 
the later considerations.
As for a comprehensive
general introduction to game theory
\index{Game theory}
the reader may be referred to Refs.\ \cite{Mye91,Schelling},
which give an exhaustive overview.
Ref.\ \cite{OtherGames} is a 
non-technical introduction to game theory,
and a text furnished with biographical remarks on
John von Neumann can be found in Ref. \cite{Pou92}.

Consider the following situation: two players,
Alice and Bob, 
 simultaneously place  a penny on a table.
The pennies can only show heads or tails up.
If the pennies match, that is, if both
heads or both tails are up, Alice
is allowed to keep both pennies, if they do not match,
Bob gets both pennies. The \index{Matching Pennies}
pay-off of each player for different outcomes
is displayed in Fig.\ 6.1.
Each player tries to maximize his or her 
final pay-off. What are they best advised to do?

\newfigure{6cm}{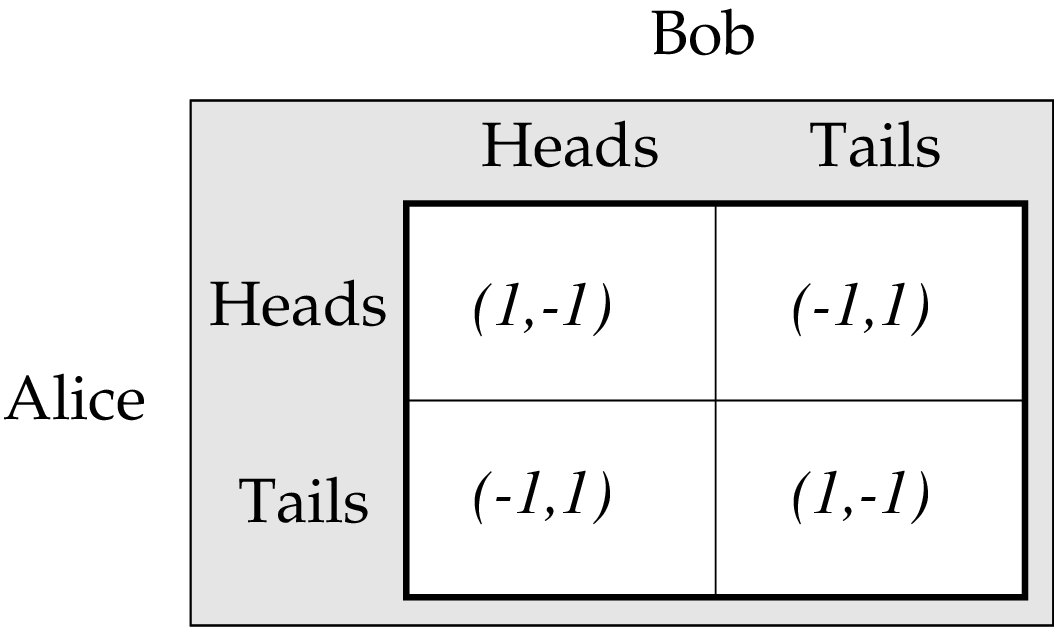}{Figure 6.1:
The pay-off table in the so-called {\it Matching Pennies game}\/.
The first entry refers to Alice's pay-off, the second
to Bob's.}

This very simple
example demonstrates \index{Player}
the typical structure of a game.
A {\it game}\/ involves several {\it players}\/ -- at least
two -- who implement one out of several
possible strategies, that is, make a decision.
Many aspects of game theory can already be understood when
only two players are involved. 
For reasons of simplicity, the following investigation
will be restricted to the case of a {\it two-player game}\/. 
A {\it strategy}\/ is a complete plan
of action. Depending on the choices of both
players they achieve a certain {\it pay-off}\/.
More formally, a (strategic-form)
\index{Two-player game}
\index{Strategic form}
 two-player game $\Gamma=(\{A,B\},S_A,S_B,u_A,u_B)$ is 
fully defined by the {\it set of players}\/, 
the {\it sets  of strategies}\/ $S_A$ and $S_B$, 
\index{Strategy}
the {\it utility functions}\/
\index{Utility function}
$u_A$ and $u_B$ 
defined on $S_A\times S_B$
specifying the pay-off for each player, 
and additional
rules of
the game consistent with the set of strategies.
The {\it pay-off}\/ provides \index{Pay-off}
a quantitative characterization of 
their individual preferences.
Both players are assumed to want to maximize
their respective pay-off, yet 
they must pick their choice without
knowing the other player's decision.

The most important solution concept for
\index{Nash equilibrium}
a game of this type
is the  {\it Nash equilibrium}\/.
A pair of strategies $(s_A,s_B)$
is said to be a (Nash) equilibrium
in {\it pure strategies}\/ if \index{Pure strategies}
\begin{eqnarray}
	u_A(s_A,s_B)&\geq& u_A(s_A', s_B),\\
	u_B(s_A,s_B)&\geq& u_B(s_A, s_B')
\end{eqnarray}
for all strategies $s_A'\in S_A$ and $s_B'\in S_B$.
Hence, in an equilibrium no player can gain by
unilaterally deviating from this equilibrium.
Given that the other player will stick to the strategy
corresponding to the equilibrium, the best
result is achieved by also playing the equilibrium solution.
Finding the equilibria of a game essentially equals
''solving'' a game.

An important class of games is the class of 
zero-sum games. A {\it zero-sum game}\/ 
is defined in that
the pay-offs of both players sum up to
\index{Zero-sum game}
zero 
for all possible pairs of strategies of
the players, that is, that
$u_A(s_A,s_B)=-u_B(s_A,s_B)$
for all possible pairs $(s_A,s_B)$, $s_A\in S_A$, $s_B\in S_B$ 
of strategies. In a zero-sum game the
interests of the players are diametrically
opposed, and the gain of one player is the
loss of the other player. 
Although not representing a typical competitive
situation, this class of games provides a suitable
object of analysis due to the strong tools of
solution that are available.
The most relevant property of such
zero-sum games is summarized in
von-Neumann's {\it min-max theorem}\/
\cite{NeuMor47}.

In the min-max theorem a value $v$ is assigned
to every 
two-person zero-sum game with
finite sets of strategies, which 
is the average pay-off that one player
can expect to win from the other player as long as
both players act rationally.
There exists at least one Nash equilibrium, while in the
case of several equilibria the pay-off will always
be represented by the same value $v$.
A particular strategy that ensures this return
$v$ to each player is called {\it min-max strategy}\/:
%
A player then adopts
a pessimistic  point of view and  
maximizes the minimal pay-off he or she may
obtain when implementing this strategy. The
equilibrium in min-max strategies is a Nash equilibrium.
\index{Min-max theorem}

\newfigure{7.67cm}{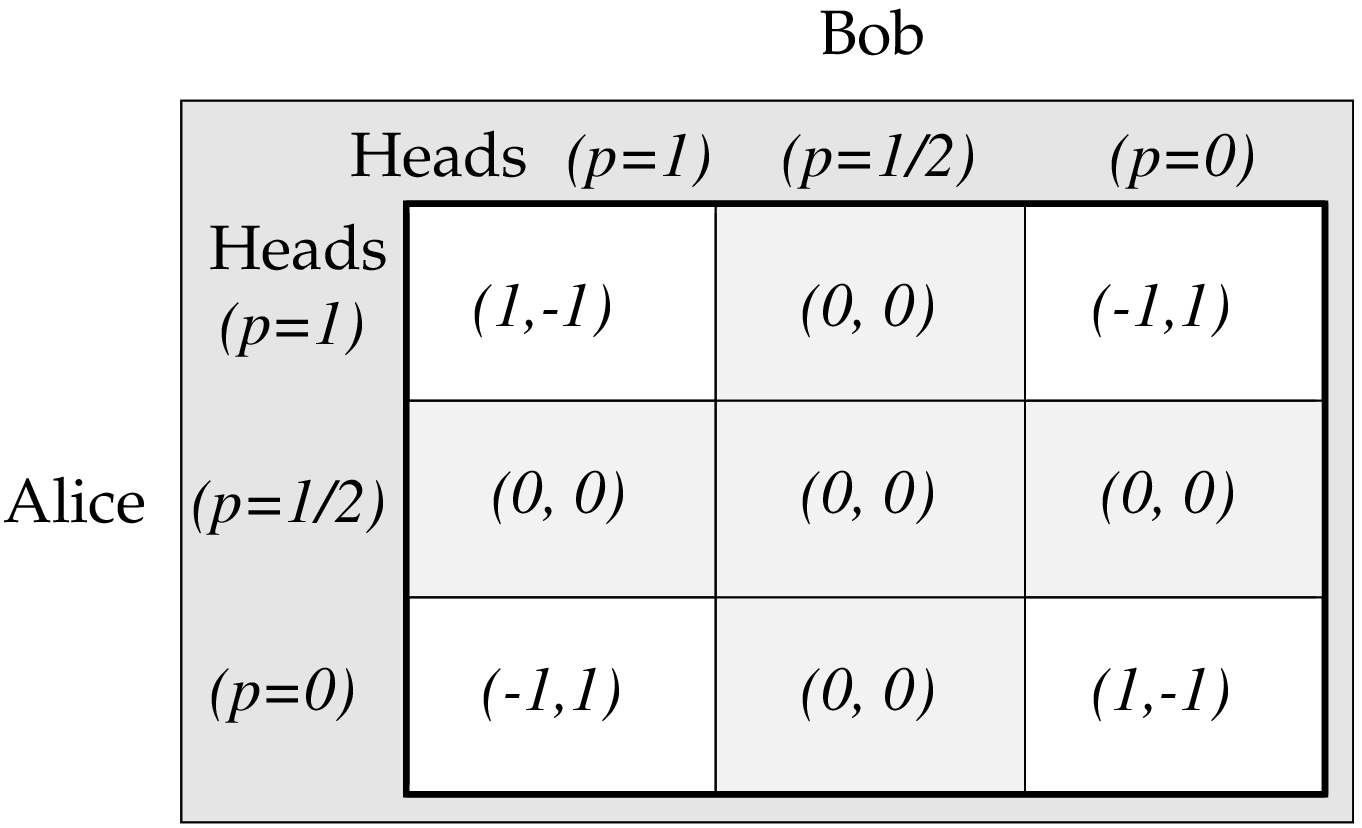}{Figure 6.2:
The pay-off matrix of the {\it Matching Pennies game}\/
including the (particular)
mixed strategy
in which a player chooses heads and tails with
probability $1/2$.}

So what would be the min-max strategy in the simple
Matching Pennies game as described above? Obviously, 
there is no pair of pure strategies which is
a Nash equilibrium. If Alice plays heads, Bob
is better off with tails, if she chooses tails,
he is best advised to play heads.
However, the players can do
more than just choosing heads or tails: they
can randomly take heads or tails with certain
probabilities (which also corresponds to how
such a game would be played in reality).

Such a strategy in which a player specifies a certain 
classical probability distribution on the
set $S_A$ or $S_B$
of pure strategies, respectively, is called
a {\it mixed strategy}\/. In such mixed strategies, the
min-max strategy can easily be identified: both
players are best advised to
play heads and tails with probability
$1/2$, as displayed in Fig.\ 6.2. 
On average, both players achieve the pay-off $0$
which is the value $v$ of this zero-sum game.

As said before, {\it non-zero-sum games}\/ 
are far more typical.
Probably the most well-known non-zero-sum game 
\index{Non-zero-sum game}
is the so-called  {\it Prisoners' Dilemma}\/ 
\cite{Tuc50}.
In this game
\index{Prisoners' Dilemma}
Alice and Bob 
have the choice between 
``cooperation'' 
and ``defection''. 
Being well aware of the 
consequences of their decisions 
the players obtain a certain pay-off
according to their respective 
strategies.

\newfigure{5.6cm}{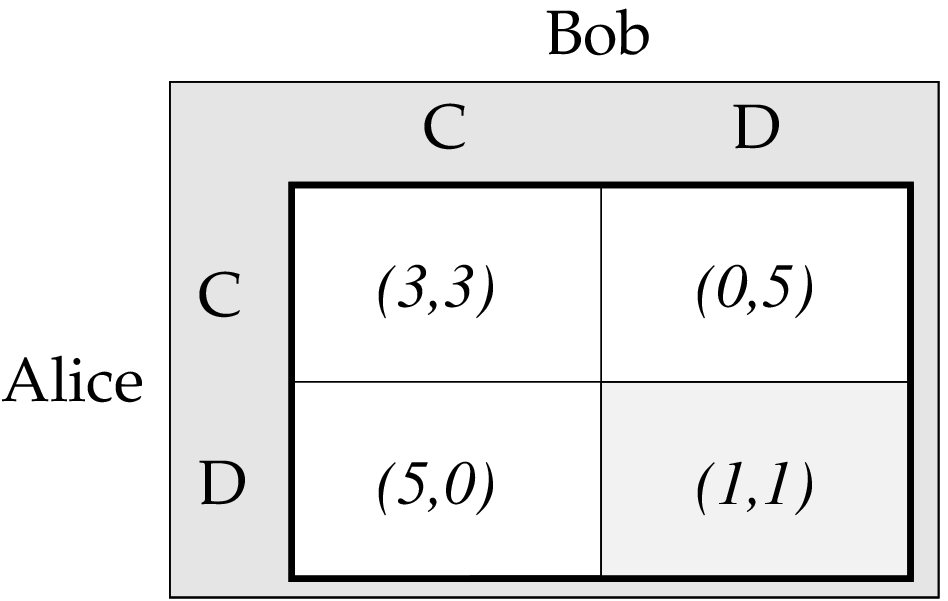}{Figure 6.3:
The pay-off matrix in the Prisoners' Dilemma game.
The first entry refers to Alice's pay-off, the second
to Bob's. If both players
cooperate, they both get $3$ units pay-off. 
If Bob defects and Alice happens
to cooperate, he obtains $5$ units, while Alice 
is in the unfortunate situation in which she does
not receive any pay-off at all. Bob faces the same
situation if he chooses to cooperate while 
Alice prefers to defect. If both defect, they 
equally get $1$ unit pay-off.}

Fig.\ 6.3
indicates the pay-off of Alice and Bob.\footnote{
For the purposes of this chapter, particular
values have been chosen for 
$A_{CC}$, $A_{CD}$, $A_{DC}$, $A_{DD}$, $B_{CC}$, $B_{CD}$, 
$B_{DC}$, and $B_{DD}$.
From a game theoretical viewpoint, any positive numbers
satisfying the symmetry conditions $A_{CC}=B_{CC}$, $A_{DD}=B_{DD}$,
$A_{CD}=B_{DC}$, $A_{DC}=B_{CD}$ and the inequalities
$A_{DC}>A_{CC}>A_{DD}>A_{CD}$ and $A_{CC}\geq (A_{CD}+A_{DC})/2$
define a (strong) Prisoners' Dilemma.
}
As Alice is better off with defection regardless
of Bob's choice, she will defect. The game being symmetric,
the same argument applies to Bob. The players face a 
{\it dilemma}\/ because
rational reasoning makes them defect, 
although they
would both benefit from mutual cooperation.

The strategy of ``defection'' is called 
{\it dominant strategy}\/: it is favorable
regardless what strategy the other party
picks.
(``defection'', ``defection'') is an equilibrium 
in dominant strategies. 
This equilibrium in dominant strategies is
the unique Nash equilibrium, and this
uniqueness also holds in mixed strategies.
It is worth noting  
that if the players agree upon
\index{Iterated game}
playing the game $N$ times, 
they will both opt for
''defection'' $N$ times \footnote{
This statement may be deduced by
backward-induction: in the last round -- the $N$-th -- 
the players will definitely not cooperate. In
the previous round they would only choose cooperation
if this behavior were rewarded in the $N$-th
round. As any agreements concerning the choice of 
the $N$-th round would not be met, the players
will stick to defection in the $N-1$-th round, and
so on until the first round.
This reasoning does not
apply to the case where the number of rounds in such an
iterated game is not known to the players in advance.
In 1980 a computer tournament was devised by Robert Axelrod 
\cite{Axelrod},
and specialists of the field were asked to
participate in a Prisoners' Dilemma tournament with
a varying number of rounds. 
A very simple strategy submitted by
Anatol Rapoport won:
\index{Tit-for-tat}
in the first round cooperation is
played, and from then on the strategy of the other
player of the previous round is copied ({\it tit-for-tat}\/).}.

A pair of strategies is called {\it Pareto-optimal}\/ 
\index{Pareto-optimal}
if
there is no outcome in which both players
do simultaneously better. 
From this perspective  
the Dilemma lies in the fact that the
unique Nash equilibrium is far from being 
Pareto-optimal. The importance
of the Prisoners' Dilemma stems
from the fact that it models a dilemma
that may arise in many situations involving
conflicting interests.




\section{Quantum Games, Strategies, and Equilibria}

This subsection deals with a ``quantum version''
of a game.
Any quantum system which can be manipulated
by two parties or more and
for which the utility
of the moves can be 
quantified in an appropriate
manner
\index{Quantum game}
may be considered a quantum game.
In the following definition, 
the physical system which serves as the underlying
setup of the game is included, as
the word ``quantum'' already 
points to the physical carrier 
of information. The quantum games
proposed in Refs.\ [E6], \cite{Mey99}, and
\cite{Vaidman}, and later in Refs.\ 
\cite{WeberGames,Hayden,MultiPlayer2,JohnsonGames,MultiPlayer3,Hall,Meyer2}
can be cast into this form; also, the
quantum cloning device as described in \cite{Wer98}
can be said to be a quantum game in this sense.

A  {\it two-player
quantum game}\/ $\Gamma=({\cal H},\rho,S_A,S_B,P_A,P_B)$
is completely specified by 
\begin{itemize}
\item[(i)] the underlying Hilbert space 
${\cal H}$ of the physical system, 
\item[(ii)]
the {\it initial state}\/ 
$\rho\in {\cal S}({\cal H})$,
\item[(iii)]
the sets $S_A$ and $S_B$ of 
permissible {\it quantum strategies}\/ of
the two players, and
\item[(iv)]
the  {\it utility functionals}\/
\index{Utility functional}
$P_A$ and $P_B$, which specify the
utility for each player.
\end{itemize}

A {\it quantum strategy}\/
\index{Quantum strategy}
${\cal E}_A\in S_A$, ${\cal E}_B\in S_B$
is a  quantum operation, that is,
\index{Completely positive map}
a completely positive trace-preserving map 
${\cal E}_A,{\cal E}_B:
{\cal S}({\cal H})\longrightarrow {\cal S}({\cal H})$.
The  definition of a quantum game
also includes certain implicit rules 
such as the order
of the implementation of the respective
quantum strategies.

As in the case of ordinary games, 
a quantum game is called {\it zero-sum game}\/ if the
\index{Zero-sum game}
expected pay-offs sum up to zero for all pairs of
strategies, that is, if 
\begin{equation}
P_A({\cal E}_A,{\cal E}_B)=-P_B({\cal E}_A,{\cal E}_B)
\end{equation}
for all ${\cal E}_A\in S_A$, ${\cal E}_B\in S_B$. Otherwise, it is 
called a {\it non-zero-sum game}\/.
\index{Non-zero-sum game}
Note that it is not required 
that a set of allowed 
strategies for a player forms a closed set.
%
Two quantum strategies of 
Alice ${\cal E}_A$ and
${\cal E}_A'$ will be called
{\it equivalent}\/, if \index{Equivalent strategy}
\begin{equation}
P_A({\cal E}_A,{\cal E}_B)=P_A({\cal E}_A',{\cal E}_B)
\,\,\,\,\,
\text{ and }
\,\,\,\,\,
P_B({\cal E}_A,{\cal E}_B)=P_A({\cal E}_A',{\cal E}_B)
\end{equation} 
for {\it all}\/ possible ${\cal E}_B$. That is, if 
${\cal E}_A$ and ${\cal E}_A'$ yield the same expected
pay-off for both players
for all allowed strategies of Bob.
Strategies ${\cal E}_B$ and ${\cal E}_B'$ 
of Bob will be identified accordingly.

A solution concept provides advice to the players 
with respect to the
action they are best advised to take. 
%
%
%
As before, a quantum strategy  of Alice
${\cal E}_A$ 
is called a {\it dominant strategy}\/
\index{Dominant strategy}
if
\begin{equation}
        P_A({\cal E}_A,{\cal E}_B')
        \geq
        P_A({\cal E}_A',{\cal E}_B')
\end{equation}
for all ${\cal E}_A'\in S_A$, ${\cal E}_B'\in S_B$.
Analogously, one can define a dominant strategy
for Bob. 
A pair $({\cal E}_A,{\cal E}_B)$ is said to be an 
\index{Equilibrium in dominant strategies}
{\it equilibrium 
in dominant strategies}\/ 
if ${\cal E}_A$ and ${\cal E}_B$ are
the players' respective dominant strategies.
A combination of strategies $({\cal E}_A,{\cal E}_B)$
is called a 
\index{Nash equilibrium}
{\it Nash equilibrium}\/ 
if
\begin{eqnarray}
        P_A({\cal E}_A,{\cal E}_B)&\geq& P_A({\cal E}_A',{\cal E}_B),\\
        P_B({\cal E}_A,{\cal E}_B)&\geq& P_B({\cal E}_A,{\cal E}_B')
\end{eqnarray}
for all ${\cal E}_A'\in S_A$, ${\cal E}_B'\in S_B$.
Again,
a pair of strategies $({\cal E}_A, {\cal E}_B)$ is called 
{\it Pareto optimal}\/, if it is not
possible to increase one player's pay-off
without lessening the pay-off of the other
player.

\section{Two-Qubit Quantum Games}

In this section specific games will be investigated
where the {\it classical version of the game}\/ is faithfully
entailed in the quantum game.
In a quantum version of a 
binary choice game,
two qubits are prepared by an {\it arbiter}\/ 
in a particular initial state, and are then
sent to the two players who dispose of 
physical instruments in order to
manipulate their qubits appropriately.
In a last step the qubits are sent back
to the arbiter who performs a measurement to evaluate
the pay-off. \footnote{By 
classical means, 
a two player binary choice game 
may be played as follows: 
An arbiter takes two coins 
and forwards one 
each to the players.
The players then receive their coin
with heads up and 
may keep it as it is (the first
pure strategy)
or turn it upside down so that
tails is up (the second strategy). 
Both players then return the coins to the
arbiter who calculates the players'
final pay-off corresponding to the combination
of strategies he obtains from the players.
Here, the coins serve as the physical
carrier of information in the game.}
 
%
For such a bi-partite quantum game
the system of interest 
is a quantum system with underlying Hilbert space
$
	{\cal H}={\cal H}_A\otimes{\cal H}_B$, 
${\cal H}_A={\cal H}_B={\mathbbm{C}}^2$,
and associated 
state space ${\cal S}({\cal H})$.
The quantum strategies of Alice and Bob
${\cal E}_A$ and ${\cal E}_B$ 
are local trace-preserving
quantum operations acting in 
${\cal H}_A$ and ${\cal H}_B$ respectively.
In other words, Alice and Bob are restricted to 
implementing 
their respective quantum strategy
${\cal E}_A$ and ${\cal E}_B$ on their qubit only.
In this step they may 
choose any quantum
strategy that is included
in the sets of strategies $S_A$ and $S_B$. They are
both aware of the sets $S_A$ and $S_B$, 
but they do not know which 
particular quantum strategy the
other party will actually implement. 
As the application of both quantum 
strategies amounts to a map
$
	{\cal E}_A\otimes {\cal E}_B:{\cal S}({\cal H})\rightarrow 
{\cal S}({\cal H})$,
the system will after execution of the moves be
in the state
\begin{equation}
	\sigma=({\cal E}_A\otimes {\cal E}_B)(\rho).
\end{equation}
The quantum strategies           
${\cal E}_A\otimes {\mathbbm{1}}_B$ and
${\mathbbm{1}}_A\otimes {\cal E}_B$
are
identified with ${\cal E}_A$ and ${\cal E}_B$, respectively.

Particular attention  will be paid to unitary
operations, which are associated
with unitary operators $U_A$ and $U_B$,
written as ${\cal E}_A\simeq U_A$ and ${\cal E}_B\simeq U_B$.
In this case  the final state $\sigma$ is  given by
\begin{equation}
	\sigma=(U_A\otimes U_B) \rho (U_A\otimes U_B)^\dagger.
\end{equation}
If not otherwise specified
both the sets of strategies of
Alice and Bob and
the pay-off functionals 
are taken to
be identical, that is,
\begin{equation}
S_A=S_B=S\,\,\,\,\text{ and }\,\,\,\,P_A=P_B=P,
\end{equation}
such 
that both parties face the same situation.

\subsection{General Setup}

Let $\rho$ be a maximally entangled state
on ${\cal H}={\mathbbm{C}}^2\otimes {\mathbbm{C}}^2$.
In order to be consistent with
Ref.\ [E6]
let $\rho=|\psi_{CC}\rangle\langle\psi_{CC}|$ with
\begin{equation}\label{psis}
	|\psi_{CC}\rangle=
		(|00\rangle+i |11\rangle)/\sqrt{2}.
\end{equation}
Any other maximally entangled state on ${\mathbbm{C}}^2\otimes 
{\mathbbm{C}}^2$
would also be appropriate.
The quantum game 
$\Gamma=({\mathbbm{C}}^2\otimes {\mathbbm{C}}^2, \rho,S,S,P,P)$ 
can be played in the following way:
The two qubits are forwarded to the arbiter who performs
a projective selective measurement
on the final state $\sigma$
with Kraus operators
$\pi_{CC}$, $\pi_{CD}$, $\pi_{DC}$, and $\pi_{DD}$,
where
\begin{eqnarray}
 	\pi_{CC}&=&|\psi_{CC}\rangle\langle\psi_{CC}|,\,\,\,\,\,
	|\psi_{CC}\rangle=(|00\rangle+i |11\rangle)/\sqrt{2},\\
		\pi_{CD}&=&|\psi_{CD}\rangle\langle\psi_{CD}|,\,\,\,\,\,
	|\psi_{CD}\rangle=(|01\rangle-i |10\rangle)/\sqrt{2},
		\label{psi12}\\
		\pi_{DC}&=&|\psi_{DC}\rangle\langle\psi_{DC}|,\,\,\,\,\,
	|\psi_{DC}\rangle=(|10\rangle-i |01\rangle)/\sqrt{2},\\
	\pi_{DD}&=&|\psi_{DD}\rangle\langle\psi_{DD}|,\,\,\,\,\,
	|\psi_{DD}\rangle=(|11\rangle+i |00\rangle)/\sqrt{2}.
\end{eqnarray}
The Kraus operators correspond to a projective
measurement associated with
the basis of ${\cal H}$ consisting of 
$|\psi_{CC}\rangle$ and three
orthonormal state vectors. On the one hand, 
due to this choice the system will be in a maximally 
entangled state when it comes
to implementing the quantum operations. On the other hand it
is guaranteed that if both players prefer to implement the
identity operation, that is, ``do nothing'', then the detector
will click in the channel with label $CC$  with certainty.

According to the outcome of the measurement, 
a pay-off of $A_{CC}$, $A_{CD}$, $A_{DC}$, or $A_{DD}$
is given to Alice, Bob receives 
$B_{CC}$, $B_{CD}$, $B_{DC}$, or $B_{DD}$.
The utility functionals, also referred to as
expected pay-off of Alice and Bob, read
\begin{eqnarray}
	P_A({\cal E}_A,{\cal E}_B)&=&A_{CC} {\text{tr}}[\pi_{CC}
	\sigma]+
	A_{CD} {\text{tr}}[\pi_{CD}
	\sigma]\nonumber\\ &+&
	A_{DC} {\text{tr}}[\pi_{DC}
	\sigma]+
	A_{DD} {\text{tr}}[\pi_{DD}
	\sigma],\label{PA}\\
	P_B({\cal E}_A,{\cal E}_B)&=&
	B_{CC} {\text{tr}}[\pi_{CC}
	\sigma]+
	B_{CD} {\text{tr}}[\pi_{CD}
	\sigma]\nonumber\\ &+&
	B_{DC} {\text{tr}}[\pi_{DC}
	\sigma]+
	B_{DD} {\text{tr}}[\pi_{DD}
	\sigma].\label{PB}
\end{eqnarray}
The Kraus operators are chosen 
in such a way that the classical game is fully entailed
in the quantum game: The
{\it classical strategies}\/ 
\index{Classical strategies}
cooperation and defection are associated
with particular unitary operations,
\begin{equation}
	C\simeq 
	\left(
	\begin{array}{cc}
	1 & 0 \\
	0 & 1 
	\end{array}
	\right),\,\,\,\,\,
	D
	\simeq
	\left(
	\begin{array}{cc}
	0 & 1 \\
	-1 & 0 
	\end{array}
	\right).
\end{equation}
$C$ does not change the state at all, $D$ implements a 
''spin-flip''. If both parties stick to these
classical strategies, Eq.\ (\ref{PA}) and Eq.\ (\ref{PB}) 
guarantee that 
the expected pay-off
is exactly the pay-off of the corresponding classical game
defined by the numbers $A_{CC}$, $A_{CD}$, $A_{DC}$, $A_{DD}$,
$B_{CC}$, $B_{CD}$, $B_{DC}$, and $B_{DD}$.

To give an example, 
if Alice plays $C$ and Bob chooses $D$,
the state $\sigma$ after implementation of the
strategies is given by
\begin{equation}
	\sigma= (C\otimes D)(\rho)= 
	|\psi_{CD}\rangle
	\langle \psi_{CD}|,
\end{equation}
such that Alice obtains $A_{CD}$ units 
and Bob $B_{CD}$ 
units pay-off.
In this way
the specificities 
of strategic moves in the quantum domain
can be adequately studied. 
The players may make use of additional
degrees of freedom which are not available
by randomization of the classical
strategies, but
they can also stick to mere
classical strategies.
This scheme can be applied to any
two player binary choice game and is canonical
to a high extent.\\

\subsection{Prisoners' Dilemma}
For the Prisoners' Dilemma, the values in the
table in Eqs. (\ref{PA})
and (\ref{PB}) are given by
(see Fig.\ 6.3),
\begin{eqnarray}
	&&A_{CC}=B_{CC}=3,\,\,\,
	A_{DD}=B_{DD}=1,\\
	&&A_{CD}=B_{DC}=0,\,\,\,
	A_{DC}=B_{CD}=5.
\end{eqnarray}
In all of the following sets of 
allowed strategies $S$  the classical 
options (to defect and to cooperate)
are included. 
Several interesting sets of strategies
and  solution concepts
will now be studied. The first three
subsections involve local unitary operations only,
while in the last subsection other quantum
operations will be considered as well.\\

\noindent
{\bf Example 6.1. -- } 
{\it One-parameter set of strategies. --}\/
The first set of strategies $S^{(6.1)}$
involves 
quantum operations ${\cal E}_A$ and ${\cal E}_B$ 
which are local
rotations with a single parameter.
The matrix representation of the
corresponding unitary operators
is taken to be
\begin{equation}
        {U}(\theta)=
        \left(
        \begin{array}{cc}
        \cos(\theta/2)&
        \sin(\theta/2)\\
        -\sin(\theta/2)
        &
		  \cos(\theta/2)
        \\
        \end{array}
        \right)
\end{equation}
with $\theta\in[0,\pi]$. Therefore,
selecting
strategies ${\cal E}_A$ and ${\cal E}_B$ amounts
in this simple case
to choosing two angles
$\theta_A$ and $\theta_B$. 
The classical pure strategies 
of defection and
cooperation can
be implemented as 
	$C\simeq 
	U(0)$,
	$D
	\simeq
	U(\pi)$.
An analysis of the expected pay-offs $P_A$ and $P_B$,
\begin{eqnarray}
	P_A(\theta_A,\theta_B)&=&
	3 |\cos(\theta_A/2)\cos(\theta_B/2)|^2
	+
	5 |\cos(\theta_B/2)\sin(\theta_A/2)|^2\nonumber\\
	&+&
	|\sin(\theta_A/2)\sin(\theta_B/2)|^2,\\
	P_B(\theta_A,\theta_B)&=&
	3 |\cos(\theta_A/2)\cos(\theta_B/2)|^2
	+
	5 |\sin(\theta_B/2)\cos(\theta_A/2)|^2\nonumber\\
	&+&
	|\sin(\theta_A/2)\sin(\theta_B/2)|^2,
\end{eqnarray}
shows that this game is nothing else but
the classical Prisoners' Dilemma game. 
The pay-off functions are identical
to the analogous functions in a Prisoners'
Dilemma with mixed, that is, randomized,
strategies, 
where cooperation is chosen with the classical
probability 
$p=\cos^2(\theta/2)$. 
The inequalities
\begin{eqnarray}
P_A(D,{\cal E}_B)&\geq& P_A({\cal E}_A,{\cal E}_B),\\
P_B({\cal E}_A,D)&\geq& P_B({\cal E}_A,{\cal E}_B)
\end{eqnarray} 
hold
for all ${\cal E}_A, {\cal E}_B\in S^{(6.1)}$, and
therefore, $(D,D)$
is an equilibrium in dominant strategies, and 
thus the unique Nash equilibrium. As explained in 
Section 6.2,
this equilibrium is far from being efficient, 
because $P_A(D,D)=P_B(D,D)=1$
instead of the Pareto optimal pay-off which would
be 3.\\

\noindent
{\bf Example 6.2. -- } 
{\it Two-parameter set of strategies. --}\/
A more general set of strategies is
the following two-parameter set $S^{(6.2)}$.
The matrix representation of 
operators corresponding to
quantum strategies from this set
is given by
\index{Pareto-optimal}
\begin{equation}
        {U}(\theta,\phi)=
        \left(
        \begin{array}{cc}
        e^{i\phi}\cos(\theta/2)&
        \sin(\theta/2)\\
        -\sin(\theta/2)
        &e^{-i\phi}\cos(\theta/2)
        \\
        \end{array}
        \right)
\end{equation}
with $\theta\in[0,\pi]$ and $\phi\in[0,\pi/2]$. 
Selecting a strategy ${\cal E}_A,{\cal E}_B$
then means choosing
appropriate angles $\theta_A,\phi_A$ and $\theta_B,\phi_B$. 
The 
classical strategies of defection and
cooperation are also included in the set of
possible strategies, as
\begin{equation}
C\simeq U(0,0)\,\,\,\, \text{ and }\,\,\,\,
D\simeq U(\pi,0).
\end{equation}
The expected pay-off for Alice explicitly
reads 
\begin{eqnarray}\label{LongPay}
	P_A(\theta_A,\phi_A,\theta_B,\theta_B)
	&=&3\left| \cos(\phi_A+\phi_B)\cos(\theta_A/2)\cos(\theta_B/2)\right|^2
	\\
		&+& 5
		\left|
			\sin(\phi_A)\cos(\theta_A/2)\sin(\theta_B/2)\right.
			-
			\left.
			\cos(\phi_B) \cos(\theta_B/2)\sin(\theta_A/2)
		\right|^2\nonumber\\
		&+& \left|
			\sin(\phi_A+\phi_B)\cos(\theta_A/2)\cos(\theta_B/2)
			\right.
			+
			\left.
			\sin(\theta_A/2)\sin(\theta_B/2)
		\right|^2\nonumber
\end{eqnarray}
and the expected pay-off of Bob is given by
\begin{eqnarray}\label{LongPay2}
	P_A(\theta_A,\phi_A,\theta_B,\theta_B)
	&=&
	3\left| \cos(\phi_A+\phi_B)\cos(\theta_A/2)\cos(\theta_B/2)\right|^2
	\\
		&+& 5
		\left|
			\sin(\phi_B)\cos(\theta_B/2)\sin(\theta_A/2)\right.
			-
			\left.
			\cos(\phi_A) \cos(\theta_A/2)\sin(\theta_B/2)
		\right|^2\nonumber\\
		&+& \left|
			\sin(\phi_A+\phi_B)\cos(\theta_A/2)\cos(\theta_B/2)
			\right.
			+
			\left.
			\sin(\theta_A/2)\sin(\theta_B/2)
		\right|^2.\nonumber
\end{eqnarray}

\newfigure{7.5cm}{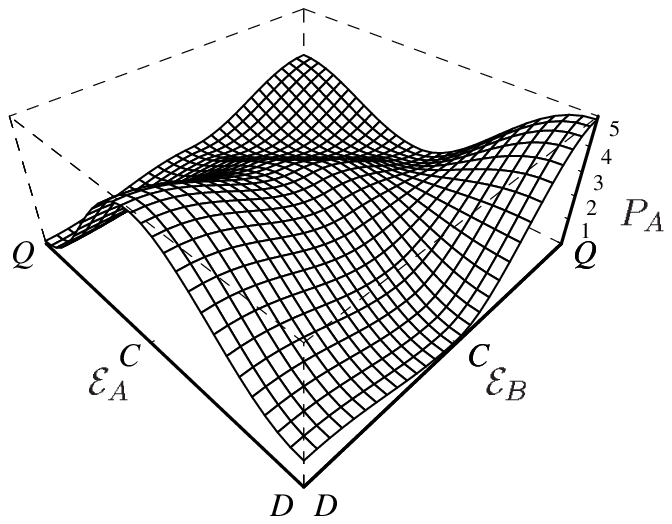}{Figure 6.4:
Alice's pay-off in the Prisoners' Dilemma game
with the set of strategies being $S^{(6.2)}$. In this
plot, a certain
parameterization has been chosen such
that the strategies ${\cal E}_A$ and ${\cal E}_B$ each depend
on a
single parameter $t\in[-1,1]$: 
${\cal E}_A\simeq U(t \pi, 0)$ for $t\in[0,1]$ and
${\cal E}_A\simeq U(0,-t\pi/2)$ for $t\in[-1,0)$, and
analogously for Bob.
Defection $D$
corresponds to the value $t=1$, cooperation $C$
to $t=0$, and $Q$ is represented by $t=-1$.}

\newfigure{5cm}{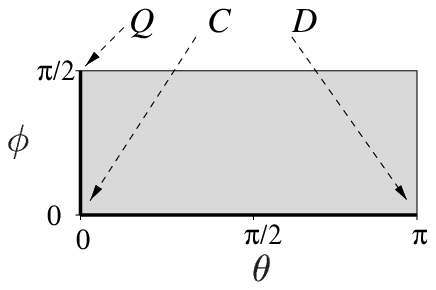}{Figure 6.5: 
This figure shows the image of the parametric curve
chosen in the previous figure.}

It becomes evident 
that the previous Nash equilibrium
$(D,D)$ of $S^{(6.1)}$
is no longer an equilibrium solution, as
both players may benefit from deviating from
the strategy $D$. At the same time a new Nash
equilibrium has emerged which will be given by
$(Q,Q)$. This strategy is not accessible in a 
classical game with mixed strategies.
The strategy $Q$ corresponds to a matrix
\begin{equation}
	Q\simeq U(0,\pi/2)=
   \left(\begin{array}{cc} i & 0 \\ 0 &  
	-i\end{array}\right)\, .
\end{equation}
For all $\theta_A\in[0,\pi]$ and $\phi_B\in[0,\pi/2]$
\begin{equation}
	P_A({\cal E}_A,Q)=
	\cos^2(\theta_A/2)
	\left(
		3\sin^2 (\phi_A)+ \cos^2 (\phi_A)
	\right)\leq 3.
\end{equation} 
In particular,
\begin{eqnarray}
	P_A({\cal E}_A,Q)\leq P_A(Q,Q)&=&3\\
	P_B(Q,{\cal E}_B)\leq P_B(Q,Q)&=&3
\end{eqnarray}
for all and all ${\cal E}_A\in S_A$ and all ${\cal E}_B\in S_B$,
such that no player can gain from unilaterally
deviating from $Q$.
This Nash equilibrium is unique   and serves
as the only acceptable solution of the game.

The astonishing fact is that $P_A(Q,Q)=P_B(Q,Q)=3$
(instead of $1$)
so that the Pareto optimum is realized. 
No player could
gain without lessening the other player's expected pay-off.
In this sense one can say that the Dilemma of
the original game has fully disappeared.
In the classical game only 
mutual cooperation is
Pareto optimal, but this pair of
strategies does not correspond
to a Nash equilibrium. If the players may resort
to quantum strategies, they can escape the Dilemma.\\

\noindent
{\bf Example 6.3. --} 
{\it General unitary operations. --}\/
One can generalize the previous setting to the
case where Alice and Bob can
implement operations ${\cal E}_A$ and ${\cal E}_B$ taken
from $S^{(6.3)}$, where $S^{(6.3)}$ is the set 
of general local 
unitary operations.
%
It might be suspected that 
the solution becomes
more efficient the larger the
sets of allowed operations are. 
But on the contrary, 
the previous Pareto optimal
unique Nash equilibrium $(Q,Q)$ ceases to be
an equilibrium solution if the set is enlarged: 
\footnote{This has already been reported in Ref.
[E6] and in greater detail in Ref.\ \cite{Comment}.
In fact, it is an open question 
whether there exist quantum games 
with (i) a unique 
equilibrium in {\it pure} 
strategies which is more efficient
than the equilibrium 
of the corresponding
classical game and (ii) in which 
the full group of local unitary operations 
is available to all players.
In this context, the 
somewhat ambiguous term solution
means either a 
unique Nash equilibrium
or an equilibrium which is clearly
distinguished from all the other
equilibria. For example, refinement
concepts like
\index{Perfect equilibrium}
{\it perfect, proper, and persistent
equilibria}\/  \cite{Mye91} 
or the {\it focal point effect}\/ \cite{Schelling}
may be employed
to eliminate all but one Nash equilibria.
Multi-player games like the elegant
scheme proposed in Ref.\ \cite{Hayden}
(see also Ref. \cite{MultiPlayer2})
may point towards a resolution of
this issue, although all equilibria
of the game of Ref.\ \cite{Hayden}
are fully symmetric, and the above
refinement concepts do not lead to a
particular 
distinguished equilibrium.
However, it is the subject of the remainder
of this example
and Example 6.3 
to show that in mixed strategies
there still exist efficient solutions of the 
quantum game.}
For any strategy ${\cal E}_B\in S^{(6.3)}$
there exists an optimal answer 
\index{Optimal answer}
${\cal E}_A\in S^{(6.3)}$ 
resulting in
\begin{equation}\label{Well}
	({\cal E}_A\otimes {\cal E}_B)(\rho)=
	|\psi_{DC}\rangle\langle \psi_{DC}|,
\end{equation}
with $\rho$ given in Eq.\ (\ref{psis}).
%
That is, for any strategy of Bob ${\cal E}_B$ 
there is a strategy ${\cal E}_A$ of Alice such that
\begin{equation}
	P_A({\cal E}_A,{\cal E}_B)=5\,\, \text{ and }\,\,
	P_B({\cal E}_A,{\cal E}_B)=0:
\end{equation}
Take
\begin{eqnarray}
		{\cal E}_A\simeq
		\left(
		\begin{array}{cc}
		a & b\\
		c & d 
		\end{array}
		\right),\,\,\,
		{\cal E}_B\simeq 
		\left(
		\begin{array}{cc}
		-i b & a\\
		-d & -ic 
		\end{array}
		\right),
	\end{eqnarray}
	where $a,b,c,d$ are appropriate complex numbers.
Given that Bob plays the strategy ${\cal E}_B$ associated
with a particular Nash equilibrium $({\cal E}_A,{\cal E}_B)$, Alice
can always apply the {\it optimal answer}\/ 
${\cal E}_A$ to achieve
the maximal possible pay-off. However, the resulting
pair of quantum strategies cannot be an equilibrium
since again, the game being symmetric,
Bob can improve his pay-off by
changing his strategy to his optimal answer ${\cal E}_B'$.
Hence, there is no pair $({\cal E}_A,{\cal E}_B)$
of pure strategies with the property that the
players can only lose from unilaterally
deviating from this pair of strategies. 

Yet, there remain 
Nash equilibria in mixed
strategies \index{Mixed strategies}
which are much more efficient than the classical 
outcome of the equilibrium in dominant strategies 
$P_A(D,D)=P_B(D,D)=1$. 
In a mixed strategy of Alice, say,
she selects a particular
quantum strategy ${\cal E}_A$ (which 
is then conceived as 
 pure strategy)
from the set of strategies ${\cal E}_A$
with a certain classical probability.
%
%
%
%
That is,
mixed strategies of Alice and Bob
are associated with
maps of the form 
\begin{equation}\label{doubly}
	\rho\longmapsto \sigma=\sum_{i,j=1}^n p^{(i)}_A p^{(j)}_B
	(U_A^{(i)}\otimes U_B^{(j)}) 
	\rho  
	(U_A^{(i)}\otimes U_B^{(j)})^\dagger, 
\end{equation}
$p^{(i)}_A, p^{(i)}_B \in [0,1]$, $i,j=1,2,...,n$, with
\begin{equation}
\sum_{i=1}^n p^{(i)}_A =\sum_{j=1}^n p^{(j)}_B =1.
\end{equation}
$U_A^{(i)}$
and $U_B^{(j)}$ are local unitary operators
corresponding to pure strategies ${\cal E}_A^{(i)}$
and ${\cal E}_B^{(j)}$. 

The map given by 
Eq.\ (\ref{doubly}) acts in ${\cal H}_A$ 
and  ${\cal H}_B$ 
\index{Doubly stochastic map}
as a doubly stochastic map, that is, as
a completely positive unital 
\index{Unital map}
map \cite{Majo2}. 
As a result, 
the final reduced states 
${\text{tr}}_B[\sigma]$ and 
${\text{tr}}_A[\sigma]$
must be more mixed 
\index{More mixed relation}
than the
reduced initial states 
${\text{tr}}_B[\rho]$ 
and ${\text{tr}}_A[\rho]$ 
in the sense of 
majorization theory \cite{Majo}.
\index{Majorization} 
The initial state $\rho$ is a maximally entangled state,
and therefore, the reduced 
states of Alice and Bob are initially 
maximally mixed. It follows that 
all accessible states after application of a mixed strategy
of Alice and Bob are locally identical to the 
\index{Maximally mixed state}
maximally mixed
state ${\mathbbm{1}_A}/2$ and ${\mathbbm{1}_B}/2$, respectively.

For example, 
the following construction
yields an equilibrium in mixed quantum strategies:
Allow Alice to choose from two strategies
${\cal E}_A^{(1)}$ and ${\cal E}_A^{(2)}$ with probabilities
$p^{(1)}_A=1/2$ and $p^{(2)}_A=1/2$,
while Bob may
take
${\cal E}_B^{(1)}$ or ${\cal E}_B^{(2)}$ with according 
probabilities, where
\begin{eqnarray}\label{PartNash}
	{\cal E}_A^{(1)}\simeq 
	\left(
	\begin{array}{cc}
        1 & 0 \\
	0 & 1 
        \end{array}
        \right),\,\,\,&&
	{\cal E}_A^{(2)}\simeq 
	\left(\begin{array}{cc}
        -i & 0 \\
	0 & i 
        \end{array}
        \right),\\
	{\cal E}_B^{(1)}\simeq 
	\left(\begin{array}{cc}
        0 & 1 \\
	-1 & 0 
        \end{array}
        \right),\,\,\,&&
	{\cal E}_B^{(2)}\simeq 
	\left(\begin{array}{cc}
        0 & -i \\
	-i & 0 
        \end{array}
        \right).\label{PartNash2}
\end{eqnarray}
The quantum strategies  of Eq.\ (\ref{PartNash}) and Eq.\ (\ref{PartNash2})
are mutually optimal answers
and have the property
\begin{eqnarray}\label{satisfaction}
	P_A({\cal E}_A^{(i)},{\cal E}_B^{(i)})=0,\,\,\,\,
	&&P_B({\cal E}_A^{(i)},{\cal E}_B^{(i)})=5,\\
	P_A({\cal E}_A^{(i)},{\cal E}_B^{(3-i)})=5,\,\,\,\,
	&&P_B({\cal E}_A^{(i)},{\cal E}_B^{(3-i)})=0,\label{satisfaction2}
\end{eqnarray}
for $i=1,2$. 
Due to the particular constraints
of Eq.\ 
(\ref{satisfaction}) and Eq.\ 
(\ref{satisfaction2}) no other mixed strategy will
entail a better pay-off for Bob than the above mixed strategy,
given that Alice sticks to the equilibrium
strategy.

This will become apparent in the following argument:
Let Alice use this particular
mixed quantum strategy as above and let Bob use
any mixed quantum strategy 
\begin{equation}\label{dono}
{\cal E}_B^{(1)},..., {\cal E}_B^{(n)}
\end{equation}
together with
 $p_A^{(1)},..., p_A^{(n)}$. The final state  $\sigma$
after application of the strategies
is given by the convex combination 
\begin{equation}
	\sigma=\sum_{i=1,2}\sum_{j=1}^n p^{(i)}_A p^{(j)}_B
	({\cal E}_A^{(i)}\otimes {\cal E}_B^{(j)})
	(\rho),
\end{equation}
This convex combination can only
lead to a smaller expected pay-off for Bob than 
the optimal pure strategy ${\cal E}_B^{(k)}$ in Eq.\ (\ref{dono}), 
$k\in \{1,...,n\}$. 
Such optimal pure strategies
are given by ${\cal E}_B^{(1)}$ and ${\cal E}_B^{(2)}$ as in Eq.\ (\ref{PartNash2})
and lead 
to an expected pay-off for Bob of $P_B({\cal E}_A,{\cal E}_B)=2.5$.
There are no pure strategies which achieve a larger
expected pay-off. While both pure strategies 
${\cal E}_B^{(1)}$ and ${\cal E}_B^{(2)}$
do not correspond to an equilibrium, the mixed strategy where
${\cal E}_B^{(1)}$ and ${\cal E}_B^{(2)}$ are chosen with $p_B^{(1)}=1/2$
and $p_B^{(2)}=1/2$ actually does. Nash equilibria consist
of pairs of mutually optimal answers, and only for this choice
of Bob the original mixed quantum strategy of Alice is
{\it her}\/ optimal choice. The game being symmetric,
the same argument applies
also to her.

However, this Nash equilibrium is 
not the only one. 
There exist other four-tuples
of matrices than the ones presented
in  Eq.\ (\ref{PartNash}) and  Eq.\ (\ref{PartNash2}) 
that satisfy Eq.\ (\ref{satisfaction})
and  Eq.\ (\ref{satisfaction2}). 
Such matrices can be found by appropriately
rotating the matrices of Eq.\ (\ref{PartNash}) and  
Eq.\ (\ref{PartNash2}).
As this means that there is more than
one equilibrium,
it is not obvious which Nash equilibrium the players will 
realize. 
It is at first not even 
evident whether a Nash equilibrium will be played at
all. But the game theoretical
concept of the {\it focal point effect}\/
\index{Focal point effect}
\cite{Schelling,Mye91}
helps to resolve this issue. 

In order to explore the
general structure of any Nash equilibrium
in mixed strategies, let  
\begin{equation}\label{Mono}
U_A^{(1)},..., U_A^{(n)}
\end{equation} 
together with
 $p_A^{(1)},..., p_A^{(n)}$ specify the mixed strategy
pertinent to a Nash equilibrium
of Alice. Then there is a mixed strategy
$U_B^{(1)},..., U_B^{(n)}$, $p_B^{(1)},..., p_B^{(n)}$
rewarding 
Bob with the best achievable pay-off, 
given that Alice plays this mixed strategy.
Yet, the pair of mixed strategies associated with
\begin{equation}\label{Dual}
	Q U_A^{(1)} Q^\dagger ,..., Q U_A^{(n)} Q^\dagger,
	\,\,\,\,\,\,
	Q U_B^{(1)} Q^\dagger ,..., Q U_B^{(n)} Q^\dagger
\end{equation}
with 
$p_A^{(1)},..., p_A^{(n)}$, 
$p_B^{(1)},..., p_B^{(n)}$ represents 
another Nash equilibrium. This equilibrium leads to the same
expected pay-off for both players and is fully
symmetric to the previous one.
Doubly applying $Q$ as
$QQU_A^{(1)} Q^\dagger
Q^\dagger ,..., Q Q U_A^{(n)} Q^\dagger Q^\dagger$
results again in a situation with equivalent
strategies as the original ones.
For a given Nash equilibrium as in Eq.\ (\ref{Mono})
the one specified by Eq.\ (\ref{Dual})
will be called dual equilibrium.

There is a single Nash equilibrium $(R,R)$
which is identical with its dual equilibrium: it is the simple map
\begin{equation}\label{SimpleMap}
	\rho\longmapsto \sigma=\mathbbm{1}/4.
\end{equation}
There exist probabilities
$p_A^{(1)},...,p_A^{(n)}$
and unitary operators $U_A^{(1)},...,U_A^{(n)}$
such that 
\begin{equation}
	\sum_i p_A^{(i)}
	(U_A^{(i)}\otimes {\mathbbm{1}}_B) \rho  (U_A^{(i)}
	\otimes {\mathbbm{1}}_B)^\dagger=
	\mathbbm{1}/4.
\end{equation}
\cite{Majo}.
If Alice has already selected ${\cal E}_A=R$, the application
of ${\cal E}_B=R$ will not change the state of the 
quantum system any more.

Assume that Eq.\ (\ref{Mono}) and Eq.\ (\ref{Dual})
are associated with equivalent quantum strategies.
This means that they have to produce the same
expected pay-off for all quantum strategies ${\cal E}_B$
of Bob. 
If Alice and Bob apply ${\cal E}_A\otimes {\cal E}_B$ they
get an expected pay-off according to 
Eq.\ (\ref{PA}) and Eq.\ (\ref{PB});
if Alice after implementation of ${\cal E}_A$ manipulates
the quantum system by applying the local unitary
operator $Q\otimes {\mathbbm{1}}_B$, they obtain 
\begin{eqnarray}
	P'_A({\cal E}_A,{\cal E}_B)&=&A_{DD} {\text{tr}}[\pi_{CC}
	\sigma]+
	A_{DC} {\text{tr}}[\pi_{CD}
	\sigma]\nonumber\\&+&
	A_{CD} {\text{tr}}[\pi_{DC}
	\sigma]+
	A_{CC} {\text{tr}}[\pi_{DD}
	\sigma],\\
	P'_B({\cal E}_A,{\cal E}_B)&=&
	B_{DD} {\text{tr}}[\pi_{CC}
	\sigma]+
	B_{DC} {\text{tr}}[\pi_{CD}
	\sigma]\nonumber\\&+&
	B_{CD} {\text{tr}}[\pi_{DC}
	\sigma]+
	B_{CC} {\text{tr}}[\pi_{DD}
	\sigma].
\end{eqnarray}
The only ${\cal E}_A$ with the property that 
$P'_A({\cal E}_A,{\cal E}_B)=P_A({\cal E}_A,{\cal E}_B)$
and $P'_B({\cal E}_A,{\cal E}_B)=P_B({\cal E}_A,{\cal E}_B)$
for all ${\cal E}_B$ is the map given by Eq.\ (\ref{SimpleMap}).

In principle, any Nash equilibrium may become a {\it self-fulfilling
prophecy}\/ if the particular Nash equilibrium is expected by
both players. It has
been pointed out that
in a game with more than one equilibrium, anything that 
attracts the players' attention towards one of the equilibria
may make them expect and therefore realize it
\cite{Schelling}. 
The corresponding {\it focal equilibrium}\/ \cite{Mye91}
\index{Focal equilibrium}
is the one which is 
distinguished from the other Nash equilibria.
There is indeed one Nash equilibrium which is
different from all the others: it is the one that 
is equivalent to its dual equilibrium, the
simple mapping of
the initial state on the maximally mixed state.
For all other expected pay-offs both players are ambivalent 
between (at least) two symmetric equilibria. 
The expected pay-off in this
focal equilibrium,
\begin{equation}
P_A(R,R)=P_B(R,R)=2.25,
\end{equation} 
is not fully
Pareto optimal, but it is again much more efficient than the 
classically achievable outcome of 1. 
Of course, in the classical game, too, both 
players could play
$C$ and $D$ with probabilities $p=1/2$ yielding the
same expected pay-off of $2.25$. However,
this pair of mixed strategies would be no 
equilibrium solution, 
as any player could benefit from simply 
choosing the dominant strategy $D$.\\

\newfigure{6.3cm}{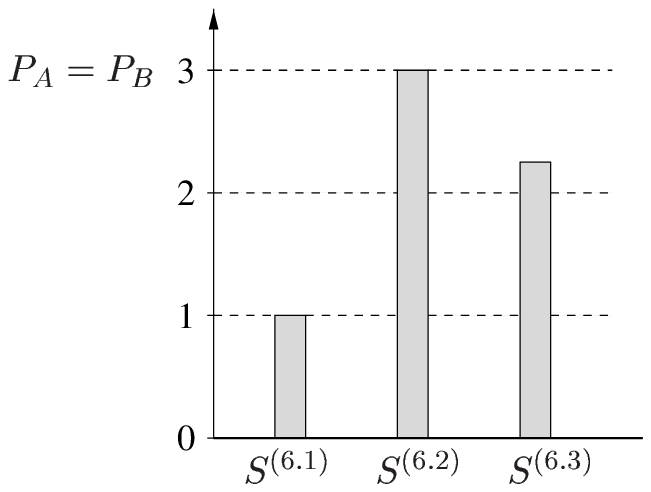}{Figure 6.6: 
The expected pay-off $P_A$ and $P_B$ in the distinguished Nash
equilibria in the sets of quantum strategies $S^{(6.1)}$,$S^{(6.2)}$, and  $S^{(6.3)}$.}

This study shows that the 
efficiency of the equilibrium 
the players can reach in
this game depends on the actions the players may
take. One feature, however, is present in
both of the sets $S^{(6.2)}$ and $S^{(6.3)}$:
both players can increase their expected
pay-offs drastically by resorting to quantum 
strategies. $S^{(6.2)}$ is equivalent with the situation in the
classical game with mixed strategies.
Fig.\ 6.6 is a schematic representation
of the achievable expected pay-off in the three 
sets of allowed strategies
\begin{equation}
S^{(6.1)}
\subset S^{(6.2)}
\subset S^{(6.3)}.
\end{equation}
In $S^{(6.2)}$ the unique Nash equilibrium is 
Pareto optimal and hence maximally efficient,
in other sets the solution is less efficient.

In the subsequent  
last two examples for the quantum version of the 
Prisoners' Dilemma
a different setting is studied: the case where
the parties face an unfair situation. In Example 6.4
one party only is restricted to classical strategies,
in Example 6.5 again such an asymmetric setting is 
investigated, but with an initial state with
varying entanglement.\\
\index{Unfair game}

\noindent 
{\bf Example 6.4. --} {\it Classical versus
quantum strategies. --}\/
An investigation of an unfair situation 
is particularly interesting in the case of 
the two-parameter strategic space $S^{(6.2)}$
introduced in Example 6.2. 
What happens if both parties do not
have access to the same strategic space?
Alice may use a 
quantum strategy, i.e., her strategic space
is still $S^{(6.2)}$, 
while Bob is restricted to apply only 
classical pure strategies $C$ or $D$ or
classically mixed strategies. 
In this case Alice is well advised to play
\begin{equation}
	M
	\simeq
	\frac{1}{\sqrt{2}}
	\left(\begin{array}{cc} i& 1 \\ -1 & -i \end{array}
	\right)\,,
\end{equation}
which corresponds to $\phi_A=\pi/2$ and $\theta_A=\pi/2$.
This strategy will ensure that she gets 
$P_A(M,{\cal E}_B)=3$
as pay-off, for all allowed mixed strategies of Bob;
he may implement $C$ and $D$ with arbitrary
probabilities $p_B$ and $1-p_B$, respectively,
with $p_B\in[0,1]$.  Hence if in an unfair game Alice can be sure that Bob plays
a mixed classical strategy, 
she may always choose $M$ as her
preferred strategy in an iterated game. This certainly out-performs
tit-for-tat, but one must keep in mind that the assumed
asymmetry is essential for this argument.\\

\noindent
{\bf Example 6.5. --} {\it Advantange in an unfair game dependent on
the entanglement of the initial state. --}
In the above considerations the initial state of the game was
fixed, $\rho=|\psi\rangle\langle\psi|$ with
$|\psi\rangle=(|00\rangle+i|00\rangle)/\sqrt{2}$. 
In this example the initial state will be varied.
Depending on an entanglement parameter $\gamma$,
$\gamma\in[0,\pi/2]$,
$|\psi\rangle$ will be taken to be
\begin{equation}\label{psis}
	|\psi\rangle=
		J |00\rangle, \,\,\,\,J=\exp 
		\left(i\gamma D \otimes
D 
		\right) .
\end{equation}
The case $\gamma=\pi/2$ corresponds to the above
case with a maximally entangled initial state, for
$\gamma=0$ the state $\rho$ is a product state.
The entanglement of the pure state $\rho$ for the
values  $\gamma\in[0,\pi/2]$ is given by
$
	E(\rho)=
	-\sin^2 (\gamma) \log_2 (\sin^2 (\gamma))
	-\cos^2 (\gamma) \log_2 (\cos^2 (\gamma)).
$
Accordingly, the Kraus operators of the measurement are
changed to 
\begin{eqnarray}
 	\pi_{CC}&=&|\psi_{CC}\rangle\langle\psi_{CC}|,\,\,\,\,\,
	|\psi_{CC}\rangle=
(C\otimes
C) J |00\rangle ,\\
		\pi_{CD}&=&|\psi_{CD}\rangle\langle\psi_{CD}|,\,\,\,\,\,
	|\psi_{CD}\rangle=(C\otimes
D )J |00\rangle,
		\label{psi12}\\
		\pi_{DC}&=&|\psi_{DC}\rangle\langle\psi_{DC}|,\,\,\,\,\,
	|\psi_{DC}\rangle=(D\otimes
C )J |00\rangle,\\
	\pi_{DD}&=&|\psi_{DD}\rangle\langle\psi_{DD}|,\,\,\,\,\,
	|\psi_{DD}\rangle=(D\otimes
D) J |00\rangle,
\end{eqnarray}
such that again, the classical strategies 
cooperate and defect correspond to $C$ and $D$ with
the matrix representations
\begin{equation}
	C\simeq 
\left(\begin{array}{cc} 1& 0 \\ 0 & 1 \end{array}
	\right),\,\,\,\,\,
D\simeq 
\left(\begin{array}{cc} 0& 1 \\ -1 & 0 \end{array}
	\right).
\end{equation}
How does the comparative advantage of
Alice depend on the parameter $\gamma$ in a game
in which 
Alice may implement any strategy from $S^{(6.2)}$,
whereas Bob chooses $C$ or $D$ with a certain classical probability?

For any value of $\gamma\in[0,1]$, 
the minimal expected pay-off $m$ Alice can always
attain by choosing an appropriate strategy ${\cal E}_A$ is given by
\begin{equation}\label{eq:4}
        m=\max_{{\cal E}_A\in{S^{(6.2)}}}
	\min_{{\cal E}_B: {\cal E}_B=p_B C+ (1-p_B) D}
	P_A({\cal E}_A,{\cal E}_B).
\end{equation}
Definitely,
Alice will not settle for anything less than this quantity.
Considering $m$ a function of the entanglement parameter
$\gamma$ it is clear that $m(0)=1$ (since in this case the
dominant strategy $D$ is the optimal choice) while for maximal
entanglement one finds $m(\pi/2)=3$ which is achieved by playing
$M$.  Fig. 6.7 shows $m$ as a function of the
entanglement parameter $\gamma$.  As a matter of fact, 
$m$ is a
monotone increasing function of $\gamma$, and the maximal advantage is
only accessible for maximal entanglement.  Furthermore, Alice should
deviate from the strategy $D$ if and only if the degree of
entanglement exceeds a certain threshold value 
\begin{equation}
	\Gamma=\arcsin(1/\sqrt{5})= 0.4636.
\end{equation} 
At the threshold she should discontinuously
change her strategy from $D$ to $Q$.\\

\newfigure{5.0cm}{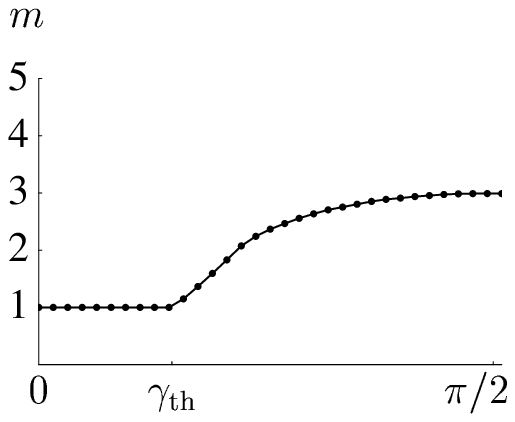}{
Figure 6.7:
The minimal pay-off Alice can always attain on average as a function of $\gamma$
in the unfair game of Example 6.5.}

\subsection{A Game With Two Equilibria}

In the (classical)
Prisoners' Dilemma the Dilemma of the players
consists of the fact that the unique Nash equilibrium is
not Pareto-optimal. Indeed, 
an unambiguous solution can be specified 
consisting of this Nash equilibrium, however, the solution is
not efficient and hence not satisfactory to
the players. In the so-called
\index{Chicken game}
{\it Chicken game}\/\footnote{The name ``Chicken'' is inspired by a
game mentioned in the
1955 movie called ``Rebel Without a Cause''.}
\cite{Mye91,Pou92}
the players
face a different Dilemma: the classical
game has multiple
Nash equilibria. The situation of the
players in the Chicken game, 
\begin{eqnarray}
        &&A_{CC}=B_{CC}=6,\,\,\,
        A_{CD}=B_{DC}=8,\\
        &&A_{DC}=B_{CD}=2,\,\,\,
        A_{DD}=B_{DD}=0,
\end{eqnarray}
can be described by the matrix of Fig.\ 6.8.

\newfigure{5.9cm}{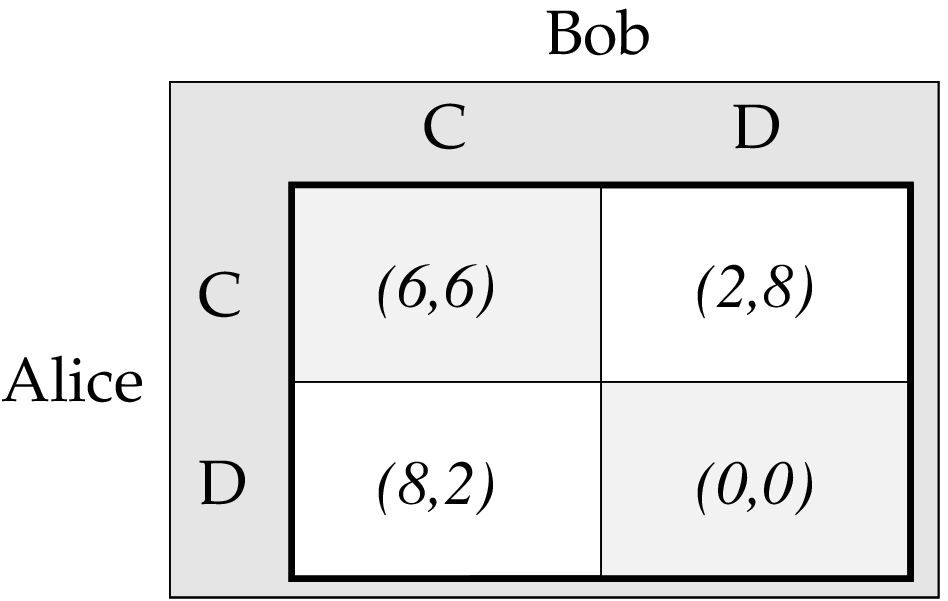}{Figure 6.8:
The pay-off matrix of the  Chicken game.}

\noindent 
This game
has two Nash
equilibria in pure strategies, namely
$(C,D)$ and $(D,C)$, and
it is not clear how to anticipate
what the players' decision would be. In
addition to the
two Nash equilibria in pure strategies there is an
equilibrium in mixed strategies, yielding an expected
pay-off $4$ \cite{Mye91}. In this equilibrium in 
mixed strategies both players choose each pure strategy
with probability $1/2$.
In order to investigate the new features of the
game if superpositions of classical strategies 
are allowed for, three set of strategies are
briefly discussed:\\

\noindent
{\bf Example 6.6. --} 
{\it One-parameter set of strategies. --}\/
The first set of strategies is again the set
$S^{(6.1)}$ of
one-dimensional rotations. 
Strategies
${\cal E}_A$ and ${\cal E}_B$ are associated with local
unitary operators
\begin{equation}
        {U}(\theta)=
        \left(
        \begin{array}{cc}
        \cos(\theta/2)&
        \sin(\theta/2)\\
        -\sin(\theta/2)
        &
                  \cos(\theta/2)
        \\
        \end{array}
        \right)
\end{equation}
with $\theta\in[0,\pi]$,
\begin{equation}
        C\simeq 
        U(0)=\left(
        \begin{array}{cc}
        1 & 0 \\
        0 & 1 
        \end{array}
        \right),\,\,\,\,\,
        D
        \simeq
        U(\pi)=\left(
        \begin{array}{cc}
        0 & 1 \\
        -1 & 0 
        \end{array}
        \right).
\end{equation}
Then as before, the 
quantum game yields the same expected pay-off
as the classical game in randomized strategies. 
This means that still two Nash equilibria in pure
strategies are present.\\

\noindent
{\bf Example 6.7. --} 
{\it Two-parameter set of strategies. --}\/
The players can actually take advantage of
an additional degree of freedom which is not accessible
in the classical game. If they may apply
unitary operations from  $S^{(6.2)}$ of the type
\begin{equation}
        {U}(\theta,\phi)=
        \left(
        \begin{array}{cc}
        e^{i\phi}\cos(\theta/2)&
        \sin(\theta/2)\\
        -\sin(\theta/2)
        &e^{-i\phi}\cos(\theta/2)
        \\
        \end{array}
        \right)
\end{equation}
with $\theta\in[0,\pi]$ and $\phi\in[0,\pi/2]$
the situation is quite different than with $S^{(6.1)}$.
$(C,D)$ and $(C,D)$ with
$C\simeq U(0,0)$ and $D\simeq U(\pi,0)$ are no
longer equilibrium solutions. E.g., given that
${\cal E}_A=D$ the pair of strategies
$(D,Q)$ with $Q\simeq U(0,\pi/2)$ yields a better
expected pay-off for Bob than $(D,C)$, that is
to say $P_B(D,Q)=8$, $P_B(D,C)=2$. 
In fact
$(Q,Q)$ is now the unique Nash equilibrium with
$P_A(Q,Q)=P_B(Q,Q)=6$, which follows from an
investigation of the actual expected 
pay-offs of Alice and Bob analogous to Eq.\ (\ref{LongPay}).
This solution is not only the 
unique acceptable solution of the game, but
it is also an equilibrium that is Pareto optimal.
This contrasts very much with the situation
in the classical game, where the two
equilibria were not that efficient.\\

\noindent
{\bf Example 6.8. --} 
{\it General unitary operations. --}\/
As in the considerations
concerning the Prisoners' Dilemma game,
more than one Nash equilibrium is present, 
if both players can take quantum strategies from the
set  $S^{(6.3)}$,
and all Nash equilibria emerge at least in pairs
as above. The focal equilibrium is given by $(R,R)$,
resulting in a pay-off of 
$P_A(R,R)=P_B(R,R)=4$, which is the same
as the mixed strategy of the classical game.

\section{Concluding Remarks}

In this chapter the idea of implementing
quantum operations as strategic moves in 
a game has been explored. First,
the notion of a game has been introduced
where strategic moves 
are identified with quantum operations.
In detail, a certain model has been 
investigated
which could be conceived as a generalization
into the quantum domain of a two player
binary choice game. As a toy model for
more complex scenarios quantum
games have been studied
in which the efficiency of the equilibria that are
attainable when using quantum strategies could be
contrasted with the efficiency of solutions in the
corresponding classical game. 

The nature of a game is determined by
the rules of the game. In particular,
the appropriate solution concept depends on
the available strategic moves. Obviously, a
player cannot make a meaningful choice without
knowing the options at his or her disposal.
So it comes to no surprise that also 
the actual achievable pay-off in such a 
game depends on the set of allowed strategies.
Roughly speaking, one can say that the possibility
of utilizing strategies which are not feasible
in the analogous classical game implicates
a significant advantage. In the models studied
in detail two kinds of ``dilemmas'' were
``resolved'': 
\begin{itemize}
\item[(i)]
On the one hand 
there are quantum games with an efficient 
unambiguous solution,
while in the classical analogue only an inefficient
equilibrium can be identified. By taking advantage
of appropriate quantum strategies much 
more efficient equilibria
could be reached. In certain sets of strategies
even a maximally efficient solution  -- the Pareto
optimum -- was attainable.

\item[(ii)] On the other hand, there exist quantum games with
a unique solution with a classical equivalent which
offers two Nash equilibria of the same quality. 
\end{itemize}
Also, 
\begin{itemize}
\item[(iii)] the performance of classical versus
quantum strategies has been compared.
It has been found [E6] 
(see also \cite{Mey99})
that the allowed quantum
strategies outperform the classical strategies
by far.
\end{itemize}

These investigations show that new features
emerge in a game in which strategies
are quantum operations applied on
quantum mechanical carriers of information.
%
In this chapter it has been the main
emphasis to examine
how such quantum generalizations 
are different from their classical equivalent.
This work can be extended in many ways.
In Refs. \cite{Hayden,JohnsonGames,MultiPlayer2}, 
e.g., multi-player
generalizations of quantum games have been
studied. Moreover, first steps in the
analysis
of evolutionary games have been taken \cite{Iqbal}.
\index{Iterated game}
{\it Iterated quantum games}\/ have, however, 
not yet been studied in detail; it is 
not even fully clear how to introduce 
the notion of such an iterated game. 
It is the actual 
hope -- at least according to the conception
of the author of this thesis -- 
that these investigations lead
to a better understanding of
competitive structures 
in a game theoretical
sense in applications of quantum 
information
theory. It seems that finding
such more pragmatic applications 
is the real challenge in the next step of 
decision theoretical investigations
in the quantum domain.

\chapter{Summary and Outlook}

This thesis was concerned with quantum entanglement. 
In the center of interest
was the resource character of entanglement
in applications of quantum information 
theory. It was the intention of this thesis to
clarify the theoretical possibilities in 
entanglement manipulation.
The subsequent list summarizes the major
contributions of this thesis to the academic 
debate on this subject:

\begin{itemize}
\item[]{\bf Chapter 2. Quantification of Quantum Entanglement.}

This chapter studied proper measures of quantum entanglement.
New measures  were  introduced,  
their properties were investigated, and they were 
compared to already known measures. 
In particular, a fully
additive entanglement monotone was proposed, and in a
numerical investigation the value for the regularized relative
entropy of entanglement 
was evaluated -- that is, the average degree of entanglement
of infinitely many copies of a certain state with high symmetry.
The last section of the chapter dealt
with multi-particle quantum systems. For such systems a
general measure of entanglement was 
defined, and the implications were investigated.

\item[]{\bf Chapter 3.  Entanglement Transformations.} 

The emphasis in this chapter was on transformation
criteria for mixed states. The underlying question was: 
what tasks can in principle be performed 
by using only local operations and classical communication
when starting from a given 
mixed state. Abstract criteria were presented 
for the possibility of a 
transformation from a single copy of a mixed state
to another mixed state. A particular example 
was studied in more detail, 
which turned out to be useful in the context of 
entanglement-assisted
transformations: In  entanglement-assisted
transformations   an auxiliary  
bi-partite quantum
system in an entangled state is borrowed, 
the desired
quantum operation is performed 
locally, and then
the auxiliary  system is returned
while leaving it in exactly the same state. 
It was shown in this chapter that such operations
are more powerful than ordinary LOCC also in the mixed state-domain.
Several aspects were addressed, such as the possibility to
increase the proportion of a certain pure state in a mixture by 
using entanglement-assisted
transformations, or  small transformations.

\item[]{\bf Chapter 4.  Non-Local Implementation of Joint Unitary Operations.} 

This chapter focused on the realization of quantum gates on remotely
located nodes. In order to enable the implementation of such gates,
both the exchange of classical information and initial entanglement
are necessary. Protocols for a non-local implementation
were presented for several quantum gates: these are the
quantum CNOT gate, the two-party control-U gate, the state swapper, the 
three-party Toffoli gate,
the three-party   control-U gate, 
and an the $N$-party control-U gate. 

\item[]{\bf Chapter 5.  Entanglement and Classical Information.} 

The amount of usable entanglement is related to how mich is known
about the state of the system. In this chapter a 
relationship between the loss of distillable entanglement and
the loss of classical information was established, 
in case that the loss of information is due to the 
loss of information about the identity
of the quantum systems. After presenting simple
examples, the chapter focused on proving
that the change in distillable entanglement is bounded
by the loss of classical information for arbitrarily 
many copies of pairs of qubits in any pure state. The proof made
use of group theoretical methods. Finally, a more general
result of a relation between entanglement and classical
information was derived when the relative entropy of entanglement
is taken as the measure of entanglement.

\item[]{\bf Chapter 6.  Quantum Information and Game Theory.} 

Chapter 6 concentrated on the connection between quantum 
information theory and the theory of games. 
The main idea of this chapter was to associate 
game theoretical strategies with quantum operations.
A general framework of the notion of a quantum game
was introduced. Simple toy models were investigated
in detail in order to clarify which 
new features may emerge
in such a situation compared to a classical game.

\end{itemize}

Needless to say, none of the research topics that have been
addressed in this thesis 
have exhaustively been represented. 
The field
of quantum information theory in general, and 
the theory
of quantum entanglement in particular, is a young
field of research, and many interesting questions
are still waiting to be resolved.  Several issues
studied in detail lead directly to challenging new 
research topics,
especially those 
related to mixed-state entanglement of bi-partite systems 
and pure-state entanglement in the multi-particle 
domain.
At the root of the open questions in multi-particle entanglement 
is the MREGS-problem. In the absence of a general framework 
of investigation it seems appropriate
to concentrate on practically motivated examples in the near
future.

Symmetries can help to greatly 
simplify many problems in entanglement theory and 
quantum information theory while preserving
the characteristic 
features of the original
problem. The idea of systematically exploring symmetries
is not yet fully exhausted. For example, 
in the context of the asymptotic limit of many copies of a state 
one can expect to yield  
useful results in further research.
Another problem amenable for the approach of utilizing symmetries
is the problem of evaluating quantum 
channel capacities.

Finally, a possibly fruitful line of thought 
could be the idea of applying the mathematical
tools of the theory of entanglement to problems which are
typically conceived not to be part of quantum information 
theory. An example could be the investigation of
the entanglement between a quantum system and its 
environment in the context of decoherence, dissipation, 
and environment-induced selection \cite{PT,Dec}. A
promising model is the so-called 
{\it quantum Browian
motion model}\/.\index{Quantum Brownian motion model} 
In this model one considers the dynamics of a quantum
harmonic oscillator that is linearly 
coupled to a heat bath. This heat bath  likewise  consists
of harmonic oscillators, which
are initially in a state corresponding to the canonical
distribution associated with a certain temperature. 
This model -- also known as Caldeira-Leggett-model \cite{CalLeg} -- 
is a frequent starting point of inquiries 
of the  ``emergence'' of classical properties of 
quantum systems, partly because 
the model is analytically solvable \cite{Hu}. 
As for Gaussian states the typical questions related
to entanglement are no more difficult to answer than
those related to small finite-dimensional systems, 
this model might be well-suited for a scrutiny of 
the entanglement with the 
tools of quantum information theory.


\newpage

\pagenumbering{Roman}

\pagestyle{fancyplain}


\chapter*{Appendix A: The von Neumann Entropy and the Relative Entropy 
Functional}

\lhead[\fancyplain{}{\bfseries\thepage}]%
   {\fancyplain{}{}}
	
\rhead[\fancyplain{}{}]%
   {\fancyplain{}{\bfseries\thepage}}%
\cfoot{}

\addcontentsline{toc}{chapter}{\protect\numberline{}Appendix 
A: The von-Neumann Entropy and the Relative Entropy Functional}

\chapterbreak

The notion of quantum entropy is of major importance
in quantum information theory. Originally,
the entropy of the state of a physical system is 
a concept from phenomenological thermodynamics
and statistical mechanics -- the respective
notions have been shaped mainly by R.\ Clausius, L.\ Boltzmann, and J.W.\ 
Gibbs
in the 19th century.
Later,
entropy was applied and extended
to an information theoretical context by C.E.\ Shannon 
\cite{Shannon,Cover}.
Quite naturally, in quantum information theory
quantum entropy became one of the most fundamental concepts.
In this appendix the major properties
of the von Neumann entropy and the relative
entropy for quantum states will be summarized.

\index{Shannon entropy} 
The {\it Shannon entropy}\/ \cite{Shannon,Cover}
is a quantity associated with a probability
distribution.
It is defined as
\begin{equation}
	H(p_1,...,p_n)=-\sum_{i=1}^n p_i \log_2(p_i).
\end{equation}	
Since $p_1,...,p_n$ is a probability distribution, it is
required that
$0\leq p_i\leq 1$ for $i=1,...,n$ and $\sum_{i=1}^n p_i =1$.
The Shannon entropy 
expresses the average information one gains when
learning about the value of a random variable $X$
which takes the value $x_i$ with
the respective probability $p_i$. 

The intuition behind the Shannon entropy can be
explained as follows:
Consider a source that produces strings $x^{(1)}, x^{(2)},...$,
where the symbols
$x^{(i)}$ are taken from the set $\{x_1,...,x_n\}$ for $i=1,2,...$\, . 
Each $x^{(i)}$
is a realization of the random variable $X$ with the above
probability distribution. So one may ask: 
what is the minimal number of bits
that are needed in order to store the information produced 
by the source, in the sense that the produced string can later
be recovered? The answer to this question is provided by  
{\it Shannon's noiseless coding theorem}\/ \cite{Shannon}:
the minimal
number of bits per source symbol is given by $H(p_1,...,p_n)$,
the Shannon entropy of the probability distribution associated
with the source.
\index{Shannon's noiseless coding theorem}

\index{Von Neumann entropy}
The ``quantum analogue'' of the Shannon entropy is the
von Neumann entropy \cite{Neu}.
The {\it von Neumann entropy}\/ 
is a measure for the degree of ``mixedness of a quantum state''.
It vanishes for pure states, and for maximally
mixed states \index{Maximally mixed state}
of the form
$\rho=\mathbbm{1}/{{\rm dim[{\cal H}]}}$ it attains its maximal
value, where ${\cal H}$ is the
underlying Hilbert space. \index{Choquet simplex}
The state space
of a quantum system is no Choquet simplex, and thus, a generic 
mixed state $\rho$ has many decompositions
of the type 
$\rho=\sum_i p_i|\psi_i\rangle\langle\psi_i|$.
As in such a decomposition the 
projections are not necessarily orthogonal,
a different Shannon information $H=-\sum_i p_i \log_2 p_i$ 
belongs to each probability
distribution  $p_1,p_2,...$, each one
associated with a different ``degree of mixing''.
The von Neumann entropy of a given state $\rho$ 
is then taken to be the 
minimal value of such a Shannon entropy of a decomposition
of $\rho$. This infimum corresponds to 
the case where the projections in the above decomposition
are pairwise orthogonal (and therefore could be conceived as
disjoint events).
In this spirit
the von Neumann entropy $S(\rho)$ of a state 
$\rho$ is defined as \cite{Neu,Ohya,LongEntropy}
\begin{equation}
	S(\rho)=-{\text{tr}}[\rho\log\rho].
\end{equation}
If $\rho=\sum_{i=1}^N p_i|\psi_i\rangle\langle\psi_i|$ is the spectral
decomposition of $\rho$, then 
\begin{equation}
	S(\rho)= -\sum_{i=1}^N p_i \log_2 (p_i).
\end{equation}
Among the important properties are the following:
Let
$\sigma,\rho$ and $\rho_i,\sigma_i$, $i=1,2,...$,
 be states
taken from ${\cal S}({\cal H})$. For a composite system
with Hilbert space ${\cal H}^{(1)}\otimes {\cal H}^{(2)}$
let $\rho^{(1)}, \sigma^{(1)}\in {\cal S}({\cal H}^{(1)})$ and
$\rho^{(2)},\sigma^{(2)}\in {\cal S}({\cal H}^{(2)})$. 
\begin{enumerate}
\item[\nb] Positivity: $S(\rho)\geq 0$.
\item[\nb] Symmetry: If $\sigma=U\rho U^\dagger$ for all unitary
operators $U$, then
$
	S(\sigma)=S(\rho)$.	
\item[\nb] Concavity:
$
	S(\lambda \rho_1 +(1-\lambda) \rho_2)\geq
	\lambda S(\rho_1)+(1-\lambda) S(\rho_2)
$
for any	$\lambda\in[0,1]$.
\item[\nb] Additivity: 
$S(\rho^{(1)}\otimes \rho^{(2)})= 
S(\rho^{(1)})+S(\rho^{(2)})$.
\index{Subadditivity}
\item[\nb] Let $\rho^{(1)}={\text{tr}}_2[\rho]$ and 
$\rho^{(2)}={\text{tr}}_1[\rho]$ be the reduced states 
of $\rho$. Then 
$
	S(\rho)\leq S(\rho^{(1)})+S(\rho^{(2)}).
$
\index{Lower semi-continuity}
\item[\nb] Lower semi-continuity: If $\lim_{n\longrightarrow \infty}
\|\rho_n-\rho\|= 0$, then
\begin{equation}
	S(\rho)\leq \lim_{n\longrightarrow\infty}\inf S(\rho_n).
\end{equation}
Here and in the following $\| .\|$ denotes the trace
norm.
\index{Monotonicity property}
\item[\nb] Monotonicity property: 
$S({\cal E}(\rho))\geq S(\rho)$
for any completely positive
unital (that is, doubly stochastic) map ${\cal E}$.
\index{Unital map}
\end{enumerate}
\index{Doubly stochastic map}
There is a close connection between the concept of 
von Neumann entropy and the theory of majorization.
\index{Majorization}
As stated in Chapter 3
a state $\rho$ is called {\it more mixed than}\/ $\sigma$,
if there exists a doubly stochastic map ${\cal E}$ such that
$\rho={\cal E}(\sigma)$. This more mixed relation --
abbreviated as $\prec$ -- implies a partial order
on the state space. 
In terms of the eigenvalues of the states this
relation can be expressed as follows:
Let $p_1,...,p_N$ and 
$q_1,...,q_N$ be the lists of eigenvalues
of states $\rho$ and $\sigma$, respectively, 
satisfying $1\geq p_1\geq...\geq p_N\geq 0$
and $1\geq q_1\geq...\geq q_N\geq 0$. 
$\rho\prec\sigma$ is equivalent with the statement that
\begin{equation}
	\sum_{i=1}^k p_i \leq \sum_{i=1}^k q_i
	\,\, \text{ for all }k=1,...,N.
\end{equation}
A connection between the relation $\prec$ and the von-Neumann
entropy is that
\begin{equation}
	\rho\prec\sigma\,\,\,\text{ implies that }
	\,\,\,
	S(\rho)\leq S(\sigma).
\end{equation}
Such a statement holds also under more general circumstances,
in that $\rho\prec\sigma$ implies that ${\text{tr}}[f(\rho)]\leq
{\text{tr}}[f(\sigma)]$ for any convex function $f:\mathbbm{R}^+
\longrightarrow \mathbbm{R}$.
For quantum systems with Hilbert space 
${\cal H}=\mathbbm{C}^2\otimes \mathbbm{C}^2$
and states $\sigma,\rho\in{\cal S}({\cal H})$
the relation
$\rho\prec\sigma$ is equivalent with $S(\rho)\leq S(\sigma)$.


The von Neumann entropy can be regarded as a special case of
another functional, the {\it relative entropy functional}\/. 
\index{Relative entropy functional}
In its simplest
version it is defined for two states $\rho$ and $\sigma$ as
\cite{Ohya,LongEntropy,AxiomaticEntropy,Lindblad,SchuRel}
\begin{equation}\label{relent}
	S(\sigma\|\rho)= {\text{tr}}[\sigma (\log_2 \sigma-\log\rho)].
\end{equation}
This relative entropy of $\sigma$ with respect to $\rho$ 
gives a measure of how different $\sigma$ is from $\rho$ in the
sense of statistical distinguishability \cite{QuantLong}. 
If $\sigma$ and $\rho$ are
identical, the relative entropy vanishes, the larger the
value of relative entropy is, the more information can -- roughly
speaking -- be
obtained from a measurement discriminating between $\sigma$ and $\rho$.
The expression given by Eq.\ (\ref{relent})
has to be interpreted in a similar way as above. Let
$\rho=\sum_{i=1}^N p_i|\psi_i\rangle\langle\psi_i|$ 
and $\sigma=\sum_{j=1}^N q_j |\phi_j\rangle\langle \phi_j|$ 
be the two spectral decompositions of $\rho$ and $\sigma$, respectively.
Then
\begin{equation}
	S(\sigma\|\rho)=
	\sum_{i,j=1}^N \bigl(
	q_j \log_2 q_j -
	q_j \log_2 p_i 
	\bigr)|\langle\psi_i|\phi_j\rangle|^2.
\end{equation}
Among the fundamental properties of the relative entropy
are the following:
\begin{enumerate}
\item[\nb] Positivity: 
$S(\sigma||\rho)\geq 0$.
\item[\nb] Nilpotence property: $S(\rho\|\rho)=0$.\index{Nilpotence}
\item[\nb] Joint convexity:
\begin{equation}
S(\lambda \sigma_1 + (1-\lambda)\sigma_2||
\lambda \rho_1 + (1-\lambda)\rho_2)\leq
\lambda S(\sigma_1||\rho_1)+(1-\lambda) S(\sigma_2||\rho_2)
\end{equation} 
for any $\lambda\in[0,1]$.
\item[\nb] Additivity: 
$
S(\sigma^{(1)}\otimes \sigma^{(2)}||\rho^{(1)}\otimes \rho^{(2)})
=S(\sigma^{(1)}||\rho^{(1)})+ S(\sigma^{(2)}||\rho^{(2)})$.

\item[\nb] Lower semi-continuity:\index{Lower semi-continuity}
If $\lim_{n\longrightarrow\infty}\| \sigma_n-\sigma\|=0$ and
$\lim_{n\longrightarrow\infty}\| \rho_n-\rho\|=0$,
then $S(\sigma\|\rho)\leq \lim_{n\longrightarrow\infty}\inf 
S(\sigma_n\|\rho_n)$.
If there exists a positive number $\lambda$ satisfying
$\sigma_n\leq \lambda \rho_n$, then
$\lim_{n\longrightarrow\infty} S(\sigma_n\|\rho_n)=
S(\sigma\|\rho)$.
\item[\nb] Monotonicity property: For any completely positive unital map ${\cal E}$
(that is, for any doubly stochastic map)
\index{Monotonicity property}
$
	S({\cal E}(\sigma)\| {\cal E}(\rho) )\leq S(\sigma\|\rho)$.
\item[\nb] Direct sum property: For all $\lambda\in[0,1]$ 
\begin{equation}
S(\lambda \sigma_1+(1-\lambda) \sigma_2\|
\lambda\rho_1 + (1-\lambda) \rho_2)
= S(\sigma_1\|\rho_1)+
S(\sigma_2\|\rho_2),
\end{equation}
if $\sigma_1 \sigma_2 = \rho_1 \rho_2=\sigma_1 \rho_2= \sigma_2\rho_1=0$.
\index{Direct sum property}
\item[\nb] Invariance property: for every unitary $U$
\index{Invariance property}
\begin{equation}
	S( U \sigma U^\dagger \| U \rho U^\dagger)=
	S(\sigma\| \rho).
\end{equation}
\end{enumerate}
It should be noted that the relative entropy functional is no
metric. In particular, it is not invariant under interchange of its
arguments, i.e., in general 
$S(\sigma\|\rho)\neq S(\rho\|\sigma)$
for two states $\rho$ and $\sigma$.


\chapter*{Appendix B: Numerical Evaluation of the Optimal PPT States}

\addcontentsline{toc}{chapter}{\protect\numberline{}Appendix 
B: Numerical Evaluation of the Optimal PPT States}

In Chapter 2 the quantities
$B_R(\sigma^{\otimes n}||\rho_n)$
have been evaluated, $n=1,2,...$ . The states $\rho_n$ 
are given by
\begin{equation}
	\rho_n=\sum_{k=0}^{n} 
	\frac{p_{k}}{\left(
	\begin{array}{c}
		n\\
		k 
	\end{array}\right)}
	\sum_{\pi\in S_n} (\pi\otimes \pi) \left(
	\sigma_a^{\otimes k}
		\sigma_s^{\otimes(n-k)}\right)(\pi\otimes \pi),
\end{equation}
where 
$p_0,...,p_n$ is a probability distribution.
The first seven probability
distributions can be evaluated as
\begin{eqnarray}
n=1&:&\,\,
p_0=\frac{1}{2},\,\,
p_1=\frac{1}{2}\\
n=2&:&\,\,
p_0=\frac{2}{3},\,\,
p_1=0,\,\,
p_2=\frac{1}{3},\,\,\\
n=3&:&\,\,
p_0=\frac{4}{5},\,\,
p_1=0,\,\,
p_2=0,\,\,
p_3=\frac{1}{5},\,\,\\
n=4&:&\,\,
p_0=\frac{3}{8},\,\,
p_1=\frac{1}{2},\,\,
p_2=0,\,\,
p_3=0,\,\,
p_4=\frac{1}{8},\,\,\\
n=5&:&\,\,
p_0=\frac{33}{106},\,\,
p_1=\frac{45}{106},\,\,
p_2=\frac{10}{53},\,\,
p_3=0,\,\,
p_4=0,\,\,
p_5=\frac{4}{53},\\
n=6&:&\,\,
p_0=\frac{23}{87},\,\,
p_1=\frac{30}{87},\,\,
p_2=\frac{30}{87},\,\,
p_3=0,\,\,
p_4=0,\,\,
p_5=0,\,\,
p_6=\frac{4}{87},\\
n=7&:&\,\,
p_0=\frac{7}{24},\,\,
p_1=\frac{7}{72},\,\,
p_2=\frac{7}{12},\,\,
p_3=0,\nonumber\\
&&p_4=0,\,\,
p_5=0,\,\,p_6=0,\,\,
p_7=\frac{1}{36}.
\end{eqnarray}
For example,
\begin{equation}
	\rho_1=\frac{1}{2}\sigma_a+\frac{1}{2}\sigma_s,\,\,\,\,
	\rho_2=\frac{1}{3}\sigma_a^{\otimes 2}+\frac{2}{3}\sigma_s^{\otimes 2},\,\,\,\,
	\rho_3=\frac{1}{5}\sigma_a^{\otimes 3}+\frac{4}{5}\sigma_s^{\otimes 3}.
\end{equation}


\chapter*{Appendix C: The Trace Norm Measure}

\addcontentsline{toc}{chapter}{\protect\numberline{}Appendix 
C: The Trace Norm Measure}

In this appendix a further measure of entanglement will be 
proposed which involves the trace norm. This measure is defined
in exactly the same way as the modified relative entropy
of entanglement $E_M$, except that the relative entropy
functional is replaced by the trace norm difference. In the
notation of Chapter 2 this measure $E_T$ is defined as
\begin{equation}
	E_T(\sigma)=\min_{\rho\in{\cal D}_\sigma({\cal H})}\|\sigma-\rho\|.
\end{equation}
\index{Trace norm measure}\index{Trace norm}
The subsequent proposition shows 
that $E_T$ has all the properties
of an entanglement monotone. 
It does not -- however -- coincide with
$S(\text{tr}[|\psi\rangle\langle\psi|])$ 
for pure states $|\psi\rangle\langle\psi|$.\\

\noindent{\bf Proposition C.1. --} 
{\it $E_T$ is an entanglement monotone.}
\probreak
\proof
Clearly, $E_T(\rho)=0$ for a separable state $\rho$. The set ${\cal D}_{\sigma}(\cal H)$
is a compact and convex set for each $\sigma\in{\cal S}({\cal H})$, and hence,
$E_T$ is also a convex functional by virtue of the triangle inequality. 
The remaining task is to show that in  a local generalized measurement
associated with Kraus operators $A_1,...,A_K$ the value of $E_T$
may only decrease on average, that is, 
$\sum_{i=1}^K  
p_i E_T( \sigma_i)\leq E_T(\sigma)$ for all states $\sigma$, 
where $p_i=\text{tr}[A_i \sigma A_i^\dagger]$ and $\sigma_i=A_i \sigma A_i^\dagger /p_i$
(see Chapter 2). The first ingredient to the proof is the fact that
$\text{tr}[A_i \rho A_i^\dagger]=\text{tr}[A_i \sigma A_i^\dagger]=p_i$
for all $\rho\in {\cal D}_{\sigma}({\cal H})$, and hence,
\begin{eqnarray}
	\sum_{i=1}^K  p_i E_T( \sigma_i) &=&
	\sum_{i=1}^K p_i 
	\min_{\rho_i \in{\cal D}_{\sigma_i}({\cal H}) } 
	\|A_i \sigma A_i^\dagger/p_i - \rho_i\|\nonumber\\
	&\leq& \sum_{i=1}^K p_i 
	\min_{\rho \in{\cal D}_\sigma({\cal H}) } 
	\frac{\|A_i\sigma A_i^\dagger - A_i\rho A_i^\dagger\|}{p_i}\nonumber\\
	&=&\min_{\rho \in{\cal D}_\sigma({\cal H}) } \sum_{i=1}^K  
		\|A_i (\sigma-\rho) A_i^\dagger\|.
\end{eqnarray}	
The second ingredient is the statement of Lemma C.2,
\begin{eqnarray}
	\sum_{i=1}^K  p_i E_T( \sigma_i)  
		&\leq&\min_{\rho \in{\cal D}_\sigma({\cal H}) }  \sum_{i=1}^K 
		\|A_i^\dagger A_i |\sigma-\rho|\, \|\nonumber\\
		&=& \min_{\rho \in{\cal D}_\sigma({\cal H}) } \|\sigma-\rho\|=E_T(\sigma).
\end{eqnarray}
Hence, $E_T$ is an entanglement monotone.
\proofend

\smallskip
It should be noted that the weaker
condition $E_T({\cal E}(\sigma)) \leq  E_T( \sigma)$
for all LOCC ${\cal E}$ and all states $\sigma$  
follows immediately from the 
fact that the trace norm fulfils
\begin{equation}
	\| {\cal E}(\sigma)- {\cal E}(\rho)\|\leq \|\sigma-\rho\|
\end{equation}
for all quantum operations  ${\cal E}$ and all states $\sigma,\rho$ \cite{Ruskai}.\\

\noindent{\bf Lemma C.2. --} {\it Let $A,B$ be $n\times n$ matrices,
and assume that $B=B^\dagger$. Then
\index{Unitarily invariant norm}\index{Normal matrix}
\begin{equation}
	\| A B A^\dagger \|\leq \| A^\dagger A |B| \|\label{Helpful}
\end{equation}
holds.}
\probreak

\proof
The trace norm is a unitarily invariant norm, 
and $A B A^\dagger$ is a normal matrix.\footnote{
An $n\times n$-matrix $M$ is called {\it normal}\/, 
if $M M^\dagger = M^\dagger M$. A norm
$\|.\|$ on $n\times n$-matrices is said to be 
{\it unitarily invariant}\/, if $\| U M V\|= \|M\|$ 
for all unitary $U,V$ \cite{Bhatia}.} 
Hence \cite{Bhatia},
\begin{equation}
	\| A (B A^\dagger) \|\leq  \|  (B A^\dagger) A\|.
\end{equation}
Eq.\ (\ref{Helpful}) is a consequence of
\begin{equation}
\|  (B A^\dagger) A\|= \text{tr}[(A^\dagger A  B^\dagger B A^\dagger A )^{(1/2)}]
= \text{tr}[(A^\dagger A |B|^2 A^\dagger A )^{(1/2)}]=
\| A^\dagger A |B| \|.
\end{equation}
\proofend


\pagestyle{fancyplain}



\addcontentsline{toc}{chapter}{\protect\numberline{}Bibliography}

\bibliographystyle{myprsty}

\input{Bibl.bbl}

\chapter*{Acknowledgements}
\addcontentsline{toc}{chapter}{\protect\numberline{}Acknowledgements}

It is my pleasure to thank all those who have
enabled me to accomplish this thesis. First and foremost,
I need to express my very sincere gratitude to the one person
who has provided me with the academic and institutional 
framework for my
research: by offering me the chance to work in his group, 
Martin Wilkens has been the focal point and the
backbone of my academic career.
He is one of the few academics who
have preserved their vigorous scientific curiosity
as well as their personal 
accessibility against the daily odds of administrative
duties. Working with him has been both extraordinarily 
enjoyable and academically 
challenging. I am most grateful to him for inspiring
new work, for enthusiastic discussions,
for maintaining the group spirit, and, most
importantly, for always giving me his full support in
every possible way. I could be sure to find
an open door as well as an open ear at any time.

Secondly, my academic progress has benefited 
enormously from 
co-operating with Martin Plenio. 
It is no accident that his
name appears so many times in this thesis. 
I have a lot of respect for his work and his deep
knowledge of the field. Throughout the two and a half years
of work for this thesis he has been a
second mentor, a patient co-worker, and a friend.
I also gratefully acknowledge the
hospitality of Peter Knight at  Imperial College.

Hans Briegel is one of those researchers who
has an outstanding ability to captivate
people with his boundless enthusiasm. 
During a single hike in the mountains I learned more
than by studying literature for hours. 
I owe a lot to him, both scientifically and 
personally.

I would like to thank my co-workers Maciek Lewenstein,
Polykarpos Papadopoulos, and Kurt Jacobs for inspiring
conversations and a fruitful collaboration.
Many thanks also to the members of the Quantum Theory
Group in Potsdam, who made
the stay in Potsdam so delightful: Timo Felbinger,
Simon Gardiner, Carsten Henkel, Fabrizio Illuminati, 
Meret Kr{\"a}\-mer, and in particular
Alexander Albus and Kim Bostr{\"o}m 
who helped me with constructive comments
on the manuscript.

I have had the chance to have interesting discussions  
about the topic of the thesis
with many colleagues and friends, and
each of them has in some way contributed to
my work, among them
Hans Aschauer,
Howard Barnum,
Almut Beige, 
Charles Bennett,
Thomas Beth,
Sogato Bose,
Dik Bouwmeester,
Daniel Braun,
Heinz-Peter Breuer,
Dagmar Bru{\ss},
Andreas Buchleitner,
Tony Chefles,
Daniel Collins,
Tom Cover,
David DiVincenzo,
Wolfgang D{\"u}r,
Tilo Eggeling,
Artur Ekert,
Ernesto F.\ Galvao,
Markus Grassl,
Lucien Hardy,
Patrick Hayden,
K.E.\ Hellwig,
Pawel Horodecki,
Susana Huelga,
A.\ Iqbal,
Jens G.\ Jensen,
Daniel Jonathan,
Sinisa Karnas,
Julia Kempe,
Michael Keyl,
Hoi-Kwong Lo,
Wolfgang Mathis,
David A.\ Meyer,
Thomas M{\"u}ller,
Arnold Neumaier,
Mike Nielsen,
Matteo G.A.\ Paris,
D{\'e}nes Petz,
Robert Raus\-sen\-dorf,
Massimiliano F.\ Sacchi,
Anna Sanpera,
Dirk Schlingemann,
Tapio Schneider,
Chris\-toph Simon,
Joel Sobel,
Dirk Sondermann,
Barbara Terhal,
Mathias Trucks,
Armin Uhlmann,
Onay Urfal{\i}o$\overline{\mathrm g}$lu,
Lieven M.K.\ Vandersypen,
Vlatko Vedral,
Guifr{\'e} Vidal,
Shash Virmani,
Karl Gerd Vollbrecht,
Reinhard F.\ Werner,
Andreas Winter, and
Christopher Witte.\\

I would like to express  my gratitude to  
my parents Ulla and 
Manfred Eisert for their support and 
encouragement during my work.
Finally, I would like to express my warmest thanks to  
my  closest companion 
Uta Simon for her love,
support, patience, and for her thoughtful
review of the manuscript of the  thesis.
%

\bigskip\vfill

This work has been 
supported by the Deutsche Forschungsgemeinschaft (''Schwerpunktprogramm 
Quanteninformationsverarbeitung'') coordinated by
G.\ Leuchs, and the project EQUIP
(IST-1999-11053) 
of the ``Proactive Initiative:  Quantum Information Processing and Communications (QIPC)''
of the European Commission.


\chapter*{Notations}
\addcontentsline{toc}{chapter}{\protect\numberline{}Notations}

\chapterbreak

\begin{tabular}{ll}
${\cal H}$ & Hilbert space\\
${\cal S}({\cal H})$ & State space associated  with a Hilbert 
space ${\cal H}$\\
${\cal D}({\cal H})$ & Subset of ${\cal S}({\cal H})$ of separable 
states\\
${\cal P}({\cal H})$ & Subset of ${\cal S}({\cal H})$ of PPT
states\\
${\cal D}_\sigma({\cal H})$ & Subset of ${\cal D}({\cal H})$ of separable states
which are locally identical to $\sigma$\\
${\cal P}_\sigma({\cal H})$ & Subset of ${\cal P}({\cal H})$ of PPT states
which are locally identical to $\sigma$\\
& \\ 
$\text{dim}$ & Dimension of a vector space\\
$\text{range}$ & Range of a linear operator\\
$\text{tr}_A$ & Partial trace with respect to system $A$\\
$||.||$& Trace norm\\
$\otimes $ & Tensor product\\
$\oplus$ & Direct sum\\
$\prec$ & Majorization relation\\
${\cal E}$ & Completely positive linear map\\
$E_1,...,E_K$ & Kraus operators\\
$A_1,...,A_K$ & Kraus operators of a local quantum operation in system $A$\\
$A_1(A_2 A_3)$ & A particular split of a system consisting of parts 
$A_1$, $A_2$, and $A_3$\\
& \\ 
$B_R$&  Relative entropy of entanglement with respect to PPT states\\
$B_R^\infty$& Regularized relative entropy of entanglement with respect to PPT states\\
$B_M$ & Modified relative entropy of entanglement with respect to PPT 
states\\
$D_C$ & Distillable entanglement with respect to the class of operations $C$\\
$D_{\leftrightarrow}$ & Distillable entanglement with respect to LOCC operations\\
$D_{\rightarrow}$ & Distillable entanglement with respect to one-local operations\\
$E_C$ & Concurrence\\
$E_F$ & Entanglement of formation\\
$E_G$ & General distance measure based on the relative entropy\\
$E_F^\infty$ & Regularized entanglement of formation\\
$E_R$ & Relative entropy of entanglement\\
$E_N$& Negativity\\
$E_\mu$ & Entanglement monotone defined in Proposition 2.6\\
$E_M$ & Modified relative entropy of entanglement\\
$E_S$ & Schmidt measure\\
$E_S^{A_1(A_2 A_3)}$ & Schmidt measure with respect to the split $A_1 (A_2 
A_3)$\\
$E_T$ & Trace norm measure\\
$E_X$ & Minimal asymptotic preparation cost of a multi-particle state\\
$I_A$ & Coherent information with respect to system $A$\\
\end{tabular}

\begin{tabular}{ll}
$(\sigma_n)_{n\in{\mathbbm{N}}}$ & Series of states\\
$\rho^{T_A}$ & Partial transpose of $\rho$ with respect to a certain basis 
of \\ & system 
$A$\\
$\rho^{\otimes n}$ & $n$-fold tensor product of a state $\rho$\\
$\sigma_a$ & Antisymmetric state of a bi-partite quantum system\\
$\sigma_s$ & Symmetric state of a bi-partite quantum system\\
$\rho_W$ & Werner state\\
$|\psi^+\rangle,|\psi^-\rangle,|\phi^+\rangle,|\phi^-\rangle$ & State 
vectors of the four Bell states\\
$|\text{GHZ}\rangle$ & State vector of the GHZ state of three qubits\\
$|\text{W}\rangle$ & State vector of the W state of three qubits\\
$\rho_M$ & Molecule state\\
& \\ 
$F(\sigma,\rho)$ & Fidelity of $\sigma$ with respect to $\rho$\\
$H(p_1,...,p_n)$ & Shannon entropy of the probability distribution 
$p_1,...,p_n$\\
$S(\rho)$ & von Neumann entropy of $\rho$\\
$S(\sigma||\rho)$ & Relative entropy functional of $\sigma$ with respect to 
$\rho$\\
& \\ 
$\sigma\rightarrow \rho$ \text{ under LOCC}& The state $\sigma$ can be transformed 
into $\rho$ under local \\ & operations with  classical communication\\
$\sigma\rightarrow \rho$ \text{ under ELOCC }& The state $\sigma$ can be transformed 
into $\rho$ under  entanglement-\\
 &assisted  local operations with classical communication\\
$\Delta D_{\leftrightarrow}$ &  Change in distillable entanglement\\
$\Delta E_R$ &  Change in relative entropy of entanglement\\
$\Delta I$ & Change in classical information\\
& \\ 
$S_n$ & Symmetric group of degree $n$\\
$\pi$ & Permutation and corresponding unitary\\
$SU(2)$ & Special unitary group of degree 2\\
$D^{(j)}$ & Spin-$j$ irreducible representation of $SU(2)$\\
& \\ 
$S_A$ & Set of quantum strategies of Alice\\
$P_A$ & Alice's utility functional\\
$u_A$ & Alice's utitily function\\
$\Gamma=(\{A,B\}, S_A, S_B, u_A, u_B)$ & (Classical) game\\
$\Gamma=({\cal H},\rho,S_A,S_B,P_A,P_B)$ & Quantum game\\

\end{tabular}



\newpage

\addcontentsline{toc}{chapter}{\protect\numberline{}Index}

\input{Index}

\end{document}

%% file: Index.tex
\begin{theindex}

  \item $K$-distillability, 17
  \item $k$-split, 43

  \indexspace

  \item Additivity of entanglement monotones, 30
  \item ALOCC operation, 50
  \item Ancilla, 10
  \item Angular momentum, 87
  \item Antisymmetric subspace, 33
  \item Approximate transformation, 54, 59
  \item Asymptotic limit, 14, 22, 33, 38, 50
  \item Asymptotically reducible, 50

  \indexspace

  \item Bell basis, 28
  \item Bell diagonal state, 28
  \item Bell state, 28
  \item Bell's inequality, 1, 15
  \item Best separable approximation, 15, 48
  \item Bit, 8
  \item Bloch sphere, 8
  \item Bound entangled state, 17
  \item Bound entanglement, 72

  \indexspace

  \item C$^{\ast }$-algebra, 12, 32
  \item Caratheodory's theorem, 15
  \item Catalysis, 60
  \item Chicken game, 113
  \item Choquet simplex, I, 9
  \item Class of operations, 13, 22
  \item Classical communication, 13, 78
  \item Classical record, 85
  \item Classical strategies, 105
  \item Cluster state, 50
  \item CNOT gate, 76
  \item Coherent information, 98
  \item Commutant, 34
  \item Complete measurement, 10
  \item Completely positive map, 11, 103
  \item Concurrence, 26
  \item Conditional expectation property, 32
  \item Continuity, 46
  \item Control-U gate, 79
  \item Convergent series, 35
  \item Convex roof extension, 23, 44

  \indexspace

  \item Decoherence, 10, 75
  \item Dense coding, 4, 79
  \item Direct sum decomposition, 91
  \item Direct sum property, III
  \item Distillable entanglement, 22, 85
  \item Distillation, 16, 22
  \item Distributed quantum computer, 75
  \item Dominant strategy, 103
  \item Doubly stochastic map, II, 11, 109
  \item Dynamical map, 9

  \indexspace

  \item Entangled state, 15
  \item Entanglement monotone, 20, 44
  \item Entanglement of formation, 23
  \item Entanglement swapping, 81
  \item Entanglement-assisted transformation, 60
  \item Equilibrium in dominant strategies, 103
  \item Equivalent strategy, 103
  \item Erasure of classical information, 85

  \indexspace

  \item Fannes' inequality, 38
  \item Focal equilibrium, 110
  \item Focal point effect, 109
  \item Free entangled state, 17
  \item Fully additive, 30
  \item Fully inseparable state, 43
  \item Fully separable state, 44, 49

  \indexspace

  \item Game theory, 100
  \item Games of incomplete information, 99
  \item Generalized measurement, 12
  \item Genuinely mixed state, 57
  \item GHZ-state, 44, 47, 51

  \indexspace

  \item Haar measure, 34
  \item Hadamard gate, 76
  \item Hashing inequality, 98
  \item Hypergeometric function, 36

  \indexspace

  \item Invariance property, III
  \item Irreversibility of entanglement manipulations, 38
  \item Isomorphism, 72
  \item Iterated game, 102, 115

  \indexspace

  \item Joint unitary operations, 75
  \item Jordan decomposition, 25

  \indexspace

  \item Kraus operator, 11

  \indexspace

  \item Linear programming, 36
  \item Local operation, 13
  \item Local state, 44
  \item Local unitary operation, 21
  \item Locally distinguishable, 13, 56
  \item LOCC operation, 14, 75
  \item Log negativity, 26, 37
  \item Lower semi-continuity, II, III

  \indexspace

  \item Majorization, II, 54, 55, 109
  \item Matching Pennies, 100
  \item Maximally mixed state, I, 27, 109
  \item Measure of entanglement, 20
  \item Measurement problem, 10
  \item Min-max theorem, 101
  \item Mixed state, 8
  \item Mixed strategies, 108
  \item Mixing, 8, 21, 25, 72
  \item Modified relative entropy of entanglement, 27
  \item Molecule state, 49
  \item Monotonicity property, II, III
  \item More mixed relation, 109
  \item MREGS, 50
  \item Multi-particle entanglement, 42
  \item Multi-party control-U, 83
  \item Multiplicity space, 91

  \indexspace

  \item N-party GHZ state, 46
  \item Nash equilibrium, 101, 103
  \item Negative eigenvalue measure, 25
  \item Negativity, 24, 37
  \item Nilpotence, III
  \item Noisy quantum channel, 16
  \item Non-locality, 75
  \item Non-PPT bound entangled state, 17, 38
  \item Non-selective measurement, 10
  \item Non-zero-sum game, 102, 103
  \item Normal matrix, VIII

  \indexspace

  \item One-local operation, 13
  \item One-system bi-separable state, 43, 49
  \item Optimal answer, 108
  \item Optimal purification, 93
  \item Ordered list, 55

  \indexspace

  \item Pareto-optimal, 102, 106
  \item Partial transposition, 16
  \item Pay-off, 101
  \item Peres-Horodecki-criterion, 16
  \item Perfect equilibrium, 108
  \item Permutation operator, 33
  \item Persistency of entanglement, 50
  \item Player, 100
  \item Posterior state, 27
  \item PPT state, 16
  \item Prisoners' Dilemma, 102
  \item Probabilistic transformation, 54, 68
  \item Product state, 9, 15, 41
  \item Pure state, 7, 8
  \item Pure strategies, 101
  \item Purification, 68

  \indexspace

  \item Quantum Brownian motion model, 118
  \item Quantum capacity, 98
  \item Quantum channel, 86
  \item Quantum circuit, 78
  \item Quantum computer, 2, 75
  \item Quantum game, 103
  \item Quantum gate, 2, 76
  \item Quantum operation, 12
  \item Quantum strategy, 103
  \item Qubit, 8

  \indexspace

  \item Reduction criterion, 16
  \item Regularized entanglement monotone, 31
  \item Relative entropy functional, III
  \item Relative entropy of entanglement, 22
  \item Restricted normal form, 37
  \item Reversible entanglement generating set, 50

  \indexspace

  \item Schmidt coefficient, 9
  \item Schmidt decomposition, 9, 43
  \item Schmidt measure, 43
  \item Schmidt number, 58
  \item Schmidt rank, 9, 44
  \item Schr{\"o}dinger equation, 9
  \item Schur's second lemma, 92
  \item Selective measurement, 10
  \item Separability structure, 43
  \item Separable operation, 14
  \item Separable state, 15
  \item Shannon entropy, I
  \item Shannon's noiseless coding theorem, I
  \item Simplex method, 37
  \item SLOCC operation, 68
  \item Spin coupling, 92
  \item State space, 8
  \item State swapper, 79
  \item Stinespring dilation, 12
  \item Strategic form, 100
  \item Strategy, 100
  \item Strong continuity, 38, 40
  \item Subadditivity, II, 31
  \item Symmetric group, 35, 91
  \item Symmetric subspace, 33
  \item System, 7

  \indexspace

  \item Teleportation, 4, 83
  \item Three-party control-U gate, 81
  \item Three-system bi-separable state, 43, 49
  \item Tit-for-tat, 102
  \item Toffoli gate, 77, 80
  \item Trace norm, VII, 21, 39
  \item Trace norm measure, VII
  \item Two-player game, 100
  \item Two-qubit gate, 77
  \item Two-system bi-separable state, 43, 49
  \item Two-way distillable entanglement, 22

  \indexspace

  \item Unfair game, 111
  \item Unique measure of entanglement, 21
  \item Uniqueness theorem for entanglement measures, 21, 39, 40, 46
  \item Unit ray, 7
  \item Unital map, II, 11, 109
  \item Unitarily invariant norm, VIII
  \item Unitary irreducible representation, 92
  \item Unlocking bound entanglement, 73
  \item Upper bound for distillable entanglement, 27
  \item Utility function, 100
  \item Utility functional, 103

  \indexspace

  \item Von Neumann entropy, I, 11, 38

  \indexspace

  \item W-state, 44, 47, 51
  \item Weakly additive, 21
  \item Weakly continuous, 21, 39
  \item Werner state, 26, 34, 48

  \indexspace

  \item Zero-sum game, 101, 103

\end{theindex}